\documentclass{dissertation}

\usepackage[dvipsnames]{xcolor}
\definecolor{azure}{rgb}{0.0, 0.5, 1.0}
\definecolor{darkblue}{rgb}{0.15,0.35,0.7}
\definecolor{reddish}{rgb}{0.65, 0.2, 0.2}
\definecolor{brandeisblue}{rgb}{0.0, 0.44, 1.0}
\definecolor{ceruleanblue}{rgb}{0.16, 0.32, 0.75}
\usepackage[linktocpage=true]{hyperref}
\hypersetup{
colorlinks=true,
citecolor=ceruleanblue,
linkcolor=ceruleanblue,
urlcolor=ceruleanblue
}

\usepackage{subfigure,epsfig,amsfonts}
\usepackage[sort&compress]{natbib}
	\setcitestyle{square,numbers,comma}	
\usepackage{amsmath}
\usepackage{chngcntr}
\counterwithout{footnote}{chapter}
\usepackage{amssymb}
\usepackage{amsthm}
\usepackage{tensor}
\usepackage{ulem}
\usepackage{physics}

\providecommand{\phantomsection}{}
\providecommand{\hypersetup}[1]{}

\usepackage[T1]{fontenc} 
\usepackage[utf8]{inputenc} 
\usepackage{caption} 	
\usepackage{ragged2e}	
\usepackage{bm}		
\usepackage{dsfont}		
\usepackage{tikz} 
\usepackage{accents} 
\usepackage{booktabs}

\def\bes #1\ees{\begin{split}#1\end{split}}

\newcommand{\be}{\begin{equation}}
\newcommand{\ee}{\end{equation}}

\newcommand{\bma}{\begin{pmatrix}}
\newcommand{\ema}{\end{pmatrix}}

\newcommand{\bea}{\begin{eqnarray}}
\newcommand{\eea}{\end{eqnarray}}
\newcommand{\bsubeq}{\begin{subequations}}
\newcommand{\esubeq}{\end{subequations}}
\newcommand{\bsubea}{\begin{subequations}\bea}
\newcommand{\esubea}{\eea\end{subequations}}

\newcommand{\overbar}[1]{\mkern 1.5mu\overline{\mkern-1.5mu#1\mkern-1.5mu}\mkern 1.5mu}

\def\TT{{T\overline{T}}}
\def\CL{{\mathcal{L}}}
\newcommand{\cL}{{\mathcal L}}

\newcommand{\zbar}{{\overbar{z}}}
\newcommand{\CO}{{\mathcal O}}

\newcommand{\Tbar}{\overbar{T}}

\newcommand{\wbar}{{\overbar{w}}}

\newcommand{\ad}{{\dot\alpha}}

\newcommand{\pa}{\partial}
\newcommand{\qb}{{\bar{\theta}}}
\newcommand{\hf}{\frac12}
\newcommand {\cY}{{\cal Y}}
\newcommand {\cS}{{\cal S}}
\newcommand {\cN}{{\cal N}}
\newcommand {\cU}{{\cal U}}
\newcommand {\cV}{{\cal V}}
\newcommand {\cVb}{{\overline{\cal V}}}

\newcommand{\cO}{{\mathcal O}}
\newcommand{\cA}{{\mathcal A}}

\newcommand{\cF}{{\mathcal F}}
\newcommand{\cQ}{{\mathcal Q}}
\newcommand{\cJ}{{\mathcal J}}

\renewcommand{\be}{\begin{equation}}
\renewcommand{\ee}{\end{equation}}
\newcommand{\bpm}{\begin{pmatrix}}
\newcommand{\epm}{\end{pmatrix}}

\newcommand{\EV}[1]{\langle #1 \rangle}
\newcommand{\beqn}{\begin{eqnarray}}
\newcommand{\eeqn}{\end{eqnarray}}

\usepackage{slashed}
\newcommand{\cD}{{\mathcal{D}}}

\newcommand{\cR}{{\mathcal{R}}}

\newcommand{\sfD}{{\sf D}}
\newcommand{\sfF}{{\sf F}}

\newcommand{\ta}{\theta}

\newcommand{\tab}{\bar\theta}

\newcommand{\cX}{\mathcal X}

\let\non\nonumber

\newcommand{\C}[1]{$(\ref{#1})$}

\newfont{\goth}{ygoth.tfm scaled 1200}                   

 \numberwithin{equation}{section}

\def\1{{(1)}}
\def\2{{(2)}}
\def\3{{(3)}}

\newcommand{\p}{\partial}
\def\s{\sigma}

\def\a{\alpha}
\def\b{\beta}

\def\d{\delta}

\def\g{\gamma}

\def\l{\lambda}
\def\m{\mu}
\def\n{\nu}

\def\q{\theta}

\def\s{\sigma}

\def\x{\xi}
\def\z{\zeta}

\def\F{\Phi}

\def\L{\Lambda}
\def\O{\Omega}

\def\S{\Sigma}

\newcommand\Tb{\overbar{T}}
\newcommand\Lb{\overbar{L}}
\newcommand\calT{\mathcal{T}}
\newcommand\calTb{\overbar{\mathcal{T}}}
\newcommand\calY{\mathcal{Y}}
\newcommand\calYb{\overbar{\mathcal{Y}}}
\newcommand\Sb{\overline{S}}
\newcommand\Wb{\overbar{W}}
\newcommand\Fb{\overline{F}}
\newcommand\xb{\overline{x}}
\newcommand\Mb{\overline{M}}

\newcommand\chib{\overline{\chi}}

\newcommand\alphad{\dot{\alpha}}
\newcommand\Phib{\overline{\Phi}}
\newcommand\bbPhi{\pmb{\Phi}}
\newcommand\bbPhib{\overline{\pmb{\Phi}}}
\newcommand\psib{\overline{\psi}}
\newcommand\Db{\overbar{D}}
\newcommand\thetab{\overline{\theta}}

\newcommand\nablab{\overline{\nabla}}

\newcommand\phib{\overbar{\phi}}
\newcommand\dps{ | D \phi |^2 }
\newcommand\dpf{ | D \phi |^4 }
\newcommand\dpsmall{ | \partial \phi |^2 }

\newcommand{\ppp}{{+++}} 
\newcommand{\mmm}{{---}}
\newcommand{\mmp}{{--+}}
\newcommand{\ppm}{{++-}}
\newcommand{\pppp}{{++++}}
\newcommand{\mmmm}{{----}}
\newcommand{\ppmm}{{++--}}
\newcommand{\mmpp}{{--++}}
\newcommand{\pp}{{++}}
\newcommand{\mm}{{--}}
\newcommand{\tp}{\theta^+}
\newcommand{\tm}{\theta^-}
\newcommand{\ww}{{\pm\pm}}
\newcommand{\www}{{\pm\pm\pm}}

\newcommand{\xp}{x^\prime}
\newcommand{\lrb}{\left(}
\newcommand{\rrb}{\right)}
\newcommand{\ld}{\left\langle}
\newcommand{\rd}{\right\rangle}
\newcommand{\T}{\mathcal{T}}

\usepackage{color}
   \definecolor{r}{RGB}{255,0,0}
   \definecolor{R}{RGB}{255,0,0}
   \definecolor{lr}{RGB}{254,109,113}
   \definecolor{b}{RGB}{0,95,251}
   \definecolor{lb}{RGB}{0,240,240}
   \definecolor{db}{RGB}{0,0,221}
   \definecolor{g}{RGB}{23,255,17}
   \definecolor{dg}{RGB}{2,172,53}
   \definecolor{w}{rgb}{1,1,1}
   \definecolor{gray}{gray}{0.6}

\title{Supersymmetry and Irrelevant Deformations}
\author{Christian Ferko}
\department{Physics}
\division{Physical Sciences}
\degree{Doctor of Philosophy}
\date{August 2021}

\dedication{To my parents}
\epigraph{
\vspace{3ex}\begin{minipage}{0.75\textwidth}\justifying 
 		Get confused. Solve problems. Repeat. The universe is waiting for you. \\[-0.5cm] 
 		\begin{flushright} 
 			--- Mark Eichenlaub
 		\end{flushright}
 		\end{minipage}
 		\vspace{7ex}

		}

\begin{document}


\pdfsuppressptexinfo=1
\pdftrailerid{}

\hypersetup{pageanchor=false}		
\maketitle
\hypersetup{pageanchor=true}		

%

\makecopyright
\makededication
\makeepigraph

\tableofcontents

\acknowledgments

First, I would like to thank my advisor, Sav Sethi. Our discussions throughout my graduate school career have shaped my intellectual development and taught me about the process of doing research. I have learned that physics is an intrinsically social enterprise, driven as much by conversations as calculations, and as such I am very grateful to Sav for generously giving his time to talk physics with me over the past few years.

Similarly, I want to thank my other colleagues who collaborated with me on the papers described in this manuscript and on unrelated projects. I have greatly enjoyed discussing physics with Chih-Kai, Gabriele, Alessandro, Hongliang, Daniel, Emil, Gautam, Stephen, Hao-Yu, Zhengdi, and Sam. The opportunity to collaborate with them has enriched my thinking and exposed me to many new perspectives on physics.

I would also like to thank my parents, for supporting my interest in science from my childhood until the present day. I've dedicated this thesis to them because they have always encouraged me to challenge myself and pursue my goals. Their love and unshakeable confidence in my abilities has been a stabilizing influence throughout my academic journey.

I am indebted to the broader theory group at the University of Chicago for providing stimulating conversations over lunch and whiskey, and for making this such an exciting place to do physics. Outside of research, I am grateful to the many students and friends who have spoken to me about science and, through our discussions, helped me to understand many subjects more deeply.

Finally, I acknowledge funding support throughout my PhD from the Physical Sciences Division and the Department of Physics at the University of Chicago; from  NSF Grants No. PHY1720480 and PHY201419; and currently from U.S. Department of Energy grant DE-SC0009999 and funds from the
University of California.

\abstract

The $\TT$ operator provides a universal irrelevant deformation of two-dimensional quantum field theories with remarkable properties, including connections to both string theory and holography beyond $\mathrm{AdS}$ spacetimes. In particular, it appears that a $\TT$-deformed theory is a kind of new structure, which is neither a local quantum field theory nor a full-fledged string theory, but which is nonetheless under some analytic control. On the other hand, supersymmetry is a beautiful extension of Poincar\'e symmetry which relates bosonic and fermionic degrees of freedom. The extra computational power provided by supersymmetry renders many calculations more tractable. It is natural to ask what one can learn about irrelevant deformations in supersymmetric quantum field theories. \vspace{10pt}

\noindent In this thesis, we describe a presentation of the $\TT$ deformation in manifestly supersymmetric settings. We define a ``supercurrent-squared'' operator, which is closely related to $\TT$, in any two-dimensional theory with $(0, 1)$, $(1, 1)$, or $(2, 2)$ supersymmetry. This deformation generates a flow equation for the superspace Lagrangian of the theory, which therefore makes the supersymmetry manifest. In certain examples, the deformed theories produced by supercurrent-squared are related to superstring and brane actions, and some of these theories possess extra non-linearly realized supersymmetries. We then show that $\TT$ defines a new theory of both abelian and non-abelian gauge fields coupled to charged matter, which includes models compatible with maximal supersymmetry. In analogy with the Dirac-Born-Infeld (DBI) theory, which defines a non-linear extension of Maxwell electrodynamics, these models possess a critical value for the electric field.\vspace{10pt}

\noindent Most of this thesis is adapted from previous publications. However, the final chapter presents new results on $\TT$-like deformations of general gauge theories in an arbitrary number of spacetime dimensions; this analysis has not appeared in past work.


\mainmatter
\normalem

\chapter{Introduction} \label{CHP:intro}
A central goal of modern theoretical physics, broadly speaking, is to better understand the space of all quantum field theories and string theories. One familiar way to explore this space is to begin with a well-understood theory and deform it, for instance by adding an integrated local operator to the Lagrangian. It is convenient to adopt a Wilsonian perspective and organize such deformations into classes according to the dimension of the perturbing operator, as one does in effective field theory. If the scaling dimension of the deforming operator is $\Delta$ and the spacetime dimension is $d$, then there are three cases.

\begin{enumerate}
    \item If $\Delta < d$, then we say that the operator is \textit{relevant}. This class of deformations triggers a conventional renormalization group flow, which modifies the behavior of the theory in the infrared. A classic example is adding a mass term $m^2 \phi^2$ to the theory of a free scalar field, which is relevant in any dimension. The term ``relevant'' refers to the fact that such operators become more important at low energies (and conversely, become negligible at high energies).
    
    \item A (classically) \textit{marginal} operator does not become more nor less important as one flows to low energies. In a general QFT, quantum corrections can introduce a weak scale dependence to such an operator which makes it either \textit{marginally relevant} or \textit{marginally irrelevant}. If the seed theory is conformal, however, there may exist certain exactly marginal operators whose scaling dimension $\Delta = d$ matches their mass dimension. In this  case, addition of a marginal operator generates motion on the conformal manifold, and the coupling constants of these marginal operators parameterize the moduli space of the theory. For example, the kinetic term $\partial^\mu \phi \partial_\mu \phi$ is marginal in any dimension; in the $2D$ CFT of a single compact boson, the addition of this operator is interpreted as a change in the radius of the boson's target-space circle.
    
    \item An \textit{irrelevant deformation} is one for which $\Delta > d$, like the square of the kinetic term $( \partial^\mu \phi \partial_\mu \phi )^2$. Such an operator grows more important at high energies and less important at low energies, which means that this class of deformations modifies the ultraviolet behavior of the theory. Because the process of flowing from high to low energies is lossy (there can be many UV theories in the same universality class), it is not possible in general to uniquely reverse a renormalization group trajectory by adding an irrelevant operator. Mathematically, adding such an operator will generically turn on infinitely many other operators which leads to a loss of predictive power. For this reason, deformation by irrelevant operators is more difficult to understand.
\end{enumerate}
In view of the technical and conceptual challenges associated with addition of an irrelevant operator, it is especially surprising that there exists a well-defined irrelevant deformation of any two-dimensional field theory with the property that the deformed theory remains under analytic control, in a sense which we will describe shortly. This observation was first made by Zamolodchikov in \cite{Zamolodchikov:2004ce}, where he studied the deformation of two-dimensional Euclidean quantum field theories by the combination
\begin{align}
    \det ( T_{\mu \nu} ) = T_{00} T_{11} - T_{01}^2 \, , 
\end{align}
where $T_{\mu \nu}$ is the stress-energy tensor of the theory defined in Euclidean signature with a flat metric $g_{\mu \nu} = \delta_{\mu \nu}$. The determinant can be written in a manifestly Lorentz-invariant way using the property of $2 \times 2$ matrices that
\begin{align}\label{stress-tensor-bilinear}
    \det ( T_{\mu \nu} ) = \frac{1}{2} \left( \left( \tensor{T}{^\mu_\mu} \right)^2 - T^{\mu \nu} T_{\mu \nu} \right) \, .
\end{align}
The expression (\ref{stress-tensor-bilinear}) involves products of stress tensor operators, and products of local operators are generally divergent in quantum field theory. However, \cite{Zamolodchikov:2004ce} showed that this particular combination gives a finite, well-defined local operator as the insertion points of the stress tensors are taken coincident (up to total derivative terms, which are not relevant inside of one-point functions or integrals over spacetimes without boundary). It is therefore possible to deform a quantum field theory by the local operator defined by $\det ( T_{\mu \nu} )$ in this coincident point limit; this operator is commonly referred to as $\TT$ due to its expression in complex coordinates for a conformal field theory.

Because $T_{\mu \nu}$ has mass dimension $d$ in $d$ dimensions, a bilinear in the stress tensor such as (\ref{stress-tensor-bilinear}) -- and, therefore, the local $\TT$ operator defined by their coincident point limit -- has dimension $2d$, and is irrelevant in any number of spacetime dimensions. As we saw above, deformation of quantum field theories by irrelevant operators is typically difficult to understand. However, for the very special case of deformations by the $\TT$ operator in two-dimensional quantum field theories, a flow equation of the form
\begin{align}\label{TT_flow_det}
    \frac{\partial \mathcal{L}^{(\lambda)}}{\partial \lambda} = \det \left( T_{\mu \nu}^{(\lambda)} \right)
\end{align}
leads to a controlled one-parameter family of theories in which certain quantities can still be computed. One can think of the equation (\ref{TT_flow_det}) as defining a curve in the space of two-dimensional quantum field theories. At each point along the curve, the ``tangent vector'' (or the local operator by which we deform to move along the curve) is the determinant of the stress-energy tensor $T_{\mu \nu}^{(\lambda)}$ computed from the Lagrangian $\mathcal{L}^{(\lambda)}$ of the theory at that point (not the stress-energy tensor of the seed theory at $\lambda = 0$).

The proof of well-definedness for the $\TT$ operator, as well as the derivations of the flow equation for the cylinder energy levels and the deformed Lagrangian for a free scalar, will be reviewed in more detail in Chapter \ref{CHP:TT}. For now, we will simply quote some results by way of motivation. Consider a family of $\TT$-deformed quantum field theories defined on a cylinder of radius $R$, whose energy levels are denoted by $E_n ( R , \lambda )$. These energies satisfy an equation of inviscid Burgers' type, namely
\begin{align}\label{burgers}
    \frac{\partial E_n}{\partial \lambda} = E_n \frac{\partial E_n}{\partial R} + \frac{1}{R} P_n^2 \, , 
\end{align}
where $P_n = P_n ( R )$ is the momentum (which does not flow). If the seed theory is a CFT, then on dimensional grounds the undeformed energies $E_n^{(0)} \equiv E_n ( R, 0 )$ must scale as $E_n^{(0)} \sim \frac{1}{R}$. In this case, one can solve the differential equation (\ref{burgers}) in closed-form to obtain
\begin{align}\label{square_root_energies}
    E_n ( R , \lambda ) = \frac{R}{2 \lambda} \left( \sqrt{ 1 + \frac{4 \lambda E_n^{(0)}}{R} + \frac{4 \lambda^2 P_n^2}{R^2} } - 1 \right) \, .
\end{align}
This is an example of what we mean by saying that the deformed theory ``remains under analytic control.'' More concretely, one can make precise statements about quantities in the deformed theory in terms of the corresponding quantities in the undeformed theory; here, the deformed energies $E_n ( R, \lambda )$ are functions of the undeformed energies $E_n^{(0)}$. There are analogous statements about other properties of the deformed theory such as its torus partition function, $S$-matrix, and -- at least perturbatively in $\lambda$ -- its correlation functions.

It turns out that the square-root structure of the energy levels (\ref{square_root_energies}) implies a surprising result about the theory's asymptotic density of states. Recall we have assumed that the UV behavior of the  undeformed theory is described by a CFT. This implies that the high energy density of states has Cardy behavior
\be\label{cardy}
\rho(E^{(0)}_n)\sim \exp \left( {\sqrt{\frac{c}{3}E^{(0)}_n
}} \right) ~. 
\ee
On the other hand the high energy behavior ($E_n \gg \frac{R}{\lambda}$) of the deformed energy scales as 
\be
E_n ( R , \lambda ) \sim \sqrt{\frac{R E_n^{(0)}}{\lambda}} \quad \implies \quad E_n^{(0)}\sim \frac{\lambda E_n^2}{R}.
\ee
It can be shown -- by first demonstrating that a $\TT$-deformed CFT enjoys modular invariance, and then by using the modular $S$ transformation to relate the high-temperature and low-temperature limits of the partition function -- that the high energy density of states in the deformed theory grows as
\be\label{hagedorn}
\rho(E_n)\sim \exp \left( {\sqrt{\frac{c\lambda}{3R}} E_n} \right)~.
\ee
A theory with the asymptotic behavior (\ref{hagedorn}) cannot be a local quantum field theory, which would necessarily exhibit the Cardy growth (\ref{cardy}). Rather, this Hagedorn growth is more characteristic of a string theory.

This result is the first hint that a $\TT$-deformed theory is some new and mysterious intermediate structure, neither a local quantum field theory nor a full-fledged string theory which would include dynamical gravity. This Hagedorn behavior was interpreted in \cite{Giveon:2017nie} -- at least for the so-called ``single trace'' version of $\TT$ -- by exploring its connection to little string theory and asymptotically linear dilaton spacetimes. In particular, it seems that adding the single-trace $\TT$ to a boundary conformal field theory dual to a bulk $\mathrm{AdS}$ spacetime corresponds to a deformation of the bulk which changes the asymptotics to linear dilaton. Because asymptotically $\mathrm{AdS}$ spacetimes are qualitatively different from asymptotically linear dilaton spacetimes, which behave more like flat space, this result suggests that $\TT$ may be a promising avenue for understanding holography beyond $\mathrm{AdS}$.

A more direct way to see that $\TT$-deformed theories share some properties of string theories is to directly deform the seed theory of a single free boson. Beginning from the undeformed Lagrangian
\begin{align}
    \mathcal{L}_0 = \partial^\mu \phi \partial_\mu \phi \, ,
\end{align}
it was shown in \cite{Cavaglia:2016oda} that applying $\TT$ to this theory leads to a deformed Lagrangian corresponding to a Nambu-Goto string in static gauge,
\begin{align}
    \mathcal{L}_\lambda = \frac{1}{2 \lambda} \left( \sqrt{1 + 4 \lambda \partial_\mu \phi \partial^\mu \phi } - 1 \right) \, .
\end{align}
This gives another piece of evidence, in addition to the Hagedorn density of states, to suggest that $\TT$ is related to string theory.

On the other hand, it has been known since the 1970s that the introduction of supersymmetry resolves several unsatisfying properties of bosonic string theory, such as the presence of a tachyon and the absence of fermions in the spectrum. Given the tantalizing connections between $\TT$-deformed field theories and (bosonic) string theories, it is natural to ask whether some supersymmetric presentation of $\TT$ might similarly be related to the superstring.

In one sense, the interplay between $\TT$ and supersymmetry is trivial: the finite-volume spectrum of a $\TT$-deformed theory obeys a differential equation whereby each deformed energy level is determined in terms of the corresponding undeformed energy level. This means that any degeneracies of the undeformed energies -- such as a pairing between the energy levels of bosonic and fermionic states, characteristic of supersymmetric theories -- will also persist in the deformed theory. From this perspective, it is obvious that applying the $\TT$ deformation to a supersymmetric seed theory produces a deformed theory which is also supersymmetric.

However, the analytic control offered by supersymmetry is most powerful when the symmetry is made manifest, for instance by a superspace construction which geometrizes the supersymmetry transformations. Although the ordinary $\TT$ deformation preserves the supersymmetry of the seed theory, the supersymmetry transformations will generically flow, so that the action of the supersymmetry generators on the fields of the theory must be corrected order-by-order in the $\TT$ parameter $\lambda$. It is desirable to present a formulation of $\TT$ where the supersymmetry of the deformed theory acts in a simple way, for instance by formulating a flow equation for a superspace Lagrangian. This is one motivation for the present work, and we will propose such superspace flow equations for theories with various amounts of supersymmetry in Chapters \ref{CHP:SC-squared-1} and \ref{CHP:SC-squared-2}.

Another motivation is the possibility of finding $\TT$-like deformations for higher-dimensional theories. We will briefly explain why the point-splitting definition of the local $\TT$ operator fails in $D>2$ spacetime dimensions. In any $D$, there is some operator product expansion
\begin{align}\label{general_tt_ope}
    T^{\mu \nu} ( x ) T_{\mu \nu} ( y ) - \frac{1}{D-1} \tensor{T}{^\mu_\mu} ( x ) \tensor{T}{^\nu_\nu} ( y ) = \sum_{\alpha} A_\alpha \left( | x - y |^2 \right) \mathcal{O}_{\alpha} ( y ) \, .
\end{align}
The argument of \cite{Zamolodchikov:2004ce}, to be reviewed in Chapter \ref{CHP:TT}, showed that that in $D = 2$ the conservation equation $\partial^\mu T_{\mu \nu} = 0$ implies $A_\alpha$ is independent of coordinate unless the corresponding $\mathcal{O}_{\alpha}$ is a total derivative. This is sufficient to define a local operator by taking $x \to y$, modulo total derivative terms. When $D > 2$, however, the conservation equation imposes fewer constraints on this operator product expansion, and there can be additional non-derivative divergences on the right side of (\ref{general_tt_ope}) which obstruct a universal definition of $\TT$.

However, in theories with supersymmetry, the stress tensor sits in a multiplet that contains additional fields which are related to $T_{\mu \nu}$. This multiplet of operators satisfies a larger collection of constraints than conservation, which correspondingly imposes more structure on the operator product expansions of fields in the multiplet. One might hope that, with a sufficiently large amount of supersymmetry (likely maximal), a particular combination of supercurrent bilinears might satisfy a property similar to that of $\TT$ in two dimensions, and therefore provide a universal irrelevant deformation of any sufficiently supersymmetric theory. Although there is no proof of this statement, we will discuss some classical properties of supersymmetric $\TT$-like deformations for four-dimensional theories in Chapter \ref{CHP:nonlinear} and speculate about the possibility of defining a quantum operator in Section \ref{concludingthoughts}.

We close this introductory section by reiterating that irrelevant current-type deformations lie at an exciting intersection of many interesting directions in theoretical physics. Here we have focused on the relationship of $\TT$ to string theory and on tantalizing hints of holography beyond $\mathrm{AdS}$. However, $\TT$ and related deformations also have connections to integrability, random geometry, topological gravity, the uniform light-cone gauge, and many other areas of interest. For an informative overview of the subject, see \cite{Jiang:2019epa}.

The layout of the remainder of this dissertation is as follows.

\begin{itemize}
    \item In Chapter \ref{CHP:TT}, we will review some facts about the ordinary $\TT$ deformation, present an introduction to supersymmetry, and lay out our conventions. This chapter is a review of previous results, primarily from \cite{Zamolodchikov:2004ce} and \cite{Cavaglia:2016oda}.
    
    \item Next, Chapter \ref{CHP:SC-squared-1} presents a manifestly supersymmetric version of the $\TT$ operator for theories with $(0, 1)$ or $(1, 1)$ supersymmetry. This chapter is based on \cite{Chang:2018dge}.
    
    \item In Chapter \ref{CHP:SC-squared-2}, we extend this definition to theories with $(2, 2)$ supersymmetry and study some deformed models in detail. Here we follow the treatment in \cite{Chang:2019kiu}.
    
    \item Chapter \ref{CHP:nonlinear} then identifies certain non-linearly realized symmetries of the $(2, 2)$ deformed models that we first introduced in Chapter \ref{CHP:SC-squared-2}. We also discuss how these theories are related to certain models with $\mathcal{N} = 1$ supersymmetry in four dimensions. This chapter is based on the paper \cite{Ferko:2019oyv}.
    
    \item Chapter \ref{CHP:nonabelian} uses the ordinary $\TT$ deformation to define a new theory of a non-abelian gauge field coupled to scalars in two dimensions; this deformed theory is compatible with maximal supersymmetry. The discussion of this chapter follows \cite{Brennan:2019azg}.
    
    \item Chapter \ref{CHP:new_chapter} continues the study of deformations in gauge theories. We present flow equations for the $\TT$ deformation of a general abelian gauge theory in arbitrary dimension, and show that it has a solution of Born-Infeld type only in $D=4$. We also comment on the (more subtle) deformation of non-abelian gauge theories. This chapter contains new material which has not appeared in any previously published work.
\end{itemize}
The details of various calculations which we use in the main body of the thesis are presented in Appendix \ref{details}.

\chapter{The \texorpdfstring{$\TT$}{TT} Deformation and Preliminaries} \label{CHP:TT}
In this chapter, we will review some features of the ordinary $\TT$ deformation and present an overview of general aspects of supersymmetry. It is our hope that these sections might provide an accessible introduction for students. We also set our conventions for the various superspaces that will be used in later chapters.

Nothing in this chapter is new. Our goal in Section \ref{TTb-review} is to present results from earlier works on $\TT$ for context and motivation, focusing on \cite{Zamolodchikov:2004ce} and \cite{Cavaglia:2016oda}. Another useful review of this content is \cite{Jiang:2019epa}. Likewise, Section \ref{sec:susy_review} contains only textbook material on supersymmetry; we will provide a list of standard references on supersymmetry at the end of that section.

\section{\texorpdfstring{Review of $T \Tb$}{Lg}}\label{TTb-review}

In this section, we will provide more detailed derivations of a few results concerning the ordinary $\TT$ deformation which were quoted in the introductory chapter.

\subsection{Definition by point-splitting}\label{sec:tt_defn_splitting}

As we mentioned in the introduction, the first remarkable property of $\TT$ is that it can be unambiguously defined by point-splitting (up to total derivatives), despite involving products of local operators. Here we will review the proof of this fact from the original paper of Zamolodchikov \cite{Zamolodchikov:2004ce}. There he considered a Euclidean QFT with the following properties:

\begin{enumerate}
    \item Local translation and rotation symmetry. This property implies the existence of a stress energy tensor $T_{\mu\nu}$ which is conserved, 
    \begin{align}
        \partial_\mu T^{\mu\nu} = 0 \, , 
    \end{align}
    and symmetric, $T_{\mu \nu} = T_{\nu \mu}$. In the $2D$ parametrization $T_{zz}=T,~\bar{T}=T_{\bar{z}\bar{z}},~\Theta=T_{z\bar{z}}$, this can be written
    \be
    \partial_{\zbar} T(z)=\partial_z \Theta(z)~,\quad \quad \partial_z \bar{T}(z)=\partial_\zbar \Theta(z)~. 
    \ee
    
    \item\label{global_trans} Global translation symmetry. This property implies that any $1$-point function is independent of position
    \be
    \langle \CO_i(z)\rangle=\langle \CO_i(0)\rangle~, 
    \ee
    and that any $2$-point function 
    \be
    \langle \CO_i(z)\CO_j(z')\rangle=G_{ij}(z-z')~,
    \ee
    depends only on the distance $| z - z' |$ between the insertion points. 
    
    \item Clustering. That there exists some direction $x$ of infinite length such that 
    \be
    \lim_{x\to \infty}\langle \CO_i(x)\CO_j(0)\rangle=\langle \CO_i\rangle\langle \CO_j\rangle~. 
    \ee
    
    \item UV CFT. The seed QFT which we begin with is assumed to be described by a CFT at short distances.
\end{enumerate}
These conditions then require that we consider a theory on the flat plane or cylinder. Using these assumptions one can show that the coincident point limit
\be
\TT:=\lim_{z'\to z}\left(T(z')\bar{T}(z)-\Theta(z')\Theta(z)\right) 
\ee
defines a local operator. 

First, note that the conservation of the stress tensor implies 
\begin{align}\begin{split}\label{deriv}
\partial_\zbar (T(z)\bar{T}(z')-\Theta(z)\Theta(z'))=(\partial_z+\partial_{z'})\Theta(z)\bar{T}(z')-(\partial_\zbar+\partial_{\zbar'})\Theta(z)\Theta(z')~,\\
\partial_z(T(z)\bar{T}(z')-\Theta(z)\Theta(z'))=(\partial_z+\partial_{z'})T(z)\Tbar(z')-(\partial_\zbar+\partial_{\zbar'})T(z)\Theta(z')~.
\end{split}\end{align}
Now using the operator product expansions
\begin{align}\begin{split}
    T(z)\Theta(z')=\sum_i A_i(z-z')\CO_i(z')~,\quad\quad\Theta(z)\Theta(z')=\sum_i C_i(z-z')\CO_i(z')~,\\
    \Theta(z)\Tbar(z')=\sum_i B_i(z-z')\CO_i(z')~,\quad\quad T(z)\Tbar(z')=\sum_i D_i(z-z')\CO_i(z')~,
\end{split}\end{align}
the equations \eqref{deriv} imply 
\begin{align}\begin{split}
&\sum_i \partial_\zbar F_i(z-z')\CO_i(z')=\sum_i \Big(B_i(z-z') \partial_{z'}\CO_i-C_i (z-z')\partial_{\zbar'}\CO_i\Big)~,\\
&\sum_i \partial_z F_i (z-z')\CO_i(z')=\sum_i\Big(D_i(z-z') \partial_{z'}\CO_i(z')-A_i(z-z')\partial_{\zbar'}\CO_i(z')\Big)~, 
\end{split}
\end{align}
where 
\be
F_i(z-z')=D_i(z-z')-C_i(z-z')~. 
\ee
This implies that any operator arising in the OPE
\be
T(z)\Tbar(z')-\Theta(z)\Theta(z')=\sum_i F_i(z-z')\CO_i(z')~,
\ee
must either have a coordinate independent coefficient function $F_i(z-z')$ or is itself the derivative of another local operator:
\be
T(z)\bar{T}(z')-\Theta(z)\Theta(z')=\CO_{\TT}(z')+{\rm derivative~ terms}~.
\ee
This allows us to define the composite operator 
\be
\TT(z):=\CO_{\TT}(z)~. 
\ee
Note that  we have only defined $\TT$ up to derivative terms, but by the assumption of global translation symmetry above, any one-point function of a total derivative vanishes.

Although it will not be used in this thesis, we note that assumption (2) of global translation symmetry can be weakened to the existence of a transitive global isometry. This can be used to define the $\TT$ deformation in $\mathrm{AdS}_2$, as is done in \cite{Brennan:2020dkw}. In that case, one can show that the $\TT$ operator obeys a factorization property in the $SL(2, \mathbb{R})$-invariant ground state.

\subsection{Deformed Lagrangian for free scalar}\label{subsec:flow_eqn_review}

In this subsection, we will review the solution of the $\TT$ flow equation for the deformed Lagrangian $\mathcal{L} ( \lambda )$ of a free scalar field $\phi$. We stress that this is a purely classical result, unrelated to the preceding argument that the $\TT$ operator is well-defined by point-splitting. Indeed, the explicit solution for the deformed Lagrangian of scalars coupled to an arbitrary background metric was already written down in \cite{Bonelli:2018kik}, which follows from the analysis in \cite{Cavaglia:2016oda}.

Consider a general $\lambda$-dependent Lagrangian for a real scalar $\phi$ coupled to a background metric $g_{\mu \nu}$. For simplicity, we assume that the Lagrangian reduces to the usual free kinetic Lagrangian for $\phi$ at $\lambda = 0$:
\begin{align}
    \mathcal{L} ( \lambda = 0 ) = \frac{1}{2} g^{\mu \nu} \partial_\mu \phi \partial_\nu \phi .
    \label{initial_condition}
\end{align}
The $\TT$ flow will not introduce dependence on the bare (undifferentiated) field $\phi$, as one would have in a potential energy term, so the finite-$\lambda$ Lagrangian can only depend on the scalar quantity $g^{\mu \nu} \partial_\mu \phi \partial_\nu \phi$ and on the parameter $\lambda$. To ease notation, we define $\x = g^{\mu \nu} \partial_\mu \phi \partial_\nu \phi$ so that
\begin{align}
    \mathcal{L} ( \lambda ) = f ( \lambda, g^{\mu \nu} \partial_\mu \phi \partial_\nu \phi ) \equiv f ( \lambda, \x ) .
\end{align}
We now compute the components of the stress tensor,
\begin{align}
\label{Tdef}
    T_{\mu \nu}^{(\lambda)} = - \frac{2}{\sqrt{-g}} \frac{\delta S^{(\lambda)}}{\delta g^{\mu \nu}} , 
\end{align}
where $S^{(\lambda)}$ is the effective action of the deformed theory
\begin{align}
    S^{(\lambda)} = \int \, d^2 x \, \sqrt{-g} \, \mathcal{L} ( \lambda ) .
\end{align}
Taking the variation, one finds
\begin{align}
    T_{\mu \nu}^{(\lambda)} = g_{\mu \nu} f - 2 \frac{\partial f}{\partial \x} \partial_\mu \phi \, \partial_\nu \phi  \, .
    \label{stress_tensor_components}
\end{align}
Written in a manifestly diffeomorphism invariant form appropriate for a general background metric, the flow equation for the Lagrangian is
\begin{align}
    \frac{\partial \mathcal{L}}{\partial \lambda} = \frac{1}{2} \left( \left( g^{\mu \nu} \tensor{T}{_\mu_\nu} \right)^2 - g^{\mu \rho} g^{\nu \sigma} T_{\mu \nu} T_{\rho \sigma} \right) .
    \label{det_T_expression}
\end{align}
Note that, when $g_{\mu \nu} = \delta_{\mu \nu}$, the right side of (\ref{det_T_expression}) reduces to the ordinary determinant of a $2 \times 2$ matrix.

Evaluating (\ref{det_T_expression}) with the components (\ref{stress_tensor_components}), one finds
\begin{align}
    \frac{1}{2} \left( \left( g^{\mu \nu} \tensor{T}{_\mu_\nu} \right)^2 - g^{\mu \rho} g^{\nu \sigma} T_{\mu \nu} T_{\rho \sigma} \right) = f^2 - 2 f \x \frac{\partial f}{\partial \x} .
\end{align}
Thus the differential equation for the deformed Lagrangian becomes
\begin{align}
    \frac{d f}{d \lambda} = f^2 - 2 f \x \frac{\partial f}{\partial \x} .
\end{align}
To solve this differential equation, we begin by noting that the solution can depend only on the dimensional parameter $\lambda$ and the dimensionless combination $\lambda \xi$. Since $\frac{1}{\lambda}$ has the same mass dimension as the Lagrangian, it must be consistent to make an ansatz of the form
\begin{align}
    f ( \lambda , \xi ) &= \frac{1}{\lambda} \cdot \tilde{f} ( \lambda \xi ) \, \nonumber \\
    &= \frac{1}{\lambda} \cdot \tilde{f} ( x ) \, ,
\end{align}
where we have defined $x = \lambda \xi$. With this ansatz, we have
\begin{align}
    \frac{df}{d \lambda} &= \frac{1}{\lambda^2} \left( x \tilde{f}' ( x ) - \tilde{f} ( x ) \right) \, , \nonumber \\
    \frac{\partial f}{\partial \x} &= \tilde{f}' ( x )  \, ,
\end{align}
so the differential equation becomes
\begin{align}
    \tilde{f}' ( x ) = \frac{ \tilde{f} ( x ) + \tilde{f} ( x )^2 }{x ( 1 + 2 \tilde{f} ( x ) ) } \, .
\end{align}
The result is now an ordinary differential equation which can be solved by separation of variables. After replacing $x = \lambda \x = \lambda g^{\mu \nu} \partial_\mu \phi \partial_\nu \phi$ and imposing the initial condition (\ref{initial_condition})), the result is
\begin{align}
    \mathcal{L} ( \lambda ) = \frac{1}{2 \lambda} \left( \sqrt{1 + 2 \lambda g^{\mu \nu} \partial_\mu \phi \partial_\nu \phi } - 1 \right) ~.
    \label{equation:Nambu-Goto}
\end{align}
Expanding about $\lambda=0$, one finds
\begin{align}
    \mathcal{L} ( \lambda ) = \frac{1}{2} g^{\mu \nu} \partial_\mu \phi \partial_\nu \phi - \frac{\lambda}{4} \left( g^{\mu \nu} \partial_\mu \phi \partial_\nu \phi \right)^2 + \mathcal{O} ( \lambda ^2 ) ~.
    \label{leading_deformed_lagrangian}
\end{align}
We emphasize that this is a purely classical result that is true for any conformally flat metric $g_{\mu \nu}$. The additional power which comes from working in (Lorentzian or Euclidean) flat space, with metric $\eta_{\mu \nu}$ or $\delta_{\mu \nu}$ is that one can unambiguously define the local $\TT$ operator up to total derivatives, as described in Section \ref{sec:tt_defn_splitting}. With a flat metric, one can also obtain a flow equation for the finite-volume spectrum, which we turn to next.\footnote{As we mentioned earlier, one can also obtain a flow equation for the $\mathrm{AdS}$-invariant ground state for $\TT$-deformed theories in $\mathrm{AdS}_2$, as shown in \cite{Brennan:2020dkw}.}

\subsection{Factorization and inviscid Burgers' equation}

We conclude the review of $\TT$ by deriving the flow equation (\ref{burgers}) for the cylinder energy levels of a $\TT$-deformed quantum field theory, which has square-root type solutions of the form (\ref{square_root_energies}) when the seed theory is conformal.

The first step is to show that the expectation value of the point-split $\TT$ operator is actually independent of the distance between the insertion points. Define the function
\begin{align}
    C ( z, w ) = \langle T ( z ) \Tbar ( w ) \rangle - \langle \Theta ( z ) \Theta ( w ) \rangle \, , 
\end{align}
where as in Section \ref{sec:tt_defn_splitting} we use the notation $T = T_{zz}$, $\Tbar = T_{\zbar \zbar}$, $\Theta = T_{z \zbar}$. Following the steps we used in (\ref{deriv}) above, we take a $z$ derivative to write
\begin{align}
    \partial_{\zbar} C ( z, w ) = \langle \partial_{\zbar} T ( z ) \Tbar ( w ) \rangle - \partial_{\zbar}  \langle \Theta ( z ) \Theta ( w ) \rangle
\end{align}
and use the conservation equation $\partial_\zbar T = \partial_z \Theta$ to write this as
\begin{align}
    \partial_{\zbar} C ( z, w ) = \partial_{z}  \langle \Theta ( z ) \Tbar ( w ) \rangle - \partial_\zbar \langle \Theta ( z ) \Theta ( w ) \rangle \, .
\end{align}
On the other hand, by the assumption of global translation invariance, the two-point function $\langle \Theta ( z ) \Theta ( w ) \rangle$ can depend only on the separation $|z - w|^2$, which means that
\begin{align}
    \partial_\zbar \langle \Theta ( z ) \Theta ( w ) \rangle = - \partial_\wbar \langle \Theta ( z ) \Theta ( w ) \rangle \, .
\end{align}
Similarly in the first term,
\begin{align}
    \partial_{z} \langle \Theta ( z ) \Tbar ( w ) \rangle = - \partial_{w}  \langle \Theta ( z ) \Tbar ( w ) \rangle \, .
\end{align}
Therefore we conclude that
\begin{align}
    \partial_{\zbar} C ( z, w ) = - \langle \Theta ( z ) \partial_{w}  \Tbar ( w ) \rangle + \langle \Theta ( z ) \partial_\wbar \Theta ( w ) \rangle = 0 \, , 
\end{align}
where we have again used the conservation equation $\partial_w \Tbar = \partial_\wbar \Theta$.

Since $C(z, w)$ can depend only on $|z - w|^2$, and since $\partial_z C ( z, w ) = 0$, it follows that $C(z,w) = C$ is a constant. Thus we may evaluate it for any choice of $z$ and $w$. On the one hand, when $z$ and $w$ are taken coincident, we have
\begin{align}
    \lim_{z \to w} C ( z, w ) = \langle \TT \rangle \, , 
\end{align}
since we have established in Section \ref{sec:tt_defn_splitting} that this limit defines a local operator up to total derivative terms (which vanish inside of one-point functions). But on the other hand, for a theory defined on the cylinder, we may take the points $z, w$ to be infinitely separated alone the non-compact cylinder direction:
\begin{align}
    C = \lim_{| z - w | \to \infty} C(z, w) = \langle T \rangle \langle \Tbar \rangle - \langle \Theta \rangle^2 \, , 
\end{align}
where in the last step we have used the fact that vacuum two-point functions cluster decompose into products of one-point functions at infinite separation. We therefore conclude that
\begin{align}
    \langle TT \rangle = \langle T \rangle \langle \Tbar \rangle - \langle \Theta \rangle^2 \, , 
\end{align}
which is referred to as the factorization property of $\TT$.

The above derivation establishes factorization of the $\TT$ operator in the vacuum state, where cluster decomposition holds. To show that it also factorizes in any energy eigenstate, one can insert a complete set of states to write
\begin{align}\label{complete_set_sum}
    \langle n \mid T ( z ) \Tbar ( w ) \mid n \rangle = \sum_m &\langle n \mid T ( z ) \mid m \rangle \, \langle m \mid \Tbar ( w ) \mid n \rangle \nonumber \\ &\quad \cdot \exp \Big( ( E_n - E_m ) | \Im z - \Im w | + i ( P_n - P_m ) | \Re z - \Re w | \Big) \, ,
\end{align}
and likewise for $\langle n \mid \Theta ( z ) \Theta ( w ) \mid n \rangle$. Because the exponential factors contain explicit dependence on the coordinates $z, w$, but the function $C ( z, w )$ is a constant when the correlation functions are taken in \textit{any} energy eigenstate, all of the terms in the sum (\ref{complete_set_sum}) must vanish except when $m = n$. In the coincident point limit, this implies that
\begin{align}
    \langle n \mid TT \mid n \rangle = \langle n \mid T \mid n \rangle \langle n \mid \Tbar \mid n \rangle - \langle n \mid \Theta \mid n  \rangle^2 \, , 
\end{align}
which is the statement that $\TT$ factorizes in all energy eigenstates, not just the vacuum.

Once the factorization property of $\TT$ is established, the flow equation for the cylinder spectrum follows almost immediately. The interpretation of the components of the stress tensor for a theory on a cylinder of radius $R$ is
\begin{align}\label{stress_component_interp}
    \langle n \mid T_{yy} \mid n \rangle &= - \frac{1}{R} E_n ( R , \lambda ) \, , \nonumber \\
    \langle n \mid T_{xx} \mid n \rangle &= - \partial_R E_n ( \lambda, R ) \, , \nonumber \\
    \langle n \mid T_{xy} \mid n \rangle &= i P_n ( R ) \, .
\end{align}
Here $x \sim x + R$ is the circular direction of the cylinder and $y$ is the non-compact direction. Because the Euclidean Lagrangian density is the Hamiltonian density, and we are deforming by adding the $\TT$ operator to the Lagrangian, the flow equation for the energy levels is
\begin{align}
    \partial_\lambda E_n ( \lambda, R ) &= - R \langle n \mid \det ( T_{\mu \nu} ) \mid n \rangle \, \nonumber \\
    &= - R \left( \langle n \mid T_{xx} \mid n \rangle \langle n \mid T_{yy} \mid n \rangle - \langle n \mid T_{xy} \mid n \rangle^2 \right) \, ,
\end{align}
where in the second step we have used factorization. Expressing the stress tensor components in terms of energies and momenta according to (\ref{stress_component_interp}), we find
\begin{align}
    \partial_\lambda E_n = E_n \partial_R E_n + \frac{1}{R} P_n^2 \, .
\end{align}
This is the inviscid Burgers' equation (\ref{burgers}) which we referred to in the introduction. Given any seed theory with energies $E_n$, this flow equation determines the energy levels of the $\TT$ deformed theory at finite $\lambda$, although for a general starting theory we cannot solve the differential equation in closed form. Matters simplify if the undeformed theory is a CFT, since in this case the undeformed energies and momenta satisfy
\begin{align}
    E_n &= \frac{1}{R} \left( n + \overbar{n} - \frac{c}{12} \right) \, , \nonumber \\
    P_n &= \frac{1}{R} \left( n - \overbar{n} \right) \, ,
\end{align}
where $c$ is the central charge. Because of the especially simple $R$-dependence of the undeformed energies, in this case we can solve the flow equation to write
\begin{align}
    E_n ( R , \lambda ) = \frac{R}{2 \lambda} \left( \sqrt{ 1 + \frac{4 \lambda E_n^{(0)}}{R} + \frac{4 \lambda^2 P_n^2}{R^2} } - 1 \right) \, ,
\end{align}
which is equation (\ref{square_root_energies}) that was quoted in the introduction.

One generic feature of $\TT$-deformed CFTs is apparently from the square-root structure of the energy levels. Let's restrict to the ground state, $n = \overbar{n} = 0$, for which the flow equation becomes
\begin{align}
    E_0 ( R, \lambda ) = \frac{R}{2 \lambda} \left( \sqrt{ 1 - \frac{4 \lambda c}{R^2} } - 1 \right)
\end{align}
When $\lambda$ exceeds $\frac{R^2}{4 c}$, the argument of the square root becomes negative and the energies are complex. This signals that there is a maximum allowed value of the flow parameter $\lambda$, after which the theory appears to suffer some pathology. A complete understanding of this complex-energy behavior is still lacking, but it might be related to the non-locality of $\TT$ deformed theories; a possible interpretation is that this maximum value of $\lambda$ represents the point at which the non-locality scale becomes comparable to the radius of the cylinder, and that this large non-locality causes the theory to become ill-defined \cite{Chakraborty:2019mdf}.

\section{Review of Supersymmetry}\label{sec:susy_review}

Much of the content of this thesis (especially Chapters \ref{CHP:SC-squared-1}, \ref{CHP:SC-squared-2}, and \ref{CHP:nonlinear}) concerns the interplay between the $\TT$ deformation and supersymmetry. To set the stage, we will present a short introduction to the ideas and formalism of supersymmetry in this section. Our discussion will necessarily be incomplete, but we will attempt to strike a balance between being brief, pedagogical, and sufficiently comprehensive so that this thesis is fairly self-contained.

It is well-established that leveraging the symmetries of a problem in physics is almost always an effective strategy for finding the solution. Motivated by this observation, one might attempt to introduce additional symmetries into a quantum field theory to make the theory more amenable to analysis. A given QFT will typically have spacetime symmetries, such as Poincar\'e, and perhaps a collection of internal symmetries -- for instance, a theory of $n$ real scalar fields $\phi_i$ might have an $O(n)$ rotational symmetry if the potential depends only on $| \vec{\phi} |^2 = \sum_{i=1}^{n} \phi_i \phi_i$. Therefore, in order to construct toy models which are more symmetric, one might attempt to combine these spacetime and internal symmetries in some way which yields a richer symmetry structure.

One obstruction to obtaining such a richer symmetry structure, however, is a no-go result known as the Coleman-Mandula theorem \cite{PhysRev.159.1251}. This theorem states that, under mild assumptions, the symmetry group of any consistent quantum field theory may only be a trivial direct product of the spacetime symmetry group and the internal symmetry group. In particular, there is no way to combine the two types of symmetries in a more interesting way, such as some kind of semi-direct product.

On the face of it, this result seems to suggest that there is no hope for enriching the symmetry structure of a quantum field theory in order to make it more analytically tractable. As with any no-go theorem, though, one can attempt to evade the result by violating one of its assumptions. One of the assumptions of the Coleman-Mandula theorem is that the symmetry structure of the quantum field theory is encoded in a Lie group. In particular, the generators of the corresponding Lie algebra are bosonic objects. One might therefore evade the theorem by considering symmetries with Grassmann (fermionic) generators, which are associated with a Lie \emph{super}algebra rather than a Lie algebra. These fermionic generators are usually assumed to transform in spinor representations of the Lorentz group, as one might expect from the spin statistics theorem (although this is not the only option).

We will write these additional fermionic symmetry generators as $Q_\alpha$, which we refer to as \emph{supercharges}. These $Q_\alpha$ necessarily send bosonic states to fermionic states, and vice-versa; we denote this by
\begin{align}
    Q_\alpha | b \rangle = | f \rangle \, , \qquad Q_\alpha | f \rangle = | b \rangle \, .
\end{align}
A supersymmetric theory must therefore possess both fermions and bosons. In any quantum field theory containing fermions, we may define a fermion number operator $(-1)^F$ which distinguishes between bosonic and fermionic states as
\begin{align}
    (-1)^F | b \rangle = | b \rangle \, , \qquad ( - 1 )^F | f \rangle = - | f \rangle \, .
\end{align}
This operator induces a $\mathbb{Z}_2$ grading of the Hilbert space $\mathcal{H}$ of our theory, which means that this Hilbert space splits into a direct sum of the form
\begin{align}
    \mathcal{H} = \mathcal{H}_b \oplus \mathcal{H}_f \, , 
\end{align}
and where $\mathcal{H}_b, \mathcal{H}_f$ contain eigenstates of $(-1)^F$ with eigenvalues $+1, -1$, respectively. Because the operators $Q_\alpha$, send bosons to fermions and vice-versa, one has $Q_\alpha : \mathcal{H}_b \to \mathcal{H}_f$ and $Q_\alpha : \mathcal{H}_f \to \mathcal{H}_b$.

We must postulate an algebra of these new symmetry generators $Q_\alpha$. The most common assumption is that this algebra takes the form
\begin{align}\label{susy_algbra}
    \{ Q_\alpha , Q_\beta^\dagger \} = c^M_{\alpha \beta} P_M + Z_{\alpha \beta} \, ,
\end{align}
where $c^M_{\alpha \beta}$ and $Z_{\alpha \beta}$ are constants, $P_M$ is the momentum operator, and the index $M$ runs over all spacetime directions. For instance, in four spacetime dimensions it is typical to take $c^M_{\alpha \beta}$ to be related to the Pauli matrices and $Z_{\alpha \beta} = 0$, so that
\begin{align}
    \{ Q_\alpha , Q^\dagger_{\dot{\beta}} \} = 2 \sigma^\mu_{\alpha \dot{\beta}} P_\mu \, ,
\end{align}
where we have distinguished between left-handed and right-handed spinors using undotted and dotted indices, as is convenient in this four-dimensional case.

The existence of such operators $Q_\alpha$ in a theory has far-reaching consequences. We will begin by mentioning some immediate implications for the spectrum; although these results are more general, for simplicity we will restrict to the case of a theory with $0$ space dimensions and one time dimension (i.e. quantum mechanics). In this case, the momentum operator has only a single component $P_0 = H$, the Hamiltonian of the theory. In this case, the upshot of (\ref{susy_algbra}) is that there exists at least one operator $Q$ with the property that $\{ Q, Q^\dagger \} \sim H$. By re-scaling the operator, we can choose the constant of proportionality to be $2$, which is the conventional choice:
\begin{align}\label{sqrt_ham}
    \{ Q, Q^\dagger \} = 2 H \, .
\end{align}
Schematically, this equation $Q Q^\dagger + Q^\dagger Q = 2 H$ looks like $Q^2 = H$, although this is of course not literally true since $Q$ is anti-commuting and therefore satisfies $Q^2 = 0$. But morally, at least, we think of this operator $Q$ as the ``square root'' of $H$. The intuition that $H$ has a square root might lead us to believe that it is non-negative, and this can be made precise. The expectation value of the Hamiltonian in any state $| \psi \rangle$ can be written
\begin{align}\label{positive_energy}
    \langle \psi \mid H \mid \Psi \rangle &= \frac{1}{2} \langle \psi \mid \{ Q, Q^\dagger \} \mid \psi \rangle \nonumber \\
    &= \frac{1}{2} \langle \psi \mid Q Q^\dagger + Q^\dagger Q \mid \psi \rangle \nonumber \\
    &= \frac{1}{2} \left( \langle Q^\dagger \psi \mid Q^\dagger \psi \rangle +\langle Q \psi \mid Q \psi \rangle \right) \, .
\end{align}
The quantity in parentheses on the last line of (\ref{positive_energy}) is the sum of the norms of two states, each of which is non-negative definite. Therefore we conclude
\begin{align}
    \langle \psi \mid H \mid \psi \rangle \geq 0
\end{align}
for any state $| \psi \rangle$ in any supersymmetric theory. Furthermore, from looking at the last line of (\ref{positive_energy}), we see that the only states which saturate this inequality are those states $| \psi \rangle$ for which $Q | \psi \rangle = 0$ and $Q^\dagger | \psi \rangle = 0$. 

The second immediate observation follows from the algebra (\ref{sqrt_ham}) and the fact that $Q$ is nilpotent, so that $Q^2 = 0$. One then has
\begin{align}
    [ H, Q ] &= \frac{1}{2} \left[ \{ Q, Q^\dagger \} , Q \right] \nonumber \\
    &= \frac{1}{2} \left( Q Q^\dagger Q + Q^\dagger Q Q - Q Q Q^\dagger - Q Q^\dagger Q \right) \nonumber \\
    &= 0 \, , 
\end{align}
where the second and third terms are identically zero since $Q^2 = 0$ and the first term cancels the fourth. 

Thus we see that the operator $Q$ necessarily commutes with the Hamiltonian, which means that it is conserved and therefore generates a symmetry of the theory. This justifies our usage of the terms super\emph{symmetry} and super\emph{charges} when referring to this structure. 

It may be useful to have a more explicit example. We now leave the discussion of supersymmetric quantum mechanics ($0+1$ dimensions) and move to field theory. A familiar and important supersymmetric theory is that of the Ramond-Neveu-Schwarz (RNS) superstring, whose worldsheet Lagrangian is
\begin{align}\label{rns_action}
    S = - \frac{1}{2 \pi} \int d^2 \sigma \left( \partial_\alpha X^\mu \partial^\alpha X_\mu - 2 i \left( \psi^\mu_- \partial_{++} \psi_{\mu -} + \psi_+^\mu \partial_{--} \psi_{\mu +} \right) \right) \, ,
\end{align}
Here $\sigma^\alpha = ( \tau, \sigma )^\alpha$ are the two worldsheet coordinates, $\partial_{\pm \pm}$ are derivatives with respect to light-cone coordinates $\sigma^{\pm \pm} = \tau \pm \sigma$ (where the notation, which uses two $+$'s or two $-$'s, will be explained in the next section), and $\mu$ is a spacetime index which runs from $0$ to $9$. 

We first observe that the action (\ref{rns_action}) is invariant (up to a total derivative term) under the transformations
\begin{align}\label{RNS_susy}
    \delta X^\mu = i \left( \epsilon_+ \psi^\mu_- - \epsilon_- \psi^\mu_+ \right) \, , \qquad \delta \psi_-^\mu = - 2 \partial_{--} X^\mu \epsilon_+ \, , \qquad \delta \psi^\mu_+ = 2 \partial_{++} X^\mu \epsilon_- \, , 
\end{align}
where $\epsilon_{\pm}$ are two infinitesimal Grassmann (anti-commuting) parameters. If we view these transformations as arising from the action of some operator $Q$, so that schematically one has $\delta X^\mu = [ \epsilon Q , X^\mu ]$ and likewise for $\delta \psi_{\pm}^\mu$, then this operator $Q$ is a fermionic symmetry generator of the form that we have discussed above. The existence of such a fermionic symmetry signals that this theory is supersymmetric, and the action of this supercharge $Q$ induces a pairing between the bosonic and fermionic degrees of freedom.

The supersymmetry of this theory is quite powerful. For instance, one learns in a first course on string theory that the bosonic string (which lacks supersymmetry) has an undesirable tachyonic state, whereas the supersymmetry of (\ref{RNS_susy}) removes the tachyon from the spectrum. But despite the usefulness of this supersymmetry, it has been presented in a way which is unsatisfying for a few reasons:

\begin{enumerate}

    \item By considering two successive supersymmetry transformations $\delta_1, \delta_2$ associated with parameters $\epsilon_{\pm}^{(1)}$, $\epsilon_{\pm}^{(2)}$, one can show that the commutators $[\delta_1, \delta_2] X^\mu$ and $[ \delta_1, \delta_2 ] \psi^\mu$ can each be expressed as a translation along the worldsheet plus a term which is proportional to one of the equations of motion. That is, the commutator of two supersymmetries is only equal to another symmetry (namely a translation) when the equations of motion are satisfied. For this reason, we say that the supersymmetry algebra only closes on-shell. It is desirable to have a different formulation of the supersymmetry transformations such that the algebra closes without using the equations of motion.
    
    \item\label{non_manifest} Although it is straightforward to verify that the transformations (\ref{RNS_susy}) leave the action (\ref{rns_action}) invariant, it is not clear how one would have guessed the form of this transformation \emph{a priori}. Said differently, the supersymmetry of (\ref{rns_action}) is not manifest.
    
    For other familiar symmetries, it is usually possible to combine the relevant fields into an object on which the symmetry action has a clear interpretation. For instance, in the example from earlier in this section, a theory of $n$ real scalar fields $\phi$ with an $O(n)$ symmetry is naturally interpreted by combining the scalars into a vector $\vec{\phi}$. In this case, $\vec{\phi}$ transforms in the vector representation of the symmetry group $O(n)$, so one has a clear intuitive picture of these symmetries as merely rotating the vector $\vec{\phi}$ in field space. If the dynamics only depend on the length $| \vec{\phi} |$ of this vector, then the theory is invariant under such rotations.
    
    In a sense, this re-packaging of the $n$ scalars $\phi_i$ into a vector $\vec{\phi}$ has ``geometrized'' the symmetry, by presenting a concrete geometric object (a vector) on which the symmetries act in a natural way and which makes it easy to check whether a Lagrangian respects that symmetry. In the same way, it is desirable to geometrize the supersymmetry transformations. We would like to package the bosonic fields $X^\mu$ and fermionic fields $\psi^\mu_{\pm}$ into an object, similar to a vector, on which the action of the supersymmetry generators is natural and which helps us to make supersymmetry manifest.
\end{enumerate}
We can remedy these two unsatisfying properties simultaneously by presenting the theory in \emph{superspace}.\footnote{It is also possible to fix the first problem but not the second, by introducing an auxiliary field but continuing to write the action in components. We will see this in (\ref{rns_components}).} Rather than considering a theory defined on the ordinary worldsheet manifold described by the coordinates $\sigma^\alpha$, we will define a theory on a \emph{supermanifold} which has two anti-commuting coordinates $\theta^+, \theta^-$, in addition to the two commuting coordinates $\tau, \sigma$. In particular, since the $\theta^A$ coordinates are anti-commuting, each of them squares to zero and therefore any Taylor series in these variables truncates. This means that a general function $\Phi ( \sigma^\alpha, \theta^A )$ can be expanded as
\begin{align}\label{superfield_example}
    \Phi ( \sigma^\alpha , \theta^A ) = \phi ( \sigma^\alpha ) + i \theta^+ \psi_+ ( \sigma^\alpha ) + i \theta^- \psi_- ( \sigma^\alpha ) + \theta^+ \theta^- f ( \sigma^\alpha ) \, ,
\end{align}
where $\phi, \psi_{\pm}$, and $f$ are ordinary fields which are functions only of the bosonic coordinates $\sigma^\alpha$. A function of the form (\ref{superfield_example}) is called a \emph{superfield}, and can be thought of as a convenient container which packages a collection of ordinary fields.

The advantage of this formalism is that we may use an explicit presentation of the supercharges for this theory, namely
\begin{align}
    Q_{\pm} = \frac{\partial}{\partial \theta^{\pm}} - \theta^{\pm} \partial_{\pm \pm} \, .
\end{align}
Here $\frac{\partial}{\partial \theta^{\pm}}$ act as the usual partial derivatives with respect to the anti-commuting coordinates, so that $\frac{\partial}{\partial \theta^{\pm}} \theta^\pm = 1$ and $\frac{\partial}{\partial \theta^{\pm}} \theta^\mp = 0$. A supersymmetry transformation $\delta$ is implemented via
\begin{align}\label{susy_delta}
    \delta = \epsilon^+ Q_+ + \epsilon^- Q_- \, .
\end{align}
For example, the action of the supersymmetry transformations on the Grassmann coordinates $\theta^A$ is
\begin{align}
    \delta \theta^+ &= [ \epsilon^+ Q_+ + \epsilon^- Q_- , \theta^+ ] = \epsilon^+ \, , \nonumber \\
    \delta \theta^- &= [ \epsilon^+ Q_+ + \epsilon^- Q_- , \theta^- ]  = \epsilon^- \, .
\end{align}
This suggests a geometrical interpretation which we asked for in point \ref{non_manifest} above: the action of the supersymmetry transformations is now a \emph{translation} in the anti-commuting directions $\theta^{\pm}$! From this perspective, the statement that our theory is supersymmetric is merely the statement that the dynamics are invariant under shifts in superspace, just as familiar non-supersymmetric examples such as free theories are invariant under spacetime translations.

We can also study the action of the supersymmetry transformations on superfields. For concreteness, let us attempt to package the degrees of freedom of the RNS superstring (\ref{rns_action}) into superfields $Y^\mu$ as
\begin{align}\label{y_expansion}
    Y^\mu  (\sigma^\alpha, \theta^A ) = X^\mu ( \sigma^\alpha ) + i \theta^+ \psi_+^\mu ( \sigma^\alpha ) + i \theta^- \psi_-^\mu ( \sigma^\alpha ) + \theta^+ \theta^- f^\mu ( \sigma^\alpha ) \, .
\end{align}
We may act on $Y^\mu$ with the supersymmetry variation (\ref{susy_delta}) and then interpret the variation $\delta Y^\mu$ as arising from individual variations $\delta X^\mu$, $\delta \psi^\mu_{\pm}$, and $\delta f^\mu$ of the component fields. Doing this, one finds
\begin{align}\label{RNS_susy_superfield}
    \delta X^\mu &= i \left( \epsilon_+ \psi^\mu_- - \epsilon_- \psi^\mu_+ \right) \, , \nonumber\\
    \delta \psi_-^\mu &= - 2 \partial_{--} X^\mu \epsilon_+ + f^\mu \epsilon_+ \, , \nonumber \\
    \delta \psi^\mu_+ &= 2 \partial_{++} X^\mu \epsilon_- + f^\mu \epsilon_- \, , \nonumber \\
    \delta f^\mu &= - 2 i \epsilon_+ \partial_{--} \psi^\mu_+ - 2 i \epsilon_- \partial_{++} \psi^\mu_- \, .
\end{align}
For any theory which is invariant under superspace translations, the transformations (\ref{RNS_susy_superfield}) will be an \emph{exact} symmetry. This is in contrast to the component supersymmetry transformations (\ref{RNS_susy}) we saw above, which were only a symmetry of the theory (\ref{rns_action}) on-shell (i.e. when the equations of motion are satisfied). Therefore, one side effect of formulating our theory in a superspace is that we arrive at an off-shell version of the supersymmetry transformations. We note that this formulation (\ref{RNS_susy_superfield}) required the introduction of an auxiliary field $f^\mu$ which was not present in the original RNS worldsheet Lagrangian. As we will see shortly, when we write a superspace version of the Lagrangian (\ref{rns_action}), the equations of motion will force $f^\mu = 0$. Therefore, after imposing the equations of motion, our off-shell supersymmetry transformations (\ref{RNS_susy_superfield}) reduce to the on-shell version (\ref{RNS_susy}) that we saw earlier, as they should. The formulation with an auxiliary field is mathematically more palatable, although it is physically equivalent to the earlier component formulation.

We now turn to the question of writing a manifestly supersymmetric Lagrangian for the superfield $Y^\mu$. Just as the usual Lagrangian (\ref{rns_action}) is built using the spacetime derivatives $\partial^\alpha$, we would like to define the appropriate notion of derivative in superspace. It turns out that a convenient choice is
\begin{align}\label{supercovariant_deriv_defn}
    D_{\pm} = \frac{\partial}{\partial \theta^{\pm}} + \theta^{\pm} \partial_{\pm \pm} \, .
\end{align}
We refer to $D_{\pm}$ as the \emph{supercovariant derivative}. One can check by direct computation that
\begin{align}
    \{ D_A , Q_B \} = 0 \, ,
\end{align}
so that any supercharge acts on the supercovariant derivative of a superfield in the same way as it acts on the superfield itself. 

Now consider the Lagrangian
\begin{align}\label{susy_11_superspace_lag_example}
    S = \int d^2 \sigma \, d^2 \theta \, D_+ Y^\mu D_- Y_\mu \, .
\end{align}
Here the integral over a Grassmann coordinate is defined in the usual way, namely
\begin{align}\label{grassman_integration}
    \int d \theta \, \left( a + b \theta \right) = b \, ,
\end{align}
which satisfies linearity and translation invariance (this integral operator is, in fact, identical to the differential operator $\frac{\partial}{\partial \theta}$). 

The Lagrangian (\ref{susy_11_superspace_lag_example}) is invariant under the action of the supersymmetry generators $Q_{\pm}$ since these are the sum of two terms, one of which is a total Grassman derivative and the other of which is a total spacetime derivative. Therefore, so long as boundary terms can be neglected, $\delta S = 0$ and this theory is supersymmetric. The same argument holds for any Lagrangian which can be written as an integral over all of superspace.

By using the component expansion (\ref{y_expansion}) of the superfield $Y^\mu$, along with the rules (\ref{grassman_integration}) of Grassman integration, one can perform the superspace integral in (\ref{susy_11_superspace_lag_example}) explicitly to find
\begin{align}\label{rns_components}
    S \sim \int d^2 \sigma \left( \partial_\alpha X^\mu \partial^\alpha X_\mu - 2 i \left( \psi^\mu_- \partial_{++} \psi_{\mu -} + \psi_+^\mu \partial_{--} \psi_{\mu +} \right) - f^\mu f_\mu \right) \, .
\end{align}
We note that this action has the same form as (\ref{rns_action}), except for the inclusion of the auxiliary field $f^\mu$. Eliminating $f^\mu$ from the theory by imposing its equation of motion $f^\mu = 0$ recovers our original formulation of the theory and collapses the off-shell supersymmetry transformations (\ref{RNS_susy_superfield}) to their on-shell versions (\ref{RNS_susy}).

We will now mention, in passing, some extensions to theories with differing amounts of supersymmetry, i.e. a different collection of operators $Q_i$, $i = 1, \cdots, \mathcal{N}$. The example of the RNS superstring which we considered above is said to have $\mathcal{N} = (1, 1)$ SUSY, where this notation refers to the fact that we have one left-moving and one-right moving supersymmetry. The chirality of our supercharges in this theory is determined by the $+$ or $-$ indices on $Q_{\pm}$. Similarly, one could consider a theory with $(0,1)$ supersymmetry in which there exists an operator $Q_+$ but no operator $Q_-$, or a theory with $(1,0)$ supersymmetry which has a supercharge $Q_-$ with no accompanying $Q_+$.

One can also consider theories with more supercharges. Of particular interest are $\mathcal{N} = (2, 2)$ theories in two spacetime dimensions, which have a collection of four supercharges $Q_{\pm}, \overbar{Q}_{\pm}$. This example has a richer structure than the $(1,1)$ theory we have considered here. For example, in such a theory we may have constrained superfields which sit in some reduced representation of the supersymmetry algebra. One common example is a so-called \emph{chiral} superfield $\Phi$, which has the property that $\overbar{D}_{\pm} \Phi = 0$, where $D_{\pm}$ and $\overbar{D}_{\pm}$ are the appropriate supercovariant derivatives defined analogously to (\ref{supercovariant_deriv_defn}). 

We could then continue to increase the amount of supsymmetry from $4$ supercharges in a $(2,2)$ theory, to $8$ supercharges, to $16$ supercharges. As the number of supercharges increases, it becomes more difficult to find a useful superspace presentation which makes all of the supersymmetry manifest. There is also an upper bound on the number of supercharges in a theory. For a theory with ``rigid'' supersymmetry -- i.e. one where the supersymmetries are global rather than gauged, which requires that there is no dynamical gravity within the theory -- this maximal quantity is $16$ supersymmetries. For theories of supergravity, the upper bound is $32$ supersymmetries.

The reason for this bound is physical rather than mathematical. If we view the supersymmetry algebra as a purely mathematical object, one can construct a collection of supercharges $Q_\alpha$, for $i = 1 , \cdots , \mathcal{N}$, satisfying the required algebra for any $\mathcal{N}$. However, because the action of a supercharge on a field changes its spin, in a theory with $16$ supercharges one could repeatedly act with the various $Q_\alpha$ to eventually produce a spin-$2$ field, which must necessarily be the graviton. Therefore, in a theory without dynamical gravity, we can have $16$ supercharges at most. A theory of supergravity \emph{does} contain a dynamical graviton, and such a theory may have up to $32$ supercharges. However, by similar reasoning, in a theory with more than $32$ supercharges, we could repeatedly act with the $Q_\alpha$ and construct an interacting spin-$3$ field. There is a general theorem \cite{WEINBERG198059} that no consistent, interacting quantum field theory\footnote{This theorem is evaded in string theory, which has an infinite tower of particle excitations at all integer and half-integer spins.} can be constructed with fields that have spin greater than $2$. Schematically, the reason for this result is that higher spin fields must be coupled to conserved currents, and there are only a handful of conserved currents in quantum field theories: vector currents associated with internal symmetries (which couple to gauge fields like the photon); the stress-energy tensor $T_{\mu \nu}$ which couples to the graviton; and the spin-$\frac{3}{2}$ supercurrent. The absence of such conserved currents beyond the stress tensor forbids us from introducing particles with spin greater than $2$.

This concludes our brief and rather superficial introduction to supersymmetry. For a more complete treatment, we refer the reader to one of the many textbooks on supersymmetry and supergravity \cite{Wess:1992cp,Buchbinder:1998qv,Gates:1983nr,freedman_van_proeyen_2012}, or to the pedagogical set of lecture notes \cite{clay_notes}.

\section{Conventions}

We now outline the various conventions that will be used in the later chapters of this thesis.

When we consider two-dimensional field theories in Lorentzian signature with coordinates $(x^0, x^1)$, it will be convenient to change coordinates to light-cone variables, defining
\begin{align}\label{lc_coordinates}
    x^{\pm \pm} = \frac{1}{\sqrt{2}} \left( x^0 \pm x^1 \right) .
\end{align}
Here we have used the bi-spinor conventions, where coordinates and vector quantities are written with a pair of $\pm$ indices ($x^{\pm \pm}$) rather than a single index ($x^{\pm})$.

The derivatives corresponding to the coordinates (\ref{lc_coordinates}) are written
\begin{align}
    \partial_{\pm \pm} = \frac{1}{\sqrt{2}}(\partial_0 \pm \partial_1)
\end{align}
In these conventions, we have $\partial_{\pm \pm} x^{\pm \pm} = 1$ and $\partial_{\pm \pm} x^{\mp \mp} = 0$.

Spinors in two dimensions carry a single index which is raised or lowered as follows:
\begin{align}
    \psi^+ = - \psi_-, \qquad   \psi^- = \psi_+ . 
\end{align}
The advantage of writing all vector indices as pairs of spinor indices is that it allows us to more easily compare terms in equations which involve a combination of spinor, vector, spinor-vector, and tensor quantities. For instance, in this notation the derivatives with respect to light-cone coordinates carry two indices $\partial_{\pm \pm}$, whereas spinor quantities $\psi_+$ carry only a single index, which allows us to distinguish between spin-$\frac{1}{2}$ and spin-$1$ objects. Similarly, the supercurrent has components $S_{+++}, S_{---}, S_{+--}$, and $S_{-++}$, which we can immediately identify as a spinor-vector because it has three indices. The stress-energy tensor carries two vector indices so its components will be written as $T_{++++}, T_{----}, T_{++--} = T_{--++}$.

\subsubsection*{\uline{\it{(1, 1) and (0, 1) Superspace}}}

When we consider $(1,1)$ supersymmetric theories, we will introduce anti-commuting coordinates $\theta^{\pm}$. The corresponding supercovariant derivatives are given in our conventions by
\begin{align}
    D_{\pm} = \frac{\partial}{\partial \theta^{\pm}} + \theta^{\pm} \partial_{\pm \pm} ,
    \label{D-def}
\end{align}
which satisfy $D_{\pm} D_{\pm} = \partial_{\pm \pm}$ and $\left\{ D_+ , D_- \right\} = 0$. There are also two supercharges $Q_{\pm}$,
\begin{align}
    Q_{\pm} = \frac{\partial}{\partial \theta^{\pm}} - \theta^{\pm} \partial_{\pm \pm},
\end{align}
which satisfy $Q_{\pm} Q_{\pm} = - \partial_{\pm \pm}$. The Lagrangian for a field theory is written as an integral over this $(1,1)$ superspace,
\begin{align}
    \mathcal{L} = \int d^2 \theta \, \mathcal{A} ( \Phi ) \, , 
\end{align}
where $\mathcal{A}$ is the superspace Lagrangian, $\Phi$ represents some collection of $(1, 1)$ superfields, and $d^2 \theta = d \theta^- \, d \theta^+$.

We will briefly describe theories in $(0, 1)$ superspace by truncating the conventions for $(1, 1)$ superspace described above. Such a superspace has a single anti-commuting coordinate $\theta^+$ along with the associated supercovariant derivative $D_+$ and supercharge $Q_+$. The Lagrangian for a $(0, 1)$ theory can be written as an integral
\begin{align}
    \mathcal{L} = \int d \theta^+ \, \mathcal{A}_+ ( \Phi ) \, , 
\end{align}
where now the superspace Lagrangian $\mathcal{A}_+$ must carry spin to ensure that the bosonic Lagrangian density $\mathcal{L}$ is a Lorentz scalar.

\subsubsection*{\uline{\it{(2, 2) Superspace}}}

When we study two-dimensional theories with $4 = 2 + 2$ supercharges, the four anti-commuting coordinates will be written $\theta^{\pm}$ and $\thetab^{\pm}$. It will sometimes be convenient to collectively denote the superspace coordinates by $\z^M=(x^\mu,\,\q^\pm,\,\qb^\pm)$. 
In Chapter \ref{CHP:SC-squared-2}, we will use the supercovariant derivatives, collectively denoted by $D_A=(\pa_a,\,D_\pm,\Db_\pm)$, given by
\begin{align}
    D_{\pm} = \frac{\partial}{\partial \theta^{\pm}} - \frac{i}{2} \thetab^{\pm} \partial_{\pm \pm} 
    ~, ~~~~~~
    \Db_{\pm} = - \frac{\partial}{\partial \thetab^{\pm}} + \frac{i}{2} \theta^{\pm} \partial_{\pm \pm} ~, 
\end{align}
and which satisfy
\begin{align}
    \{ D_{\pm} , \Db_{\pm} \} &= i \partial_{\pm \pm} , 
\end{align}
with all other (anti-)commutators vanishing.

The supersymmetry transformations for an $\cN=(2,2)$ superfield  $\cF({\z})=\cF(x^{\pm\pm},\q^\pm,\qb^\pm)$  are given by
\bea
\d_Q \cF
:=
i\epsilon^+ \cQ_+ \cF
+i\epsilon^- \cQ_- \cF
-i\bar\epsilon^+ \overline{\cQ}_+ \cF
-i\bar\epsilon^- \overline{\cQ}_- \cF
~,
\label{susySuperfield22}
\eea
where the action of supercharges on superfields is represented by the differential operators
\begin{align}
    \cQ_{\pm} = \frac{\partial}{\partial \theta^{\pm}} + \frac{i}{2} \thetab^{\pm} \partial_{\pm \pm} 
    ~, ~~~~~~
    \overline{\cQ}_{\pm} = - \frac{\partial}{\partial \thetab^{\pm}} - \frac{i}{2} \theta^{\pm} \partial_{\pm \pm} ~, 
\end{align}
satisfying
\begin{align}
    \{ \cQ_{\pm} , \overline{\cQ}_{\pm} \} &= -i \partial_{\pm \pm} 
    ~, 
\end{align}
and commuting with the covariant derivatives $D_A$.

In Chapter \ref{CHP:nonlinear}, to more easily facilitate comparison between $\mathcal{N} = (2,2)$ in $2D$ and $\mathcal{N} = 1$ theories in $4d$, we will use a slightly different convention for the supercovariant derivative which differs only by relative constant factors. There we will use supercovariant derivatives
\begin{align}\label{different_(2,2)_conventions}
    D'_\pm = \frac{\p} {\p \theta^\pm}+i \bar \theta^\pm \p_{\pm\pm} \, , \qquad \Db'_\pm = -\frac{\p} {\p\bar  \theta^\pm}  -  i  \theta^\pm \p_{\pm\pm} \, , 
\end{align}
which satisfy the rescaled algebra $\{ D'_\pm , \Db'_\pm \}=-2i \p_{\pm\pm}$.

When we write integrals over $(2, 2)$ superspace, we use the notation $d^2\q = d\q^- d\q^+$, $d^2 \bar\q = d\bar\q^+d \bar\q^-$ and  $d^4\q = d^2\q d^2\qb$.

\subsubsection*{\uline{\it{Four Dimensional Theories}}}

When we study theories in four spacetime dimensions, we will mostly follow the conventions of Wess and Bagger \cite{Wess:1992cp}. The $D=4$, $\cN=2$ superspace is parametrised by bosonic coordinates $x^\mu$ and anti-commuting coordinates $(\theta^\a,\,\bar{\theta}^\ad)$, $(\tilde \theta^\a,\,\bar{\tilde \theta}^\ad)$. To discuss chirality constraints in these theories, it is convenient to introduce the coordinates
\begin{align}
    y^\mu=x^\mu+ i \ta \sigma ^\mu \tab + i \tilde \ta \sigma ^\mu \bar{\tilde\theta}
\end{align}
In terms of the $y^\mu$, the supercovariant derivatives are given by
\be
D_\alpha =\frac{\p}{\p \ta^\alpha}+2i \sigma^\mu_{\alpha \dot \alpha}\tab^{\dot \alpha} \frac{\p}{\p y^\mu},\qquad \bar D_{\dot\alpha} =-\frac{\p}{\p \bar\ta^{\dot\alpha}}
~,
\ee
and similarly for $\tilde D_\alpha,  \bar{ \tilde{D}}_{\ad}$.

Our conventions differ from those of \cite{Wess:1992cp} in the conversion between vector and bi-spinor indices. We will use the normalization
\begin{align}
    v_{\alpha\dot \alpha}=-2\sigma^\mu_{\alpha\dot \alpha} v_\mu,  \; 
   v_\mu =\frac14 \bar \sigma^{\alpha\dot \alpha}v_{\alpha\dot \alpha} \, .
\end{align}
In these conventions, for instance, one has
 \be
\cJ_{\alpha\dot \alpha}=-2 \sigma^\mu_{\alpha\dot \alpha} \cJ_\mu, \qquad
\cJ^\mu = \frac{1}{4} \cJ_{\alpha \dot \alpha} \bar \sigma^\mu{}^{\dot \alpha \alpha}, \qquad
\cJ^2\equiv \eta^{\mu\nu} J_\mu J_\nu=-\frac18  \epsilon^{\alpha \beta}\epsilon^{\dot \alpha \dot \beta}
 \cJ_{\alpha\dot \alpha}J_{\beta\dot \beta}
~.
\ee

\chapter{\texorpdfstring{$(1, 1)$}{(1, 1)} and \texorpdfstring{$(0, 1)$}{(0, 1)} Supercurrent-Squared Deformations} \label{CHP:SC-squared-1}
In this chapter, we propose a solvable, irrelevant deformation that is built from bilinears in currents and which manifestly preserves supersymmetry. Just as the remarkable properties of the $\TT$ deformation follows from continuity equations, in the supersymmetric case, we will describe analogous relations based on the conservation laws in superspace. This chapter is primarily based on the paper ``Supersymmetry and $\TT$ Deformations'' \cite{Chang:2018dge}, except for Section \ref{sec:on-shell} which is new.

\section{\texorpdfstring{$T \Tb$}{TTbar} and Supersymmetry}\label{section:tt_and_susy}

\subsection{Supercurrent-squared}\label{section:supercurrent-squared}

Because the usual $T \Tb$ deformation discussed in section (\ref{TTb-review}) is built from the Noether current for spatial translations, we will generalize this construction by writing a manifestly supersymmetric Noether current associated with translations in superspace. For concreteness, we work in the $(1,1)$ theory, but a similar calculation in $(0,1)$ will be described in section \ref{section:(0,1)}.

Consider a supersymmetric Lagrangian which is written as an integral over $(1,1)$ superspace as $\mathcal{L} = \int d^2 \theta \, \mathcal{A}$. We allow $\mathcal{A}$ to depend on a superfield $\Phi$ and a particular set of its derivatives listed below:
\be \mathcal{A} = \mathcal{A}\left( \Phi, D_+ \Phi, D_- \Phi, \partial_{++} \Phi, \partial_{--} \Phi, D_+ D_- \Phi \right).
\ee 
The supercovariant derivatives $D_{\pm}$ are defined in (\ref{D-def}). The superspace equation of motion associated with this Lagrangian is
\begin{align}\label{superspace_eom_general}
	\frac{\delta \mathcal{A}}{\delta \Phi} & =  D_+ \left( \frac{\delta \mathcal{A}}{\delta D_+ \Phi} \right) + D_- \left( \frac{\delta \mathcal{A}}{\delta D_- \Phi} \right) + \partial_{++} \left( \frac{\delta \mathcal{A}}{\delta \partial_{++} \Phi} \right) \cr & + \partial_{--} \left( \frac{\delta \mathcal{A}}{\delta \partial_{--} \Phi} \right) - D_+ D_- \left( \frac{\delta \mathcal{A}}{\delta D_+ D_- \Phi} \right) . 
\end{align}
As in the derivation of the usual stress tensor $T$, we now consider a spatial translation of the form $\delta x^{\pm \pm} = a^{\pm \pm}$ for some constant $a^{\pm \pm}$. The variation $\delta \mathcal{A}$ of the superspace Lagrangian is given by
\begin{equation}\label{general_superspace_variation} 
\begin{split}
	\delta \mathcal{A} &= D_+ \left( \delta \Phi \frac{\delta \mathcal{A}}{\delta D_+ \Phi} \right) + D_- \left( \delta \Phi \frac{\delta \mathcal{A}}{\delta D_- \Phi} \right) + \partial_{++} \left( \delta \Phi \frac{\delta \mathcal{A}}{\delta \partial_{++} \Phi} \right) \cr
    &+ \partial_{--} \left( \delta \Phi \frac{\delta \mathcal{A}}{\delta \partial_{--} \Phi} \right) + \frac{1}{2} \left( D_+ \left( \frac{\delta \mathcal{A}}{\delta D_+ D_- \Phi} D_- \delta \Phi \right) + D_- \left( \delta \Phi D_+ \frac{\delta \mathcal{A}}{\delta D_+ D_- \Phi} \right) \right) \cr
    &- \frac{1}{2} \left( D_- \left( \frac{\delta \mathcal{A}}{\delta D_+ D_- \Phi} D_+ \delta \Phi \right) + D_+ \left( \delta \Phi D_- \frac{\delta \mathcal{A}}{\delta D_+ D_- \Phi} \right)  \right) \cr
    &- \delta \Phi \left( - \frac{\delta \mathcal{A}}{\delta \Phi} + D_+ \frac{\delta \mathcal{A}}{\delta D_+ \Phi} + D_- \frac{\delta \mathcal{A}}{\delta D_- \Phi} + \partial_{++} \frac{\delta \mathcal{A}}{\delta \partial_{++} \Phi} + \partial_{--} \frac{\delta \mathcal{A}}{\delta \partial_{--} \Phi}  \right. \cr & \left. - D_+ D_- \frac{\delta \mathcal{A}}{\delta D_+ D_- \Phi} \right) . 
\end{split} 
\end{equation}
Here we have chosen to symmetrize the term involving $D_+ D_- \frac{\delta \mathcal{A}}{\delta D_+ D_- \Phi}$ using $\left\{ D_+ , D_- \right\} = 0$.

The last two lines of (\ref{general_superspace_variation}) are the superspace equation of motion; we now specialize to the case of on-shell variations, for which this last term vanishes. Further, the left side of (\ref{general_superspace_variation}) is $\delta \mathcal{A} = a^{++} \partial_{++} \mathcal{A} + a^{--} \partial_{--} \mathcal{A}$, which is a total derivative. We use $\partial_{\pm \pm} = D_{\pm} D_{\pm}$ to express (\ref{general_superspace_variation}) in the form
\begin{align}
\begin{split}
	0 &= a^{++} D_+ \Big[ \partial_{++} \Phi \frac{\delta \mathcal{A}}{\delta D_+ \Phi} + D_+ \left(  \partial_{++} \Phi \frac{\delta \mathcal{A}}{\delta \partial_{++} \Phi} \right) + \frac{1}{2} \frac{\delta \mathcal{A}}{\delta D_+ D_- \Phi} D_- \left( \partial_{++} \Phi  \right) \label{superspace-T-conservation} \\
	&\qquad \qquad \qquad - \frac{1}{2} \partial_{++} \Phi D_- \left( \frac{\delta \mathcal{A}}{\delta D_+ D_- \Phi} \right) - D_+ \mathcal{A} \Big]  \\
    &+ a^{++} D_- \Big[ \partial_{++} \Phi \frac{\delta \mathcal{A}}{\delta D_- \Phi} + D_- \left(  \partial_{++} \Phi  \frac{\delta \mathcal{A}}{\delta \partial_{--} \Phi} \right)  - \frac{1}{2} \frac{\delta \mathcal{A}}{\delta D_+ D_- \Phi} D_+ \left( \partial_{++} \Phi \right)  \\
    &\qquad \qquad \qquad + \frac{1}{2} \partial_{++} \Phi D_+ \left( \frac{\delta \mathcal{A}}{\delta D_+ D_- \Phi} \right) \Big]  \\
	&+ a^{--} D_+ \Big[ \partial_{--} \Phi \frac{\delta \mathcal{A}}{\delta D_+ \Phi} + D_+ \left( \partial_{--} \Phi \frac{\delta \mathcal{A}}{\delta \partial_{++} \Phi} \right) + \frac{1}{2} \frac{\delta \mathcal{A}}{\delta D_+ D_- \Phi} D_- \left( \partial_{--} \Phi \right)  \\
	&\qquad \qquad \qquad - \frac{1}{2} \partial_{--} \Phi D_- \left( \frac{\delta \mathcal{A}}{\delta D_+ D_- \Phi} \right) \Big]  \\
    &+ a^{--} D_- \Big[  \partial_{--} \Phi \frac{\delta \mathcal{A}}{\delta D_- \Phi} + D_- \left( \partial_{--} \Phi \frac{\delta \mathcal{A}}{\delta \partial_{--} \Phi} \right) - \frac{1}{2} \frac{\delta \mathcal{A}}{\delta D_+ D_- \Phi} D_+ \left( \partial_{--} \Phi \right)  \\
    &\qquad \qquad \qquad + \frac{1}{2} \partial_{--} \Phi D_+ \left( \frac{\delta \mathcal{A}}{\delta D_+ D_- \Phi} \right) - D_- \mathcal{A} \Big] .
\end{split}
\end{align}
This equation gives a conservation law for a superfield $\mathcal{T}$ which we define by
\begin{align}
	\mathcal{T}_{++-} &= \partial_{++} \Phi \frac{\delta \mathcal{A}}{\delta D_+ \Phi} + D_+ \left(  \partial_{++} \Phi \frac{\delta \mathcal{A}}{\delta \partial_{++} \Phi} \right) + \frac{1}{2} \frac{\delta \mathcal{A}}{\delta D_+ D_- \Phi} D_- \left( \partial_{++} \Phi  \right)  \nonumber \\
	&\quad - \frac{1}{2} \partial_{++} \Phi D_- \left( \frac{\delta \mathcal{A}}{\delta D_+ D_- \Phi} \right) - D_+ \mathcal{A}  , \nonumber \\
    \mathcal{T}_{+++} &= \partial_{++} \Phi \frac{\delta \mathcal{A}}{\delta D_- \Phi} + D_- \left(  \partial_{++} \Phi  \frac{\delta \mathcal{A}}{\delta \partial_{--} \Phi} \right)  - \frac{1}{2} \frac{\delta \mathcal{A}}{\delta D_+ D_- \Phi} D_+ \left( \partial_{++} \Phi \right) \nonumber \\
    &\quad + \frac{1}{2} \partial_{++} \Phi D_+ \left( \frac{\delta \mathcal{A}}{\delta D_+ D_- \Phi} \right) , \nonumber \\
    \mathcal{T}_{---} &= \partial_{--} \Phi \frac{\delta \mathcal{A}}{\delta D_+ \Phi} + D_+ \left( \partial_{--} \Phi \frac{\delta \mathcal{A}}{\delta \partial_{++} \Phi} \right) + \frac{1}{2} \frac{\delta \mathcal{A}}{\delta D_+ D_- \Phi} D_- \left( \partial_{--} \Phi \right) \label{final_ttbar_general} \\
    &\quad - \frac{1}{2} \partial_{--} \Phi D_- \left( \frac{\delta \mathcal{A}}{\delta D_+ D_- \Phi} \right) , \nonumber \\
    \mathcal{T}_{--+} &=  \partial_{--} \Phi \frac{\delta \mathcal{A}}{\delta D_- \Phi} + D_- \left( \partial_{--} \Phi \frac{\delta \mathcal{A}}{\delta \partial_{--} \Phi} \right) - \frac{1}{2} \frac{\delta \mathcal{A}}{\delta D_+ D_- \Phi} D_+ \left( \partial_{--} \Phi \right) \nonumber \\
    &\quad + \frac{1}{2} \partial_{--} \Phi D_+ \left( \frac{\delta \mathcal{A}}{\delta D_+ D_- \Phi} \right) - D_- \mathcal{A} \nonumber . 
\end{align}
Equation (\ref{superspace-T-conservation}) implies that, when the equations of motion are satisfied, $\mathcal{T}$ obeys the superspace conservation laws:
\begin{align}\label{equation:conservation in superspace}
    D_+ \mathcal{T}_{++-} + D_- \mathcal{T}_{+++} &= 0 , & D_+ \mathcal{T}_{---} + D_- \mathcal{T}_{--+} &= 0 .
\end{align}
We are now in a position to propose the supercurrent-squared deformation. Consider a one-parameter family of superspace Lagrangians labeled by $t$, which satisfy the ordinary differential equation
\begin{align}
	\frac{\partial}{\partial t} \mathcal{A}^{(t)} = \mathcal{T}_{+++}^{(t)} \mathcal{T}_{---}^{(t)} - \mathcal{T}_{--+}^{(t)} \mathcal{T}_{++-}^{(t)} ,
    \label{ttbar_general_flow}
\end{align}
where $\mathcal{T}^{(t)}$ is the supercurrent superfield (\ref{final_ttbar_general}) computed from the superspace Lagrangian $\mathcal{A}^{(t)}$. This uniquely defines the supercurrent-squared deformation of an initial Lagrangian $\mathcal{A}^{(0)}$ at finite deformation parameter $t$. We note that one can perform a similar Noether procedure in a $(0+1)$-dimensional theory and define an analogous deformation constructed from current bilinears in supersymmetric quantum mechanics \cite{us:susy}.

\subsection{Reduction to components for a free theory}\label{section:free_components}

To illutrate the relationship between the flow equation (\ref{ttbar_general_flow}) and the usual $T \Tb$ operator, let us explicitly compute the components of the supercurrent-squared deformation for a free $(1,1)$ superspace Lagrangian
\begin{align}
    \mathcal{A} = D_+ \Phi D_- \Phi ,
\end{align}
where $\Phi$ is a superfield with component expansion
\begin{align}
    \Phi = \phi + i \theta^+ \psi_+ + i \theta^- \psi_- + \theta^+ \theta^- f .
\end{align}
The entries of $\mathcal{T}$, defined by (\ref{final_ttbar_general}), for the free theory are
\begin{align}
\begin{split}
    \mathcal{T}_{++-} &= \partial_{++} \Phi D_- \Phi - D_+ \left( D_+ \Phi D_- \Phi \right) , \label{free_superspace} \\
    \mathcal{T}_{+++} &= - \partial_{++} \Phi D_+ \Phi ,  \\
    \mathcal{T}_{---} &= \partial_{--} \Phi D_- \Phi ,  \\
    \mathcal{T}_{--+} &= - \partial_{--} \Phi D_+ \Phi - D_- \left( D_+ \Phi D_- \Phi \right)  .
\end{split}
\end{align}
In components, (\ref{free_superspace}) is
\begin{align}
    \mathcal{T}_{++-} =& - i \psi_+ f + \theta^+ \left( - f \partial_{++} \phi + \psi_+ \partial_{++} \psi_- \right) + \theta^- \left( - f^2 - \psi_+ \partial_{--} \psi_+ \right) \nonumber \\
    & + i \theta^+ \theta^- \left( - \partial_{++} \phi \partial_{--} \psi_+ - \partial_{++} \psi_+ \partial_{--} \phi - f \partial_{++} \psi_- + \psi_- \partial_{++} f + \partial_{++} \left( \psi_+ \partial_{--} \phi - \psi_- f \right) \right) , \nonumber \\
    \mathcal{T}_{+++} =& - i \psi_+ \partial_{++} \phi - \theta^+ \left( \psi_+ \partial_{++} \psi_+ + \left( \partial_{++} \phi \right)^2 \right) - \theta^- \left( f \partial_{++} \phi + \psi_+ \partial_{++} \psi_- \right) \nonumber \\
    & - i \theta^+ \theta^- \left( 2 \partial_{++} \phi \partial_{++} \psi_- + \psi_+ \partial_{++} f - f \partial_{++} \psi_+ \right) , \nonumber \\
    \mathcal{T}_{---} =&  \, i \psi_- \partial_{--} \phi + \theta^+ \left( \psi_- \partial_{--} \psi_+ - f \partial_{--} \phi \right) + \theta^- \left( \psi_- \partial_{--} \psi_- + \left( \partial_{--} \phi \right)^2 \right) \nonumber \\
    & + i \theta^+ \theta^- \left( \psi_- \partial_{--} f - f \partial_{--} \psi_- - 2 \partial_{--} \phi \partial_{--} \psi_+ \right) , \label{free_superspace_components}  \\
    \mathcal{T}_{--+} =& - i\psi_- f + \theta^+ \left( f^2 + \psi_- \partial_{++} \psi_- \right) + \theta^- \left( -f \partial_{--} \phi - \psi_- \partial_{--} \psi_+ \right) \nonumber \\
    & + i \theta^+ \theta^- \left( -\partial_{--} \phi \partial_{++} \psi_- + f \partial_{--} \psi_+ - \partial_{--} \psi_- \partial_{++} \phi - \psi_+ \partial_{--} f + \partial_{--} \left( \psi_+ f + \psi_- \partial_{++} \phi \right) \right). \nonumber
\end{align}
To compare with the non-superspace $T \Tb$ deformation, we identify the components of the usual stress tensor $T$ for the theory of a free boson $\phi$ and fermions $\psi_{\pm}$ which one obtains by performing the integrals over $\theta^{\pm}$. In our conventions, these take the form:
\begin{align}
\begin{split}
    T_{++++} &= \left( \partial_{++} \phi \right)^2 + \psi_+ \partial_{++} \psi_+ , \\
    T_{----} &= \left( \partial_{--} \phi \right)^2 + \psi_- \partial_{--} \psi_- .
\end{split}
\end{align}
We will also drop terms involving the auxiliary field $f$, since in the bosonic part of the supercurrent-squared deformation, these terms vanish after integrating out $f$ using its equation of motion. Then the bilinears appearing in our flow equation (\ref{ttbar_general_flow}) are
\begin{align}
\begin{split}
    \mathcal{T}_{+++} \mathcal{T}_{---} &= \psi_+ \psi_- \partial_{++} \phi \partial_{--} \phi + i \theta^+ \left( \psi_+ \psi_- \partial_{++} \phi \partial_{--} \psi_+ - T_{++++} \psi_- \partial_{--} \phi \right) \label{sc-square-no-f} \\
    &\quad + i \theta^- \left( \psi_+ \partial_{++} \phi T_{----} + \psi_+ \psi_- \partial_{++} \psi_- \partial_{--} \phi \right) - \theta^+ \theta^- \big( T_{++++} T_{----} \\
    &\quad + 2 \partial_{++} \phi \partial_{--} \phi \left( \psi_+ \partial_{--} \psi_+ +  \psi_- \partial_{++} \psi_- \right) - \psi_- \partial_{++} \psi_- \psi_+ \partial_{--} \psi_+ \big) , \\
    \mathcal{T}_{++-} \mathcal{T}_{--+} &= - 2 \theta^+ \theta^- \left( \psi_+ \partial_{--} \psi_+ \psi_- \partial_{++} \psi_- \right) .
\end{split}
\end{align}
The superspace integral of the deformation $\mathcal{T}_{+++} \mathcal{T}_{---} + \mathcal{T}_{++-} \mathcal{T}_{--+}$ picks out the top component, which is
\begin{align}
\begin{split}
    \int &d^2 \theta \, \left( \mathcal{T}_{+++} \mathcal{T}_{---} + \mathcal{T}_{++-} \mathcal{T}_{--+} \right) = \label{sc-square-integrated} \\
    &- T_{++++} T_{----} - 2 \partial_{++} \phi \partial_{--} \phi \left( \psi_+ \partial_{--} \psi_+ + \psi_- \partial_{++} \psi_- \right) - \psi_- \partial_{++} \psi_- \psi_+ \partial_{--} \psi_+ .
\end{split}
\end{align}
We see that (\ref{sc-square-integrated}) contains the usual $T \Tb$ deformation, given in our bi-spinor notation by $-T_{++++} T_{----}$, along with extra terms which are all proportional to the fermion equations of motion, $\partial_{\pm \pm} \psi_{\mp} = 0$. These added terms vanish on-shell, which means that the supercurrent-squared deformation is on-shell equivalent to the usual bosonic $\TT$ deformation. In the next subsection, we will verify that these additional terms do not affect the energy levels of the theory deformed by supercurrent-squared, which means that the same inviscid Burgers' relation between the deformed and undeformed energy levels holds for our supersymmetric deformation as for the ordinary $\TT$ flow.

\subsection{Relationship with the $\mathcal{S}$-multiplet}

The $(1,1)$ superfield $\mathcal{T}$ contains the conserved stress-energy tensor $T_{\mu \nu}$ and the supercurrent $S_{\mu \alpha}$. Such current multiplets have received much attention in the literature; the first construction for four-dimensional theories was the FZ multiplet \cite{Ferrara:1974pz}, which was later shown to be a special case of the more general $\mathcal{S}$-multiplet \cite{dumitrescuSupercurrentsBraneCurrents2011a}.

For the two-dimensional theories we consider here, it is known that the $\mathcal{S}$-multiplet is the most general multiplet containing the stress tensor and supercurrent, subject to assumptions that the multiplet be indecomposable\footnote{A representation is \emph{indecomposable} if it is not isomorphic to a direct sum of representations. All irreducible representations are indecomposable, but not all indecomposable represenations are irreducible.} and contain no other operators with spin greater than one. Since our supercurrent superfield $\mathcal{T}$ satisfies these properties, it must be equivalent to the $\mathcal{S}$-multiplet. As we will show, the four superfields contained in $\mathcal{T}$ are identical to the four superfields of the $\mathcal{S}$-multiplet, up to terms which vanish on-shell and therefore do not affect the conservation equations for the currents.

The $\mathcal{S}$-multiplet is a reducible but indecomposable set of two superfields $\mathcal{S}$ and $\chi$ satisfying the constraints
\begin{align}
\begin{split}
    D_{\mp} \mathcal{S}_{\pm \pm \pm} &= D_{\pm} \chi_{\pm} , \label{s-multiplet-constraints} \\
    D_- \chi_+ &= D_+ \chi_- . 
\end{split}
\end{align}
In components, the $\mathcal{S}$-multiplet for $(1,1)$ theories contains the usual stress tensor $T_{\mu \nu}$, the supercurrent $S_{\mu \alpha}$, and a vector $Z_\mu$ which is associated with a scalar central charge:
\begin{align}
\begin{split}
    \mathcal{S}_{+++} &= S_{+++} + \theta^+ T_{++++} + \theta^- Z_{++} - \theta^+ \theta^- \partial_{++} S_{-++} , \label{s-multiplet} \\
    \mathcal{S}_{---} &= S_{---} + \theta^+ Z_{--} + \theta^- T_{----} + \theta^+ \theta^- \partial_{--} S_{+--} , \\
    \chi_+ &= S_{-++} + \theta^+ Z_{++} + \theta^- T_{++--} - \theta^+ \theta^- \partial_{++} S_{+--} , \\
    \chi_- &= S_{+--} + \theta^+ T_{++--} + \theta^- Z_{--} + \theta^+ \theta^- \partial_{--} S_{-++} . 
\end{split}
\end{align}
In terms of these component fields, the constraints (\ref{s-multiplet-constraints}) give conservation equations for the currents:
\begin{align}
\begin{split}
    \partial_{++} T_{----} + \partial_{--} T_{++--} &= 0 = \partial_{++} T_{--++} + \partial_{--} T_{++++} , \\
    \partial_{++} S_{+--} + \partial_{--} S_{+++} &= 0 = \partial_{++} S_{---} + \partial_{--} S_{-++} , \\
    \partial_{++} Z_{--} + \partial_{--} Z_{++} &= 0 .
\end{split}
\end{align}

We claim that the components (\ref{free_superspace_components}) of our superspace supercurrent are the same as those in the two superfields $\mathcal{S}$ and $\chi$ appearing in the $(1,1)$ $\mathcal{S}$-multiplet (\ref{s-multiplet}), up to signs and terms which vanish on-shell. In particular, after discarding terms which are proportional to the equations of motion, we find the identifications:
\begin{align}
\mathcal{S}_{\pm \pm \pm} = \mp \mathcal{T}_{\pm \pm \pm}, \qquad \chi_+ = \mathcal{T}_{++-}, \qquad \chi_- = - \mathcal{T}_{--+}. 
\end{align}
We will check this explicitly for the free theory, $\mathcal{A} = D_+ \Phi D_- \Phi$, for which we computed the components of $\mathcal{T}$ in section (\ref{section:free_components}). Writing only those terms that survive when the component equations of motion $f=0$, $\partial_{++} \psi_- = 0 = \partial_{--} \psi_+$, and $\partial_{++} \partial_{--} \phi = 0$ are all satisfied, (\ref{free_superspace_components}) becomes
\begin{align}
\begin{split}
    \mathcal{T}_{++-} &\overset{\text{on-shell}}{=} 0 , \label{free_superspace_on_shell} \\
    \mathcal{T}_{+++} &\overset{\text{on-shell}}{=} - i \psi_+ \partial_{++} \phi - \theta^+ \left( \psi_+ \partial_{++} \psi_+ + \left( \partial_{++} \phi \right)^2 \right) , \\
    \mathcal{T}_{---} &\overset{\text{on-shell}}{=} i \psi_- \partial_{--} \phi + \theta^- \left( \psi_- \partial_{--} \psi_- + \left( \partial_{--} \phi \right)^2 \right) , \\
    \mathcal{T}_{--+} &\overset{\text{on-shell}}{=} 0 .
\end{split}
\end{align}
For the free $(1,1)$ superfield considered here, the supercurrent is given in our conventions by
\begin{align}
\begin{split}
    S_{+++} &= \psi_+ \partial_{++} \phi , \\
    S_{---} &= \psi_- \partial_{--} \phi , \\
    S_{+--} &= 0 = S_{-++} ,
\end{split}
\end{align}
while the stress tensor components are as in (\ref{bos-T-free}). To find expressions for the scalar central charge current $Z_{\pm \pm}$, we use the supersymmetry algebra implied by the $\mathcal{S}$-multiplet constraints, which gives
\begin{align}
\begin{split}
    \left\{ Q_{\pm} , S_{\pm \pm \pm} \right\} &= T_{\pm \pm \pm \pm} , \label{S-algebra} \\
    \left\{ Q_{\pm} , S_{\pm \mp \mp} \right\} &= T_{\pm \pm \mp \mp} ,  \\
    \left\{ Q_{\pm} , S_{\mp \pm \pm} \right\} &= Z_{\pm \pm} ,  \\
    \left\{ Q_{\pm} , S_{\mp \mp \mp} \right\} &= Z_{\mp \mp} . 
\end{split}
\end{align}
Note that the $\mathcal{S}$-multiplet constraints only hold when the conservation equations for the currents hold, so the relations (\ref{S-algebra}) should be viewed as an on-shell algebra. Acting with the supercharges $Q_{\pm}$ on the stress tensor and supercurrent components, one finds that $Z_{--} \sim \psi_- \partial_{--} \psi_+$ and $Z_{++} \sim \psi_+ \partial_{++} \psi_-$, both of which vanish when the fermion equations of motion are satisfied.

Thus, after imposing the equations of motion, we can write our supercurrent superfield components as
\begin{align}\begin{split}
    \mathcal{T}_{++-} = \chi_+ = 0 \, , &\qquad \mathcal{T}_{--+} = 0 = - \chi_{-}, \nonumber \\
    \mathcal{T}_{+++} = - S_{+++} - \theta^+ T_{++++} = - \mathcal{S}_{+++} \, , &\qquad \mathcal{T}_{---} = S_{---} + \theta^- T_{----} = \mathcal{S}_{---} .
\end{split}\end{align}
Since terms which vanish on-shell do not affect conservation equations, one can view $\mathcal{T}$ as an improvement transformation of the $\mathcal{S}$-multiplet. The constraint equation $D_{\mp} \mathcal{S}_{\pm \pm \pm} - D_{\pm} \chi_{\pm} = 0$ is expressed by our conservation equations (\ref{equation:conservation in superspace}).

\subsection{Equivalence to $T \overline{T}$ for Spectrum}\label{section:solvable}
In this section we prove the theory deformed by (\ref{ttbar_general_flow}) satisfies the same flow equation for the finite-volume energy levels as the usual $\TT$ deformation. In this sense, the supercurrent-squared deformation has the same solvability properties as $\TT$.

We begin with the superspace conservation equation given in (\ref{equation:conservation in superspace}). It is straightforward to solve these constraints in components by using the conservation of the stress energy tensor:
\begin{align}
\begin{split}
    &\T_\ppp = H_\ppp - \tp T_\pppp - \tm W_\pp + \tp\tm G_\ppp, \\
    &\T_\mmm = H_\mmm + \tm T_\mmmm + \tp W_\mm - \tp\tm G_\mmm, \\
    &\T_\mmp = H_\mmp - \tp T_\mmpp - \tm W_\mm + \tp\tm G_\mmp, \\
    &\T_\ppm = H_\ppm + \tm T_\ppmm + \tp W_\pp - \tp\tm G_\ppm .
\end{split}
\end{align}
Here $(H_\www,H_{\mp\mp\pm})$ are the lowest components of $\T$ while $(G_\www,G_{\mp\mp\pm})$ are its highest components. The conservation law in (\ref{equation:conservation in superspace}) implies the following constraints on $G$ and $H$:
\begin{align}\label{equation:Conservation G H}
\begin{split}
    &G_{\mp\mp\pm} = \partial_{\pm\pm} H_{\mp\mp\mp} , \\
    &G_{\pm\pm\pm} = \partial_{\pm\pm} H_{\mp\pm\pm} .
\end{split}
\end{align}
In terms of these components, the deformation in (\ref{ttbar_general_flow}) becomes
\begin{align}\label{reduce_to_ttbar}
    \frac{\partial}{\partial t} \mathcal{L}^{(t)} =
    & -\int d^2\theta \lrb \T_\ppp\T_\mmm + \T_\ppm\T_\mmp \rrb \nonumber \\
    = & -\lrb T_\pppp T_\mmmm - T_\ppmm T_\mmpp \rrb \\
    &\qquad + \lrb H_\ppp G_\mmm - G_\ppp H_\mmm - H_\ppm G_\mmp + G_\ppm H_\mmp \rrb . \nonumber
\end{align}
The first term $-\lrb T_\pppp T_\mmmm - T_\ppmm T_\mmpp \rrb$ of (\ref{reduce_to_ttbar}) is the usual $T\bar T$ deformation. To understand how the second bracket changes the spectrum, we consider the two-point correlation function:
\begin{align}
    \mathcal{C} = \ld H_\ppp(x)G_\mmm(\xp) \rd 
    & - \ld G_\ppp(x)H_\mmm(\xp) \rd \\
    & - \ld H_\ppm(x)G_\mmp(\xp) \rd + \ld G_\ppm(x)H_\mmp(\xp) \rd  . \nonumber
\end{align}
Up to contact terms that vanish at separated points, we can replace $G$ by using the conservation equation (\ref{equation:Conservation G H}):
\begin{align}
    \mathcal{C} = \ld H_\ppp(x)\partial^\prime_\mm H_{\mmp}(\xp) \rd 
    &- \ld \partial_\pp H_\ppm(x)  H_\mmm(\xp) \rd \\
    &- \ld H_\ppm(x)\partial^\prime_\pp H_\mmm(\xp) \rd + \ld \partial_\mm H_\ppp(x)H_\mmp(\xp) \rd . \nonumber
\end{align}
Here $\partial^\prime$ means the derivative with respect to the coordinate $\xp$. Now we can use translational invariance to move the derivative from $\xp$ to $x$. Then the first term cancels the fourth term and the third term cancels the second one, because both $H$ and $G$ are fermionic, so that $\mathcal{C}$ vanishes at separated points. This implies that the extra term can have no effect on the spectrum -- the presence of this term is only to make the action supersymmetric. The theory remains solvable, like the usual $\TT$ deformation, with the same relation between deformed and undeformed energy levels.

\section{Theories with \texorpdfstring{$(1,1)$}{Lg} Supersymmetry}\label{section:(1,1)}

In this section, we consider the supercurrent-squared deformation of a theory involving a single $(1,1)$ superfield $\Phi$, both in the free case and with a superpotential.

\subsection{Free \texorpdfstring{$(1,1)$}{Lg} superfield}

First consider an undeformed superspace Lagrangian $\mathcal{A}^{(0)} = D_+ \Phi D_- \Phi$. We make the following ansatz for the deformed Lagrangian at finite $t$:
\begin{align}
    \mathcal{A}^{(t)} = F \left( \, t \partial_{++} \Phi \partial_{--} \Phi , \, t \left( D_+ D_- \Phi \right)^2 \right) D_+ \Phi D_- \Phi .
\end{align}
Here $F$ may only depend on the two dimensionless combinations which we define by 
\begin{align}
   x = t\, \partial_{++} \Phi \partial_{--} \Phi, \qquad  y = t \left( D_+ D_- \Phi \right)^2.
\end{align} In order to reduce to the free theory as $t \to 0$, we must also impose the boundary condition $F(0,0) = 1$.

After computing the components of the supercurrent-squared deformation and simplifying, the flow equation (\ref{ttbar_general_flow}) yields
\begin{align}
\begin{split}
    \frac{\partial}{\partial t} F &= \left( \left( D_+ D_- \Phi \right)^2 - \partial_{++} \Phi \partial_{--} \Phi \right) F^2 \\
    &\quad - 2 F \left( \partial_{++} \Phi \partial_{--} \Phi \right) \left( \partial_{++} \Phi \partial_{--} \Phi + \left( D_+ D_- \Phi \right)^2 \right) \frac{\partial F}{\partial x} .
    \label{free_pde}
\end{split}
\end{align}
In terms of the dimensionless variables $x$ and $y$, equation (\ref{free_pde}) becomes
\begin{align}
    \frac{\partial F}{\partial x} x + \frac{\partial F}{\partial y} y = (y-x)F^2 - 2 F \frac{\partial F}{\partial x} x ( x + y ) .
    \label{free_pde_nondim}    
\end{align}
Supplemented with the boundary condition $F(0,0) = 1$, the partial differential equation (\ref{free_pde}) uniquely determines the deformed Lagrangian at finite $t$.

As a check, we would like to verify that the bosonic structure of the solution to (\ref{free_pde}) reduces to the known results for the $T \Tb$-deformed theory of a free boson. We will argue that, in fact, it suffices to set $y = 0$ in (\ref{free_pde}) and note that the result agrees with the flow equation obtained in the purely bosonic case \cite{cavagliaBarTDeformed2D2016}.

Indeed, let us write the components of the superfield $\Phi$ as $\Phi = \phi + i \theta^+ \psi_+ + i \theta^- \psi_- + \theta^+ \theta^- f$. To probe the bosonic structure, it suffices to set $\psi_{\pm} = 0$, perform the superspace integration, and then integrate out the auxiliary field $f$ using its equation of motion. Thus consider an arbitrary superspace integral of the form
\begin{align}\label{arbitrary_superspace_integral_11}
    \mathcal{L}^{(t)} = \int d^2 \theta \, F^{(t)} (x, y) D_+ \Phi D_- \Phi .
\end{align}
The lowest component of the superfield $y = t ( D_+ D_- \Phi )^2$ is to $t f^2$, and the higher components will not contribute to the bosonic part because they come multiplying $D_+ \Phi D_- \Phi$, which is already proportional to $\theta^+ \theta^-$ after setting the fermions to zero.

Thus the purely bosonic piece of the physical Lagrangian associated with a superspace Lagrangian $\mathcal{A}^{(t)} = F^{(t)} ( x, y) D_+ \Phi D_- \Phi$ is
\begin{align}
    \mathcal{L}^{(t)} = F^{(t)} \left( t \partial_{++} \phi \partial_{--} \phi , t f^2 \right) \left( f^2 + 4 \partial_{++} \phi \partial_{--} \phi \right) .
\end{align}
The equation of motion for the auxiliary field $f$ is
\begin{align}
    2 t f \frac{\partial F}{\partial y} \left( f^2 + 4 \partial_{++} \phi \partial_{--} \phi \right) + 2 f F = 0 ,
\end{align}
which admits the solution $f=0$. The Lagrangian for the bosonic field $\phi$ is then
\begin{align}
    \mathcal{L}^{(t)} = 4 F^{(t)} \left( t \partial_{++} \phi \partial_{--} \phi , 0 \right) \partial_{++} \phi \partial_{--} \phi .
\end{align}
Therefore, to determine the terms in the Lagrangian which involve only $\phi$, we may solve the simpler partial differential equation
\begin{align}\label{equation:free_pde_y=0}
    \frac{\partial F}{\partial x} x = - x F^2 - 2 F x^2 \frac{\partial F}{\partial x},
\end{align}
which holds upon setting $y=0$ in (\ref{free_pde_nondim}). But this is precisely the equation discussed in section \ref{TTb-review}, whose solution is equation (\ref{equation:Nambu-Goto}):
\begin{align}
    \mathcal{L}^{(t)} = \frac{\sqrt{1 + 4 t \partial_{++} \phi \partial_{--} \phi} - 1}{2 t} . 
\end{align}
We see that the supercurrent-squared deformation of the free superfield is indeed a generalization of the $T \Tb$ deformation of a free boson, in the sense that it yields the same modification to the purely bosonic terms in the action but also includes additional terms which affect only the fermions.

\subsection{On-Shell Simplification of Flow Equation}\label{sec:on-shell}

In the previous subsection, we investigated the purely bosonic part of the physical Lagrangian along the supercurrent-squared flow, after imposing the equation of motion for the auxiliary field $f$. This allowed us to simplify the flow equation and present a closed-form solution for the Lagrangian of the boson $\phi$, which reproduces the known square-root result.

In fact, a similar procedure can be performed in superspace rather than in components. We will now demonstrate that, if we are willing to go ``partially on-shell'' in the superspace Lagrangian -- by imposing one implication of the superspace equations of motion, rather than the full equations -- then the superspace flow equation will also simplify in a similar way. We can then solve this simplified flow equation and write an exact superspace Lagrangian. Since we have imposed part of the equations of motion, the resulting expression is only equivalent to the true solution on-shell (i.e. up to field redefinitions), but if we are willing to accept this shortcoming, this procedure allows for an explicit solution along the flow.

We begin with the general superspace Lagrangian (\ref{arbitrary_superspace_integral_11}). The function $F^{(t)} ( x, y )$ can be expanded as a power series in $y$ to write
\begin{align}
    \label{y_taylor_exp}
    \mathcal{L}^{(t)} = \int d^2 \theta \, \left[ F^{(t)} ( x, 0 ) + F^{(t)}_y ( x, 0 ) y + \cdots \right] \, D_+ \Phi D_- \Phi \, .
\end{align}
On the other hand, the superspace equation of motion (\ref{superspace_eom_general}) for this Lagrangian is
\begin{align}\label{special_11_eom}
    0 &= D_+ \left[ F^{(t)} ( x, y ) \, D_- \Phi \right] - D_- \left[ F^{(t)} ( x, y ) \, D_- \Phi \right] + \partial_{++} \left[ t  \frac{\partial F^{(t)}}{\partial x} \partial_{--} \Phi D_+ \Phi D_- \Phi \right] \nonumber \\ &\qquad  + \partial_{--} \left[ t \frac{\partial F^{(t)}}{\partial x} \partial_{++} \Phi D_+ \Phi D_- \Phi \right] + D_+ D_- \left[ 2 t \frac{\partial F^{(t)}}{\partial y} \left( D_+ D_- \Phi \right) D_+ \Phi D_- \Phi \right] \, . 
\end{align}
Now suppose that we multiply both sides of (\ref{special_11_eom}) by the product $D_+ \Phi D_- \Phi$. Both $D_+ \Phi$ and $D_- \Phi$ are fermionic, so any term which was already proportional to either of those factors will vanish upon such a multiplication. The only terms which survive are
\begin{align}
    0 &= F^{(t)} ( x, y ) D_+ \Phi D_- \Phi \left( D_+ D_- \Phi \right) - 2 t D_+ \Phi D_- \Phi \frac{\partial F^{(t)}}{\partial y} \left( D_+ D_- \Phi \right)^3 \nonumber \\
    &= D_+ \Phi D_- \Phi \left( D_+ D_- \Phi \right) \cdot \left[ F^{(t)} (x, y) - 2 t \frac{\partial F^{(t)}}{\partial y} ( D_+ D_- \Phi )^2 \right] \, .
\end{align}
Thus the equations of motion require that either $D_+ \Phi D_- \Phi \left( D_+ D_- \Phi \right) = 0$ on-shell or $F^{(t)} (x, y) - 2 t \frac{\partial F^{(t)}}{\partial y} ( D_+ D_- \Phi )^2 = 0$ on-shell. But the latter equation cannot hold identically since, at least for $t$ small enough, it would require that $F^{(t)} ( x, y ) = 0$ and thus the action vanishes. Since $y \sim ( D_+ D_- \Phi)^2$, we conclude that
\begin{align}\label{y_eom_result}
    y D_+ \Phi D_- \Phi = 0 
\end{align}
on-shell. This means that, when the equations of motion are satisfied, we may replace the Taylor expansion (\ref{y_taylor_exp}) with
\begin{align}
    \mathcal{L}^{(t)} = \int d^2 \theta \, F^{(t)} ( x, 0 ) \, D_+ \Phi D_- \Phi \, ,
\end{align}
or said differently, we may simply set $y = 0$ whenever it appears in a superspace expression multiplying $D_+ \Phi D_- \Phi$. In particular, since the supercurrent-squared flow equation (\ref{free_pde_nondim}) appears inside such a superspace integral, we may set $y = 0$ there to find
\begin{align}
    x \frac{\partial F}{\partial x} + x F^2 + 2 x^2 F \frac{\partial F}{\partial x} = 0 \, . 
\end{align}
The general solution to this differential equation, as we have seen, is
\begin{align}
    F(x) = \frac{1}{2x} \left( -1 + \sqrt{ 1 + 4 c_1 x } \right) \, , 
\end{align}
and the initial condition $F(0) = 1$ fixes $c_1 = 1$. Thus we obtain a solution to the flow equation
\begin{gather}
    \mathcal{L}^{(t)} = \int d^2 \theta \, \frac{1}{2x} \left( -1 + \sqrt{ 1 + 4 x } \right) D_+ \Phi D_- \Phi \, , \nonumber \\
    x = t \partial_{++} \Phi \partial_{--} \Phi \, ,
\end{gather}
which is on-shell equivalent to the true solution to the flow equation (involving both $x$ and $y$). The advantage of this approach is that it allows us to see the square-root structure of the solution to the supercurrent-squared flow directly in superspace. We will see that a similar on-shell analysis is possible in the case of the $(2,2)$ supercurrent-squared deformation, where the analysis leading up to equation (\ref{y_eom_result}) is replaced by the calculation in Appendix (\ref{appendix:on-shell}).

\subsection{Interacting \texorpdfstring{$(1,1)$}{Lg} superfield}

Next, we consider the case with a superpotential: that is, we begin from the undeformed superspace Lagrangian
\begin{align}
    \mathcal{A}^{(0)} = D_+ \Phi D_- \Phi + h ( \Phi ) , 
\end{align}
where $h(\Phi)$ is an arbitrary function (it need not give rise to a theory with infinitely many integrals of motion). After performing the superspace integral, the physical Lagrangian is
\begin{align}
    \mathcal{L}^{(0)} = \int d^2 \theta \, \mathcal{A}^{(0)} = \partial_{++} \phi \partial_{--} \phi + \psi_+ \partial_{--} \psi_+ + \psi_- \partial_{++} \psi_- + f^2 + h'(\phi) f .
\end{align}
Integrating out the auxiliary field using its equation of motion $f = - \frac{1}{2} h'(\phi)$, we see that the physical potential $V$ is given by $V = - \frac{1}{4} h'(\phi)^2$.

We might expect that both the kinetic and potential terms are modified by a finite supercurrent-squared deformation, which would lead us to make the ansatz
\begin{align}
    \mathcal{A}^{(t)} = F(x, y) D_+ \Phi D_- \Phi + G(t, \Phi) ,
    \label{potential_ansatz}
\end{align}
where $G$ is a new function to be determined, and $x = t \partial_{++} \Phi \partial_{--} \Phi$, $y = t \left( D_+ D_- \Phi \right)^2$ as above. However, the deformation does not induce any change in the potential $h$, so in fact we may put $G = h$ for all $t$. To see this, we can write down the supercurrent-squared deformation associated with the ansatz (\ref{potential_ansatz}), which gives
\begin{align}
	& \frac{\partial}{\partial t} F ( x, y ) D_+ \Phi D_- \Phi + \frac{\partial}{\partial t} G (t,\Phi) = \label{interacting_deformation} \\
	&\frac{1}{t} \Bigg( \left( y - x \right) F^2 - 2 F x ( x + y ) \frac{\partial F}{\partial x} + \left( G' \right)^2 + 2 G' \sqrt{y} \left( x \frac{\partial F}{\partial x} - F \right) - 2 \sqrt{y} x G' \frac{\partial F}{\partial y} \Bigg) D_+ \Phi D_- \Phi . \non
\end{align}
The details of the calculation leading to (\ref{interacting_deformation}) are discussed in Appendix \ref{11_appendix}. We see that deformation is proportional to $D_+ \Phi D_- \Phi$, so it does not source any change in the potential $h(\Phi)$; thus we may take $G(h, \Phi) = h(\Phi)$ in our ansatz. This leaves us with a single partial differential equation for $F$, namely
\begin{align}
	 x \frac{\partial F}{\partial x} + y \frac{\partial F}{\partial y} &= \left( y - x \right) F^2 - 2 F x ( x + y ) \frac{\partial F}{\partial x} + \left( h' \right)^2 + 2 h' \sqrt{y} \left( x \frac{\partial F}{\partial x} - F \right) - 2 \sqrt{y} x h' \frac{\partial F}{\partial y} .
    \label{potential_pde}
\end{align}	
In the second line, we have used the constraint that $F$ can depend only on the dimensionless combinations $x = t \partial_{++} \Phi \partial_{--} \Phi$ and $y = t \left( D_+ D_- \Phi \right)^2$.

As in the free case, we would like to study the purely bosonic terms in the physical Lagrangian resulting from (\ref{potential_pde}) and compare them to known results. Here the auxiliary will play a more important role since $f=0$ is no longer a solution. 

We can expand both the Lagrangian $\mathcal{L} = \int d^2 \theta \left( F(x,y) D_+ \Phi D_- \Phi + h ( \Phi ) \right)$ and the auxiliary field $f$ as power series in $t$:
\begin{align}
    \mathcal{L} = \sum_{j=0}^{\infty} t^j \mathcal{L}^{(j)}, \qquad f = \sum_{j = 0}^{\infty} t^j f^{(j)} , 
\end{align}
and then integrate out the auxiliary order-by-order in $t$. Doing so to order $t^3$, we arrive at 
\begin{align}
\begin{split}
    \mathcal{L}  &= -\frac{1}{4} h'(\phi )^2 + \frac{x}{t} +t \left(\frac{1}{16} h'(\phi )^4 - \left( \frac{x}{t} \right)^2\right) +t^2 \left(-\frac{1}{4} \left( \frac{x}{t} \right)^2 h'(\phi )^2-\frac{1}{64} h'(\phi )^6+2 \left( \frac{x}{t} \right)^3\right) \label{our_result_expanded} \\
    &\quad + t^3 \left(\left( \frac{x}{t} \right)^3 h'(\phi )^2+\frac{1}{256} h'(\phi )^8-5 \left( \frac{x}{t} \right)^4\right) + \mathcal{O} ( t^4 ), 
\end{split}
\end{align}
after setting the fermions to zero. Up to conventions, this matches the Taylor expansion of the known result \cite{cavagliaBarTDeformed2D2016,bonelliBarDeformationsClosed2018} for the $T \Tb$ deformation of a boson with a generic potential $V$, which is given in our conventions as
\begin{align}
    \mathcal{L}^{(t)} = -\frac{1}{2t} \frac{1-2 t V}{1-t V} + \frac{1}{2t} \sqrt{\frac{ t \left( 4 V + \partial_{++} \phi \partial_{--} \phi \right) }{1-t V}+\frac{(1-2 t V)^2}{(1-t V)^2}} .
    \label{closed_form_pot}
\end{align}
Again the physical potential $V$ is related to $h$ via $V = - \frac{1}{4} h'(\phi)^2$. We have checked explicitly that the bosonic part of the series solution to the PDE (\ref{potential_pde}) matches the Taylor expansion of (\ref{closed_form_pot}) up to $\mathcal{O} ( t^7 )$.

\section{Theories with \texorpdfstring{$(0,1)$}{Lg} Supersymmetry}\label{section:(0,1)}

In this section we study the deformation of a theory with chiral $(0,1)$ supersymmetry; a $(0,1)$ scalar superfield $\Phi$ consists of a scalar and a real fermion, $\Phi = \phi + i \tp \psi_+$. The Lagrangian in superspace is a function of $D_+ \Phi$, $\partial_\pp\Phi$, $\partial_\mm\Phi$, as well as $\Phi$ itself.

\subsection{Free $(0,1)$ superfield}

The free theory is defined by the Lagrangian,
\begin{align}
\begin{split}
    \mathcal{L} &= \int d\tp\ D_+\Phi\partial_\mm\Phi, \label{equation:(0,1) free} \\
    &= \partial_{++} \phi \partial_{--} \phi + \psi_+ \partial_{--} \psi_+ .
\end{split}
\end{align}
Following the approach of section \ref{section:supercurrent-squared}, we first look for conservation laws for a given superspace Lagrangian $\mathcal{A}$. They take the form, 
\begin{align}
\begin{split}
	& \partial_\mm \mathcal{S}_\ppp + D_+ \T_\ppmm = 0, \\
	& \partial_\mm \mathcal{S}_\mmp + D_+ \T_\mmmm = 0,
\end{split}
\end{align}
where $\mathcal{S}_{\pm\pm+}$ and $\T_{\ww\mm}$ are superfields given by:
\begin{align}\label{equation:(0,1) supercurrent}
\begin{split}
    & \mathcal{S}_\ppp = \frac{\delta\mathcal{A}}{\delta\partial_\mm \Phi}\partial_\pp \Phi, \\
    & \mathcal{S}_\mmp = \frac{\delta\mathcal{A}}{\delta\partial_\mm \Phi}\partial_\mm \Phi - \mathcal{A}, \\
    & \T_\ppmm = \frac{\delta\mathcal{A}}{\delta D_+ \Phi} \partial_\pp \Phi + D_+ \left( \frac{\delta\mathcal{A}}{\delta\partial_\pp \Phi} \partial_\pp \Phi \right) -D_+ \mathcal{A}, \\
    & \T_\mmmm = \frac{\delta\mathcal{A}}{\delta D_+ \Phi} \partial_\mm \Phi + D_+ \left( \frac{\delta\mathcal{A}}{\delta\partial_\pp \Phi} \partial_\mm  \Phi \right).
\end{split}
\end{align}
We define the supercurrent-squared deformation as follows: 
\begin{align}\label{equation:(0,1) flow equation}
    \frac{\partial}{\partial t} \mathcal{A}^{(t)} = \mathcal{S}_\ppp \T_\mmmm - \mathcal{S}_\mmp\T_\ppmm .
\end{align}

To understand what the deformation (\ref{equation:(0,1) flow equation}) does to a $(0,1)$ theory, consider an undeformed Lagrangian in superspace
\begin{align}
    \mathcal{A}^{(0)} = g(\Phi) D_+\Phi\partial_\mm\Phi, 
\end{align}
where $g(\Phi)$ is an arbitrary differentiable function of the superfield. A free theory corresponds to a constant $g(\Phi)$. To find the deformed theory $\mathcal{A}^{(t)}$, we first make a general ansatz for the deformed  Lagrangian 
\begin{align}
    \mathcal{A}^{(t)} = f(t \partial_\pp\Phi\partial_\mm\Phi)D_+\Phi\partial_\mm\Phi,
\end{align} 
where $f(x)$ is some differentiable function. Using the expression for the supercurrents given in (\ref{equation:(0,1) supercurrent}) and imposing the initial condition $f(x\rightarrow 0) = g(\Phi)$, we find the function $f(x)$ satisfies the same differential equation found in (\ref{equation:free_pde_y=0}). Its solution is given by
\begin{align}
    f(x) = \frac{\sqrt{1+4x g(\Phi)}-1}{2x}.
\end{align}

\subsection{Reduction of $(1,1)$ to $(0,1)$}

Any theory with $(1,1)$ global supersymmetry can also be viewed as a theory with $(0,1)$ global supersymmetry. Up to possible field redefinitions, we should therefore be able to relate the $(1,1)$ theory deformed by the supercurrent-squared deformation defined in (\ref{final_ttbar_general}) to the $(0,1)$  of (\ref{equation:(0,1) flow equation}), which we would have used if we had simply restricted to $(0,1)$ supersymmetry. 

To be more precise, consider a $(1,1)$ theory whose physical Lagrangian $\mathcal{L}$ is given by the integral of a superspace Lagrangian $\mathcal{A}^{(1,1)}$ over $(1,1)$ superspace. We can also view this as a $(0,1)$ theory, 
\begin{align}
    \mathcal{L} = \int d^2 \theta \, \mathcal{A}^{(1,1)} = \int d \theta^+ \, \mathcal{A}^{(0,1)} . 
    \label{11-to-01}
\end{align}
The flow equation defining the supercurrent-squared deformation of $\mathcal{A}^{(1,1)}$ is $\partial_{t} \mathcal{A}^{(1,1)} = \mathcal{T}_{+++} \mathcal{T}_{---} - \mathcal{T}_{--+} \mathcal{T}_{++-}$. By performing the integral over $\theta^-$, this induces a flow for $\mathcal{A}^{(0,1)}$ due to (\ref{11-to-01}), namely
\begin{align}
    \frac{\partial}{\partial t} \mathcal{A}^{(0,1)} &= \int d \theta^+ \, \left( \mathcal{T}_{+++} \mathcal{T}_{---} - \mathcal{T}_{--+} \mathcal{T}_{++-} \right) .
    \label{reduced_flow}
\end{align}
For instance, let us consider the deformation of the free theory $\mathcal{A}^{(1,1)} = D_+ \Phi^{(1,1)} D_- \Phi^{(1,1)}$. This can be written as an integral over $(0,1)$ superspace as
\begin{align}
\begin{split}
    \int d^2 \theta \, D_+ \Phi^{(1,1)} D_- \Phi^{(1,1)} &= \int d \theta^+ \Big( - i \psi_+ \partial_{--} \phi - i \psi_- f - \theta^+ \big( f^2 + \partial_{++} \phi \partial_{--} \phi \label{11-reduced} \\
    &\hspace{70pt}  + \psi_+ \partial_{--} \psi_+ + \psi_- \partial_{++} \psi_- \big) \Big) \\
    &= - \int d \theta^+ \, \left(D_+ \Phi^{(0,1)} \partial_{--} \Phi^{(0,1)} + \Psi_- D_+ \Psi_- \right)  .
\end{split}
\end{align}
Here we have written the integrand on the right side of (\ref{11-reduced}) as a superspace Lagrangian $\mathcal{A}^{(0,1)} \left( \Phi^{(0,1)}, \Psi_- \right)$ for a superfield $\Phi^{(0,1)} = \phi + i \theta^+ \psi_+$ of the form discussed above, along with an extra Fermi superfield $\Psi_- = i \psi_- + \theta^+ f$:
\begin{align}
    \mathcal{A}^{(0,1)} \left( \Phi^{(0,1)}, \Psi_- \right) = D_+ \Phi^{(0,1)} \partial_{--} \Phi^{(0,1)} + \Psi_- D_+ \Psi_- .
\end{align}

For comparison, we compute the supercurrent-squared deformation to leading order in $t$; that is, we compute the tangent vector $\frac{\partial \mathcal{A}^{(1,1)}}{\partial t} \vert_{t=0}$ to the free theory along the flow and compare it to that of the free $(0,1)$ theory with an extra fermion.

The components of the supercurrent superfield associated with the free theory, after integrating out the auxiliary using $f=0$, are given in equation (\ref{sc-square-no-f}). Using these and performing the integral over $\theta^-$, the reduced flow equation (\ref{reduced_flow}) at $t=0$ becomes
\begin{align}
\begin{split}
    \frac{\partial}{\partial t} \mathcal{A}^{(0,1)} \vert_{t=0} &=  i \left( \psi_+ \partial_{++} \phi T_{----} + \psi_+ \psi_- \partial_{++} \psi_- \partial_{--} \phi \right) + \theta^+ \Big( T_{++++} T_{----} \label{bos-T-free} \\
    &\quad + 2 \partial_{++} \phi \partial_{--} \phi \left( \psi_+ \partial_{--} \psi_+  +  \psi_- \partial_{++} \psi_- \right) + \psi_- \partial_{++} \psi_- \psi_+ \partial_{--} \psi_+ \Big), 
\end{split}
\end{align}
where we have used $T_{\pm \pm \pm \pm} = \left( \partial_{\pm \pm} \phi \right)^2 + \psi_\pm \partial_{\pm \pm} \psi_\pm$.

We know that the solution to (\ref{bos-T-free}) must represent a solvable deformation of the original $(0,1)$ theory because it descends from a solvable deformation in the parent $(1,1)$ theory. On the other hand, one can construct the flow equation (\ref{equation:(0,1) flow equation}) directly in the $(0,1)$ theory. This must also yield a solvable deformation since it is built out of currents which satisfy a superspace conservation equation of the same form as in the $(1, 1)$ supercurrent multiplet. One might suspect that these two deformations should be the same, up to field redefinitions which do not affect the spectrum. To check this, let us compare the leading-order deformations for these two cases in components. After including the contributions $\partial_{\pm \pm} \Psi_- \frac{\delta \mathcal{A}}{\delta D_{\pm} \Psi_-} $ to $\mathcal{T}_{\pm \pm - -}$ due to the fermion $\Psi_-$, the currents (\ref{equation:(0,1) supercurrent}) for this theory are
\begin{align}
\begin{split}
    \mathcal{S}_\ppp &= D_+ \Phi \partial_{++} \Phi \\
    &= i \psi_+ \partial_{++} \phi + \theta^+ \left( \psi_+ \partial_{++} \psi_+ + \left( \partial_{++} \phi \right)^2 \right) , \\
    \mathcal{S}_\mmp &= - \Psi_- D_+ \Psi_- \\
    &= - i \psi_- f - \theta^+ \left( f^2 + \psi_- \partial_{++} \psi_- \right) , \\
    \T_\ppmm &= \partial_{--} \Phi \partial_{++} \Phi + \Psi_- \partial_{++} \Psi_- - D_+ \left( D_+ \Phi \partial_{--} \Phi + \Psi_- D_+ \Psi_- \right)  \\
    &=  - \psi_+ \partial_{--} \psi_+ - f^2 + i \theta^+ \left( \partial_{++} \phi \partial_{--} \psi_+ - \psi_+ \partial_{++} \partial_{--} \phi \right), \\
    \T_\mmmm &= \left( \partial_{--} \Phi \right)^2 - \Psi_- \partial_{--} \Psi_- \\
    &= \left( \partial_{--} \phi \right)^2 + \psi_- \partial_{--} \psi_- + i \theta^+ \left( - f \partial_{--} \psi_- + \psi_- \partial_{--} f + 2 \partial_{--} \phi \partial_{--} \psi_+ \right) .
\end{split}
\end{align}
The bilinears appearing in the $(0,1)$ deformation are
\begin{align}
\begin{split}
    \mathcal{S}_{+++} \mathcal{T}_{----} &= i \psi_+ \partial_{++} \phi T_{----} + \theta^+ \big( T_{++++} T_{----} \\
    &\hspace{70pt} + \psi_+ \partial_{++} \phi \left( \psi_- \partial_{--} f - f \partial_{--} \psi_- + 2 \partial_{--} \phi \partial_{--} \psi_+ \right) \big) , \\
    \mathcal{S}_{--+} \mathcal{T}_{++--} &= i \psi_- f \left( \psi_+ \partial_{--} \psi_+ + f^2 \right) + \theta^+ \big( \left( f^2 + \psi_- \partial_{++} \psi_- \right) \left( f^2 + \psi_+ \partial_{--} \psi_+ \right) \\
    &\hspace{70pt} + \psi_- f \left( \psi_+ \partial_{++} \partial_{--} \phi - \partial_{++} \phi \partial_{--} \psi_+ \right) \big),
\end{split}
\end{align}
and thus the $(0,1)$ flow equation at $t=0$ is given by
\begin{align}
    \frac{\partial}{\partial t} \mathcal{A}^{(0,1)} \vert_{t=0} &= \mathcal{S}_{+++} \mathcal{T}_{----} - \mathcal{S}_{--+} \mathcal{T}_{++--} \label{01_free_flow} \cr
    &= i \psi_+ \partial_{++} \phi T_{----} - i \psi_- f \left( \psi_+ \partial_{--} \psi_+ + f^2 \right) \\
    &+ \theta^+ \Big( T_{++++} T_{----} + \psi_+ \partial_{++} \phi \left( \psi_- \partial_{--} f - f \partial_{--} \psi_- + 2 \partial_{--} \phi \partial_{--} \psi_+ \right) \cr
    &- \left( f^2 + \psi_- \partial_{++} \psi_- \right) \left( f^2 + \psi_+ \partial_{--} \psi_+ \right)  - \psi_- f \left( \psi_+ \partial_{++} \partial_{--} \phi - \partial_{++} \phi \partial_{--} \psi_+ \right) \Big). \non
\end{align}
The deformations (\ref{bos-T-free}) and (\ref{01_free_flow}) agree up to terms proportional to the equations of motion $f=0$ and $\partial_{++} \psi_- = 0$. At this order in $t$, such terms can be removed by making a field redefinition involving $f$ and $\psi_-$. If this can be repeated order-by-order in $t$, as we suspect should be the case, then the two flows are genuinely equivalent and give rise to deformed theories with the same energies.

\chapter{\texorpdfstring{$(2, 2)$}{(2, 2)} Supercurrent-Squared Deformations} \label{CHP:SC-squared-2}
In this chapter, we will extend the supercurrent-squared deformation of Chapter \ref{CHP:SC-squared-1} to theories with more supersymmetry, focusing on the case of $(2,2)$ theories. Our discussion is based on the paper ``$\TT$ Flows and $(2,2)$ Supersymmetry'' \cite{Chang:2019kiu}.

\section{\texorpdfstring{$D=2 \;\, \cN=(2,2)$}{2D N=(2,2)} Supercurrent Multiplets} \label{sec:supercurrents}

As in the case of $(1,1)$ theories and $(0,1)$ theories considered in the previous chapter, the manifestly supersymmetric construction of $T \Tb$ for $(2,2)$ theories is built from bilinears in fields of the supercurrent multiplet. In this section we review the structure of such multiplets in a general $D=2$ theory with $\mathcal{N} = (2,2)$ supersymmetry.

\subsection{The \texorpdfstring{$\cS$}{S}-multiplet}
\label{section-S-multiplet}

For Lorentz invariant supersymmetric theories, there is an essentially unique 
supermultiplet which contains the stress-energy tensor $T_{\mu \nu}$, the supercurrent $S_{\mu \alpha}$, and no other operators 
with spin larger than one, under the assumption that the multiplet be indecomposable~\cite{Dumitrescu:2011iu}. That is, though in general such a multiplet is not an irreducible representation of the supersymmetry algebra, it cannot be separated into a direct sum of supersymmetry multiplets. By ``essentially unique,'' we mean that this $\cS$-multiplet 
is unique up to improvement terms which preserve the superspace constraint equations. The $\cS$-multiplet can be defined in any theory with $D=2$ 
$\cN=(2,2)$ supersymmetry. 

For two-dimensional theories with $(2,2)$ supersymmetry, the $\cS$-multiplet consists of superfields 
$\mathcal{S}_{\pm \pm}$, $\chi_{\pm}$, and $\mathcal{Y}_{\pm}$ which satisfy the constraints:
\bsubeq
\label{conservation-S}
\begin{gather}
    \Db_{\pm} \mathcal{S}_{\mp \mp} 
    = \pm \left( \chi_{\mp} + \mathcal{Y}_{\mp} \right)~ , \\
    \Db_{\pm} \chi_{\pm} = 0~ , 
    \qquad 
    \Db_{\pm} \chi_{\mp} = \pm C^{(\pm)} ~, 
    \qquad 
    D_+ \chi_- - \Db_- \chib_+ = k ~, 
    \\
    D_{\pm} \mathcal{Y}_{\pm} = 0~ ,
    \qquad 
    \Db_{\pm} \mathcal{Y}_{\mp} = \mp C^{(\pm)}~, 
    \qquad 
    D_+ \mathcal{Y}_- + D_- \mathcal{Y}_+ = k' ~.
\end{gather}
\esubeq
Here $k$ and $k'$ are real constants and $C^{(\pm)}$ is a complex constant.
The~$\cS$-multiplet contains~$8+8$ independent real component operators and the constants~$k, k', C^{(\pm)}$ 
\cite{Dumitrescu:2011iu}.
The expansion in components of $\cS_{\pm\pm}$,
$\chi_\pm$, and $\calY_{\pm}$ are given 
for convenience in Appendix
\ref{components-S}.

Among the various component fields 
it is important to single out 
the complex supersymmetry current $S_{\a}{}_\mu$
 and the energy-momentum tensor $T_{\mu\nu}$.
 The complex supersymmetry current, associated to $S_{+\pm\pm}$ and $S_{-\pm\pm}$, is conserved: $\pa^\mu S_{\a}{}_\mu=0$. 
 The energy-momentum tensor,
 associated with $T_{\pm\pm\pm\pm}$ and $T_{++--}=T_{--++}$, is real, conserved ($\pa^\mu T_{\mu\nu}=0$), 
 and symmetric ($T_{\mu\nu}=T_{\nu\mu}$). 
 In light-cone notation the conservation equations are given by
\bsubeq
\label{conserT}
\bea
 \pa_{++} S_{+--}(x) &=&\, - \pa_{--} S_{+++}(x)\,,\\
 \pa_{++} \bar{S}_{+--}(x) &=&\, - \pa_{--} \bar{S}_{+++}(x)\,,\\
 \pa_{++} T_{----}(x) &=&\, -\pa_{--} \Theta(x) \,,\\
 \pa_{++} \Theta(x) &=&\, - \pa_{--} T_{++++}(x) \,,
\eea
\esubeq
where we have defined as usual 
\bea
\Theta(x):=T_{++--}(x)=T_{--++}(x)
~.
\eea

To conclude this subsection, let us describe the ambiguity in the form of the $\cS$-multiplet which is parametrized by a choice of 
improvement term. 
If $\cU$ is a real superfield, we are free to modify the $\cS$-multiplet superfields as follows
\bsubeq    \label{S_mult_improvement}
\bea
    \cS_{\pm \pm} &\to \cS_{\pm \pm} + [ D_{\pm} , \Db_{\pm} ] \cU~ , \\
    \chi_{\pm} &\to \chi_{\pm} - \Db_+ \Db_- D_{\pm} \cU ~, \\
    \mathcal{Y}_{\pm} &\to \mathcal{Y}_{\pm} - D_{\pm} \Db_+ \Db_- \cU ~,
\eea
\esubeq
which keeps invariant the conservation equations \eqref{conservation-S}.
In general the $\cS$-multiplet is a reducible representation of supersymmetry and  some of its component can consistently be set 
to zero by a choice of improvement.
The reduced Ferrara-Zumino supercurrent multiplet, which plays a central role in this chapter, is described next.

\subsection{The Ferrara-Zumino (FZ) multiplet and old-minimal supergravity}

If there exists a well-defined superfield $\cU$ such that $\chi_{\pm} = \Db_+ \Db_- D_{\pm} \cU$,  then we may 
use the transformation (\ref{S_mult_improvement}) to set $\chi_{\pm} = 0$ in the $\cS$-multiplet.  If in addition $k =  C^{(\pm)}=0$, 
then we will rename $\mathcal{S}_{\pm \pm}$ to $\mathcal{J}_{\pm \pm}$; then the fields $\mathcal{J}_{\pm \pm}$ and $\mathcal{Y}_{\pm}$ satisfy the  constraints
\bsubeq    \label{FZ_constraint}
\bea
    \Db_{\pm} \mathcal{J}_{\mp \mp} &=& \pm \mathcal{Y}_{\mp}~ , \\
    D_{\pm} \mathcal{Y}_{\pm} &=& 0 ~, \\
    \Db_{\pm} \mathcal{Y}_{\mp} &=& 0~ , \\
    D_+ \mathcal{Y}_- + D_- \mathcal{Y}_+ &= &k' ~. 
\eea
\esubeq
These are the defining equations for the Ferrara-Zumino (FZ) multiplet 
$(\mathcal{J}_{\pm \pm} , \mathcal{Y}_{\pm})$. In this case, $\mathcal{J}_{\pm \pm}$ turns out to be associated to the axial $U(1)_A$ $R$-symmetry current and satisfies the conservation equation
\be
\pa_{--}{\cal J}_{++}
-\pa_{++}{\cal J}_{--}
=0
~.
\ee
This multiplet, which has $4+4$ real components,
 is the dimensionally-reduced version of the $D=4$ $\cN=1$ FZ-multiplet \cite{Ferrara:1974pz}; 
 see Appendix \ref{components-S}\ for more details. All of the models we consider in section \ref{sec:models} have the property that 
 $\chi_{\pm}$ can be improved to zero;  that is, they all have a well-defined FZ-multiplet. 

In Chapter \ref{CHP:SC-squared-1}, we obtained the components of the supercurrent superfield by finding the Noether currents associated with translations in superspace. One could use a similar Noether procedure in the case $(2, 2)$ superspace, as is done for $4 D \, \mathcal{N} = 1$ in \cite{Magro:2001aj}. However, we will find it more convenient to avoid the Noether procedure in this chapter and instead use the supersymmetrized version of the Hilbert definition of the stress tensor. Just as the bosonic Hilbert stress tensor $T_{\mu \nu}$ represents the response function of the Lagrangian to a linearized perturbation $h_{\mu \nu}$ of the metric, the supercurrent multiplets correspond to linearized couplings to supergravity. Therefore, we can compute the components of the supercurrent by coupling our $(2, 2)$ theories to one of the formulations of $2D$ supergravity, which we now briefly review.

Different formulations of off-shell supergravity couple to different supercurrent multiplets. If a theory has a well-defined FZ-multiplet, as is the case for all the examples found in section \ref{sec:models}, then the theory can be consistently coupled to the old-minimal supergravity prepotentials $H^{\pm \pm}$ and $\sigma$. The nomenclature ``old-minimal''  is again inherited from $D=4$ $\cN=1$ supergravity; see \cite{Gates:1983nr,Buchbinder:1998qv} for pedagogical reviews and references. Here $H^{\pm \pm}$ is the conformal supergravity prepotential---the analogue of the traceless part of the metric---and $\sigma$ is a chiral conformal compensator. 

We refer the reader to \cite{Grisaru:1994dm, Grisaru:1995dr, Grisaru:1995kn, Grisaru:1995py, Gates:1995du} 
and references therein for an exhaustive description of  $D=2$ $\,\cN=(2,2)$ off-shell supergravity in superspace, which we will use in 
our analysis; see also Appendix \ref{appendix:supercurrent_calculation}. For the scope of this work, it will be enough to know the 
structure of linearized old-minimal supergravity. For instance, at the linearized level the gauge symmetry of the supergravity 
prepotentials $H^{\pm\pm},\,\s$ and $\bar{\s}$, can be parameterized as follows
\bsubeq\label{old_minimal_gauge}
\bea
    \delta H^{++} &=& \frac{i}{2} \left( \Db_- L^+ - D_- \Lb^+ \right) ~,\\
    \delta H^{--} &= &\frac{i}{2} \left( \Db_+ L^- - D_+ \Lb^- \right) ~,\\
    \delta \sigma &=& - \frac{i}{2} \Db_+\Db_- \left( D_+ L^+ - D_- L^- \right) ~,\\
    \delta \bar{\sigma} &=& - \frac{i}{2} D_-D_+ \left( \Db_+ \Lb^+ - \Db_- \Lb^- \right) 
    ~,
\eea
\esubeq
in terms of  unconstrained spinor superfields $L^{\pm}$ and their complex conjugates. 

The conservation law (\ref{FZ_constraint}) for the FZ-multiplet can be derived by using the previous gauge transformations.
The linearized supergravity couplings for a given model are written as
\begin{align}
    \mathcal{L}_{\text{linear}} &= \int d^4 \theta \, \left( H^{++} \mathcal{J}_{++} + H^{--} \mathcal{J}_{--} \right)
    - \int d^2 \theta \, \sigma \, \mathcal{V}
        - \int d^2 \bar{\theta} \, \bar{\sigma} \, \overline{\mathcal{V}}
~,
\label{linearized_couplings_D_term}
\end{align}
with $\cV$ a chiral superfield and $\cVb$ its complex conjugate. Assuming the matter superfields satisfy their equations of motion, 
the change in the Lagrangian \eqref{linearized_couplings_D_term}  under the gauge transformation (\ref{old_minimal_gauge}) is
\begin{align}
    \delta \mathcal{L}_{\text{linear}} &= \int d^4 \theta \, \left( \delta H^{++} \mathcal{J}_{++} 
    + \delta H^{--} \mathcal{J}_{--} 
\right)
  - \int d^2 \theta \, \d\sigma \, \mathcal{V}
        - \int d^2 \bar{\theta} \, \d\bar{\sigma} \, \overline{\mathcal{V}}
     \nonumber \\
    &= \frac{i}{2} \int d^4 \theta \, \Big\{ \left( \Db_- L^+ - D_- \Lb^+ \right) \mathcal{J}_{++} 
    + \left( \Db_+ L^- - D_+ \Lb^- \right) \mathcal{J}_{--} \nonumber \\
    &\qquad~~~~~~~~~~~~
     -  \left( D_+ L^+ - D_- L^- \right) \mathcal{V} 
    -\left( \Db_+ \Lb^+ - \Db_- \Lb^- \right) \overline{\mathcal{V}} \Big\} \nonumber \\
    &= \frac{i}{2} \int d^4 \theta \, \Big\{ L^+ \left( \Db_- \mathcal{J}_{++} +  D_+ \mathcal{V}  \right) 
    + L^- \left( \Db_+ \mathcal{J}_{--} -  D_- \mathcal{V}  \right) + \text{c.c.} \Big\} ~, 
\end{align}
where we have integrated by parts. Demanding that the variation vanishes for any gauge parameter $L^{\pm}$ gives
\be
    \Db_{-} \mathcal{J}_{++} + D_+ \mathcal{V} = 0 ~,~~~~~~
    \Db_+ \mathcal{J}_{--} - D_- \mathcal{V}  = 0 
    ~.
    \label{FZ-2_(2,2)}
\ee
This matches the constraints (\ref{FZ_constraint}) for the FZ-multiplet if we identify
\begin{align}
    \mathcal{Y}_{\pm} = D_{\pm} \mathcal{V} 
         ~,
\end{align}
and set $k'= 0$.

As we will soon see, studying $T \Tb$ deformations requires consideration of a 
composite operator constructed out of the square of the supercurrent multiplet. 
Hence to solve the $T \Tb$ flow equations we need to be able to calculate the supercurrent multiplet explicitly. 
The coupling to supergravity provides a straightforward prescription for computing the FZ-multiplet for matter models 
that can be coupled to old-minimal supergravity.%
\footnote{Though we will not need it in this thesis, it is worth mentioning that the non-minimal supergravity results of 
\cite{Grisaru:1994dm,Grisaru:1995dr,Grisaru:1995kn,Grisaru:1995py,Gates:1995du} allow the computation of the supercurrent 
multiplet for more general classes of models.} 
In particular, for a given $\cN=(2,2)$ matter theory we will:
\begin{enumerate}

    \item Begin with an undeformed superspace Lagrangian  $\mathcal{L}$ in flat $\cN=(2,2)$ superspace.
    
    \item Minimally couple $\mathcal{L}$ to the supergravity superfield prepotentials $H^{\pm\pm}$, $\sigma$ and  $\bar\sigma$.
    
    \item Extract the superfields $\mathcal{J}^{\pm\pm}$, $\mathcal{V}$ and  $\overline{\mathcal{V}}$
     which couple linearly to $H^{\pm\pm}$,
    $\sigma$ and  $\bar\sigma$, respectively, in the D- 
    and F-terms of (\ref{linearized_couplings_D_term}).
    
\end{enumerate}
Thanks to the analysis given above, the superfields $\mathcal{J}^{\pm\pm}$, $\mathcal{V}$ and  $\overline{\mathcal{V}}$ 
will automatically satisfy the FZ-multiplet constraints \eqref{FZ-2_(2,2)}.
A detailed description of the computation of the FZ-multiplet for the models relevant for our discussion is given in Appendix 
\ref{appendix:supercurrent_calculation}.


\section{The \texorpdfstring{$T \Tb$}{T Tbar} Operator and \texorpdfstring{$\cN = (2,2)$}{N=(2,2)} Supersymmetry}
 \label{secTTbar}

After having reviewed in the previous section the structure of the  $\cS$-multiplet, we are 
ready to describe $\cN=(2,2)$ $T \Tb$ deformations.

\subsection{The \texorpdfstring{$\calT \calTb$}{super T Tbar} operator}

Given a $D=2$ $\cN=(2,2)$ supersymmetric theory with an $\cS$-multiplet, 
we define the supercurrent-squared deformation of this theory, denoted $\calT \calTb$ in analogy with $T \Tb$, by the flow equation
\begin{align}
    \partial_\lambda \mathcal{L} &= - \frac{1}{8} \calT \calTb 
    ~,
\end{align}
where $\calT \calTb$ is constructed from current bilinears with
\begin{align}
     \calT \calTb &\equiv 
     -\int d^4 \theta \, \mathcal{S}_{++} \mathcal{S}_{--}
    - \left( \int d \theta^-  d \theta^+ \, \chi_+ \chi_- + \int d \thetab^-  d \theta^+ \, \calYb_+ \calY_- + \text{c.c.} \right) ~,
    \label{TTbar-S}
\end{align}
and where the factor of $\frac{1}{8}$ is chosen for later convenience.
This deformation generalizes the results we recently obtained for $D=2$ theories possessing $\cN=(0,1)$, $\cN=(1,1)$ and 
$\cN=(0,2)$ supersymmetry \cite{Chang:2018dge, Baggio:2018rpv, Jiang:2019hux} to theories with  $\cN=(2,2)$ supersymmetry.

Let us recall the form of the $T \Tb$ composite operator \cite{Zamolodchikov:2004ce}, which we denote
\bea
T \Tb(x)=T_{++++}(x)\,T_{----}(x)-\big[\Theta(x)\big]^2
~.
\label{component-TTbar-22}
\eea
An important property of the $\cN=(0,1)$, $\cN=(1,1)$ and $\cN=(0,2)$ cases is that the $T \Tb$ operator turns out to be
 the bottom component of a long supersymmetric multiplet. This is true up to both total vector derivatives ($\pa_{++}$ and $\pa_{--}$), 
 and terms that vanish upon using the supercurrent conservation equations (Ward identities). For this reason, in the supersymmetric 
 cases studied previously, the original $T \Tb$ deformation of \cite{Zamolodchikov:2004ce} is manifestly supersymmetric and 
 equivalent to the deformations constructed in terms of the full superspace integrals of primary supercurrent-squared composite 
 operators \cite{Chang:2018dge, Baggio:2018rpv, Jiang:2019hux}. 

Remarkably, despite the much more involved structure of the $(2,2)$ $\cS$-multiplet compared to theories with fewer 
supersymmetries, it is possible to prove that the following relation holds:
\begin{align}
\calT\calTb(x)
=
T \Tb(x)
+{\rm EOM's}
+\pa_{++}(\cdots)
+\pa_{--}(\cdots)
~.
\label{calTTb=TTb_(2,2)}
\end{align}
In~\eqref{calTTb=TTb_(2,2)}, we  use ${\rm EOM's}$ to denote terms that are identically zero when \eqref{conservation-S} are used.
Showing~\eqref{calTTb=TTb_(2,2)} requires using  \eqref{comp-S}--\eqref{comp-Y}, along with several cancellations, integration by parts 
and the use of the $(2,2)$ $\cS$-multiplet conservation equations
\eqref{conservation-S}. 

In fact, the specific combination of current superfields given in \eqref{TTbar-S} was chosen precisely for \eqref{calTTb=TTb_(2,2)} to hold. 
The combination \eqref{calTTb=TTb_(2,2)} is also singled out by being invariant under the improvement transformation 
\eqref{S_mult_improvement}.
The important implication of  \eqref{calTTb=TTb_(2,2)} is that the $T \Tb$ deformation for an $\cN=(2,2)$ supersymmetric quantum field 
theory is manifestly supersymmetric and equivalent to the $\calT \calTb$ deformation of eq.~\eqref{TTbar-S}. 

Note that in the $\cN=(2,2)$ case the deformation we have introduced in \eqref{TTbar-S} is conceptually different from the cases 
with less supersymmetry. Specifically, the deformation is not given by the descendant of a \emph{single} composite superfield. 
On the other hand, suppose the $\cS$-multiplet is such that $C^{(\pm)}=k=k'=0$ and it is possible to improve the superfields  
$\chi_\pm$ and $\cY_{\pm}$ to a case where
\bsubeq\label{simplified-S-multiplet}
\bea
&
\cY_\pm=D_\pm\cV
~,~~~
\overline{\cY}_\pm=\Db_\pm\overline{\cV}
~,
\\
&
\chi_+=i\Db_+\overline{\cal B}
~,~~~
\chi_-=i \Db_-{\cal B}
~,~~~
\bar{\chi}_+= - i D_+{\cal B}
~,~~~
\bar{\chi}_-= - i D_-\overline{\cal B}
~,
\eea
\esubeq
with $\cV$ chiral and ${\cal B}$ twisted-chiral:
\bsubeq
\bea
&\Db_\pm \cV=0~,~~~
D_\pm \overline{\cal V}=0
~,
\\
&\Db_+ {\cal B}=D_-{\cal B}=0~,~~~
D_+ \overline{\cal B}=\Db_-\overline{\cal B}=0
~.
\eea
\esubeq
In this case \eqref{TTbar-S} simplifies to
\bea
\calT\calTb
&=&  - \int d^4 \theta \, \mathcal{S}_{++} \mathcal{S}_{--} 
+ \left( \int d \theta^-  d \theta^+ \, \Db_+\overline{\cal B}\Db_-\overline{\cal B}
- \int d \thetab^- \, d \theta^+ \, \Db_+ \overline{\mathcal{V}} D_- \mathcal{V} + \text{c.c.} \right) 
\non \\
        &=& -\int d^4 \theta \, \left( 
        \mathcal{S}_{++} \mathcal{S}_{--} 
        - 2 \mathcal{B} \overline{\mathcal{B}}
        - 2 \mathcal{V} \overline{\mathcal{V}} \right) 
        ~,
        \label{simplifiedTTbar}
\eea
and we see that, up to EOM's,  $T \Tb(x)$ is the bottom component of a long supersymmetric multiplet. 
In this situation, once we define the composite superfield 
\bea
\cO(\z):=
- \mathcal{S}_{++}(\z) \mathcal{S}_{--}(\z)
+ 2 \mathcal{B}(\z) \overline{\mathcal{B}}(\z)
+ 2 \mathcal{V}(\z) \overline{\mathcal{V}}(\z)
~,
\label{primary-TTbar}
\eea
eq.~\eqref{calTTb=TTb_(2,2)} turns into the equivalent result%
\footnote{In the subsequent discussion by $\q=0$ we will always mean $\q^\pm=\qb^{\pm}=0$.}
\begin{align}
 \int d^4 \theta\,\cO (\z)
 =D_-D_+\Db_+\Db_-\cO(\z)|_{\q=0}
=
T\Tb(x)
+{\rm EOM's}
+\pa_{++}(\cdots)
+\pa_{--}(\cdots)
~,
\label{calTTb=TTb-2}
\end{align}
stating that the D-term of the operator $\cO(\z)$ is equivalent to the standard $T\Tb(x)$ operator. 

For a matter theory that can be coupled to old-minimal supergravity, leading to the FZ-multiplet described by \eqref{FZ-2_(2,2)},
 the operator $\cO(\z)$ further simplifies thanks to the fact that the twisted-(anti-)chiral operators ${\cal B}$ and $\overline{{\cal B}}$ 
 disappear. For these cases, the $T \Tb$ flow turns into the following equation
\be
        \partial_{\lambda} \mathcal{L} 
  = \frac{1}{8} \int d^4 \theta \, \left( \mathcal{J}_{++} \mathcal{J}_{--} - 2 \mathcal{V} \overline{\mathcal{V}} \right) 
  ~.
  \label{FZTTbar}
  \ee
This will be our starting point in analyzing  $\cN=(2,2)$ deformed models in section \ref{sec:models}.


\subsection{Point-splitting and well-definedness}
\label{point-splitting-section}

The $T \Tb (x)$ operator \eqref{component-TTbar-22} is quite magical because it is a well-defined irrelevant composite local operator, 
free of short distance divergences \cite{Zamolodchikov:2004ce}. In fact,  this property generalizes to the larger class of operators 
\begin{align}
[A_{s}(x)\, A'_{s'}(x)-B_{s+2}(x)\, B_{s'-2}(x)]
\end{align} 
where $(A_s,B_{s+2})$ and $(A'_{s'},B'_{s'-2})$ are two pairs of conserved currents with spins $s$ and $s'$. The operator 
$T \Tb (x)$ is a particular example with $s=s'=0$. As  proven in \cite{Smirnov:2016lqw}, these composite operators of 
``Smirnov-Zamolodchikov''-type have a well-defined point splitting which is free of short-distance divergences. In the case of 
$\cN = (0,1)$ and $\cN = (1, 1)$ supersymmetric $T\Tb$ deformations, the entire supermultiplet whose bottom component is 
$T \Tb (x)$ is comprised of well-defined  Smirnov-Zamolodchikov-type operators \cite{Baggio:2018rpv,Chang:2018dge}. In the 
$\cN=(0,2)$ case, the primary%
\footnote{We denote as primary operator the top component of a supersymmetric multiplet even when the theory is not 
superconformal.} 
operator whose bottom component is $T \Tb(x)$ is not of Smirnov-Zamolodchikov-type. Nevertheless, also in this case it was recently 
shown that, thanks to supersymmetry, the whole multiplet is  well-defined \cite{Jiang:2019hux}.

In the $\cN=(2,2)$ case it is clear that the situation is more complicated than any of the cases mentioned above. 
First, in the general situation, according to \eqref{TTbar-S}, the $\calT\calTb$ deformation is a linear combination of a D-term together 
with chiral and twisted-chiral F-terms contributions. Though the F-terms might be protected by standard perturbative 
non-renormalization theorems (see, for example, \cite{Gates:1983nr,Buchbinder:1998qv} for the $D=4$ $\cN=1$ case which 
dimensionally reduces to  $D=2$ $\cN=(2,2)$), the D-term associated to the $\cS_{++}\cS_{--}$ operator has no clear reason to be 
protected in general from short-distance  divergences in point-splitting regularization, and hence has no obvious reason to be 
well-defined. This indicates that there might be a clash between supersymmetry and a point-splitting procedure in the general setting. 

We will not attempt to analyze this issue in full generality in the current work; instead our aim is to describe a subclass of models 
for which the $\calT \calTb$ deformation turns out to be well-defined. A natural restriction to impose is that the $\cS$-multiplet is 
constrained by \eqref{simplified-S-multiplet} and the $\calT\calTb$ deformation is therefore described by the D-term 
\eqref{simplifiedTTbar}. By trivially extending the arguments used in \cite{Jiang:2019hux} for the $\cN=(0,2)$ case, it is not difficult to 
show that these restrictions are sufficient to imply that the multiplet described by the $\cN=(2,2)$ primary operator $\cO(\z)$ of 
\eqref{primary-TTbar} is indeed well-defined despite not being of Smirnov-Zamolodchikov-type. As in the $\cN=(0,2)$,
unbroken $\cN=(2,2)$ supersymmetry turns out to be the reason for this to happen.

Let us quickly explain how this works for the FZ-multiplet and the deformation \eqref{FZTTbar}, which are the main players in this chapter. Note, however, that the same argument extends to more general cases where both chiral and twisted-chiral 
current superfields, $\chi_{\pm}$ and $\cY_{\pm}$,
satisfying \eqref{simplified-S-multiplet}
are turned on. 
We also refer to  \cite{Jiang:2019hux} for details that we will skip in the following discussion, which are trivial extensions from the 
$(0,2)$ to the $(2,2)$ case.

A first indication of the well-definedness of the multiplet associated to $\cO(\z)$ comes by looking  at the vacuum expectation 
value of its lowest component. Define the primary composite operator
\bea
O(x):=-j_{--}(x)j_{++}(x)+2v(x)\bar{v}(x)
=  \cO(\z)|_{\q=0}
~,
\label{comp-primary}
\eea
and its point-split version
\bea
O(x,x'):=
-j_{--}(x)j_{++}(x')
+v(x)\bar{v}(x')
+\bar{v}(x)v(x')
~,
\label{comp-primary-ps}
\eea
where
\bea
j_{\pm\pm}(x):={\cal J}_{\pm\pm}(\z)|_{\q=0}
~,~~~
v(x):=\cV(\z)|_{\q=0}
~,~~~
\bar{v}(x):=\overline{\cV}(\z)|_{\q=0}
~.
\label{FZ-3}
\eea
Note that equation \eqref{FZ-2_(2,2)} implies the following relation among the component operators
\bea
\big[\overline{Q}_\pm,j_{\mp\mp}(x)\big]
=\pm\big[Q_\mp,v(x)\big]
~,~~~
\big[Q_\pm,j_{\mp\mp}(x)\big]
=\pm\big[\overline{Q}_\mp,\bar{v}(x)\big]
~,
\label{conservation-components}
\eea
with $Q_{\pm}$ and $\overline{Q}_\pm$ denoting the $\cN=(2,2)$ supercharges.%
\footnote{Given an operator $F(x)$ defined as the $\theta=0$ component of the superfield $\cF(\zeta)$, $F(x):=\cF(\z)|_{\theta=0}$, 
then its supersymmetry transformations are such that $\big[Q_\pm,F(x)\big\}=\cQ_\pm\cF(\z)|_{\theta=0}
=D_\pm\cF(\z)|_{\theta=0}$
and
$\big[\overline{Q}_\pm,F(x)\big\}
=\overline{\cQ}_\pm\cF(\z)|_{\theta=0}
=\overline{D}_\pm\cF(\z)|_{\theta=0}$.
}
By then using $\pa_{\pm\pm}=i\{Q_\pm,\overline{Q}_\pm\}$, $\{Q_+,Q_-\}=\{\overline{Q}_+,\overline{Q}_-\}=0$, 
$[\overline{Q}_\pm,v(x)]=[Q_\pm,\bar{v}(x)]=0$, 
super-Jacobi identities,
together with the conservation equations \eqref{conservation-components}, 
and the assumption that the vacuum is invariant under  supersymmetry, it is straightforward to show that vacuum expectation value
 of $O(x,x')$ satisfies
\bea
\pa_{++}\left\langle j_{--}(x)j_{++}(x')\right\rangle 
&=&
i\left\langle
\left[\{Q_+,\overline{Q}_+\},j_{--}(x)\right]
j_{++}(x')\right\rangle 
\non\\
&=&
i\left\langle
\left\{Q_+,[Q_-,v(x)]\right\}
j_{++}(x')
+\left\{\overline{Q}_+,[\overline{Q}_-,\bar{v}(x)]\right\}
j_{++}(x')
\right\rangle 
\non\\
&=&
-i\left\langle
[Q_+,v(x)]
[Q_-,j_{++}(x')]
+[\overline{Q}_+,\bar{v}(x)]
[\overline{Q}_-,j_{++}(x')]
\right\rangle 
\non\\
&=&
i\left\langle
[Q_+,v(x)]
[\overline{Q}_+,\bar{v}(x')]
+[\overline{Q}_+,\bar{v}(x)]
[Q_+,v(x')]
\right\rangle 
\non\\
&=&
\left\langle
\left[i\{\overline{Q}_+,Q_+\},v(x)\right\}
\bar{v}(x')
+\left[i\{Q_+,\overline{Q}_+\},v(x)\right\}
\bar{v}(x')
\right\rangle 
\non\\
&=&
\pa_{++}\left\langle
v(x)\bar{v}(x')+v(x)\bar{v}(x')
\right\rangle 
~,
\eea
and, after performing a similar calculation for
$\left\langle \pa_{--}j_{--}(x)j_{++}(x')\right\rangle
=-\left\langle j_{--}(x)\pa'_{--}j_{++}(x')\right\rangle $,
it is clear that the relation
\be
\pa_{\pm\pm}\left\langle O(x,x')\right\rangle 
=0
\ee
holds. 
Therefore, $\left\langle O(x,x')\right\rangle $ is independent of the positions and free of short distance divergences. 
It is worth noting that similarly to the argument showing that the two point function of two chiral or twisted-chiral operators is 
independent of the positions $x$ and $x'$, the previous analysis for $\left\langle O(x,x')\right\rangle$ necessarily relies on unbroken 
$\cN=(2,2)$ supersymmetry.

The argument given above can be generalized to a statement about operators in  superspace  in complete analogy to the 
$\cN=(0,2)$ case of \cite{Jiang:2019hux}. Let us investigate the short distance singularities in the bosonic coordinates by defining  a 
point-split version of the  $\cN=(2,2)$ primary $\calT\calTb$ operator,
 \bea
 \cO(x,x',\q)
 :=
 - {\cal J}_{--}(x,\q)\,{\cal J}_{++}(x',\q)   
 +\cV(x,\q)\, \overline{\cV}(x',\q)
 +\overline{\cV}(x,\q)\, \cV(x',\q)
 ~.
 \label{pointsplit-primary}
 \eea
We want to show that the preceding bilocal superfield is free of short distance divergences in the limit $x\to x'$. 
A straightforward calculation shows that
\bea\hspace{-12pt}
\pa_{++}\cO(x,x',\q)
&=&
- \Big\{
 i  D_+\cV(\z)\left[D'_-\cJ_{++}(\z')+\Db'_+\overline{\cV}(\z')\right]
+ i   \Db_+\overline{\cV}(\z)\left[\Db'_-\cJ_{++}(\z')+D'_+\cV(\z')\right]
\non\\
&&
+ i  (\cQ_++ \cQ'_+)
\left[(\Db_+\overline{\cV}(\z))\cV(\z')\right]
+ i  (\overline{\cQ}_++\overline{\cQ}'_+)
\left[(D_+\cV(\z))\overline{\cV}(\z')\right]
\non\\
&&
+ i  (\cQ_-+\cQ'_-)
\left[(D_+\cV(\z))\cJ_{++}(\z')\right]
+ i   (\overline{\cQ}_-+\overline{\cQ}'_-)
\left[(\Db_+\overline{\cV}(\z))\cJ_{++}(\z')\right]
\non\\
&&
+(\pa_{++}+\pa'_{++})
\left[\qb^+(\Db_+\overline{\cV}(\z))\cV(\z')
+\qb^-(D_+\cV(\z))\cJ_{++}(\z')\right]
\non\\
&&
-(\pa_{++}+\pa'_{++})
\left[\q^+ (D_+\cV(\z))\overline{\cV}(\z')
+\q^-(\Db_+\overline{\cV}(\z))\cJ_{++}(\z')\right]
\Big\}\Big|_{\q=\q'}
~.
\label{coinciding-trick}
\eea
Note that the first line in the preceding expression 
is zero because of the FZ conservation equations \eqref{FZ-2_(2,2)}, which hold up to contact terms in correlation functions. 
The other lines are either total vector derivatives or 
supersymmetry transformations of bilocal operators. A 
similar equation holds for $\pa_{--}\cO(x,x',\q)$ showing 
that the operator $\cO(x,x',\q)$ satisfies
\bea
\pa_{\pm\pm}\cO(x,x',\q)
&=&
0
\,+\,\textrm{EOM's}
\,+\,[P,\cdots]
\,+\,[ Q,\cdots]
~,
\label{paO}
\eea
where $[P,\cdots]$ and $[ Q,\cdots]$ schematically indicate a translation and supersymmetry transformation of some bilocal superfield 
operator.%
\footnote{See Appendix A of \cite{Jiang:2019hux} for the relation between the operators $(\cQ_\pm+ \cQ'_\pm)$, 
$(\overline{\cQ}_\pm+ \overline{\cQ}'_\pm)$ and the generators of supersymmetry transformations on bilocal superfields such as 
$\cO(x,x',\q)$. The extension of that analysis from $\cN=(0,2)$ to $\cN=(2,2)$ is straightforward.} 
To conclude, by employing an OPE argument completely analogous to the one originally given by Zamolodchikov in 
\cite{Zamolodchikov:2004ce} and extended to the $\cN=(0,2)$ supersymmetric case in \cite{Jiang:2019hux}, one can show that 
eq.~\eqref{paO} implies
\be
\cO(x,x',\q)
=
\cO(\z)
\,+\,
{\rm derivative~terms}
~.
\ee 
Here ``derivative terms'' indicate superspace covariant derivatives 
$D_A=(\pa_{\pm\pm},D_\pm,\Db_\pm)$ 
acting on local superfield operators while $\cO(\z)$ arises from the regular, non-derivative part of the OPE of $\cO(x,x',\q)$. 
As a result the integrated operator 
\bea
S_{\cO} 
= 
\int d^2x\,d^4\q\,
\lim_{\varepsilon\to0}\cO(x, x+\varepsilon, \theta) 
=\int d^2x\,d^4\q:\cO(x, x, \theta):
~,
\eea
which can be considered as a definition of the integrated
$\calT\calTb(x)$ operator,%
\footnote{Note that, consistently, one can show that
\be
\{Q_+,[\overline{Q}_+,\{Q_-,[\overline{Q}_-,O(x,x')]\}]\}
=T_{----}(x)T_{++++}(x')
-\Theta(x)\Theta(x')
+\textrm{EOM's}
+[P,\cdots]
\ee
implying
that the descendant of the point-split
primary operator $O(x)$  is equivalent, up to Ward identities and total vector derivatives ($\pa_{\pm\pm}$), to
the point-split version of the descendant 
$T\Tb(x)$ operator.}
is free of short distance divergences and well-defined in complete analogy to the non-supersymmetric case
 \cite{Zamolodchikov:2004ce} and the $\cN=(0,1)$, $\cN=(1,1)$, and $\cN=(0,2)$  cases 
 \cite{Baggio:2018rpv, Chang:2018dge, Jiang:2019hux}.


\section{Deformed \texorpdfstring{$(2,2)$}{(2,2)} Models } \label{sec:models}

In this section, we will apply our supercurrent-squared deformation (\ref{FZTTbar}) to a few examples of
 $\mathcal{N} = (2,2)$ supersymmetric theories for a chiral multiplet $\Phi$. The superfield $\Phi$ can be written in components as
\bea
    \Phi &=& \phi  
    + \theta^+ \psi_+ 
    + \theta^- \psi_-
    + \theta^+ \theta^- F
    - i \theta^+ \thetab^+ \partial_{++} \phi 
    - i \theta^- \thetab^- \partial_{--} \phi 
     \non\\
    &&
    - i \theta^+ \theta^- \thetab^- \partial_{--} \psi_+ 
     - i \theta^- \theta^+ \thetab^+ \partial_{++} \psi_- 
    - \theta^+ \theta^- \thetab^- \thetab^+ \partial_{++} \partial_{--} \phi
    ~,
\eea
where $\phi$ is a complex scalar field, $\psi_{\pm}$ are Dirac fermions, and $F$ is a complex auxiliary field. 
The multiplet $\Phi$ satisfies the chirality constraint $\Db_{\pm} \Phi = 0$.

We denote the physical Lagrangian by $\mathcal{L}$ and the superspace D-term Lagrangian by $\mathcal{A}$, so that
\begin{align}
    S = \int d^2 x \, \mathcal{L} = \int d^2 x \, d^4 \theta \, \mathcal{A} ~.
\end{align}
A broad class of 
two-derivative theories for a chiral superfield can be described by superspace Lagrangians of the form
\begin{align}
    \mathcal{L} = \int d^4 \theta \, K ( \Phi , \Phib) + \int d^2 \theta \, W ( \Phi ) + \int d^2 \thetab \Wb ( \Phib ) 
    ~,
\end{align}
where $K ( \Phi , \Phib )$ is a real function called the K\"ahler potential and $W ( \Phi )$ is a holomorphic function called the 
superpotential. These are $\cN=(2,2)$ Landau-Ginzburg models. In order for the kinetic terms of the component fields of $\Phi$ to 
have the correct sign, we will assume that $K_{\Phi \Phib} = \frac{\partial^2 K}{\partial \Phi \partial \Phib}$ is positive.  

Although we will not expand on this point in detail, all the results found in this section can be derived almost identically for the case of 
a generic model of a single scalar twisted-chiral superfield $\calY$, 
$\Db_+\calY=D_-\calY=0$,
and its conjugate. This is not surprising since theories containing only chiral superfields are physically equivalent to theories 
formulated in terms of twisted-chiral superfields; see,
 for example,~\cite{Grisaru:1994dm,Grisaru:1995dr,Grisaru:1995kn,Grisaru:1995py,Gates:1995du} 
 for a discussion of this equivalence in models with global and local supersymmetry. There are also many more involved $(2,2)$ 
 theories that one might also want to study involving chiral, twisted-chiral and semi-chiral superfields; see,
  for example,~\cite{Caldeira:2018ynv} for a recent discussion and references. 
  For this analysis,  we have chosen to consider only models based on a single 
chiral multiplet.

\subsection{K\"ahler potential} \label{subsec:kahler}

First we will set the superpotential $W$ to zero and begin with an undeformed superspace Lagrangian of the form
\begin{align}
    \mathcal{L} = \int d^4 \theta \, K ( \Phi , \Phib )
\end{align}
for some K\"ahler potential $K$. To leading order around this undeformed theory, the FZ supercurrents are
\bsubeq
\bea
    \mathcal{J}_{\pm \pm} &=& 2 K_{\Phi \Phib} D_{\pm} \Phi \Db_{\pm} \Phib~ , \\
    \mathcal{V} &=& 0~ ,
\eea
\esubeq
where $K_{\Phi} = \frac{\partial K}{\partial \Phi}, K_{\Phi \Phib} = \frac{\partial^2 K}{\partial \Phi \partial \Phib}$, etc. 
Therefore, at first order the supercurrent-squared deformation driven by 
$\cO=
\left(- \mathcal{S}_{++} \mathcal{S}_{--}
+ 2 \mathcal{V} \overline{\mathcal{V}}\right)$
will source a four-fermion contribution in the D-term, giving
\begin{align}
    \mathcal{L}^{(1)} = \mathcal{L}^{(0)} + \frac{1}{2} \lambda K_{\Phi \Phib}^2 D_+ \Phi \Db_+ \Phib D_- \Phi \Db_- \Phib ~.
\end{align}
Next, we would like to find the all-orders solution for the deformed theory. We make the ansatz that, at finite deformation parameter 
$\lambda$, the Lagrangian takes the form
\begin{align}
    \mathcal{L}_{\lambda} = \int d^4 \theta  \Big\{ K ( \Phi, \Phib ) 
    + f ( \lambda, x, \xb, y ) K_{\Phi \Phib}^2 D_+ \Phi \Db_+ \Phib D_- \Phi \Db_- \Phib \Big\}~ , 
    \label{kahler_ansatz}
\end{align}
where we define the combinations
\begin{align}
    x = K_{\Phi \Phib} \partial_{++} \Phi \partial_{--} \Phib
    ~, \qquad
    y = K_{\Phi \Phib} \left( D_+ D_- \Phi \right) \left( \Db_+ \Db_- \Phib \right)
    ~.
\end{align}
Using the results in Appendix 
\ref{appendix:supercurrent_calculation}, one finds that the 
superfields $\mathcal{J}_{\pm \pm}$ and $\mathcal{V}$ appearing in 
our supercurrent-squared deformation, computed for the Lagrangian
\eqref{kahler_ansatz},
are given by
\begin{align} \hspace*{-50pt}
    \mathcal{J}_{++} &= 2 K_{\Phi \Phib} D_+ \Phi \Db_+ \Phib 
    \left[ 1 + f ( x + \xb - 3 y ) 
    + x \frac{\partial f}{\partial x} ( \xb - y ) 
    + \xb \frac{\partial f}{\partial \xb} ( x - y ) 
    + y \frac{\partial f}{\partial y} ( x + \xb - 2 y ) \right]
     \nonumber \\
    &\quad 
    + 2 K_{\Phi \Phib}^2 D_- \Phi \Db_- \Phib \partial_{++} \Phi \partial_{++} \Phib
     \left[ - f 
     - x \frac{\partial f}{\partial x} 
     - \xb \frac{\partial f}{\partial \xb} 
     + y \left( \frac{\partial f}{\partial x} 
     + \frac{\partial f}{\partial \xb} \right) \right] 
      \nonumber \\
    &\quad -2 i K_{\Phi \Phib}^2 D_+ \Phi D_- \Phi \partial_{++} \Phib \Db_+ \Db_- \Phib
     \left[ - f  
     + ( x - \xb ) \frac{\partial f}{\partial \xb} 
     + (x-y) \frac{\partial f}{\partial y}  \right]  
     \nonumber \\
    &\quad 
    - 2 i K_{\Phi \Phib}^2 \Db_+ \Phib \Db_- \Phib \partial_{++} \Phi D_+ D_- \Phi 
    \left[ f 
    + ( x - \xb ) \frac{\partial f}{\partial x} 
    + ( y - \xb ) \frac{\partial f}{\partial y}  \right]   
    ~,
    \label{kahler_jpp}
\end{align}
and
\begin{align} \hspace*{-50pt}
    \mathcal{J}_{--} &= 2 K_{\Phi \Phib} D_- \Phi \Db_- \Phib 
    \left[ 1 
    + f ( x + \xb - 3 y ) 
    + x \frac{\partial f}{\partial x} ( \xb - y ) 
    + \xb \frac{\partial f}{\partial \xb} ( x - y ) 
    + y \frac{\partial f}{\partial y} ( x + \xb - 2 y ) \right] 
    \nonumber \\
    &\quad
    +2 K_{\Phi \Phib}^2 D_+ \Phi \Db_+ \Phib \partial_{--} \Phi \partial_{--} \Phib 
    \left[ - f
     - x \frac{\partial f}{\partial x}
      - \xb \frac{\partial f}{\partial \xb} 
      + y \left( \frac{\partial f}{\partial x} 
      + \frac{\partial f}{\partial \xb} \right) \right]
      \nonumber \\
    &\quad 
     - 2 i K_{\Phi \Phib}^2 D_+ \Phi D_- \Phi \partial_{--} \Phib  \Db_+ \Db_- \Phib 
     \left[ - f 
     + ( \xb - x ) \frac{\partial f}{\partial x} 
     + ( \xb - y ) \frac{\partial f}{\partial y} \right] 
      \nonumber \\
    &\quad 
    - 2 i K_{\Phi \Phib}^2 \Db_+ \Phib \Db_- \Phib \partial_{--} \Phi  D_+ D_- \Phi 
    \left[ f 
    + ( \xb - x ) \frac{\partial f}{\partial \xb} 
    + (y - x ) \frac{\partial f}{\partial y} \right]~  ,
    \label{kahler_jmm}
\end{align}
and
\begin{align}\hspace*{-25pt}
    \mathcal{V} &= 2 K_{\Phi \Phib}^2
     \left( f 
     + y \frac{\partial f}{\partial y} 
     + x \frac{\partial f}{\partial x} 
     + \xb \frac{\partial f}{\partial \xb} \right) 
     \Big[ - i \partial_{++} \Phib ( D_+ D_- \Phi ) D_- \Phi \Db_- \Phib 
     + \partial_{++} \Phib \partial_{--} \Phib D_- \Phi D_+ \Phi 
     \nonumber \\
    &\quad 
    - \Db_- \Phib \Db_+ \Phib \left( D_+ D_- \Phi \right)^2 
    - i \partial_{--} \Phib ( D_+ D_- \Phi ) D_+ \Phi \Db_+ \Phib 
    \Big] 
    ~.
    \label{kahler_v}
\end{align}
The supercurrent-squared flow then induces a differential equation for the superspace Lagrangian 
$\mathcal{A}_\lambda$ (where, again, $\mathcal{L}_\lambda = \int d^4 \theta \, \mathcal{A}_\lambda$) given by
\begin{align}
    \frac{d}{d \lambda} \mathcal{A}_\lambda &=- \frac{1}{8}\cO= 
    \frac{1}{8} \left( \mathcal{J}_{++} \mathcal{J}_{--} 
    - 2 \mathcal{V} \overline{\mathcal{V}} \right)~ .
    \label{sc2_kahler_flow}
\end{align}
Given our ansatz (\ref{kahler_ansatz}), we see that
\begin{align}
    \frac{d \mathcal{A}_\lambda}{d \lambda} = \frac{df}{d \lambda} K_{\Phi \Phib}^2 \, D_+ \Phi \Db_+ \Phib D_- \Phi \Db_- \Phib ~.
\end{align}
On the other hand, plugging in our expressions (\ref{kahler_jpp}), (\ref{kahler_jmm}), and (\ref{kahler_v}) 
for the supercurrents into the right hand side of (\ref{sc2_kahler_flow}) also gives a result proportional to 
$K_{\Phi \Phib}^2 D_+ \Phi \Db_+ \Phib D_- \Phi \Db_- \Phib$. Equating the coefficients, we find a differential equation for $f$:
\begin{align}\hspace{-15pt}
    \frac{d f}{d \lambda} 
    =&\, 
    \frac{1}{2} \Bigg\{ - \xb y \left[ f + ( \xb - x ) \frac{\partial f}{\partial \xb} 
    + (y - x) \frac{\partial f}{\partial y }  \right]^2 
    - x y \left[ f + ( x - \xb ) \frac{\partial f}{\partial x} 
    + ( y - \xb ) \frac{\partial f}{\partial y} \right]^2 
     \nonumber \\
    & 
    + 2 ( x - y ) ( y - \xb ) \left[ f + y \frac{\partial f}{\partial y} + \xb \frac{\partial f}{\partial \xb}
     + x \frac{\partial f}{\partial x} \right]^2 
     + x \xb \left[ f + ( \xb - y ) \frac{\partial f}{\partial \xb} 
     + ( x - y ) \frac{\partial f}{\partial x} \right]^2
      \nonumber \\
    &+ \left[ 1 + ( x + \xb - 3 y ) f + ( x + \xb - 2 y ) y \frac{\partial f}{\partial y} + \xb ( x - y ) \frac{\partial f}{\partial \xb} 
    + x ( \xb - y ) \frac{\partial f}{\partial x} \right]^2 \Bigg\} 
    ~.
    \label{bg_def_unsimp}
\end{align}
In particular, this shows that our ansatz (\ref{kahler_ansatz}) for the finite-$\lambda$ superspace action is consistent: 
the supercurrent-squared deformation closes on an action of this form. It could have been otherwise: the flow equation might have
 sourced additional terms proportional, say, to two-fermion combinations $D_+ \Phi \Db_+ \Phib$, or required dependence on other 
 dimensionless variables like $\lambda (D_+ D_- \Phi)^2$, but these complications do not arise in the case where the undeformed 
 theory only has a K\"ahler potential.

On dimensional grounds, $f$ must be proportional to $\lambda$ times a function of the dimensionless combinations $\lambda x$ 
and $\lambda y$. Thus, although the differential equation for $f$ determined by (\ref{bg_def_unsimp}) is complicated, 
one can solve order-by-order in $\lambda$. The solution to $\mathcal{O} ( \lambda^3 )$ is
\begin{align}
    f ( \lambda, x, \xb, y ) =&\, \frac{\lambda}{2} + \lambda^2 \left( \frac{x + \xb}{4} - \frac{3}{4} y \right)
     \cr &
      + \lambda^3 \left( \frac{x^2 + \xb^2 + 3 x \xb}{8} + \frac{37}{24} y^2 - \frac{25}{24} \left( x + \xb \right) y \right) + \cdots~ .
\label{series-1}\end{align}
We were unable to find a closed-form expression for $f$ to all orders in $\lambda$. However, the differential equation 
simplifies dramatically when we impose the equations of motion for the theory, and in this case one can write down an exact formula.
This is similar to the $T\Tb$ flow of the free action for a real $\cN=(1,1)$ scalar multiplet that was analyzed in
\cite{Baggio:2018rpv,Chang:2018dge}.

We claim that, on-shell, one may drop any terms where $y \sim ( D_+ D_- \Phi ) ( \Db_+ \Db_- \Phib )$ multiplies the
 four-fermion term $|D\Phi|^4 \equiv D_+ \Phi \Db_+ \Phib D_- \Phi \Db_- \Phib$. This is shown explicitly in Appendix 
 \ref{appendix:on-shell} and follows directly from the superspace equation of motion and nilpotency of the fermionic terms
  $D_{\pm} \Phi$ and $\Db_{\pm} \Phib$. It is also an intuitive statement associated to the 
fact that for these models, on-shell, $\cN=(2,2)$
supersymmetry is not broken.
In fact, note that the superfields $( D_+ D_- \Phi )$ and
$( \Db_+ \Db_- \Phib )$ have as their lowest components the 
auxiliary fields $F$ and $\overline{F}$. If supersymmetry is not broken, the vev of $F$ has to be zero, 
$\langle F\rangle = 0$, which implies that
the auxiliary field $F$ is on-shell at least quadratic 
in fermions and, more precisely, can be proven
to be at least linear in $\psi_\pm=D_\pm\Phi|_{\q=0}$
and $\bar{\psi}_\pm=\Db_\pm\overline{\Phi}|_{\q=0}$.
From this argument it follows that on-shell 
$( D_+ D_- \Phi )$ is at least linear in 
$D_\pm\Phi$ and $\Db_\pm\overline{\Phi}$, and then the two conditions
$( D_+ D_- \Phi )|D\Phi|^4=0$ and $y|D\Phi|^4=0$ follow.

After removing  from \eqref{bg_def_unsimp}
the $y$-dependent terms which vanish on-shell,
we find a simpler differential equation 
for the function $f$,
\begin{align}
    \frac{df}{d \lambda} &= \frac{1}{2} \left\{ - x \xb \left[ f + x \frac{\partial f}{\partial x} + \xb \frac{\partial f}{\partial \xb} \right]^2 
    + \left[ 1 + (x + \xb) f + x \xb \left( \frac{\partial f}{\partial x} + \frac{\partial f}{\partial \xb} \right) \right]^2 \right\} ~,
\end{align}
whose solution is
\begin{align}
    f( \lambda , x, \xb, y=0) = \frac{\lambda}{1 - \frac{\lambda}{2} ( x + \xb )  + \sqrt{1 - \lambda ( x + \xb )  
    + \frac{\lambda^2}{4} \left( x - \xb \right)^2 }}
    ~.
\end{align}
Thus we have shown that the supercurrent-squared deformed Lagrangian at finite $\lambda$ is equivalent on-shell 
to the following superspace Lagrangian
\begin{align}
    \mathcal{L}_{\lambda} &= \int d^4 \theta \, \left( K ( \Phi, \Phib ) 
    + \frac{\lambda K_{\Phi \Phib}^2 D_+ \Phi \Db_+ \Phib D_- \Phi \Db_- \Phib }{1 - \frac{1}{2} \lambda K_{\Phi \Phib}^2 A 
    + \sqrt{1 - \lambda K_{\Phi \Phib}^2 A  + \frac{1}{4} \lambda^2 K_{\Phi \Phib}^4 B^2 } } \right)~ ,
    \label{bg_on_shell}
\end{align}
where
\begin{align}
    A = \partial_{++} \Phi \partial_{--} \Phib + \partial_{++} \Phib \partial_{--} \Phi ~,\qquad
    B = \partial_{++} \Phi \partial_{--} \Phib - \partial_{++} \Phib \partial_{--} \Phi ~. 
\end{align}

When $K ( \Phi , \Phib ) = \Phib \Phi$, 
it is simple to show that this model represents 
an  $\mathcal{N} = (  2, 2 )$ 
off-shell supersymmetric extension of the $D=4$ Nambu-Goto string in an appropriate gauge---
often referred to as a static gauge in presence of a $B$ field, though it can be more naturally described as uniform light-cone 
gauge~\cite{Arutyunov:2004yx,Arutyunov:2005hd} (see refs.~\cite{Baggio:2018gct,Frolov:2019nrr} for a discussion of this point).
In particular, by setting various component fields to zero and performing the superspace integrals, one can show that 
(\ref{bg_on_shell}) matches the expected answer for $T \Tb$ deformations in previously known non-supersymmetric cases. 
For instance, setting the fermions to zero and integrating out the auxiliary fields 
$F$ and $\overline{F}$
gives  the $T \Tb$ deformation of the complex free boson
$\phi$, whose Lagrangian is
\begin{equation}
\label{eq:bosonTTbar22}
 {\cal L}_{\lambda,\text{bos}}= 
 \frac{\sqrt{1+2 \lambda  a +\lambda^2 b^2}-1}{4\lambda}
 =
  \frac{  a}{4 } - \lambda \frac{ \pa_{++} \phi \pa_{--}\phi \pa_{++} \bar \phi \pa_{--} \bar\phi }
 {  1+\lambda a+\sqrt{1+2 \lambda  a +\lambda^2 b^2}  } 
 ~,
\end{equation}
where
\begin{equation}
  \label{eq:xydef_(2,2)}
 a=\pa_{++} \phi \pa_{--} \bar \phi +\pa_{++}  \bar \phi \pa_{--}  \phi
 ~, \qquad 
b= \pa_{++} \phi \pa_{--} \bar \phi -\pa_{++}  \bar \phi \pa_{--}  \phi ~.
\end{equation}
The Lagrangian \eqref{eq:bosonTTbar22} indeed describes the $D=4$ light-cone gauge-fixed Nambu-Goto string model.

Alternatively, setting all the bosons to zero
in \eqref{bg_on_shell} can be shown to give the $T \Tb$ deformation of a complex free fermion. 
These calculations are similar to those in the case of the $(0,2)$ supercurrent-squared action, 
which are presented in \cite{Jiang:2019hux}. In fact, it can even be easily shown
that an $\cN=(0,2)$ truncation of \eqref{bg_on_shell}
gives precisely the $T\Tb$ deformation of a free $\cN=(0,2)$ chiral multiplet that was derived in \cite{Jiang:2019hux}.

It is worth highlighting that, unlike the $\cN=(2,2)$ case,
an off-shell 
$(0,2)$ chiral scalar multiplet contains only physical degrees
of freedom and no auxiliary fields.
Interestingly, related to this fact, it turns out that
(up to integration by parts and total derivatives)
the $\cN=(0,2)$ off-shell supersymmetric extension of the  
$D=4$ Nambu-Goto string action in light-cone gauge is unique
and precisely matches the off-shell $T\Tb$ deformation of a free 
$\cN=(0,2)$ chiral multiplet action \cite{Jiang:2019hux}.

In the $\cN=(2,2)$ case, because of the presence of the auxiliary field $F$ in the chiral multiplet $\Phi$, 
there are an infinite set of inequivalent $\cN=(2,2)$ off-shell extensions of the Lagrangian \eqref{eq:bosonTTbar22} that are all equivalent 
on-shell. 
A representative of these equivalent actions is described by
(\ref{bg_on_shell}) when $K(\F,\overline{\F})=\overline{\F}\F$.

The non-uniqueness of dynamical systems described by actions
of the form (\ref{bg_on_shell})
can also be understood by noticing that, for example,
it is possible to perform a class of redefinitions that leaves the action (\ref{bg_on_shell}) invariant on-shell. 
As a (very particular)  example, 
note that we are free to perform a shift 
of the form
\begin{align}
    D_+ \Db_- \left( \Db_+ \Phib D_- \Phi \right) \longrightarrow D_+ \Db_- \left( \Db_+ \Phib D_- \Phi \right)
     + a \left( D_+ D_- \Phi + \Db_+ \Db_- \Phib \right)^2
    \label{field_redef}
\end{align}
for any real number $a$. In terms of $A$ and $B$, (\ref{field_redef}) implements the shifts
\begin{align}\begin{split}
    A \longrightarrow A + a \left( \left( D_+ D_- \Phi \right)^2 + 2 y + \left( \Db_+ \Db_- \Phib \right)^2 \right)~,\qquad
    B \longrightarrow B 
        \label{field_redef-2}
\end{split}\end{align}
in (\ref{bg_on_shell}).
The resulting Lagrangian would enjoy the same on-shell 
simplifications described in  Appendix \ref{appendix:on-shell}
and would turn out to be on-shell equivalent to
the Lagrangian (\ref{bg_on_shell}).
In this infinite set of on-shell
equivalent actions, a particular choice would represent
an exact solution of the $T\Tb$ flow equation
\eqref{sc2_kahler_flow}--\eqref{bg_def_unsimp},
whose leading terms in a $\lambda$ series expansion are given in \eqref{series-1}.
Another representative in this on-shell equivalence class
is the simplified model described by (\ref{bg_on_shell}).

These types of redefinition 
and on-shell equivalentness 
are not a surprise, nor really new. 
In fact, they are  of the same nature as redefinitions 
that have  been studied in detail in \cite{GonzalezRey:1998kh}
(see also \cite{Kuzenko:2011tj} for a description of these 
 types of ``trivial symmetries'')
 in the context of 
$D=4$ $\cN=1$ chiral and linear superfield models possessing a non-linearly realised additional supersymmetry
\cite{Bagger:1997pi,GonzalezRey:1998kh}. 
 As in (\ref{field_redef-2}), the field redefinition in 
this context does not affect the dynamics of the physical fields---
it basically corresponds only to an arbitrariness in the definition of the 
auxiliary fields that always appear quadratically
in the action and then are set to zero (up to fermion terms
that will not contribute due to nilpotency in the action)
on-shell. 
Although here we only focused on discussing the on-shell ambiguity
of the solution of the $\cN=(2,2)$ $T\Tb$ flow,
we expect that the exact solution of the flow equations
with $y$ nonzero \eqref{sc2_kahler_flow}--\eqref{bg_def_unsimp} 
can be found by a 
field redefinition of the kind we made in the action (\ref{bg_on_shell}). 

It is also interesting to note that similar
freedoms and field redefinitions 
are also described in the construction
of $D=4$ $\cN=1$ supersymmetric Born-Infeld actions;  see, for example,~\cite{Bagger:1996wp}.
In fact, as will be analyzed in more detail elsewhere 
\cite{SUSYDBITTbar},
it can be shown that the Lagrangian \eqref{bg_on_shell}
is structurally of the type described by Bagger and Galperin
for the $D=4$ $\cN=1$ supersymmetric Born-Infeld action 
\cite{Bagger:1996wp}. 
The equivalence can be formally shown by identifying 
$W_+ = \Db_+ \Phib$, $W_- = D_- \Phi$, $W^2 = \Db_+ \Phib D_- \Phi$, and 
$D^\alpha W_\alpha = D_+ D_- \Phi + \Db_- \Db_+ \Phib$ to match their conventions. 
As a consequence, we can show that our solution for the $T\Tb$ flow possesses a second non-linearly realised $\cN=(2,2)$ supersymmetry, besides the $(2,2)$ supersymmetry which is made manifest by the superspace construction, which is discussed in much greater detail in Chapter \ref{CHP:nonlinear}. We note that the presence of a second supersymmetry is analogous to what happens in the $\cN=(0,2)$ case~\cite{Jiang:2019hux}.

\subsection{Adding a superpotential}

Now suppose we begin with an undeformed theory that has a superpotential $W(\Phi)$,
\begin{align}
    \mathcal{L}^{(0)} = \int d^4 \theta \, K ( \Phi , \Phib ) + \left( \int d^2 \theta \, W ( \Phi ) \right) 
    + \left( \int d^2 \thetab \, \overline{W} ( \Phib ) \right)~ .
    \label{lag_with_W}
\end{align}
As shown in Appendix \ref{appendix:supercurrent_calculation}, the superpotential F-term gives a contribution 
$\delta \mathcal{V} = 2 W(\Phi)$ to the field $\mathcal{V}$ which appears in supercurrent-squared. 
To leading order in the deformation parameter, the Lagrangian takes the form 
\begin{align}
    \mathcal{L}^{(0)} &\to \mathcal{L}^{(0)} + \mathcal{L}^{(1)} \nonumber \\
    &= \mathcal{L}^{(0)} 
    + \lambda \int d^4 \theta \, 
    \left( \frac{1}{2} K_{\Phi \Phib}^2 D_+ \Phi \Db_+ \Phib D_- \Phi \Db_- \Phib + W(\Phi) \Wb ( \Phib ) \right) ~.
\end{align}
In addition to the four-fermion term which we saw in section \ref{subsec:kahler}, we see that the deformation modifies the 
K\"ahler potential, adding a term proportional to $|W ( \Phi )|^2$.

Next consider the second order term in $\lambda$. For convenience, we use the combination 
$|D\Phi|^4 = D_+ \Phi \Db_+ \Phib D_- \Phi \Db_- \Phib$, which is the four-fermion combination that appeared at first order. Then
\begin{align}
    \mathcal{L}^{(2)} 
    =
    \frac{\lambda^2}{4}
    \int d^4 \theta \,
     \Big( x + \xb - 3 y 
     -2 | W' ( \Phi ) |^2 
     +W D_- D_+ +\Wb \Db_+ \Db_-\Big) |D\Phi|^4
    ~. 
    \label{superpotential_second_order}
\end{align}
The new terms involving supercovariant derivatives of $|D\Phi|^4$ will generate contributions with two fermions in the D-term.

As we continue perturbing to higher orders, the form of the superspace Lagrangian becomes more complicated. 
It is no longer true that the supercurrent-squared flow closes on a simple ansatz with one undetermined function, 
as it did in the case with only a K\"ahler potential. Indeed, the finite-$\lambda$ deformed superspace Lagrangian
 in the case with a superpotential will depend not only on the variables $x$, $\xb$, 
 and $y$ as in section \ref{subsec:kahler}, 
 but also, for example,
 on combinations like $\partial_{++} \Phi \Db_+ \Db_- \Phib$, which can appear multiplying the two-fermion term
  $D_- \Phi \Db_- \Phib$ in the superspace Lagrangian. To find the full solution, one would need to determine several functions contributing to the D-term---one multiplying the four-fermion term $|D\Phi|^4$ as in the K\"ahler case; one for the deformed K\"ahler potential which may 
  now depend on $x$, $y$, and other combinations; and four functions multiplying the two-fermion terms 
  $D_+ \Phi D_- \Phi$, $D_+ \Phi \Db_- \Phib$, etc. Each function can depend on several dimensionless combinations. 

In the presence of a superpotential, 
the situation might further be complicated by the fact that 
supersymmetry can be spontaneously broken.
This would make it impossible, for example, to use on-shell
simplifications like $y|D\Phi|^4=0$ that we employed in the section~\ref{subsec:kahler}, where 
supersymmetry is never spontaneously broken.

It should be clear that the case with a superpotential is significantly
more involved and rich than just a pure K\"ahler potential.
In this case, we have not attempted to find a solution of the 
$T\Tb$ flow equation in closed form.
However, it is evident from the form of supercurrent-squared 
eq.~\eqref{sc2_kahler_flow}---which is always written as a D-term integral of current bilinears---
that this deformation will only affect the $D$ term and not the $\cN=(2,2)$ superpotential $W$ appearing in the chiral integral. 
Therefore the superpotential, besides being protected from perturbative quantum corrections, is also protected from corrections
 along the supercurrent-squared flow.

\subsection{The physical classical potential} 

In view of the difficulty of finding the all-orders deformed superspace action for a theory with a superpotential, 
we now consider the simpler problem of finding the local-potential approximation (or zero-momentum potential) for the bosonic complex scalar 
$\phi$ contained in the superfield $\Phi$. 
We stress that our analysis here is purely classical
and we will make a couple of  
comments about possible quantum effects 
later in this section.
For simplicity, we will also restrict to 
the case in which the K\"ahler 
potential is flat, $K(\Phi,\overline{\Phi})=\overline{\Phi}\F$.
By ``zero-momentum potential'' we mean the physical potential
$V ( \phi )$ which appears in the  
Lagrangian after performing the superspace integral in the deformed theory and then setting $\partial_{\pm \pm} \phi = 0$. 
For instance, consider the undeformed Lagrangian
\begin{align}
    \mathcal{L}^{(0)} = \int d^4 \theta \, \Phib \Phi + \int d^2 \theta \, W ( \Phi ) + \int d^2 \thetab \, \Wb ( \Phib ) ~.
\end{align}
When we ignore all terms involving derivatives and the fermions $\psi_{\pm}$, the only contributions to the physical Lagrangian 
(after performing the superspace integral) come from an $|F|^2$ term from the kinetic term, plus the term 
$W(\Phi) = W ( \phi ) + W'(\phi) \theta^+ \theta^- F$. This gives us the zero-momentum, zero-fermion component action
\begin{align}
    S = \int d^2 x \, \left(|F|^2 + W' ( \phi ) F + \Wb' ( \phib) \Fb \right)~ .
    \label{superpotential_integrated}
\end{align}
We may integrate out the auxiliary field $F$ using its equation of motion $\Fb = - W' ( \phi )$, which yields
\begin{align}
    S = \int d^2 x \, \left( - | W' ( \phi ) |^2 \right)~ ,
\end{align}
so the zero-momentum potential for $\phi$ is $V = | W' ( \phi )|^2$, as expected.
Note that the previous potential might have extrema that breaks $\cN=(2,2)$ supersymmetry
while supersymmetric vacua will always set 
$\langle F\rangle=\langle W'(\phi)\rangle = 0$. We will
assume supersymmetry of the undeformed theory 
not to be spontaneously broken in our discussion.

Now suppose we deform by the supercurrent-squared operator to second order in $\lambda$, 
which gives the superspace expression (\ref{superpotential_second_order}). If we again perform the superspace 
integral and discard any terms involving derivatives or fermions, we now find the physical Lagrangian
\bea
\mathcal{L} \big\vert_{\partial_{\pm \pm} \phi = 0 } 
&=& |F|^2 + F W' + \Fb \Wb' + \lambda \left( \frac{1}{2} |F|^4 - |F|^2 |W'|^2 \right) 
+ \frac{1}{4} \lambda^2 | F |^4 \left( W' F + \Wb' \Fb \right) 
\non\\
    && - \frac{1}{2} \lambda^2 | W' |^2 | F |^4 + \frac{3}{4} \lambda^2 | F |^6 ~.
    \label{second_order_W_action}
\eea
Remarkably, the equations of motion for the auxiliary $F$
in (\ref{second_order_W_action}) admit the solution 
$F = - \Wb' ( \phib ) $, $\Fb = - W' ( \phi )$, which is the same 
as the unperturbed solution.
This for instance implies that if we  start
from a supersymmetric vacua in the undeformed theory we will 
remain supersymmetric along the $T\Tb$ flow.
On the one hand, this
is not a surprise considering that we know the $T\Tb$
flow preserves the structure of the spectrum, and in particular should leave a zero-energy supersymmetric vacuum unperturbed.
On the other hand, it is a reassuring check to see 
this property  explicitly appearing in our analysis.

Returning to \eqref{second_order_W_action} and
integrating out the auxiliary fields gives
\begin{align}
    \mathcal{L} \big\vert_{\partial_{\pm \pm} \phi = 0 } = - |W'(\phi)|^2 - \frac{1}{2} \lambda | W' ( \phi ) |^4 
    - \frac{1}{4} \lambda^2 | W' ( \phi ) |^6~ .
    \label{leading_geometric}
\end{align}
These are the leading terms in the geometric series $\frac{-|W'|^2}{1 - \frac{1}{2} \lambda |W'|^2}$.
 In fact, up to conventions for the scaling of $\lambda$, one could have predicted this outcome from the form of the
  supercurrent-squared operator and the known results for $T \Tb$ deformations of a bosonic theory with a potential 
  \cite{Cavaglia:2016oda}.
We know that, up to terms which vanish on-shell, the effect of adding supercurrent-squared to the physical Lagrangian
 is to deform by the usual $T \Tb$ operator. However, in the zero-momentum sector, we see that the $T \Tb$ deformation 
 reduces to deforming by the square of the potential:
\begin{align}
    T \Tb \Big\vert_{\partial_{\pm \pm} \phi = 0} = \mathcal{L}^2 \Big\vert_{\partial_{\pm \pm} \phi = 0} = V^2 ~.
\end{align}
Therefore, it is easy to solve for the deformed potential if we deform a physical Lagrangian
 $\mathcal{L} = f ( \lambda, \partial_{\pm \pm} \phi ) + V ( \lambda , \phi )$ by $T \Tb$, 
 since the flow equation for the potential term is simply
\begin{align}
    \partial_\lambda \mathcal{L} = \frac{\partial V}{\partial \lambda} = V^2~ ,
\end{align}
which admits the solution
\begin{align}\label{pole_potential_solution}
    V ( \lambda , \phi ) = \frac{V ( 0, \phi ) }{1 - \lambda V ( 0 , \phi ) } ~.
\end{align}
We can apply this result to the Lagrangian (\ref{superpotential_integrated}), treating the entire expression involving
 the auxiliary field $F$ as a potential (since it is independent of derivatives). The deformed theory has a zero-momentum
  piece which is therefore equivalent to
\begin{align}
    S (\lambda ) \Big\vert_{\partial_{\pm \pm} \phi = 0} = \int d^2 x \, \frac{\left(|F|^2 + W' ( \phi ) F 
    + \Wb' ( \phib) \Fb \right)}{1 - \lambda \left(|F|^2 + W' ( \phi ) F + \Wb' ( \phib) \Fb \right)}~ ,
\end{align}
at least on-shell. Integrating out the auxiliary now gives
\begin{align}
    S (\lambda ) \Big\vert_{\partial_{\pm \pm} \phi = 0} = \int d^2 x \,  \frac{- |W'(\phi)|^2 }{1 - \lambda |W'(\phi)|^2 }
    \label{deformed_superpotential}
\end{align}
as the deformed physical potential. This matches the first few terms of (\ref{leading_geometric}), 
up to a convention-dependent factor of $\frac{1}{2}$ in the scaling of $\lambda$.

Now one might ask what superspace Lagrangian would yield the physical action (\ref{deformed_superpotential}) 
after performing the $d \theta$ integrals. One candidate is
\begin{align}
    \mathcal{L} (\lambda ) \Big\vert_{\partial_{\pm \pm} \phi = 0} \sim \int d^4 \theta \, 
    \left( \Phib \Phi - \lambda | W (\Phi) |^2 \right) + \int d^2 \theta \, W ( \Phi ) + \int d^2 \thetab \, \Wb ( \Phib ) ~,
    \label{deformed_superpotential_superspace}
\end{align}
where here $\sim$ means ``this superspace Lagrangian gives an equivalent zero-momentum physical potential for the boson 
$\phi$ on-shell.''

It is important to note that (\ref{deformed_superpotential_superspace}) is \emph{not} the true solution for the deformed superspace 
Lagrangian using supercurrent-squared. The genuine solution involves a four-fermion term, all possible two-fermion terms, and more 
complicated dependence on the variable $y = \lambda ( D_+ D_- \Phi ) ( \Db_+ \Db_- \Phib )$ in the zero-fermion term. However, if 
one were to perform the superspace integral in the true solution and then integrate out the auxiliary field $F$ using its equation of 
motion, one would obtain the same zero-momentum potential for $\phi$ as we find by performing the superspace integral in 
(\ref{deformed_superpotential_superspace}) and integrating out $F$. 

The form (\ref{deformed_superpotential_superspace}) is interesting because it shows that the effect of supercurrent-squared on the 
physical potential for $\phi$ can be interpreted as a change in the K\"ahler metric, which for this Lagrangian is
\begin{align}
    K_{\Phi \Phib} = 1 - \lambda | W' ( \Phi ) |^2~ .
    \label{effective_kahler}
\end{align}
When one performs the superspace integrals in (\ref{deformed_superpotential_superspace}), the result is
\begin{align}
    \mathcal{L} \Big\vert_{\partial_{\pm \pm} \phi = 0} = K_{\Phi \Phib} | F |^2 + W' ( \phi ) F + \Wb' ( \phib ) \Fb ~, 
\end{align}
which admits the solution $F = - \frac{\Wb'(\phib)}{K_{\Phi \Phib}}$. Substituting this solution gives
\begin{align}
    \mathcal{L} \Big\vert_{\partial_{\pm \pm} \phi = 0} = \frac{ - | W' ( \phi ) |^2 }{K_{\Phi \Phib}} 
    = \frac{- | W' ( \phi )|^2 }{1 - \lambda | W' ( \phi ) |^2}~ ,
\end{align}
which agrees with (\ref{deformed_superpotential}).

As already mentioned, supersymmetric vacua of the original, undeformed, theory are associated with critical points of the 
superpotential $W(\phi)$. Any vacuum of the undeformed theory will persist in the deformed theory: near a point where 
$W' ( \phi ) = 0$, we see that the physical potential $V ( \phi ) = \frac{ | W' |^2}{1 - \lambda |W'|^2}$ also vanishes (away from the pole 
$|W'|^2 = \frac{1}{\lambda}$, the deformed potential is a monotonically increasing function of $|W'|^2$). Further, the auxiliary field $F$ 
does not acquire a vacuum expectation value because $F = - \Wb' ( \phib )$ remains a solution to its equations of motion in the 
deformed theory. Once more, this indicates that supersymmetry is unbroken along the whole $T\Tb$ flow if is in the undeformed
theory.

However, this classical analysis suggests that the soliton spectrum of the theory has changed dramatically at any finite deformation 
parameter $\lambda$. There are now generically poles in the physical potential $V(\phi)$ at points where $|W'|^2 = \frac{1}{\lambda}$ 
which might separate distinct supersymmetric vacua of the theory. 
For instance, if the original theory had a 
double-well superpotential 
with two critical points $\phi_1$, $\phi_2$ where 
$W'(\phi_i) = 0$, then this undeformed theory supports BPS 
soliton solutions which 
interpolate between these two vacua. But if the superpotential 
$W$ reaches a value of order $\frac{1}{\lambda}$ at some point 
between $\phi_1$ and $\phi_2$, then this soliton solution appears 
naively forbidden in the deformed theory because it requires 
crossing an infinite potential barrier. Another way of seeing 
this is by considering the effective K\"ahler potential (\ref{effective_kahler}), 
which would change sign at some point between the two 
supersymmetric vacua in the deformed theory and thus give rise 
to a negative-definite K\"ahler metric.

Our discussion has been purely classical. As we emphasised in the introduction, a fully quantum analysis of this problem is desirable, though subtle because of the non-local nature of the $T\Tb$ deformation.\footnote{For an analysis of correlation functions in theories deformed by supercurrent-squared operators, which may offer a first step towards understanding the quantum properties of these deformations, see \cite{Ebert:2020tuy}.} The advantage of performing such an analysis in models with extended supersymmetry is that holomorphy and associated non-renormalization theorems provide control over the form of any possible quantum corrections. For example, the superpotential for the models studied in this work is not renormalized perturbatively along the flow. It would be interesting to examine the structure of perturbative quantum corrections along the lines of~\cite{Rosenhaus:2019utc}, but in superspace with manifest supersymmetry. It should be possible to absorb any quantum corrections visible in perturbation theory by a change in the $D$-term K\"ahler potential
meaning that at least the structure  of the
supersymmetric vacua would be preserved.


\chapter{Non-Linearly Realized Symmetries} \label{CHP:nonlinear}
In this chapter, we will investigate certain additional symmetries which are present in some $\TT$ and supercurrent-squared deformed models. The treatment of this chapter follows the paper ``Non-Linear Supersymmetry and $\TT$-like Flows'' \cite{Ferko:2019oyv}.

It was briefly mentioned in Chapter \ref{CHP:SC-squared-2} that our solution (\ref{bg_on_shell}) for the supercurrent-squared deformation of a chiral multiplet possesses a second non-linearly realized supersymmetry -- which we will demonstrate in more detail shortly -- but there are already examples of such additional symmetries in the non-supersymmetric context. For instance, we reviewed in Section \ref{subsec:flow_eqn_review} that the $\TT$ deformation of a free boson is given by
\begin{align}\label{tt_deformed_free_boson}
    \mathcal{L}_\lambda = \frac{1}{2 \lambda} \left( \sqrt{1 + 2 \lambda \partial_\mu \phi \partial^\mu \phi + 1 } - 1 \right) \, .
\end{align}
Ignoring the constant $-\frac{1}{2 \lambda}$, this is of the same form as the Dirac Lagrangian which describes the transverse fluctuations $\phi$ on a brane. The Dirac action for a $p$-brane is usually written 
\begin{align}\label{general_Dp_brane}
    S_{\text{Dirac}} = - T_p \int d^{p+1} \xi \, \sqrt{ - \det ( \gamma ) } \, , 
\end{align}
where $T_p$ is the tension of the $Dp$-brane and
\begin{align}
    \gamma_{ab} = \frac{\partial X^\mu}{\partial \xi^a} \frac{\partial X^\nu}{\partial \xi^b} \eta_{\mu \nu} 
\end{align}
is the pullback of the target space metric $\eta_{\mu \nu}$ onto the brane's worldvolume. We can choose static gauge which identifies the $p+1$ worldvolume coordinates $\xi^a$ with the target space coordinates $X^a$, and the remaining worldvolume coordinates are transverse oscillations:
\begin{align}\begin{split}
    X^a &= \xi^a \, , \qquad \qquad \; \, a = 0 , \cdots , p \, , \\
    X^I &= 2 \pi \alpha' \phi^I ( \xi ) \, , \quad I = p+1 , \cdots D - 1 \, .
\end{split}\end{align}
Here $D$ is the total dimension of the target spacetime and $\alpha'$ is a parameter which is proportional to the square of the string length. In the case of a $D1$-string embedded in a three dimensional spacetime, so that $p=1$ and $D=3$, the Dirac action can be written as
\begin{align}\label{dirac_lagrangian}
    S_{\text{Dirac}} = - T_p \int d^2 \xi \sqrt{ 1 + 2 \pi \alpha' \partial^a \phi \partial_a \phi } \, , 
\end{align}
which is the same form as the $\TT$ deformed Lagrangian (\ref{tt_deformed_free_boson}). Here $\alpha'$ plays the role of the $\TT$ parameter $\lambda$.

It is well-known that the $Dp$-brane Lagrangian (\ref{general_Dp_brane}) possesses a non-linearly realized symmetry for any $p$. We will describe this symmetry in the case (\ref{dirac_lagrangian}) of a $D1$-string. In this case, for either fixed index $a \in \{ 0, 1 \}$, the action (\ref{dirac_lagrangian}) is invariant under the transformation
\begin{align}\label{nonlinear_dirac_symmetry}
    \delta \phi = \xi^a + \phi \partial^a \phi \, .
\end{align}
The interpretation of this symmetry is that the presence of the brane as an embedded hypersurface spontaneously breaks part of the Poincar\'e symmetry of the ambient spacetime. The scalars $\phi$ are Nambu-Goldstone bosons of this spontaneously broken symmetry, which is then non-linearly realized. One can view the transformation (\ref{nonlinear_dirac_symmetry}) as a rotation which rotates the transverse direction $\phi$ into the worldvolume direction $\xi^a$, breaking static gauge, along with a compensating worldvolume diffeomorphism which then restores static gauge. This symmetry, and its extensions to the Dirac-Born-Infeld action which includes a gauge field, are nicely discussed in \cite{Gliozzi:2011hj, Casalbuoni:2011fq, Maxfield:2016vpw}.

\section{\texorpdfstring{$D=2 \;\, \cN=(2,2)$}{2D N=(2,2)} Flows and Non-Linear \texorpdfstring{$\cN=(2,2)$}{N=(2,2)} Supersymmetry} 
\label{2D}

In this section, we are going to explore in detail how the non-linear supersymmetries -- which we briefly alluded to in Chapter \ref{CHP:SC-squared-2} -- arise for the simplest $\cN=(2,2)$ $\TT$ flows. The undeformed models are supersymmetrized theories of free scalars, while the deformed models are $\cN=(2,2)$ supersymmetric extensions of the $D=4$ gauge-fixed Nambu-Goto string.

\subsection{\texorpdfstring{$T \Tb$}{T Tbar} deformations with \texorpdfstring{$\cN=(2,2)$}{N=(2,2)} supersymmetry}

In Chapter \ref{CHP:SC-squared-2}, supersymmetric flows for various theories were studied; the conclusion of Section \ref{point-splitting-section} was that $T \Tb(x)$ operator of a supersymmetric theory is related to a supersymmetric 
descendant operator $\calT \calTb(x)$, 
\begin{align}
\calT\calTb(x)
=
T \Tb(x)
+{\rm EOM}
+\pa_{++}(\cdots)
+\pa_{--}(\cdots)
~.
\label{calTTb=TTb}
\end{align}
The previous equation states the equivalence of $T \Tb(x)$ and $\calT\calTb(x)$ up to total derivatives and terms that vanish on-shell, which we have indicated with ``${\rm EOM}$''.

The simplest cases of deformed models, on which we will focus in this section,  are $T\Tb$-deformed theories of free scalars, fermions and auxiliary fields. In the case of $D=2$ $\cN=(2,2)$ supersymmetry, a scalar multiplet can have several different off-shell representations 
\cite{Gates:1984nk,Buscher:1987uw,Grisaru:1997pg,Lindstrom:2005zr}. The two cases we will consider here are chiral and twisted-chiral supermultiplets, which are the most commonly studied cases. 

In $\cN=(2,2)$ superspace, parametrized by coordinates $\z^M=(x^{\pm\pm},\q^{\pm},\qb^\pm)$, let
the complex superfields $X(x,\q)$ and $Y(x,\q)$ satisfy chiral
and twisted-chiral constraints, respectively,
\bea
\bar  D_\pm X=0
~,\quad
\bar  D_+ Y= D_- Y=0
~.
\label{chiral_twisted-chiral_constraints}
\eea
In this chapter, we use slightly different conventions for the supercovariant derivatives than in Chapter \ref{CHP:SC-squared-2}. These were referred to as $D'_\pm$ and $\Db'_\pm$ in equation (\ref{different_(2,2)_conventions}), but we will drop the primes in the following discussion. For convenience, the conventions are repeated here:
\begin{gather}
 D_\pm =\frac{\p} {\p \theta^\pm}+i \bar \theta^\pm \p_{\pm\pm}
 \, , \; \;
\bar  D_\pm =-\frac{\p} {\p\bar  \theta^\pm}  -  i  \theta^\pm \p_{\pm\pm}
~ \; \; ,
\\
 Q_\pm = i\frac{\p} {\p \theta^\pm}+\bar \theta^\pm \p_{\pm\pm}
~, \; \;
\bar  Q_\pm = -i\frac{\p} {\p\bar  \theta^\pm}  -    \theta^\pm \p_{\pm\pm}
\; \; .
\end{gather}
These satisfy
\begin{gather}
 D_\pm^2 = \bar  D_\pm^2=0
~, \qquad  
\{ D_\pm , \bar  D_\pm \}=-2i \p_{\pm\pm}
~, \qquad   
[D_\pm , \p_{\pm \pm} ]=[\bar D_\pm , \p_{\pm \pm} ]=0
~, \quad
\\
 Q_\pm^2 = \bar  Q_\pm^2=0
~, \qquad  
\{ Q_\pm , \bar  Q_\pm \}=-2i \p_{\pm\pm}
~, \qquad   
[Q_\pm , \p_{\pm \pm} ]=[\bar Q_\pm , \p_{\pm \pm} ]=0
~. \quad
\end{gather}
There is one more caveat worth mentioning: in much of the $\cN=(2,2)$ literature, twisted-chiral multiplets, often denoted $\S$ in this context, naturally arise as field strengths for $\cN=(2,2)$ vector superfields $V$. The lowest component of such a superfield is a complex scalar, but the top component proportional to $\bar{\theta}^-\theta^+$ encodes the gauge-field strength along with a real auxiliary field. On the other hand, there are twisted chiral superfields denoted $Y$ whose bottom component is a complex scalar and whose top component is just a complex auxiliary field. In this work we restrict to the latter case. 
The free Lagrangians for these supermultiplets are given by
\bea
\mathcal{L}^{\text{c}}_{0} = \int d^4 \theta \, X\bar{X}
~,\qquad
\mathcal{L}^{\text{tc}}_{0} =
- \int d^4 \theta \, Y\bar{Y}
~.
\label{free-scaral-Lagrangians}
\eea

In \cite{Chang:2019kiu} it was shown that the following Lagrangian
\bsubea    \label{bgc_on_shell}
    \mathcal{L}^{\text{c}}_{\lambda} &=& \int d^4 \theta  \left( X \bar{X} 
    + \frac{\lambda  D_+ X \Db_+ \bar{X} D_-  X \Db_- \bar{X} }{1 - \frac{1}{2} \lambda  A 
    + \sqrt{1 - \lambda  A  + \frac{1}{4} \lambda^2  B^2 } } \right)~ ,
\eea
with
\bea
    A = \partial_{++}  X \partial_{--} \bar{X} + \partial_{++} \bar{X} \partial_{--}  X ~,
\quad
    B = \partial_{++}  X \partial_{--} \bar{X} - \partial_{++} \bar{X} \partial_{--}  X ~,
\esubea
is a solution of the flow equation \eqref{FZTTbar} on-shell, and hence describes the $T\Tb$-deformation of the free chiral supermultiplet Lagrangian \eqref{free-scaral-Lagrangians}.

A simple way to generate
the $T\Tb$-deformation of the free twisted-chiral theory is to remember that a twisted-chiral multiplet can be obtained from a chiral one
by acting with a $\mathbb{Z}_2$ 
automorphism
on the Grassmann coordinates 
of $\cN=(2,2)$ superspace:
\bea
\q^+\leftrightarrow\q^+
~,\quad
\q^-\leftrightarrow-\,\bar\q^-
~.
\label{Z2auto}
\eea
This leaves the $D_+$ and $\bar{D}_+$ derivatives  invariant while it exchanges $D_-$ with $\bar{D}_-$. As a result, 
the chiral and twisted-chiral differential constraints \eqref{chiral_twisted-chiral_constraints} are mapped into each others under the automorphism \eqref{Z2auto}.\footnote{In the literature this ${\mathbb Z}_2$ automorphism
\eqref{Z2auto} is often called  the ``mirror-map'' or ``mirror-image" because it exchanges the vector and axial $U(1)$ R-symmetries.}

Under the $\mathbb{Z}_2$ 
automorphism \eqref{Z2auto}, the Lagrangian \eqref{bgc_on_shell} turns into the following twisted-chiral Lagrangian
\bsubea    \label{bgtc_on_shell}
    \mathcal{L}^{\text{tc}}_{\lambda} &=& -\int d^4 \theta  \left(Y \bar{Y}
    + \frac{\lambda  D_+ Y \Db_+ \bar{Y} \Db_- Y D_- \bar{Y} }{1 - \frac{1}{2} \lambda  A 
    + \sqrt{1 - \lambda  A  + \frac{1}{4} \lambda^2  B^2 } } \right)~ ,
   \eea
   where
   \bea
    A = \partial_{++} Y \partial_{--} \bar{Y} + \partial_{++} \bar{Y} \partial_{--} Y~,
\quad
    B = \partial_{++} Y \partial_{--} \bar{Y} - \partial_{++} \bar{Y} \partial_{--} Y ~. 
\esubea
Thanks to the map \eqref{Z2auto}, by construction the Lagrangian \eqref{bgtc_on_shell} is a $T\Tb$-deformation and its superspace Lagrangian
$\cA^{\rm tc}_\l$,
$\cL^{\rm tc}_\l=\int d^4\q\,\cA^{\rm tc}_\l$, is an on-shell solution of the following flow equation
\bea
    \frac{d}{d \lambda} \mathcal{A}^{\text{tc}}_\lambda = 
    \frac{1}{8} \left( \mathcal{\cR}_{++} \mathcal{\cR}_{--} 
    - 2 \mathcal{B} \bar{\mathcal{B}} \right)~ .
    \label{sc2_flow_tc}
\eea
Here $\cR_{\pm\pm}(x,\q)$, $\mathcal{B}(x,\q)$ and its complex conjugate $\bar{\mathcal{B}}(x,\q)$
are the local operators describing the $\cR$-multiplet of currents
for $D=2$ $\cN=(2,2)$ supersymmetry 
that arise by applying \eqref{Z2auto} to the FZ multiplet of the chiral theory
\eqref{bgc_on_shell}~\cite{Chang:2019kiu}.
They satisfy the conservation equations, 
\be
    \Db_{+} \mathcal{R}_{--} 
    = i \Db_-{\cal B}
    ~,
\qquad
   D_{-} \mathcal{R}_{++} 
    = iD_+{\cal B}
    ~,\qquad
    \Db_+ {\cal B}=D_-{\cal B}=0~,
\label{conservation-R}
\ee
together with their complex conjugates.
Like the case of the FZ-multiplet, 
the supercurrent-squared operator
\be
\calT\calTb(x)
  = \int d^4 \theta \, 
  \mathcal{O}^{\cR}(x,\q)
  ~,\quad
  \cO^{\cR}(x,\q)
  :=
- \mathcal{R}_{++}(x,\q) \mathcal{R}_{--}(x,\q) + 2 \mathcal{B}(x,\q) \bar{\mathcal{B}}(x,\q)
~,
\ee
satisfies \eqref{calTTb=TTb}; namely, $\calT\calTb(x)$
is equivalent to $T\bar{T}(x)$ up to total derivatives and ${\rm EOM}$, as we showed in Chapter \ref{CHP:SC-squared-2}.

Note that the bosonic truncation of both \eqref{bgc_on_shell} and \eqref{bgtc_on_shell} give the Lagrangian
\begin{equation}
\label{eq:bosonTTbarNonlinear}
 {\cal L}_{\lambda,\text{bos}}= 
 \frac{\sqrt{1+2 \lambda  a +\lambda^2 b^2}-1}{4\lambda}
 =
  \frac{  a}{4 } - \lambda \frac{ \pa_{++} \phi \pa_{--}\phi \pa_{++} \bar \phi \pa_{--} \bar\phi }
 {  1+\lambda a+\sqrt{1+2 \lambda  a +\lambda^2 b^2}  } 
 ~,
\end{equation}
where
\begin{equation}
  \label{eq:xydef}
 a=\pa_{++} \phi \pa_{--} \bar \phi +\pa_{++}  \bar \phi \pa_{--}  \phi
 ~, \qquad 
b= \pa_{++} \phi \pa_{--} \bar \phi -\pa_{++}  \bar \phi \pa_{--}  \phi ~,
\end{equation}
and $\phi$ is either $\phi= X|_{\q=0}$ or $\phi=Y|_{\q=0}$.
This is the Lagrangian for the gauge-fixed Nambu-Goto string in four dimensions
\cite{Cavaglia:2016oda}.

The aim of the remainder of this section is to show
that the Lagrangians \eqref{bgc_on_shell} and \eqref{bgtc_on_shell}
are structurally identical to the Bagger--Galperin action
for the $D=4$ $\cN=1$ supersymmetric Born-Infeld theory 
\cite{Bagger:1996wp}, which we will analyse in detail in section \ref{BI-flows}. 
Since the Bagger--Galperin action possesses a second non-linearly realized $D=4$, $\cN=1$ supersymmetry,
we will show that the theories described by \eqref{bgc_on_shell} and \eqref{bgtc_on_shell} also possess an extra set of non-linearly realized $\cN=(2,2)$ supersymmetries.

\subsection{The \texorpdfstring{$T \Tb$}{T Tbar}-deformed twisted-chiral model and  partial-breaking}

Let us start with the twisted-chiral Lagrangian \eqref{bgtc_on_shell}
which, as we will show, is the one more directly related to the $D=4$ Bagger-Galperin action.
In complete analogy to the $D=4$ case, we are going to show that \eqref{bgtc_on_shell} is a model for a Nambu-Goldstone multiplet of $D=2$ 
$\cN=(4,4)\to \cN=(2,2)$ partial supersymmetry breaking.
The analysis is similar in spirit to the $D=4$ construction of the Bagger-Galperin action using $D=4$ $\cN=2$ superspace proposed by Ro\v{c}ek and Tseytlin  \cite{Rocek:1997hi}; see also
\cite{Kuzenko:2015rfx,Antoniadis:2017jsk,Antoniadis:2019gbd} for  more recent analysis.

To describe manifest $\cN=(4,4)$ supersymmetry we can use $\cN=(4,4)$ superspace which augments the $\cN=(2,2)$ superspace coordinates $\z^M=(x^{\pm\pm},\q^{\pm},  \bar \q^\pm)$  of the previous section with
the following additional complex Grassmann coordinates $(\eta^{\pm},  \bar \eta^\pm)$.  
The extra supercovariant derivatives and supercharges are given by
\bsubeq
\bea
 \cD_+ 
 &=&
 \frac{\p} {\p \eta^+ }+i \bar \eta^+ \p_{++ }
 ~, \quad
\bar \cD_+ =-\frac{\p} {\p \bar\eta^+ } - i  \eta^+ \p_{++ }
~,
\\
 \cQ_+ 
 &=&
 i \frac{\p} {\p \eta^+ }+  \bar \eta^+ \p_{++ }
 ~, \quad
\bar \cQ_+ =- i \frac{\p} {\p \bar\eta^+ } -    \eta^+ \p_{++ }
~,
\eea
 \esubeq
with similar expressions for $\cD_-$ and $\cQ_-$. They satisfy
\begin{gather}
 \cD_\pm^2 = \bar  \cD_\pm^2=0
~, \quad  
\{ \cD_\pm , \bar  \cD_\pm \}=-2i \p_{\pm\pm}
~, \quad   
[\cD_\pm , \p_{\pm \pm} ]=[\bar \cD_\pm , \p_{\pm \pm} ]=0
~,~~~~~~
\\
 \cQ_\pm^2 = \bar  \cQ_\pm^2=0
~, \quad  
\{ \cQ_\pm , \bar  \cQ_\pm \}=-2i \p_{\pm\pm}
~, \quad   
[\cQ_\pm , \p_{\pm \pm} ]=[\bar \cQ_\pm , \p_{\pm \pm} ]=0
~,~~~~~~
\end{gather}
while they (anti-)commute with all the usual $D_\pm$ and $Q_{\pm}$ operators.

Two-dimensional $\cN=(4,4)$ supersymmetry can also be usefully described in the language of $\cN=(2,2)$ superspace.
In this section, we will largely refer to \cite{Ivanov:2004yv} for such a description. 
In this approach from the full $(4,4)$ supersymmetry, one copy of $(2,2)$ is manifest while a second $(2,2)$ is hidden. 
For our goal of describing a model of partial supersymmetry breaking, 
we view the hidden $(2,2)$ supersymmetry as broken and non-linearly realized. 
We will derive such a description starting from $\cN=(4,4)$ superspace and describe the broken/hidden supersymmetry 
using the $\eta^{\pm}$ directions.

The hidden supersymmetry transformation of a generic $D=2$ $\cN=(4,4)$ superfield 
$U=U(x^{\pm\pm},\q^\pm,\qb^\pm,\eta^\pm,\bar\eta^\pm)$ under the hidden $(2,2)$ supersymmetry  is
 \be
 \delta U= i(\epsilon^+ \cQ_+ +\epsilon^- \cQ_- - \bar \epsilon^+  \bar \cQ_+-\bar \epsilon^- \bar \cQ_-  )U
 ~.
 \label{hidden22_1}
 \ee
The $(2,2)$ supersymmetry, generated by the $Q_\pm$ and $\bar{Q}_\pm$ operators,
will always be manifest and preserved,
so we will not bother to discuss it in detail. 
For convenience, we also introduce the chiral coordinate $y^{\pm\pm}=x^{\pm\pm} + i \eta^\pm \bar \eta^\pm $.
Using this coordinate, the spinor covariant derivatives and supercharges take the form
\bsubea
 \cD_\pm 
 &=&
 \frac{\p} {\p \eta^\pm } + 2 i  \bar\eta^\pm \frac{\p}{\p y^ {\pm\pm }}
 ~,  \quad
\bar \cD_\pm =-\frac{\p} {\p \bar\eta^\pm }
~,
\\
 \cQ_\pm &=&
 i \frac{\p} {\p \eta^\pm } ~, \quad
\bar \cQ_\pm =- i \frac{\p} {\p \bar\eta^\pm }  - 2  \eta^\pm\frac{\p}{\p y^ {\pm\pm }} 
~.
\esubea

After this technical introduction, let us turn to our main construction.
  Consider a $(4,4)$  superfield   which is   chiral under the hidden $(2,2)$ supersymmetry: 
 \be
\bar  \cD_\pm {  \bm{\mathcal X}} =0
~.
 \ee
We can  expand it in terms of   hidden fermionic coordinates, 
\be
  {\bm{\mathcal X}} =X +  \eta^+ X_++  \eta^- X_- +   \eta^+  \eta^-F
  ~,
  \label{calX}
\ee
 where $X=X(y^{\pm\pm},\q^\pm,\bar\q^\pm)$,
 $X_\pm=X_\pm(y^{\pm\pm},\q^\pm,\bar\q^\pm)$
 and $F=F(y^{\pm\pm},\q^\pm,\bar\q^\pm)$ 
 are   themselves $(2,2)$
 superfields. In the following discussion, we will keep
 the $\theta^\pm ,\,\bar \theta^\pm$ dependence implicit. 
 The hidden $(2,2)$ supersymmetry transformation rules 
 can then be straightforwardly computed using \eqref{hidden22_1} and \eqref{calX}. 
 They take the form 
 \bsubeq
 \label{susyYsfXY}
 \bea
 \delta X &=&  - \epsilon^+ X_+   - \epsilon^- X_-  
 ~,
 \label{susyYsfXY_a}
 \\
 \d X_\pm&=& \mp \epsilon^\mp F -2i \bar \epsilon ^\pm \p_{\pm \pm }X,
  \label{susyYsfXY_b}
 \\
  \delta F &=&  - 2i \bar \epsilon^-  \p_{--} X_+ + 2i \bar \epsilon^+ \p_{++} X_-
  ~.
   \label{susyYsfXY_c}
\eea
\esubeq
The $\bm{\mathcal X}$ superfield is still reducible under $\cN=(4,4)$ supersymmetry so we can put additional constraints on the $(2,2)$ superfields $X$, $X_\pm$ and $F$. 
Here we will consider $(4,4)$ twisted multiplets, and refer the reader to \cite{Gates:1984nk,Gates:1983py,Gates:1995aj,Gates:1998fr,Ivanov:1984ht,Ivanov:1984fe,Ivanov:1987mz} for a more detailed analysis.
For this discussion, we will follow the $\cN=(2,2)$ superspace description of \cite{Ivanov:2004yv}.
One type of twisted multiplet with $(4,4)$ supersymmetry can be defined by setting  
\be
X_+= \bar  D_+ \bar Y, \qquad X_-=  -\bar  D_-  Y
~,
\ee
where  $X$ and $Y$ are chiral and twisted-chiral, respectively, under the manifest $(2,2)$ supersymmetry: 
\be\label{XYchiral}
\bar  D_+ X=\bar  D_- X=\bar  D_+ Y =   D_- Y=0
~, \qquad
  D_+ \bar  X=   D_- \bar  X=   D_+ \bar  Y = \bar   D_-  \bar  Y=0
 ~.
\ee
The superfield \eqref{calX} becomes
\be
{{\bm{\mathcal X}} }=X+   \eta^+    \Db_+ \bar Y  - \eta^- \bar  D_- Y+    \eta^+ \eta^-F
~.
\ee
The supersymmetry transformation rules then become
 \bsubea\label{susyYsfXY-2}
 \delta X &=&  - \epsilon^+  \bar  D_+ \bar Y   + \epsilon^-  \bar  D_-  Y 
 ~,
 \label{susyYsfXY-2_X}
\\
  \delta F &=&  - 2i \bar \epsilon^-  \p_{--}  \bar  D_+ \bar Y - 2i \bar \epsilon^+ \p_{++}   \bar  D_-  Y
  ~,
 \esubea
 while $\d X_\pm$ remains the same as \eqref{susyYsfXY_b}.
By using the conjugation property for two fermions,  
$\overline {\chi\xi} =\bar\xi\bar \chi =-\bar\chi \bar \xi $, 
and the conjugation property $\overline{ D_+ A} =\bar  D_+ \bar A $ 
for a bosonic superfield $A$, it follows that
 \be
  \delta \bar X =    \bar \epsilon^+   D_+ Y   -  \bar \epsilon^-    D_-  \bar  Y 
  ~.
 \ee
  One can check that
\be
 \delta  \bar  D^2 \bar X =\bar  D^2 \delta  \bar X
 =\bar  D^2 \Big(    \bar \epsilon^+   D_+ Y   -  \bar \epsilon^-    D_-  \bar  Y    \Big) 
 =2i \bar\epsilon^+ \p_{++} \bar  D_- Y +2i \bar \epsilon^- \p_{--}\bar D_+ Y
 ~,
\ee
where $\bar  D^2 =\bar  D_+ \bar  D_-$. Note that in the first equality we made use of the fact that the manifest and hidden 
$(2,2)$ supersymmetries are independent.   The supersymmetry transformation rule for $-\bar  D^2 X$ is then exactly that of the auxiliary field $F$. Thus we can  consistently set
\bea
F= -\bar  D^2\bar X
~,
\eea
which is the last constraint necessary to describe a version of the $(4,4)$ twisted multiplet
in terms of a chiral and twisted-chiral $\cN=(2,2)$ superfields.
The resulting  $(4,4)$ superfield ${{\bm{\mathcal X}} }$,
expanded in terms of the hidden $(2,2)$ fermionic coordinates, takes the form
\be
{{\bm{\mathcal X}} }=X+   \eta^+    \Db_+ \bar Y  - \eta^- \bar  D_- Y  -    \eta^+ \eta^-\bar  D^2\bar X
~,
\label{calX1}
\ee
which closely resembles the expansion of a $D=4$ $\cN=2$ vector multiplet when one identifies 
the analogue of the $D=4$ $\cN=1$ chiral vector multiplet field strength $W_\a$ with the $(2,2)$ chiral superfields
$\Db_+\bar{Y}$
and
$\Db_- Y$.
Note in particular that ${{\bm{\mathcal X}} }$ turns to be $(4,4)$ chiral:
 \be
\bar  D_\pm {  \bm{\mathcal X}} =0
~,\quad
\bar  \cD_\pm {  \bm{\mathcal X}} =0
~.
 \ee
To summarize: the entire $(4,4)$ off-shell twisted multiplet is described in terms of one chiral and one twisted-chiral $(2,2)$ superfield, which 
possess the following hidden $(2,2)$ supersymmetry transformations:
 \bsubea\label{susyYsfXY-3}
 \delta X &=&  - \epsilon^+  \bar  D_+ \bar Y   + \epsilon^-  \bar  D_-  Y 
 \label{susyYsfXY-3-X}
 ~,\\
 \d Y&=& \bar \epsilon^- D_- X+\epsilon^+ \bar D_+ \bar X
  ~.
 \esubea

Let us now introduce the action for a free $\cN=(4,4)$ twisted multiplet.
Taking the square of ${{\bm{\mathcal X}} }$ in \eqref{calX1} we obtain
\be
{ {\bm{\mathcal X}} } ^2= \eta^+  \eta^-  \Big( -2X   \bar  D^2 \bar X+2  \bar  D_+ \bar Y \bar  D_- Y \Big) +\ldots
~,
\ee
where the ellipses denote terms that are not important for our analysis. 
Since ${\bm{\mathcal X}}  $ and therefore ${\bm{\mathcal X}} ^2$ are chiral superfields, we can consider the chiral integral in the hidden 
direction
\be
\int d \eta^+ d  \eta^-{{\bm{\mathcal X}} } ^2 =2X   \bar  D^2 \bar X-2  \bar  D_+ \bar Y \cdot  \bar  D_- Y
~.
\ee
 Note also that,  since $X$ and $Y$ are chiral and twisted-chiral  under the manifest supersymmetry \eqref{XYchiral}, it follows that 
\begin{align}
\int d^2x\,d\theta^+ d\theta^- d\bar \theta^+ d\bar \theta^- (X\bar X-Y \bar Y)
 &= \int d^2x\,d\theta^+ d\theta^-  \bar  D _+  \bar  D_- (X\bar X-Y \bar Y)
~,
\nonumber \\
&= \int d^2x\,d\theta^+ d\theta^-   \Big(X \bar  D _+  \bar  D_-\bar X-  \bar  D_+ \bar Y \cdot  \bar  D_- Y 
 \Big) \, , 
\end{align}
which can also be rewritten as 
\be
\int d^2x\,d\theta^+ d\theta^- d\bar \theta^+ d\bar \theta^- (X\bar X-Y \bar Y)
=
\int d^2x\,d\bar\theta^+ d\bar\theta^-   \Big(\bar X    D _+     D_-  X-    D_+  Y \cdot     D_- \bar Y  \Big) 
~.
\ee
The sum of the two equations above  yields
\be\label{action44SUSY}
4 \int d^2x\,d\theta^+ d\theta^- d\bar \theta^+ d\bar \theta^- (X\bar X-Y \bar Y) 
= \int d^2x\,d\theta^+ d\theta^-    d \eta^+ d  \eta^-{ {\bm{\mathcal X}} } ^2 +{}c.c.~.
\ee
The left-hand side has an enhanced $\cN=(4,4)$ supersymmetry as discussed in  \cite{Ivanov:2004yv}. This becomes manifest from our $(4,4)$ superspace construction on the right-hand side.

To describe $\cN=(4,4)\to\cN=(2,2)$ supersymmetry breaking
we can appropriately deform the $(4,4)$ twisted multiplet. 
Analogous to the case of a $D=4$ $\cN=2$ vector multiplet deformed by a magnetic Fayet-Iliopoulos term \cite{Antoniadis:1995vb}
(see also \cite{Antoniadis:2019gbd,Rocek:1997hi,Ivanov:1998jq,Kuzenko:2015rfx,Antoniadis:2017jsk}), 
we add a deformation parameter to the auxiliary field $F$ of ${{\bm{\mathcal X}} }$,
which is deformed to
\be
{{\bm{\mathcal X}} }_{\text{def}}=X+   \eta^+    D_+ \bar Y  - \eta^- \bar  D_- Y  -    \eta^+ \eta^- \Big( \bar  D^2\bar X+\kappa\Big)
~.
\ee
Assuming that the auxiliary field $F$ gets a VEV, $\langle F\rangle=\kappa$ or equivalently $\langle \Db^2\bar X\rangle=0$,
then by looking at the supersymmetry transformations of $X_\pm$ for the deformed multiplet
\bea
\d X_\pm
&=&
  \pm \epsilon^\mp \Big( \bar  D^2\bar X+\kappa\Big)
  -2i \bar \epsilon ^\pm \p_{\pm \pm }X
~,
\label{dYdef}
\eea
we can see the $\cN=(4,4)\to\cN=(2,2)$ supersymmetry breaking pattern arises; specifically, the hidden $\cN=(2,2)$ is spontaneously broken and non-linearly realized.
For later use, it is important to stress that, though the hidden transformations of $\d X_\pm$ are modified by the non-linear term
proportional to $\kappa$,
the hidden transformation of $X$ remains the same as in the undeformed case given in eq.~\eqref{susyYsfXY-2_X}.

In analogy to the $D=4$ case of \cite{Rocek:1997hi,Kuzenko:2015rfx,Antoniadis:2017jsk}, 
to describe the Goldstone multiplet associated to partial supersymmetry breaking
we impose the following nilpotent constraint on the deformed $(4,4)$ twisted superfield:
\be
{ {\bm{\mathcal X}} }_{\text{def}} ^2=0=-2   \eta^+  \eta^-  \Big(    X (\kappa+   \bar  D^2 \bar X )-  \bar  D_+ \bar Y  \cdot \bar  D_- Y \Big) 
+\ldots
~.
\ee
This implies the constraint 
\be
 X  \Big(\kappa +   \bar  D^2  \bar X \Big)-  \bar  D_+ \bar Y \cdot  \bar  D_- Y=0
 ~,
 \label{BG44-1}
\ee
which requires
\be\label{Xeq}
X=\frac{ \bar  D_+ \bar Y \cdot  \bar  D_- Y}{\kappa+   \bar  D^2\bar  X}  =\frac{ W^2}{ {\kappa}+   \bar  D^2\bar  X}
~,
\ee
and its conjugate 
\be\label{Xbeq}
\bar{X}=
-\frac{  D_+ Y \cdot   D_- \bar{Y}}{\kappa+   D^2X}
=-\frac{     \bar W^2 }{ {\kappa}+     D^2  X}
~.
\ee
Here $\bar  D^2 =\bar  D_+ \bar  D_-,   D^2 =-  D_+   D_-$ and we have  introduced the superfields:
\bsubea
W^2 
&=& 
   -X_+ X_-
   =   \bar  D_+ \bar Y \cdot  \bar  D_- Y 
   =\bar  D_+      \bar  D_-  (Y \bar Y)= \bar D^2 (Y\bar Y)~,
   \\
\bar W^2 &=&
   \bar{X}_+ \bar X_-
   =   -   D_+   Y \cdot   D_- \bar Y 
   = -  D_+   D_-  ( Y\bar Y) =   D^2 (Y\bar Y)
~.
\esubea
The constraint \eqref{BG44-1} is the $D=2$ analogue of the Bagger-Galperin constraint for a Maxwell-Goldstone multiplet for 
$D=4\,$ $\cN=2\to\cN=1$ supersymmetry breaking \cite{Bagger:1996wp}. 
 Combining \eqref{Xeq} and \eqref{Xbeq} gives 
\be\label{XwithY}
\kappa X=\bar D^2 (   Y\bar Y - X\bar X)=\bar D^2 \Big[ Y\bar Y - \frac{   \bar  D_+ \bar Y \cdot  \bar  D_- Y\cdot    D_- \bar Y \cdot    D_+  Y  }{
( {\kappa}+     D^2  X)( {\kappa}+   \bar  D^2  \bar X)}   \Big]
~,
\ee
which is consistent thanks to the $\kappa$ terms in the denominator.
Because of the four fermion coupling in the numerator of the last term, 
 no fermionic terms can appear in the denominator. So effectively we have the equation
\be
(\kappa + D^2 X)_{\text{eff}}  =  \Big(\kappa + D^2  \frac{ W^2}{ {\kappa}+   \bar  D^2\bar  X}\Big)_{\text{eff}}
=\kappa+\frac{ D^2 W^2}{\kappa+(\bar D^2 \bar X)_{\text{eff}}}
~,
\ee
and its conjugate 
\be
(\kappa +\bar D^2\bar X)_{\text{eff}} = \kappa+\frac{ \bar  D^2 \bar W^2}{\kappa+(   D^2 X)_{\text{eff}}}
~.
\ee
Solving them we get 
\bsubea
(   D^2  X)_{\text{eff}} &=&\frac{B-\kappa^2 +\sqrt{B^2 +2\kappa^2 A+\kappa^4}   }{2\kappa}
~, 
\\
( \bar D^2\bar X)_{\text{eff}} &=&\frac{-B-\kappa^2 +\sqrt{B^2 + 2\kappa^2 A+\kappa^4}   }{2\kappa}
~.
\esubea
Substituting these expressions into \eqref{XwithY} gives
\be
X=\frac{1}{\kappa} \bar D^2\Upsilon
~, \qquad 
\bar  X=\frac{1}{\kappa}   D^2 \Upsilon
~, \qquad 
 \Upsilon=\bar\Upsilon= Y\bar Y - \frac{2 W^2 \bar W^2}{A+\kappa^2 +\sqrt{B^2 + 2\kappa^2 A+\kappa^4}   }  
 ~,
\ee
where 
\begin{align}\label{2DAB}
A&= D^2 W^2 +\bar D^2 \bar W^2=\{  D^2, \bar  D^2 \}(Y\bar Y)
=\partial_{++} Y \partial_{--} \bar{Y} + \partial_{++} \bar{Y} \partial_{--} Y
~ ,
\\
B&=
 D^2 W^2 -\bar D^2 \bar W^2= [  D^2, \bar  D^2  ](Y\bar Y)
 =\partial_{++} Y \partial_{--} \bar{Y} - \partial_{++} \bar{Y} \partial_{--} Y
~.
\end{align}
The result is that the $\cN=(2,2)$ chiral part  $X$ of the $\cN=(4,4)$ twisted multiplet is expressed 
in terms of the $(2,2)$ twisted-chiral superfield $Y$. 
Thanks to the linearly realized construction in terms of $(4,4)$ superfields, it is straightforward
to obtain the non-linearly realized $\cN=(2,2)$ supersymmetry transformations for $Y$. In particular, 
it suffices to look at the transformations of
$D_+Y$ and $\Db_- Y$ that
can be obtained by substituting back the composite expression for $X=X[Y]$ into the 
transformations \eqref{dYdef}. By construction, these expressions ensure that $\d X$ transforms according to \eqref{susyYsfXY-2_X}.

Since $X$ is chiral under the manifest $(2,2)$ supersymmetry \eqref{XYchiral}, we can consider the chiral  integral
\bea
S_{\kappa^2}&=&
-\frac12 \kappa\int d^2 x \,d\theta^+ d\theta^- X+c.c
=-\frac12  \int d^2x\,d\theta^+ d\theta^-   \bar D^2\Upsilon +c.c.
\non\\
&=&-  \int d^2x\,d\theta^+ d\theta^- d \bar \theta^+ d\bar \theta^-   \Upsilon  
 ~.
\eea
A remarkable property of this action is that
it is invariant under the hidden non-linearly realized supersymmetry. 
Using \eqref{susyYsfXY-2_X}, we see that 
\begin{align}
\delta S_{\kappa^2} &=
-\frac12 \kappa\int d^2 x D_+ D_- \delta X\Big|_{\theta=\bar\theta=0}+c.c.~, \nonumber 
 \\
 &=
 - \frac12 \kappa\int d^2 x  \Big( -2i  \epsilon^-  \p_{--} D_+  Y 
-2i  \epsilon^+ \p_{++}     D_- \bar Y\Big) \Big|_{\theta=\bar\theta=0}+c.c.=0 \label{deriv-susy-1}
~,
\end{align}
where we used the fact that $Y$ is a twisted-chiral superfield \eqref{XYchiral}.  

Explicitly, the action reads 
\be
S_{\kappa^2}= -\int d^2x\,d\theta^+ d\theta^- d \bar \theta^+ d\bar \theta^-  
\Bigg( Y\bar Y - \frac{  2W^2 \bar W^2}{\kappa^2+A+ \sqrt{\kappa^4+2 \kappa^2 A+ B^2  }   } \Bigg)
~,
\ee
which precisely matches the model of eq.~\eqref{bgtc_on_shell} if we 
identify the coupling constants:
\be
\l =-\frac{2}{\kappa^2}
~.
\ee
This shows explicitly that the $T\Tb$-deformation of the free twisted-chiral action possesses a non-linearly realized 
$\cN=(2,2)$ hidden 
supersymmetry.

\subsection{The \texorpdfstring{$T \Tb$}{T Tbar}-deformed chiral model and  partial-breaking}

Let us now turn to the $T\Tb$ deformation of the free chiral model of eq.~\eqref{bgc_on_shell}. The construction follows the previous
subsection with the difference that we will start with a different formulation of the $(4,4)$ twisted multiplet
described in terms of $(2,2)$ superfields.
Consider again an $\cN=(4,4)$  superfield   which is   chiral under the hidden (2,2) supersymmetry: 
 \be
\bar  \cD_+ {  \bm{\mathcal Y}} =\bar \cD_-{  {\bm{\mathcal Y}} }=0
~.
 \ee
Its expansion in hidden superspace variables is  
\be
  {\bm{\mathcal Y}}=Y +   \eta^+ Y_++   \eta^- Y_- +   \eta^+  \eta^-G
  ~,
\ee
 where $Y=Y(y^{\pm\pm},\q^\pm,\bar{\q}^{\pm})$, $Y_\pm=Y_\pm(y^{\pm\pm},\q^\pm,\bar{\q}^{\pm})$  
 and $G=G(y^{\pm\pm},\q^\pm,\bar{\q}^{\pm})$ 
 are themselves superfields with manifest $(2,2)$ supersymmetry.
 The hidden $(2,2)$ supersymmetry transformation rules of the components are 
\bsubea\label{susyYsfXY_4}
 \delta Y &=&  - \epsilon^+ Y_+   - \epsilon^- Y_-  
 ~,\\
 \d Y_\pm&=&
  \mp \epsilon^\mp G -2i \bar \epsilon ^\pm \p_{\pm \pm }Y
 ~,
 \\
  \delta G &=&  - 2i \bar \epsilon^-  \p_{--} Y_+ + 2i \bar \epsilon^+ \p_{++} Y_-
  ~.
\esubea

This representation of $(4,4)$  off-shell supersymmetry is again reducible so we can impose constraints. As in the construction of the previous section, we impose
\be
Y_+= \bar  D_+ \bar X
~, \qquad Y_-=   D_-  X
~,
\ee
then
\be
  {\bm{\mathcal Y}} =Y +   \eta^+ \bar  D_+ \bar X  +   \eta^-   D_-  X  +   \eta^+  \eta^-G
  ~.
\ee
Here  $X$ and $Y$ are consistently chosen to be chiral and twisted-chiral under the manifest $(2,2)$ supersymmetry:
\be\label{XYchiral-2}
\bar  D_+ X=\bar  D_- X=\bar  D_+ Y =   D_- Y=0
~, \qquad
  D_+ \bar  X=   D_- \bar  X=   D_+ \bar  Y = \bar   D_-  \bar  Y=0
 ~. 
\ee
Then we have 
\be
\delta Y =-\epsilon^+ \bar  D_+ \bar X-\epsilon^-  D_- X
~,
\label{dY2}
\ee
as well as its conjugate 
\be
\delta \bar Y= \bar \epsilon^+  D_+ X +\bar\epsilon^- \bar  D_- \bar X
~.
\ee
Hence it follows that
\be
\delta (\bar D_+  D_- \bar Y) =\bar D_+  D_- \delta  \bar Y
=2i  \bar \epsilon^+ \p_{++}  D_- X - 2i \bar\epsilon^- \p_{--} \bar D_+ \bar X
~.
\ee
This should be compared with
\be
\delta G=2i  \bar \epsilon^+ \p_{++}  D_- X - 2i \bar\epsilon^- \p_{--} \bar D_+ \bar X
~,
\ee
showing that $\bar D_+  D_- \bar Y$ transforms exactly like the auxiliary field $G$. 
This enables us to further constrain the $(4,4)$ multiplet  by setting 
\be
G= \bar D_+  D_- \bar Y
~.
\ee
Imposing these conditions gives a $(4,4)$ twisted superfield 
\be
  {\bm{\mathcal Y}} =Y +   \eta^+ \bar  D_+ \bar X  +   \eta^-   D_-  X  +   \eta^+  \eta^-  \bar D_+  D_- \bar Y
  ~,
\ee
which by construction is twisted-chiral and chiral with respect to the manifest and hidden $(2,2)$ supersymmetries, respectively:
\be
\bar  D_+ {  \bm{\mathcal Y}} =
D_- {  \bm{\mathcal Y}} =0
~,\quad
\bar  \cD_\pm {  \bm{\mathcal Y}} =0
~.
\ee
Its free dynamical action can be easily constructed by considering  its square
\be
  {\bm{\mathcal Y}}^2 =2  \eta^+ \eta^- \Big(  Y  \bar D_+  D_- \bar Y - \bar  D_+ \bar X   \cdot   D_-  X   \Big) 
  +\ldots
  ~.
\ee
In fact, the following relations hold:
\begin{align}
\int d^2x\,d\theta^+ d\theta^- d\bar \theta^+ d\bar \theta^- (X\bar X-Y \bar Y)
&=
\int d^2x\,d\theta^+ d\bar \theta^- \bar D_+  D_- (X\bar X-Y \bar Y)~,
\nonumber \\
&= \int d^2x\,d\theta^+ d\bar \theta^-  \Big(  D_+   X \cdot  \bar  D_-  \bar X - \bar  Y  D_+ \bar  D_-  Y  \Big)  \, .
\end{align}
Alternatively, 
\begin{align}
\int d^2x\,d\theta^+ d\theta^- d\bar \theta^+ d\bar \theta^- (X\bar X-Y \bar Y)
&=
\int d^2x\,d \bar \theta^+ d \theta^-   D_+ \bar  D_- (X\bar X-Y \bar Y) \, , 
\nonumber \\
&=
\int d^2x\,d \bar \theta^+ d \theta^-  \Big(\bar D_+  \bar X \cdot  D_- X - Y\bar   D_+  D_- \bar Y  \Big)  \, .
\end{align}
These relations imply
\be 
4 \int d^2x\,d\theta^+ d\theta^- d\bar \theta^+ d\bar \theta^- (X\bar X-Y \bar Y) 
=\int d^2x\,d\theta^+ d\bar \theta^- d \eta^+d  \eta^-   \ { {\bm{\mathcal Y}} } ^2 +c.c.~.
\ee
Once again the  $(4,4)$  supersymmetry of the left hand side becomes manifest on the right hand side. 

As in the $(4,4)$ twisted multiplet considered in the previous subsection, 
we can deform this representation 
to induce the partial breaking.
The deformed multiplet is described by the following $(4,4)$ superfield:
\be
  {\bm{\mathcal Y}}_{\text{def}} =Y +   \eta^+ \bar  D_+ \bar X  +   \eta^-   D_-  X  +   \eta^+  \eta^-   \Big( \bar D_+  D_- \bar Y
  +\kappa\Big) 
  ~.
\ee
The hidden supersymmetry transformations of the component $(2,2)$ superfields can be straightforwardly computed using the arguments of the previous subsection. For the goal of this section, it is enough to mention that
$\d Y$ is the same as the undeformed case of eq.~\eqref{dY2}.

To eliminate half of the degrees of freedom of $  {\bm{\mathcal Y}}_{\text{def}} $ 
and describe a Goldstone multiplet for $\cN=(4,4)\to\cN=(2,2)$ partial supersymmetry breaking,
we again impose the nilpotent constraint
\be
{\bm{\mathcal Y}}^2_{\text{def}}=0 =2  \eta^+ \eta^- \Big(  Y  (\kappa+ \bar D_+  D_- \bar Y) - \bar  D_+ \bar X   \cdot   D_-  X   \Big) +\ldots
~.
\ee
This yields the following constraint for the $(2,2)$ superfields
\be
Y  (\kappa+ \bar D_+  D_- \bar Y) - \bar  D_+ \bar X   \cdot   D_-  X=0
~,
\label{BG44-2}
\ee
which is equivalent to
\be
Y=\frac{ \bar  D_+ \bar X   \cdot   D_-  X}{\kappa+ \bar D_+  D_- \bar Y }
=\frac{  \widetilde W^2  }{\kappa+ \overline{\widetilde  D}^2 \bar Y }
~,\qquad
\bar Y=\frac{ \bar  D_-  \bar X   \cdot   D_+  X}{\kappa+ \bar D_+  D_- \bar Y }
=\frac{ \bar{ \widetilde W}^2  }{\kappa+  \widetilde{   D}^2 \bar Y }
~.
\ee
Here $\bar{\widetilde  D}^2 =\bar  D_+  D_-,  \widetilde D^2 =-  D_+  \bar D_-$ and we have  introduced
the following bilinears:
\be
\widetilde W^2 \equiv   \bar  D_+ \bar X   \cdot   D_-  X =\bar{\widetilde  D}^2  (X\bar X) , \qquad
\bar{ \widetilde W}^2   \equiv  \bar  D_-  \bar X   \cdot   D_+  X  = \widetilde{   D}^2 (X\bar X)
~.
\ee
 
Using exactly the same tricks as before 
and inspired by the $D=4$ Bagger-Galperin model, 
 we can solve the constraints \eqref{BG44-2} to find
\be
Y=\frac{1}{\kappa}\bar{\widetilde  D}^2 \widetilde \Upsilon
~, \quad 
\bar  Y=\frac{1}{\kappa} \widetilde D^2 \widetilde  \Upsilon
~, \qquad 
\widetilde \Upsilon=\bar{\widetilde\Upsilon}= X\bar X - \frac{2 \widetilde W^2 \bar { \widetilde W}^2}{ \widetilde  A+\kappa^2 +\sqrt{ \widetilde  B^2 + 2\kappa^2 \widetilde  A+\kappa^4}   }  
~,
\ee
where 
\begin{gather}
 \widetilde A = \widetilde D^2  \widetilde W^2 +\bar{ \widetilde   D^2} \bar{ \widetilde  W^2}
 =\{ \widetilde   D^2, \bar{ \widetilde   D^2} \}(X\bar X)
 = \partial_{++}  X \partial_{--} \bar{X} + \partial_{++} \bar{X} \partial_{--}  X 
 ~ , 
 \\
 \widetilde  B = 
  \widetilde  D^2  \widetilde W^2 -\bar{ \widetilde   D}^2 \bar{ \widetilde  W}^2= [ \widetilde   D^2, \bar{ \widetilde   D^2 } ](X\bar X)
  =\partial_{++}  X \partial_{--} \bar{X} - \partial_{++} \bar{X} \partial_{--}  X 
 ~.
\end{gather}
Since $Y$ is twisted-chiral under the manifest $(2,2)$ supersymmetry \eqref{XYchiral-2}, we can consider the 
twisted-chiral  integral
\be
S_{\kappa^2}=\frac12 \kappa\int d^2x\,d \theta^+ d  \bar \theta^- Y+c.c
=\frac12  \int d^2x\, d   \theta^+ d \bar \theta^-  \bar{\widetilde  D}^2 \widetilde \Upsilon  +c.c.
=\int d^2x\,d\theta^+ d\theta^- d \bar \theta^+ d\bar \theta^-   \widetilde \Upsilon  
 ~.~~~~~~~~~
 \label{chiralYaction}
\ee
By using arguments analogous to those around eqs.~\eqref{deriv-susy-1} of the previous subsection, 
the action \eqref{chiralYaction} proves to be $\cN=(4,4)$ supersymmetric.

Explicitly, the action reads 
\bea
S_{\kappa^2}
= \int d^2x\,d\theta^+ d\theta^- d \bar \theta^+ d\bar \theta^-  
\Bigg( X\bar X - \frac{2 \widetilde W^2 \bar { \widetilde W}^2}{ \kappa^2+\widetilde  A 
+\sqrt{\kappa^4+ 2\kappa^2 \widetilde  A+ \widetilde  B^2 }   }  \Bigg)  
~,
\eea
which precisely matches the model of eq.~\eqref{bgc_on_shell} if we 
identify the coupling constants:
\be
\l =-\frac{2}{\kappa^2}
~.
\ee
This shows explicitly that the $T\Tb$ deformation of the free chiral action possesses a non-linearly realized 
$\cN=(2,2)$  
supersymmetry.


\section{\texorpdfstring{$T^2$}{T-Squared} Deformations in \texorpdfstring{$D=4$}{D=4} and Supersymmetric Extensions}\label{Tsquaredcomments} 
 
In Section~ \ref{2D} we exhibited the non-linear supersymmetry possessed by two $D=2$, $\cN=(2,2)$ 
models constructed in \cite{Chang:2019kiu} from the $T\Tb$ deformation of free actions. The striking relationship with the $D=4$ supersymmetric 
Born-Infeld (BI) theory naturally makes one wonder whether some kind of $T\Tb$ flow equation is satisfied by 
supersymmetric $D=4$ BI, 
and related actions. We will spend the rest of the chapter exploring this possibility. In this section, we start with a few general 
observations on $T^2$ or supercurrent-squared operators in $D>2$.

\subsection{Comments on the \texorpdfstring{$T^2$}{T-Squared} operator in \texorpdfstring{$D=4$}{D=4}}

In two dimensions, by $T\bar T$ we mean the operator 
 $T_{\mu\nu}T^{\mu\nu}- (T_ \mu^\mu )^2 $, which is proportional to $\det [T_{\mu\nu}]$~\cite{Zamolodchikov:2004ce,Smirnov:2016lqw,Cavaglia:2016oda}. 
 One can attempt to generalize this structure to $D>2$. 
 In general, one could consider the following  stress-tensor squared operator
 \be
 O_{T^2}^{[r]} = T^{\mu\nu}T_{\mu\nu}- r\,\Theta^2 
 ~,
 \qquad
  \Theta\equiv \tensor{T}{^\mu_\mu} 
 ~,
 \label{OT2r0}
 \ee  
with $r$ a real constant parameter.
 In two dimensions,  the  unique choice $r=1$ yields a well defined operator which  is free of short distance singularities~\cite{Smirnov:2016lqw,Zamolodchikov:2004ce}. 
 However, to the best of our knowledge, there is no analogous argument in higher dimensions that guarantees a well-defined irrelevant operator $O_{T^2}^{[r]}$ at the quantum level. 
 Nevertheless,  in a $D$-dimensional space-time, one possible extension is given by 
 $O_{T^2}^{[r]}$ with  $r=1/(D-1)$, which  reduces to the $T\bar T$ operator in two dimensions. 
 
 This operator has received some attention recently since it is motivated by a particular holographic picture in $D>2$
\cite{Taylor:2018xcy,Hartman:2018tkw}.
We will not enter into a detailed discussion of the physical properties enjoyed by  $O_{T^2}^{[1/(D-1)]}$, but simply comment that this combination is invariant under a set of improvement transformations of the stress-energy tensor.
Indeed it is easy to show that such a $T^2$ operator transforms by,
 \be
  O_{T^2}^{[1/(D-1)]}  \rightarrow  O_{T^2}^{[1/(D-1)]} +\text{total derivatives}
  ~,
 \ee
 if the (symmetric) stress-energy tensor shifts by the following improvement transformation,
  \be
 T_{\mu\nu} \rightarrow   T_{\mu\nu} +\big(\p_\mu \p_\nu-\eta_{\mu\nu}\p^2 \big)u
 ~,
  \ee
 for an arbitrary scalar field $u$.

In four dimensions, there is another choice of interest, specifically $r=1/2$. 
In fact, it was shown in \cite{Conti:2018jho} that the bosonic Born-Infeld action can be obtained by deforming the free Maxwell theory
with  the operator  $O_{T^2}^{[1/2]}$.\footnote{It is worth mentioning that another type of higher-dimensional generalization 
of $T\Tb$-deformations, specifically the operator $|\det T |^{1/(D-1)}$, was studied in \cite{Cardy:2018sdv,Bonelli:2018kik}.}
In this work, we are going to use $O_{T^2}^{[1/2]}$ as our deforming operator. Once generalized to the
supersymmetric case, we will see that this operator plays a central role for various models possessing non-linearly realized symmetries. 

One interesting property enjoyed by $O_{T^2}^{[1/2]}$ is its invariance under a shift of the Lagrangian density of the theory, or equivalently a shift of the zero point energy. This can serve as motivation for this particular combination. Under a constant  shift  of the Lagrangian 
density $\cL$, and correspondingly its stress-energy tensor,
\be
\cL \rightarrow \cL+ c
~, \qquad 
T^{\mu\nu} \rightarrow T^{\mu\nu}-c\,\eta^{\mu\nu}
~,
\ee
the composite operator $O_{T^2}^{[r]}$ transforms in the following way:
\be
 O_{T^2}^{[r]}  \rightarrow  O_{T^2}^{[r]} +2c(2r-1) \Theta+4c^2(1-r)
 ~.
\ee 
When the theory is not conformal, which is the general situation at an arbitrary point in the flow since the deformation introduces a scale, 
and $r\neq 1/2$, the operator $O_{T^2}^{[r]}$ always transforms in a non-trivial way because of the extra trace term. 
This implies that under a constant shift in the Lagrangian, the dynamics is modified which is certainly peculiar since the shift is trivial in the undeformed theory.\footnote{It is worth noting that $\TT$ in $D=2$ shares this peculiarity.}

However if $r=\frac12$,  $O_{T^2}^{[r]}$ is unaffected up to an honest field-independent cosmological constant term. The  shift of the  
 vacuum energy does not affect the dynamics of the theory, as long as the theory is not coupled to gravity. This property is especially interesting, since the $D=4$ $\cN=1$ Goldstino action, which is also given by a $\TT$ flow,
 is the low-energy description of supersymmetry breaking which can generate a cosmological 
 constant. 
For these reasons, we will study the particular operator quadratic in stress-energy tensors given by 
\be\label{Tsquare}
 O_{T^2}\equiv T^{\mu\nu}T_{\mu\nu}- \frac12\Theta^2 
 ~,
\ee 
in the remainder of this chapter. More general $T^2$ type deformations of gauge theories in any spacetime dimension will be considered in Chapter \ref{CHP:new_chapter}.

\subsection{\texorpdfstring{$\cN=1$}{N=1} supercurrent-squared operator in \texorpdfstring{$D=4$}{D=4}}

We would like to find the $\cN=1$ supersymmetric extension of the $ O_{T^2}$ operator in four dimensions. 
As reviewed in section \ref{2D},
in two dimensions the manifestly supersymmetric $T\bar T$ deformation is roughly given by the square of the supercurrent
superfields.
 One might suspect that a similar construction holds in four dimensions.

 For the remainder of this work, we will assume that the $D=4$,  $\cN=1$ supersymmetric theories under our consideration admit
 a Ferrara-Zumino (FZ)  multiplet of currents \cite{Ferrara:1974pz}. Generalizations of this case involving
 the supercurrent multiplets described in
  \cite{Komargodski:2010rb,Dumitrescu:2011iu,Gates:1981yc,Ambrosetti:2016ieg,Gates:1983nr,Dienes:2009td,Kuzenko:2010am}\ 
  might be possible, but merit separate investigation. 
 The operator content of the FZ multiplet, which has $12+12$ component fields,
 includes the conserved supersymmetry current $S_{\mu\alpha}$, its conjugate $\bar S_\mu{}^\ad$ 
 and the conserved symmetric energy-momentum tensor $T_{\mu\nu}$:
 \be
 T_{\mu\nu}=T_{\nu\mu}
 ~, \qquad \p^\mu T_{\mu\nu}=0
 ~, \qquad \p^\mu S_\mu = \p^\mu \bar S_\mu=0
 ~.
 \ee
The  FZ multiplet also includes a complex scalar field ${\sf x}$,  as well as 
 the  $R$-current vector field $j_\mu$, which is not necessarily conserved 
 \cite{Ferrara:1974pz}.
 
In $D=4$, $\cN=1$ superspace, 
the FZ multiplet is described by a vector superfield $\cJ_\mu$ and a complex scalar scalar superfield $\cX$ 
satisfying the following constraints:
 \be
 \bar D^{\dot \alpha}  \mathcal J_{\alpha \dot \alpha}=D_\alpha \mathcal X
 ~, \qquad
 \bar D_{\dot \alpha} \mathcal X=0
 ~.
 \ee
The constraints can be solved, and the FZ supercurrents expressed in terms of its $12+12$ 
independent components read\footnote{For convenience, we have rescaled the supersymmetry current compared to 
\cite{Dumitrescu:2011iu}: $ S_\mu^{\text{here}}=-i S_\mu^{\text{there}}$.}
\bea
 \mathcal J_\mu(x)  &=& j_\mu + \theta \Big(S_\mu - \frac{1} {\sqrt{2}} \sigma_\mu \bar \chi\Big) 
 +   \bar\theta \Big(\bar S_\mu + \frac{1}{\sqrt2} \bar \sigma_\mu \chi\Big) 
 +\frac{i}{2} \theta^2 \p_\mu \bar {\sf x} -  \frac{i}{2} \bar\theta^2 \p_\mu  {\sf x}
\non
\\&&
 + \theta \sigma^\nu \bar \theta 
 \Big( 2 T_{\mu\nu} -\frac23 \eta_{\mu\nu} \Theta -\frac12 \epsilon_{\nu\mu\rho\sigma} \partial^\rho j^\sigma \Big)
 \non\\
 &&
- \frac{i}{2} \theta^2\bar\theta\Big(\bar{\slashed\p}  S_\mu + \frac{1}{\sqrt2} \bar \sigma_\mu \slashed\p \bar \chi\Big) 
- \frac{i}{2} \bar\theta^2 \theta\Big( {\slashed\p} { \bar S}_\mu - \frac{1}{\sqrt2}  \sigma_\mu\bar{  \slashed \p}   \chi \Big) 
\non\\&&
+\frac{1}{2} \theta^2 \bar \theta^2 \Big(  \p_\mu\p^\nu j_\nu -\frac{1}{2} \p^2 j_\mu \Big) 
~,
\eea
and
\bsubea
\mathcal X(y)&=&{\sf x}(y)  +\sqrt2 \theta \chi(y)   +\theta^2 {\sf F}(y)
~,
~~~
\\
\chi_\alpha
&=& \frac{  \sqrt{2}   }{3}( \sigma^\mu)_{\alpha\dot\alpha}\bar S^{\dot \alpha}_\mu
~,
\qquad
{\sf F}=  \frac23 \Theta +i \partial_\mu j^\mu
  ~,
\esubea
where the chiral coordinate is defined by $y^\mu=x^\mu +i \theta\sigma^\mu \bar \theta$,
and  we used $\slashed\p=\sigma^\mu\p_\mu, \, \bar{\slashed\p}=\bar\sigma^\mu\p_\mu$.

If we seek a manifestly supersymmetric completion of the operator \eqref{Tsquare} 
by using combinations of the supercurrent superfields with dimension $4$,
it is clear that the only possibility is the full superspace
integral of a linear combination of $\cJ^2$ and $\cX\bar\cX$.
Up to total derivatives and terms that vanish by using the supercurrent conservation equations, or equivalently that vanish
on-shell, the $D$-terms of $\cJ^2$ and $\cX\bar\cX$ are given by\footnote{The composite $A$ (and analogously its conjugate $\bar{A}$)   is given by
\begin{align}
A =\Big(S_\mu - \frac{1}{\sqrt{2}} \sigma_\mu \bar \chi\Big)
\Big( {\slashed\p} { \bar S}^\mu - \frac{1}{\sqrt{2}} \sigma^\mu\bar{  \slashed \p}   \chi \Big)
=S_\mu {\slashed\p} { \bar S}^\mu  -  \bar \chi  \bar{  \slashed \p}   \chi  +\sqrt{2}  { \bar S}^\mu \p_\mu \bar\chi
+\text{total derivatives}
~.
\end{align}
The equality can be obtained with some algebra. Note that the last term drops  
after integration by parts because of the conservation equation for $S_\mu$.
}
\bsubea
\mathcal J^2 |_{\theta^2\bar\theta^2} \equiv \eta^{\mu\nu} \cJ_\mu \cJ_\nu|_{\theta^2\bar\theta^2}
&=&
-\frac12  \Big( 2 T_{\mu\nu} -\frac23 \eta_{\mu\nu} \Theta -\frac12 \epsilon_{\nu\mu\rho\sigma} \partial^\rho j^\sigma \Big)
 ^2  +j^\mu \Big(  \p_\mu\p^\nu j_\nu -\frac12 \p^2 j_\mu \Big)  
\non\\
&&
 +\frac12 \p_\mu {\sf x}\p^\mu \bar {\sf { x}}
+\frac{i}{2} \Big(A-\bar A \Big)
\\&=&
-2(T_{\mu\nu})^2   +\frac49 \Theta^2 - \frac54  \Big( \p_  \mu j^\mu \Big)^2 -\frac34 j_\mu \p^2  j^\mu + \frac12 \p_\mu {\sf \bar x}\p^\mu {\sf x}
\non\\
&&
 +i  \Big( S_\mu {\slashed\p} { \bar S}^\mu  -   \bar \chi  \bar{  \slashed \p}   \chi 
 \Big)
  +\text{total derivatives} 
  +{\rm EOM}
  ~,
\esubea
and
\bsubea
\mathcal X \bar\cX |_{\theta^2\bar\theta^2} 
&=&
  {\sf F} \bar   {\sf F} -\p_\mu {\sf x}\p^\mu\bar {\sf x}
-i \bar \chi{ \bar{\slashed \p }}  \chi  +\text{total derivatives} 
\\
&=&   \frac49 \Theta^2 +(\p_\mu j^\mu)^2-\p_\mu {\sf x}\p^\mu\bar {\sf x}
-i \bar \chi{ \bar{\slashed \p }}  \chi
 +\text{total derivatives} 
 ~.
\esubea

To get a manifestly supersymmetric extension of  $O_{T^2}=T^2 -\frac12 \Theta^2$, we have to consider the following linear 
combination 
   \be\label{superTsquare}
  \mathcal O_{T^2}= -\frac{1}{2} \Big( \eta^{\mu\nu} \mathcal J_\mu \mathcal  J_\nu+\frac54 \mathcal X\bar {\mathcal X} \Big)= 
   \frac{1}{16} \mathcal  J^{\alpha\dot\alpha}   \mathcal J_{\alpha\dot\alpha}- \frac58  \mathcal  X    \bar {\mathcal    X}
   ~.
   \ee 
In fact, the supersymmetric descendant of the supercurrent-squared operator $  \mathcal O_{T^2}$ is 
 \bsubeq
 \label{superTsquare2}
 \bea
{\sf  O}_{T^2}&=& \int d^4 \theta \,  \mathcal O_{T^2} 
\\&=&
 T^2 -\frac12\Theta^2 +\frac38 j_\mu \p^2 j^\mu +\frac38 \p_\mu {\sf x}\p^\mu\bar {\sf x}
-\frac{i}{2}  \Big( S_\mu {\slashed\p} { \bar S}^\mu  - \frac94  \bar \chi  \bar{  \slashed \p}   \chi   \Big)
\non\\
&&
 +\text{total derivatives}+{\rm EOM}
~.
 \eea
 \esubeq
This result shows that ${\sf O}_{T^2}$ is the natural supersymmetric extension of $O_{T^2}$. 
However, it is worth emphasizing that in the $D=4$ case,  the supersymmetric descendent ${\sf  O}_{T^2}$ of $\cO_{T^2}$ 
has extra non-trivial contributions from other currents. This should be contrasted with the $D=2$ case where ${\sf  O}_{T^2}=O_{T^2}$
up to ${\rm EOM}$ and total derivatives, see eq.~\eqref{calTTb=TTb}.

It actually does not seem possible to find a linear combination of $\cJ^2$ and $\cX\bar\cX$ such that an analogue 
of eq.~\eqref{calTTb=TTb}  holds in $D=4$.
This suggests that, in contrast with the $D=2$ case, deformations of a Lagrangian triggered
by the operators $O_{T^2}$ and ${\sf  O}_{T^2}$
will in general lead to different flows: one manifestly supersymmetric, while the other not.


\section{Bosonic Born-Infeld as a \texorpdfstring{$T^2$}{T-Squared} Flow}
\label{bosonicBI}

It was shown in \cite{Conti:2018jho} that the $D=4$ Born-Infeld action arises from a $D>2$ generalization of the $T\bar T$ deformation.  Specifically, the operator driving the flow equation was shown to be the $ O_{T^2}$ defined in eq.~\eqref{Tsquare} of the preceding section.
In this section we review this result  in detail as it is a primary inspiration for our supersymmetric extensions. 
   
The $D=4$ bosonic BI action on a flat background
 is given by 
\bea
S_{\text{BI }}
&=&    \frac{1}{\alpha^2} \int d^4x\;  \Big[  1-  \sqrt{-\det (\eta_{\mu\nu}+\alpha F_{\mu\nu } )}   \Big]~,
\non\\
&=&  \frac{1}{\alpha^2}\int d^4x\;  \Big[  1-  \sqrt{1+\frac{\alpha^2}{2}  F^2 -\frac{\alpha^4}{16} (F\tilde F)^2    }   \Big]~,
\non\\
&=&
 -\frac14\int d^4x\;  F^2
 +{\rm higher~derivative~terms}
~,
\label{flat_BI}
\eea
where $F_{\mu\nu}=(\pa_{\mu}v_\nu-\pa_{\nu}v_\mu)$ is the field strength for an Abelian gauge field $v_\mu$, and
\be
F^2\equiv F_{\mu\nu}F^{\mu\nu}, \qquad   F\tilde F\equiv F_{\mu\nu} \tilde F^{\mu\nu}=\frac12 \epsilon_{\mu\nu\rho\sigma}F^{\mu\nu}F^{\rho\sigma}
~.
\ee

The Hilbert stress-energy tensor
  for the BI action can be  computed straightforwardly and it reads  \cite{Rasheed:1997ns} 
\be 
T^{\mu\nu} =-\frac{  F^{\mu   \lambda} F^\nu{}_{\lambda  } +
 \frac{1}{\alpha^2}  \Big(\sqrt{1+\frac{\alpha^2}{2}  F^2 -\frac{\alpha^4}{16} (F\tilde F)^2    } -1-\frac{\alpha^2}{2}  F^2\Big) \eta^{\mu\nu}   }
 {  \sqrt{1+\frac{\alpha^2}{2}  F^2 -\frac{\alpha^4}{16} (F\tilde F)^2    }  }
 ~.
\ee
This can be written in the following useful form
\be
T^{\mu\nu}   =   \frac{  T^{\mu\nu} _{\text{Maxwell}}  } { \sqrt{1+2 A +  B^2    }  } 
+
\frac{  \eta ^{\mu\nu} }{\alpha^2\sqrt{1+2    A +   B^2    }  }
\frac{ A^2 - B^2}{ 1+  A  +\sqrt{1+2    A +  B^2    }   } 
~,
\ee
where we used the stress-energy tensor for the Maxwell theory 
\be\label{maxwell_lag_nonlinear_chapter}
T^{\mu\nu} _{\text{Maxwell}} = - F^{\mu   \lambda} F^\nu{}_{\lambda}  +\frac14 F^2  \eta ^{\mu\nu}
~,
\ee
while $A$ and $B$ are defined by
\be\label{4DAB}
A=\frac14 \alpha^2 F^2
~, \qquad 
B=\frac{i}{4}   \alpha^2 F\tilde F
~.
\ee
 It is easy to compute the trace of the stress-energy tensor 
\be
\Theta=T^{\mu\nu} \eta_{\mu\nu} = \frac{ 4}{\alpha^2\sqrt{1+2    A +   B^2    }  }
\frac{ A^2 - B^2}{ 1+  A  +\sqrt{1+2    A +  B^2    }   } 
~,
\ee
where, interestingly, the combination $(A^2-B^2)$ proves to be related to the square of $T^{\mu\nu}_{\text{Maxwell}}$. Using the identity 
 \be
 (F\tilde F)^2= \frac14 ( \epsilon_{\mu\nu\rho\sigma}F^{\mu\nu}F^{\rho\sigma})^2
 =4F_{\mu\nu}F^{\nu \rho}F_{\rho \sigma}F^{\sigma \mu} -2 (F^2)^2
 ~,
 \ee
we see that 
\be
T^2 _{\text{Maxwell}}=F_{\mu\nu}F^{\nu \rho}F_{\rho \sigma}F^{\sigma \mu} -\frac14 (F^2)^2
=\frac14 \Big(  (F^2)^2 +(F\tilde F)^2 \Big) 
=\frac{4}{\alpha^4}(A^2-B^2)
~.
\ee

Using tracelessness of the free Maxwell stress-energy tensor,
the $O_{T^2}$ operator can be easily computed:
\bsubeq\beqn\label{OT2}
O_{T^2}=T^2 -\frac12 \Theta^2&=&
  \frac{4(  A^2 -B^2)   } {\alpha^4 \sqrt{1+2 A +  B^2    }^2  } 
  \Big( 1
-
\frac{ A^2 - B^2}{ (1+  A  +\sqrt{1+2    A +  B^2    } )^2  }  \Big)~,
~~~~~~~~~
\\&=&
  \frac{4(  A^2 -B^2)   } {\alpha^4 \sqrt{1+2 A +  B^2    }^2  } 
  \Big( 1
-
\frac{ 1+  A  -\sqrt{1+2    A +  B^2    }  }{ 1+  A  +\sqrt{1+2    A +  B^2    }   }\Big)~,
\\&=&
  \frac{8(  A^2 -B^2)   } {\alpha^4 \sqrt{1+2 A +  B^2    }   } 
\frac{ 1      }{ 1+  A  +\sqrt{1+2    A +  B^2    }} ~,
\\&=&
  \frac{8(  1+  A  -\sqrt{1+2    A +  B^2 )   }} {\alpha^2 \sqrt{1+2 A +  B^2    }   } 
  ~.
\eeqn
\esubeq
The variation of the BI Lagrangian with respect to the 
parameter $\a^2$ can be readily computed, 
and it is given by
\be
\frac{\p \mathcal L_\alpha}{\p \alpha^2} = \frac{1+\frac14 \alpha^2 F^2- \sqrt{1+\frac12 \alpha^2 F^2 -\frac{1}{16} \alpha^4 (F\tilde F)^2}}{\alpha^2 \sqrt{1+\frac12 \alpha^4 F^2 -\frac{1}{16} \alpha^4 (F\tilde F)^2}}
~.
\label{der-BI}
\ee
Once we use \eqref{4DAB} it is clear that \eqref{OT2} and \eqref{der-BI} have exactly the same structure and satisfy
\be\label{one_eighth_maxwell}
\frac{\p\mathcal L_\alpha}{\p \alpha^2} =\frac{1}{8} O_{T^2}
~,
\ee
showing that the BI Lagrangian satisfies a $T^2$-flow driven by the operator $O_{T^2}$.

Before turning to $D=4$ supersymmetric analysis, it is worth mentioning that the structure  
of the computation relating  the $O_{T^2}$ operator to the bosonic BI theory, which we just reviewed,
is quite similar to what we saw in section \ref{2D} for the $D=2$ $\cN=(2,2)$ supersymmetric $T\bar T$ flows. 
For example, in the deformation of the free twisted-chiral multiplet action,  
the analogue of the $A$ and $B$ combinations of \eqref{4DAB} is given by 
\eqref{2DAB}, but the square root structure of the actions is completely analogous. 
This fact, together with the non-linearly realized supersymmetry we investigated in section \ref{2D},
naturally lead to the guess that the $D=4$ $\cN=1$ supersymmetric Born-Infeld (BI) theory 
 may also satisfy a  $T^2$ flow. 
The next section is devoted to explaining how this is the case.


\section{Supersymmetric Born-Infeld from Supercurrent-Squared Deformation}
\label{BI-flows}

In Section \ref{2D} we proved, by analogy and extension of the $D=4$ results of \cite{Bagger:1996wp},   that two $D=2$ supercurrent-squared flows possess additional non-linearly realized supersymmetry. In this section we reverse the logic. We will look at a well-studied model, namely the Bagger-Galperin construction \cite{Bagger:1996wp} of $D=4$, $\cN=1$ Born-Infeld theory \cite{Deser:1980ck,cecotti:1987}, and show that it satisfies a supercurrent-squared flow equation.

\subsection{\texorpdfstring{$D=4$}{D=4} \texorpdfstring{$\cN=1$}{N=1} supersymmetric BI  and non-linear supersymmetry}

Let us review some well known results about the $D=4$ $\cN=1$ Born-Infeld theory
\cite{cecotti:1987}, the 
Bagger-Galperin action \cite{Bagger:1996wp}, the non-linearly realized second supersymmetry, and its precise $\cN=2\to\cN=1$
supersymmetry breaking pattern. 
For more detail, we refer to the following references on the subject \cite{Bagger:1996wp,Rocek:1997hi,Kuzenko:2015rfx,Antoniadis:2017jsk,Antoniadis:2019gbd,cecotti:1987}.

We start  with the following  $\mathcal N=2$ superfield,
\be\label{n2chiral}
\mathcal  W (y, \theta, \tilde \theta)= X(y,\theta )+\sqrt{2} i \tilde  \theta  W(y,\theta ) - \tilde \theta^2 G(y,\theta )
~, 
\qquad y^\mu=x^\mu+ i \ta \sigma ^\mu \tab+ i \tilde \ta \sigma ^\mu \bar{\tilde\theta}
~,
\ee
which is chiral with respect to both supersymmetries:
\be
\bar D_\ad\mathcal W= \bar{ \tilde{D}}_\ad\mathcal W=0~.
\ee
Since we are ultimately interested in partial $\cN=2\to\cN=1$ supersymmetry breaking, 
we will mostly use $\cN=1$ superfields associated to the $\q$ Grassmann variables to describe manifest supersymmetry,
while we use the $\tilde{\q}$ variable for the hidden non-linearly realized supersymmetry.
 The $\cN=1$ superfields $X$, $W_\a$, and $G$ of eq.~\eqref{n2chiral} 
  are chiral under the manifest $\cN=1$ supersymmetry. Under the additional hidden $\cN=1$ supersymmetry, they transform as follows:
\bsubeq\beqn
\tilde \delta X&=&\sqrt2 i \epsilon W~, \\
\tilde \delta W &=&\sqrt2  \sigma^\mu \bar \epsilon  \p_\mu X+\sqrt2  i \epsilon G~, \\
\tilde \delta  G&=& - \sqrt2 \p_\mu W \sigma^\mu \bar \epsilon~.   \label{susytsfG}
\eeqn\esubeq

The superfield \eqref{n2chiral} has $16+16$ independent off-shell components and is reducible. 
It contains the degrees of freedom of an $\mathcal N=2$ vector and tensor multiplet.
To reduce the degrees of freedom and describe an irreducible $\cN=2$ off-shell vector multiplet,
we impose the following conditions on the $\cN=1$ components of $\cal W$:
\begin{itemize}
\item[{(i)}] First that $W_\a$ is the field-strength superfield of an $\mathcal N=1$ vector multiplet satisfying,
\be
D^\a W_\a-\bar D_\ad \bar W^\ad=0
~,
\ee
\item[{(ii)}] 
and that
\be
G=\frac14 \bar D^2 \bar X
~.
\label{constrW2}
\ee
\end{itemize}
The latter condition can easily be seen to be consistent since
it is straightforward to verify 
that $\frac14 \bar D^2 \bar X$ transforms in the same way  {as  $G$} given in \eqref{susytsfG}. 
Therefore 
we can impose \eqref{constrW2}
without violating $\mathcal N=2$  supersymmetry.

Since $\mathcal W$ is chiral with respect to both sets of supersymmetries, we can consider the following  Lagrangian, 
\bea
\mathcal L^{\mathcal N=2}_{ \mathcal W^2} 
&=& 
\frac14\int d^2 \theta  d^2 \tilde \theta \,\mathcal W^2+c.c.
=\frac14 \int d^2 \theta \Big(W^2-\frac12 X\bar D^2 \bar X\Big)+c.c.~.
\eea
On the other hand,  the $\mathcal N=2$  Maxwell theory written in terms of the $\mathcal N=1$ chiral superfields $X$ and $W_\a$ is  given by 
\bea\label{N2Maxwell}
\mathcal L^{\mathcal N=2}_{ \text{ Maxwell}}
&=&
\int d^2 \theta d^2 \bar \theta\, \bar X X +\frac14 \int d^2 \theta \, W^2
+\frac14 \int d^2 \theta\,  \bar W^2~,
\non\\
&=&\frac14 \int d^2 \theta \Big(W^2-\frac12 X\bar D^2 \bar X\Big)+c.c.
+{\rm total~derivative}
~.
\eea
We see that these two Lagrangians are the same, confirming that the extra constraint imposed on $\mathcal W$ is correct.  
 The  off-shell $\mathcal N=2$  vector multiplet can therefore be described in term of the following $\mathcal N=2$ superfield
\be\label{n2vector}
\mathcal  W (y, \theta, \tilde \theta)= X(y,\theta )+\sqrt{2} i \tilde  \theta  W(y,\theta ) -\frac14 \tilde \theta^2  \bar D^2 \bar X(y,\theta)~,
\ee
where $X$ and $W_\a$ are  $\mathcal N=1$ chiral and vector  multiplets, respectively. {Their component expansion reads:}
\bsubeq\beqn
W_\alpha &=& -i \lambda_\alpha +  \theta_\alpha \sfD  -  i (\sigma^{\mu\nu}\theta)_\alpha F_{\mu\nu}    +\q^2 (\sigma^\mu   \p_\mu {\bar \lambda}  )_\alpha~,
\label{N1_vector_components}
\\
X&=&x+\sqrt{2} \theta \chi -\theta^2\, \sfF~.
\eeqn
 \esubeq
 
Following    \cite{Rocek:1997hi} (see also \cite{Kuzenko:2015rfx,Antoniadis:2017jsk,Antoniadis:2019gbd}), 
we break  $\mathcal N=2$ supersymmetry by considering a Lorentz  and $\cN=1$  invariant condensate with a non-trivial dependence on the hidden Grassmann variables   $\EV{\mathcal W}=\mathcal W_{\text{def}} \propto \tilde \theta^2  \neq 0$, 
such that
\bsubea
\mathcal W&\rightarrow&  {\mathcal W_{\text{new}}}=\EV{\mathcal W}+\mathcal W= \mathcal W+\mathcal W_{\text{def}}~,
\\
{\mathcal W_{\text{new}}}
&=&
X +\sqrt{2} i \tilde  \theta  W  
 -\frac14 \tilde \theta^2 \Big(   \bar D^2 \bar X  + \frac{2}{ \kappa}\Big)
 ~.
 \label{Wnew}
\esubea
The hidden supersymmetry transformations of the $\cN=1$ components of the deformed $\cN=2$ vector multiplet turn out to be
\bsubea
\tilde \delta X
&=&
\sqrt2 i \epsilon W
\label{hidden_susy-1}
~, \\
\tilde \delta W 
&=&
\frac{i}{\sqrt2 \kappa}\epsilon 
+\frac{i}{2\sqrt2}   \epsilon \Db^2\bar{X}
+\sqrt2  \sigma^\mu \bar \epsilon  \p_\mu X
~.
\label{hidden_susy-2}
\esubea
Assuming the model under consideration preserves the manifest $\cN=1$ supersymmetry, which implies $\EV{\Db^2X}=0$,
the explicit non-linear $\kappa$-dependent term in the transformation of the fermionic $W_\a$ signals 
the spontaneous partial breaking $\cN=2\to\cN=1$ of the hidden supersymmetry.

To describe the Maxwell-Goldstone multiplet for the partial breaking $\cN=2\to\cN=1$, we impose the following nilpotent constraint on the deformed $\cN=2$ superfield ${\mathcal W_{\text{new}}}$
\cite{Rocek:1997hi}:
\be\label{N2constraint}
( {\mathcal W_{\text{new}}})^2=0~.
\ee
Once reduced to $\cN=1$ superfields, following the expansion \eqref{Wnew}, this constraint implies
the Bagger-Galperin constraint \cite{Bagger:1996wp}
\be\label{XWeq}
\frac1\kappa X=W^2-\frac12  X\bar D^2 \bar X
~,
\ee 
which can be solved to eliminate $X$ in terms of 
$W^2=W^\a W_\a$ and its complex conjugate 
$\bar{W}^2=\bar{W}_\ad \bar{W}^\ad$ as
 \be\label{XWconstraint}
 X=\kappa W^2 -\kappa^3 \bar D^2 \Big[   \frac{W^2 \bar W^2}{ 1+ \mathcal A +\sqrt{1+2 \mathcal A- \mathcal B^2}}  \Big]~ .
 \ee
Here we have introduced
 \be\label{AtBt}
 \mathcal A=\frac{\kappa^2}{2} (D^2 W^2+\bar D^2 \bar W^2 )= \overline{\mathcal A}
 ~, \qquad 
 \mathcal B= \frac{\kappa^2}{2} (D^2 W^2-\bar D^2 \bar W^2 )= - \overline{\mathcal B}~.
 \ee
 For later use, we denote the lowest components of the composite superfields $\cal A$ and $\cal B$ as
 \be
 A= \mathcal A |_{\theta=0}
 ~, \qquad  
 B= \mathcal B |_{\theta=0}~.
 \ee
We will not repeat the derivation of \eqref{XWconstraint} which can be found in the original paper
 \cite{Bagger:1996wp}, and was reviewed and slightly modified in section \ref{2D} for our analysis in two dimensions.

 The $\cN=1$ supersymmetric BI action can be constructed using the following $\cN=1$ (anti-)chiral Lagrangian linear in $X$:
\be\label{BI}
 \mathcal L_\kappa=\frac{1}{4 \kappa   } \Big(\int d^2 \ta X+\int d^2 \tab \bar X \Big)~.
 \ee
The second hidden supersymmetry eq.~\eqref{hidden_susy-1} written in terms of the 
unconstrained real vector multiplet $V$, where $W_\a=-1/4 \Db^2D_\a V$, takes the form
\be
\tilde \delta X=-\frac14 \sqrt2 i \epsilon^\alpha    \bar D^2 D_\alpha V
~.
\ee
Using the fact that $D^2\Db^2D_\a\propto\pa_{\a\ad}D^2\Db^\ad$,
one can immediately see that the supersymmetry variation of $\mathcal L_\kappa $ in \eqref{BI} is a total derivative. 
Therefore this supersymmetric BI action is invariant under the second hidden non-linear supersymmetry. 
  
Using the solution \eqref{XWconstraint}, the  supersymmetric BI Lagrangian  takes the explicit form 
  \begin{align}
 \mathcal L_\kappa&= 
 \frac{1}{4\kappa}\int d^2 \theta \Big(\kappa W^2 -\kappa^3 \bar D^2 \Big[   \frac{W^2 \bar W^2}{ 1+ \mathcal A +\sqrt{1+2 \mathcal A+ \mathcal B^2}}  \Big] \Big)
 +c.c. \, 
 \nonumber
\\
 &= \frac14 \int d^2 \theta \, W^2+ \frac14 \int d^2\bar \theta \, \bar W^2 
 +2 \kappa^2 \int d^2 \theta d^2 \bar\theta  \,\frac{W^2 \bar W^2}{ 1+ \mathcal A +\sqrt{1+2 \mathcal A+ \mathcal B^2}}  
 ~,
 \label{susyBI1}
  \end{align}
which makes it clear that  the supersymmetric BI is a non-linear deformation of the free $\cN=1$ Maxwell theory.  
  This supersymmetric extension of BI was first constructed by Bagger and Galperin in
\cite{Bagger:1996wp}. In this work when we refer to the supersymmetric BI theory, we will always
mean the Bagger-Galperin action.
  
We can easily calculate the flow under the $\kappa^2$ coupling constant,
\be\label{BIflow}
\frac{\p \mathcal L_\kappa}{\p \kappa^2}=2 \int d^2 \theta d^2\bar \theta\,
\frac{  W^2 \bar W^2   }{ 1+\mathcal A  +\sqrt{ 1+2 \mathcal A+\mathcal  B^2    }   } \frac{1}{\sqrt{ 1+2 \mathcal A+\mathcal  B^2    } }
~.
\ee
Our goal is now to show that the right hand side of this flow equation on-shell
is the specific supercurrent bilinear \eqref{superTsquare}  that we introduced earlier. This will establish a supercurrent-squared flow for the supersymmetric BI action.

Before turning to the core of this analysis
let us recall that at the leading order in $\kappa^2$, the fact that $D=4$ $\cN=1$ BI 
satisfies a supercurrent-squared flow was already noticed in \cite{cecotti:1987}.
This result was also highlighted recently in the introduction of \cite{Chang:2018dge}.
In fact, note that in the free limit $\alpha =\kappa=0$,  the Lagrangian \eqref{susyBI1} becomes 
the $\cN=1$ supersymmetric Maxwell theory. 
Its supercurrent multiplet is
\be
\cJ_{\alpha \dot \alpha}=-4W_\alpha \bar W_{\dot \alpha}
~, \qquad  
\cX=0
~,
\ee
where $\cX=0$ because super-Maxwell theory is scale invariant.  
The supersymmetric $T^2$ deformation operator  \eqref{superTsquare} is then simply given by
\be
\mathcal O_{T^2}=   \frac{1}{16}   \cJ_{\alpha\dot \alpha}\cJ^{\alpha\dot \alpha}  -\frac58 \cX \bar \cX=  W^2 \bar W^2
~,
\ee 
and to leading order  \eqref{BIflow} turns into \cite{cecotti:1987}
\be
\frac{\p \mathcal L_\kappa}{\p \kappa^2}=\int d^2 \theta d^2\bar \theta\,  W^2 \bar W^2 +\mathcal O(\kappa^2)
=\int d^2 \theta d^2\bar \theta\,   \mathcal O_{T^2}+\mathcal O(\kappa^2)
~.
\ee
This shows that the supercurrent-squared flow equation 
is satisfied at this order.
The rest of this section is
devoted to demonstrating the  full non-linear
extension of this result.
First, we are going to look at the bosonic truncation of \eqref{susyBI1} and \eqref{BIflow}.

\subsection{Bosonic truncation}
\label{bosonicBI-2}

  In the pure bosonic case
  the gauginos are set to zero in \eqref{N1_vector_components}, $\lambda=\bar \lambda=0$, and  $W^2, \bar W^2$ only have $\theta^2, \bar \theta^2$ components, so $\mathcal A, \mathcal B$ can only contribute to the lowest components:
\be
 A=\mathcal A|_{\theta=0}= 2 \kappa^2  \Big( F^2 - 2 \sfD^2 \Big)
 ~,\qquad
 B=\mathcal  B|_{\theta=0}=  2    \kappa^2  i F \tilde F
 ~.
\ee
Therefore the supersymmetric BI Lagrangian reduces to 
 \be
 \mathcal L=  
 \frac{1}{8\kappa^2} \Big[1- \sqrt{1+4 \kappa^2  \Big( F^2 - 2\sfD^2 \Big) -4  \kappa^4  \Big( F \tilde F  \Big)^2
}   \Big] 
~.
 \ee
The auxiliary field can be set to $\sfD=0$ using its ${\rm EOM}$, 
and the Lagrangian is then equivalent to the bosonic BI Lagrangian \eqref{flat_BI} with the identification $\alpha^2=8 \kappa^2$. 
This immediately implies that on-shell the bosonic truncation of the supersymmetric BI
satisfies a $T^2$ flow equation driven by the $O_{T^2}$ operator \eqref{Tsquare}, as we discussed in \eqref{OT2}.
A similar conclusion holds for the complete supersymmetric model of \eqref{susyBI1} and \eqref{BIflow}.

\subsection{Supersymmetric Born-Infeld as a supercurrent-squared flow}

The supercurrent for the supersymmetric BI action 
\eqref{susyBI1} was computed in \cite{Kuzenko:2002vk}  for  $\kappa^2=\frac12$. 
To simplify notation, we will also consider the special case  $\kappa^2=\frac12$ in our intermediate computations.  
The $\kappa$-dependence can be restored easily and will appear in the final formulae. 

We can straightforwardly use the results of \cite{Kuzenko:2002vk} for our supercurrent-squared flow analysis. 
The FZ multiplet was computed for a class of 
models described by the following Lagrangian,
\bea
\cL 
&=& 
\frac14 \int d^2 \theta \, W^2
+ \frac14 \int d^2\bar \theta \, \bar W^2 
+\frac{1}{4} \int d^2 \theta d^2 \bar\theta  \, W^2 \bar W^2\L(u,\bar{u})
 ~,
 \label{GeneralBI}
\eea
where
\be
u=\frac18 D^2 W^2
~, \qquad \bar u=\frac18 \bar D^2 \bar W^2
~.
\ee
The action \eqref{susyBI1} turns out to be given by the following choice of $\L(u,\bar{u})$
\be
\Lambda(u,\bar u)=\frac{4}{  1+ \mathcal A +\sqrt{1+2 \mathcal A+ \mathcal B^2}  }
~,
\label{LambdaBI}
\ee
where
\be
\mathcal A=2(u+\bar u)
~ , \qquad
\mathcal B=2(u-\bar u)
~.
\ee 
Following \cite{Kuzenko:2002vk}, we also introduce the composite superfields
\be
\Gamma(u,\bar u)=\frac{\p(u\Lambda)}{\p u}
~, \qquad
\bar \Gamma(u,\bar u)=\frac{\p(\bar u\Lambda)}{\p \bar u}
~,
\ee
which, in the case of interest to us where \eqref{LambdaBI} holds, satisfy
\bsubeq
\beqn \label{gammau}
\Gamma +\bar \Gamma -\Lambda&=&\frac{4}{\Big( 1+ \mathcal A +\sqrt{1+2 \mathcal A+ \mathcal B^2} \Big) \sqrt{1+2 \mathcal A+ \mathcal B^2} }
~,
\\
\bar u \Gamma+ u \bar \Gamma
&=& 1 -\frac{1}{  \sqrt{1+2 \mathcal A+ \mathcal B^2}  }  
~.
\eeqn
\esubeq
The supercurrents will also be functionals of the following composite
\bsubeq
\beqn
i M_\alpha &=&W_\alpha \Bigg[ 1-\frac14 \bar D^2 \Bigg(\bar W^2\Big(\Lambda +\frac18 D^2(W^2 \frac{\p \Lambda}{\p u})\Big)  \Bigg)   \Bigg]~,  \\
&=&
W_\alpha \Big( 1- 2 \bar u\Gamma \Big)+  W \bar W (\cdots)+   W^2 (\cdots)
~,
\eeqn
\esubeq
where $W \bar W (\cdots)$ denotes terms which are proportional to $W_\alpha  \bar W_{\dot \alpha}$, 
while $  W^2 (\cdots)$ denotes terms  proportional to $W^2$.
We will use similar notation with ellipses  denoting quantities  with bare fermionic terms that will not contribute to the calculation because of nilpotency conditions.

 With the ingredients introduced above, the FZ multiplet for 
 the supersymmetric BI action is given by  
\cite{Kuzenko:2002vk} 
\bsubeq
\beqn
\mathcal X&=&  \frac16 W^2 \bar D^2 \Big( \bar W^2(\Gamma+\bar \Gamma -\Lambda  ) \Big) 
~,
\\
\mathcal J_{\alpha \dot{\alpha}} &=&  -2 i M_\alpha \bar W_{\dot \alpha}+2i W_\alpha \bar M_{\dot \alpha}
+\frac{1}{12} [ D_\alpha, \bar  D_{\dot \alpha}]  \Big(W^2 \bar W^2 \Big) \cdot \Big(\Gamma +\bar \Gamma -\Lambda \Big)
\non\\
&&
 +W^2 \bar W  (\cdots)+ \bar  W^2  W (\cdots)
 ~.
\eeqn
\esubeq
For our purposes, the superfields $X$ and $\cJ_{\a\ad}$ can be further simplified as follows:
\bsubeq
\bea
\mathcal X&=&\frac16  W^2 \bar D^2  \bar W^2\cdot  \Big(     \Gamma+\bar \Gamma -\Lambda  \Big) +W^2 \bar W (\cdots)~,
\\
&=&
\frac{2W^2 \bar D^2  \bar W^2}{3 \Big( 1+\mathcal A  +\sqrt{ 1+2 \mathcal A+\mathcal  B^2    } \Big)  }+W^2 \bar W (\cdots)
~,
\label{X2}
\eea
\esubeq
and
\bsubeq
\bea
\mathcal J_{\alpha \dot{\alpha}} &= &
 -4W_\alpha \bar W_{\dot \alpha} (1- \bar u \Gamma  -   u\bar\Gamma)
+\frac{1}{12 } [ D_\alpha, \bar D_{\dot \alpha}]  \Big(W^2 \bar W^2 \Big) \cdot \Big(\Gamma +\bar \Gamma -\Lambda \Big)
\non\\
&&
 +W^2 \bar W  (\cdots)+ \bar  W^2  W(\cdots) ~,
\\&= & -\frac{ 4W_\alpha \bar W_{\dot \alpha}  }{\sqrt{1+2 \mathcal A+ \mathcal B^2}}
+\frac{ 2 D_\alpha W^2  \cdot \bar  D_{\dot \alpha}  \bar W^2  }{3\Big( 1+ \mathcal A +\sqrt{1+2 \mathcal A+ \mathcal B^2} \Big) \sqrt{1+2 \mathcal A+ \mathcal B^2} }
\non\\
&&
 +W^2 \bar W  (\cdots)+ \bar  W^2  W (\cdots)
 ~,
\label{trace}
\eea
\esubeq
where we used \eqref{gammau}. 

The computation of $\mathcal  X \bar \cX $ is trivial and receives contributions only from the square of the first term in \eqref{X2}.
The computation of $\mathcal J^2$ is less trivial. 
It is obvious that the last two complicated terms in the second line of \eqref{trace} make no contribution since all the terms are 
proportional to $W\bar W$, and we have the nilpotency property $W_\alpha W_\beta W_\gamma=0$.
The square of the first term is easy to compute, and it is proportional to $W^2\bar{W}^2$.
Next we consider the product between the first and second term in  \eqref{trace} which leads to the relation: 
\be
W_\alpha \bar W_{\dot \alpha} \cdot D^\alpha W^2  \cdot \bar  D^{\dot \alpha}   \bar W^2
= W^2   (DW) \cdot \bar W^2(\bar D\bar W)=0
~.
\ee
Remarkably, this cross term vanishes, as shown in Appendix \ref{appendix:EoMBI}. Therefore, we have the on-shell relation that
\be
W^2 \overbar{W}^2 DW=0
~.
\label{on-shell_susy-condition}
\ee
A simple physical interpretation of this condition is that the manifest 
supersymmetry is preserved on-shell,  implying that the   auxiliary field $\sfD\propto D^\a W_\a|_{\q=0}$ 
has no vev,  and is at least linear in gaugino fields $\l_\a\propto W_{\a}|_{\q=0}$.
The vanishing of this cross term can be compared with the pure bosonic  case where the cross terms in $T^2$ 
vanish because of the tracelessness property of the free Maxwell stress tensor; see section \ref{bosonicBI}. 
 Finally, we  compute the square of the second term in \eqref{trace} which includes the following structure:
\be
 D^\alpha W^2  \cdot \bar  D^{\dot \alpha}  \bar W^2 \cdot  D_\alpha W^2  \cdot \bar  D_{\dot \alpha}  \bar W^2
= W^2 \bar W^2 D^2 W^2 \bar D^2 \bar W^2
~.
\ee
Here we have used $ (D_\alpha W_\beta)( D^\alpha W^\beta)  =  -\frac12 D^2 W^2  +W^\beta D^2 W_\beta$
to simplify the result.

In summary, on-shell the contributions to the supercurrent-squared operator $\cO_{T^2}$
defined in eq.~\eqref{superTsquare}
are given by    
\begin{align}
\mathcal J^2 &=
-\frac18  \Bigg\{ 
   \frac{16W^2\bar W^2}{\sqrt{1+2 \mathcal A+ \mathcal B^2}^2}+\frac{4W^2\bar W^2D^2 W^2 \bar D^2 \bar W^2 }{9 \sqrt{1+2 \mathcal A+ \mathcal B^2}^2  \Big( 1+\mathcal A  +\sqrt{ 1+2 \mathcal A+\mathcal  B^2    } \Big)^2 } 
   \Bigg\} \, , 
\\
\mathcal  X \bar \cX &=\frac49     \frac{W^2\bar W^2 D^2 W^2 \bar D^2 \bar W^2 }
{  \sqrt{1+2 \mathcal A+ \mathcal B^2}  ^2\Big( 1+\mathcal A  +\sqrt{ 1+2 \mathcal A+\mathcal  B^2    } \Big)^2} \, .
\end{align}
Adding these results gives the supersymmetric $T^2$ primary operator $\cO_{T^2}$: 
\begin{align}
  \mathcal O_{T^2}= -\frac{1}{2} \Big(    \mathcal J^2  +\frac54 \mathcal X \bar{\mathcal  X} \Big)
&=  \frac{ W^2\bar W^2}{ \sqrt{1+2 \mathcal A+ \mathcal B^2}^2}
 \Bigg( 1-  \frac{ D^2 W^2 \bar D^2 \bar W^2 }
 { 4 \Big( 1+\mathcal A  +\sqrt{ 1+2 \mathcal A+\mathcal  B^2    } \Big)
^2 }   \Bigg) \, , \nonumber 
\\&=  \frac{W^2 \bar W^2}{   \sqrt{1+2 \mathcal A+ \mathcal B^2} ^2 } 
\Big( 1-  \frac{\mathcal A^2- \mathcal B^2 }{ (1+\mathcal A  +\sqrt{ 1+2 \mathcal A+\mathcal  B^2    })^2}  \Big) \, \nonumber
\\&=  \frac{2 W^2 \bar W^2}{ \sqrt{ 1+2 \mathcal A+\mathcal  B^2    }  \Big({1+\mathcal A  +\sqrt{ 1+2 \mathcal A+\mathcal  B^2 }  \Big)  }}  \, .
\label{OT2-susyBI}
\end{align}
It is worth noting that the simplifications occurring in constructing $\cO_{T^2}$ from the supercurrents 
are very similar to the bosonic case of \eqref{OT2}.

Comparing with \eqref{BIflow}, we see that
eq.~\eqref{OT2-susyBI} proves that the supersymmetric BI action \eqref{susyBI1}
is an on-shell solution of the flow equation 
\bsubeq
\bea
\frac{\p  \mathcal L_\kappa}{\p \kappa^2} 
&=& 
 \int d^2 \theta d^2\bar \theta^2
 \,\frac{2 W^2 \bar W^2}{ \sqrt{ 1+2 \mathcal A+\mathcal  B^2    }  \Big({1+\mathcal A  +\sqrt{ 1+2 \mathcal A+\mathcal  B^2 }  
\Big)  }}~, \nonumber
\\
&=& \int d^2 \theta d^2\bar \theta^2\, \mathcal O_{T^2} +\text{total derivatives} +{\rm EOM}
~.
\eea
\esubeq
It therefore describes a supercurrent-squared deformation
of the $\cN=1$ free Maxwell Lagrangian. 
This result establishes a relationship between non-linearly realized supersymmetry and 
supercurrent-squared flow equations in $D=4$. We note that this analysis can be extended to the supersymmetric ModMax theory, whose supercurrent-squared deformation is the corresponding supersymmetric ModMax-BI theory \cite{Ferko:2022iru}.

Before closing  
 this section, we should make a few comments regarding the on-shell condition \eqref{on-shell_susy-condition}
 used in establishing the 
supercurrent-squared flow equation for the $D=4$ $\cN=1$ BI action.
First it is important to stress that the flow equation is not satisfied by the supersymmetric BI action off-shell. 
Second, we note that the specific combination of $ \mathcal J^2$ and $\mathcal X \bar{\mathcal  X}$ studied
is the unique choice for which \eqref{susyBI1} satisfies a supercurrent-squared flow equation, even if only on-shell.

Such a non-trivial condition satisfied by the on-shell supersymmetric BI action is intriguing and hints at the existence
of appropriate (super)field redefinitions which might
lead to a different supersymmetric extension of BI that satisfies the flow equation
off-shell. For example, it is know that the dependence of the off-shell extension on the  auxiliary field $\sfD$
can be modified by appropriate (super)field redefinitions, as well as redefinitions of the full superspace Lagrangian. 
We refer to \cite{Cecotti:1986gb, GonzalezRey:1998kh,Kuzenko:2011tj,Bagger:1997pi} for a list of relevant papers on this subject.
Under field redefinitions, the hidden supersymmetry will be modified but will remain a non-linearly realized symmetry of the 
theory. The existence of an off-shell solution of the supercurrent-squared flow is an interesting question for future research.

\section{\texorpdfstring{$D=4$}{D=4} Goldstino Action From Supercurrent-Squared Deformation}
\label{Goldstino-Flow}

In section \ref{BI-flows} we showed that the Bagger-Galperin action for the $D=4$, $\cN=1$ supersymmetric BI
theory satisfies a supercurrent-squared flow. It is known that the truncation of this model to fermions describes 
a Goldstino action for $D=4$, $\cN=1$ supersymmetry breaking; see, for example, \cite{Hatanaka:2003cr,Kuzenko:2005wh,Kuzenko:2011tj}. 
The $\cN=1$ non-linearly
realized supersymmetry arises as the non-linearly realized part of the $\cN=2\to\cN=1$ 
breaking of the supersymmetric BI.
We have shown in sections \ref{bosonicBI} and \ref{bosonicBI-2} 
that the bosonic truncation of the supersymmetric BI satisfies a $T^2$ flow equation.
The same should be true for the fermionic truncation.
More generally, one might argue that $D=4$ $\cN=1$ Goldstino models could satisfy a sort of flow equation that organizes their 
expansion in the supersymmetry breaking scale parameter. 

Note that in the $D=2$ case, the intuition is similar. If we consider the actions analyzed in section \ref{2D} 
that describe Goldstone models for partial $D=2$ $\cN=(4,4)\to\cN=(2,2)$ supersymmetry breaking,
one can immediately argue that their fermionic truncation describes Goldstino actions possessing non-linearly realized $D=2$
$\cN=(2,2)$ supersymmetry. These, by construction, are expected to satisfy a $T\Tb$-flow equation.
In fact,
such an argument is in agreement 
with the very nice recent analysis of \cite{Cribiori:2019xzp} 
where a $D=2$ Goldstino model possessing $\cN=(2,2)$ non-linearly realized supersymmetry
was shown to satisfy the supercurrent-squared flow equation for the FZ multiplet, which was given in \eqref{FZTTbar} of Chapter \ref{CHP:SC-squared-2}.\footnote{We refer to 
\cite{Farakos:2016zam} for a discussion of various models possessing non-linearly realized $(2,2)$ supersymmetry.}
The model analyzed in \cite{Cribiori:2019xzp} 
is the analogue of the $D=4$ model  of \cite{Casalbuoni:1988xh,Komargodski:2009rz} and related on-shell to 
the Goldstino model of \cite{Rocek:1978nb}.\footnote{Note that the Goldstino models of
\cite{Casalbuoni:1988xh,Komargodski:2009rz,Rocek:1978nb} 
were shown in \cite{Kuzenko:2005wh,Kuzenko:2011tj} to be identical to the 
fermionic truncation of the supersymmetric BI action up to a field redefinition of the Goldstino.} 
This section is devoted to  showing that these $D=4$ $\cN=1$ Goldstino models 
satisfy a supercurrent-squared flow driven by the operator $\cO_{T^2}$ of the supersymmetric BI, in agreement with
the arguments given above.

 \subsection{\texorpdfstring{$D=4$}{D=4} Goldstino actions}  
   
The Volkov-Akulov (VA) action is the low energy description of supersymmetry breaking. There are several representations of the  
Goldstino action that are equivalent to the Volkov-Akulov form; see \cite{Kuzenko:2011tj,Cribiori:2016hdz} for  comprehensive discussions. 
Here we will focus on two models, but we start by reviewing a few general features of Goldstino actions.

The original VA action was obtained by requiring its invariance under the  the non-linear supersymmetry
 transformation~\cite{Volkov:1973ix}
 \be
 \delta_\xi \lambda^\alpha =\frac{1}{\kappa} \xi^\alpha -i \kappa(\lambda \sigma^m \bar \xi  -\xi \sigma^m \bar \lambda)\p_m \lambda^\alpha
~. \ee
 Explicitly, the original  Lagrangian was proven to be
 \be\label{VAaction}
 \mathcal L_{\rm VA}= -\frac{1}{2\kappa^2} \det A =
 -\frac{1}{2\kappa^2} -\frac{i}{2} 
 ( \lambda \sigma^m\p_m \bar \lambda-\p_m\lambda \sigma^m \bar \lambda  )
 + \text{interactions}
 ~,
 \ee  
 where 
 \be
 A_m{}^a= \delta_m{}^a -i \kappa^2 \p_m \lambda \sigma^a \bar \lambda +i \kappa^2 \lambda \sigma^a \p_m \bar \lambda
 ~.
 \ee
 
The alternative representation of the Goldstino action 
that interests us 
was originally introduced by Casalbuoni {\it et al.} in \cite{Casalbuoni:1988xh},
and later rediscovered and made fashionable
 by Komargodski and Seiberg \cite{Komargodski:2009rz}.
This model, which following recent literature we will call the KS model,
was constructed by imposing  nilpotent superfield constraints as a generalization of Ro\v{c}ek's 
seminal ideas for the Goldstino model described in \cite{Rocek:1978nb}. 
After integrating out an auxiliary field in the KS model, described in more detail in the next section, 
the  explicit form of the Lagrangian  is given by 
the following very simple combination of terms:
    \be\label{KSaction}
 \mathcal L_{\text{KS}}  = -f^2
-\frac{i}{2} 
 ( \psi \sigma^m\p_m \bar \psi-\p_m\psi \sigma^m \bar \psi  )
-  \frac{1}{4f^2} \p^\mu\bar \psi^2 \p_\mu \psi^2 -\frac{1}{16f^6} \psi^2 \bar\psi^2 \p^2 \psi^2 \p^2 \bar\psi^2
~.
   \ee
   The action is invariant under a quite involved non-linearly realized supersymmetry transformation whose
   explicit form can be found in \cite{Kuzenko:2010ef,Kuzenko:2005wh}.
 The Goldstino actions described by \eqref{VAaction} and \eqref{KSaction} 
 prove to be equivalent off-shell up to a field redefinition \cite{Kuzenko:2010ef,Kuzenko:2005wh}.

\subsection{\texorpdfstring{$D=4$}{D=4} KS Goldstino model as a supercurrent-squared flow}

 The goal in the rest of this section is to straightforwardly generalize   the analysis of \cite{Cribiori:2019xzp} to $D=4$ and to show how the  KS action satisfies a flow equation 
arising from a $T^2$ deformation of the free fermion action.

\subsubsection{\uline{KS model}}
   
Let us start by reviewing the Goldstino model of  \cite{Casalbuoni:1988xh,Komargodski:2009rz}.
Consider the following Lagrangian 
\be\label{KSlag}
\mathcal  L_{\rm KS}=\int d^4 \theta\, \bar \Phi \Phi
+ \int d^2 \theta  \Big( f  \Phi+ \frac12 \Lambda \Phi^2\Big) 
+ \int d^2 \bar\theta \Big( f \bar \Phi+   \frac12   \bar\Lambda \bar \Phi^2\Big)
~,
\ee
where $\Phi,\, \bar \Phi$ are $D=4$ $\cN=1$ chiral and anti-chiral superfields, satisfying the constraints $\bar D_{\dot \alpha  }\Phi=D_{\alpha} \bar \Phi=0$. The constant parameter $f$, which describes the supersymmetry breaking scale,
 is real. The superfields $\Lambda,\,\bar \Lambda$ are chiral and anti-chiral
  Lagrange multipliers whose ${\rm EOM}$ yield the nilpotent constraints
\be\label{nilpotent}
\Phi^2=\bar \Phi^2=0
~.
\ee
The equation of motion for $\Phi$ is
   \be\label{eomPhi}
   \frac14 \bar D^2 \bar \Phi=\Lambda \Phi+f
   ~, \qquad    \frac14  D^2  \Phi= \bar\Lambda \bar \Phi+f
   ~.
   \ee
As a consequence, we also have 
   \be\label{eomPhiDPhi}
   \Phi  \bar D^2 \bar \Phi =4f \Phi
   ~,\qquad      \bar \Phi  D^2   \Phi =4f \bar\Phi
   ~,
   \ee
 where the nilpotent properties of  \eqref{nilpotent} are used. 
 Note that the constraints \eqref{nilpotent} and \eqref{eomPhiDPhi} are the ones originally used by Ro\v{c}ek to define 
 his Goldstino model \cite{Rocek:1978nb}. These observations make manifest the on-shell equivalence of the KS model with Ro\v{c}ek's Goldstino model in a simple superspace setting. 
 The off-shell equivalence of all these Golstino models up to field redefinitions, including the VA action, was proven in 
 \cite{Kuzenko:2011tj}.

 The  Lagrange  multiplier  in  \eqref{KSlag} imposes the nilpotent constraint $\Phi^2=0$ on the chiral superfield $\Phi$. 
 This condition can be solved in terms of the spinor field $\psi$ and the auxiliary field $F$ of the chiral multiplet,  
 \cite{Casalbuoni:1988xh,Komargodski:2009rz}: 
    \be
   \Phi=\frac{\psi^2}{2F}+\sqrt{2} \theta \psi+\theta^2 F
   ~,
   \ee
   which is sensible assuming that $F\ne 0$.
Substituting back into \eqref{KSlag} gives a Lagrangian expressed  in terms of  $\psi $ and the auxiliary field $F$,
\be
\cL_{\rm KS}
=
-\frac{i}{2}\psi \s^\mu\pa_\mu\bar\psi
+\hf\bar{F}F
+\frac{1}{8}\frac{\bar\psi^2}{\bar{F}}\pa^2\Big(\frac{\psi^2}{F}\Big)
+fF
+c.c.~.
\label{KSaction2}
\ee
The auxiliary field can then be eliminated using its equation of motion, which can be solved in closed form
\be
F=-f\Bigg(
1
+\frac{\bar\psi^2}{4f^4}\pa^2\psi^2
-\frac{3}{16f^8}\psi^2\bar\psi^2\pa^2\psi^2\pa^2\bar\psi^2
\Bigg)
~,
\label{Fsolution}
\ee
together with the complex conjugate expression for $\bar{F}$.
Plugging \eqref{Fsolution} into \eqref{KSaction2}
gives the Goldstino action \eqref{KSaction}
 \cite{Casalbuoni:1988xh,Komargodski:2009rz}.

\subsubsection{\uline{$D=4$ Goldstino action as a supercurrent-squared flow}}
   
One advantage of using the KS model compared to other Goldstino actions is the relatively simple form of the action, thanks to the Lagrange multiplier, which makes the computation of its supercurrent easier. 
The FZ multiplet resulting from the action \eqref{KSlag} is 
\begin{align}
   \cJ_{\alpha\dot\alpha} &=
   2 D_\alpha\Phi \cdot \bar D_{\dot\alpha}   \bar \Phi -\frac23 [D_\alpha, \bar D_{\dot\alpha}] (\Phi \bar \Phi)
   =
   \frac23 D_\alpha \Phi  \cdot \bar D_{\dot\alpha} \bar \Phi 
-     \frac{2i}{3} \Big(    \Phi \p_{\alpha\dot\alpha} \bar \Phi -\bar \Phi \p_{\alpha\dot\alpha} \Phi \Big) \, , 
   \\
\cX   &= 4\Big(f\Phi +\frac12\Lambda \Phi^2 \Big) 
 -\frac 13 \bar D^2 (\Phi \bar\Phi)
= \frac83 f\Phi +2\Lambda \Phi^2 \, .
\end{align}
The composite operators $   \cJ^{\alpha\dot\alpha}   \cJ_{\alpha\dot\alpha}  $
and 
$\cX  \bar\cX$
are then
   \be
   \cJ^{\alpha\dot\alpha}   \cJ_{\alpha\dot\alpha}  = \frac{64}{9} f^2 \Phi\bar \Phi 
   + \text{total derivatives}
   +{\rm EOM}
   ~,
      \ee
and 
\be
\cX  \bar\cX  =\frac{64}{9}f^2 \Phi\bar\Phi
   +{\rm EOM}
~,
\ee   
where we used \eqref{nilpotent} and \eqref{eomPhi}.
The supercurrent-squared operator   \eqref{superTsquare}    then takes the form
   \be
     \mathcal O_{T^2}=  \frac{1}{16}  \cJ^{\alpha\dot\alpha}   \cJ_{\alpha\dot\alpha}- \frac58 \cX\bar \cX
     =-4f^2 \Phi\bar \Phi+ {\rm EOM}+  \text{total derivatives}
     ~.
   \ee
The descendant operator ${\sf O}_{T^2}$ of eq.~\eqref{superTsquare2} becomes
   \be
  {\sf O}_{T^2}  =\int d^2 \theta d^2 \bar \theta\;     \mathcal O_{T^2}
   =-4f^2\int d^2 \theta d^2 \bar \theta\;    \Phi\bar \Phi
   =2 f^3 \int d^2 \theta\, \Phi+2 f^3 \int d^2  \bar \theta\, \bar \Phi
   ~,
   \ee
  where we used \eqref{eomPhiDPhi} in the last equality.

  From \eqref{KSlag}, it is easy to see that the following relation holds:
  \be
  \frac{ \p\mathcal L_{\rm KS} }{\p f}=   \int d^2 \theta \,\Phi+  \int d^2  \bar \theta\, \bar \Phi
  ~.
 \ee
By identifying the coupling constants, 
   \be
\gamma=-\frac{1}{4   f^2 }
~,
   \ee
it follows immediately that the KS action,  
\be\label{SKS}
S_{\gamma}=\int d^4 x \, \mathcal L_{\rm KS}
~,
\ee
satisfies the flow equation
 \be\label{TTflow}
  \frac{ \p S_{\gamma} }{\p \gamma}=    \int d^4 x d^2 \theta\, \Phi+  \int d^4 x d^2   \bar \theta \,\bar \Phi 
  =\int d^4 x   d^2 \theta d^2\bar\theta  \;  \mathcal O_{T^2}
  = \int d^4 x \,  {\sf  O}_{T^2} 
  ~.
   \ee
   This proves that
   \eqref{KSlag} satisfies a supercurrent-squared flow
   (or $T^2$ flow) equation. 
  Because   on-shell the actions   \eqref{KSaction} and    \eqref{KSlag} are equivalent,
 and   the equation 
 $$\int  d^2 \theta d^2\bar\theta  \;  \mathcal O_{T^2}  =  {\sf  O}_{T^2} $$
 holds,
   eq.~\eqref{TTflow} proves that the $D=4$ $\cN=1$ Goldstino action arises from a supercurrent-squared deformation.\footnote{The careful reader may find that the flow can also be satisfied by other supercurrent-squared operators, $\cJ^2 - r \bar\cX\cX$, with arbitrary $r$ because of the linearity between $\cJ^2 $ and $\bar \cX \cX$. It is worth pointing out the same thing happens in $D=2$  \cite{Cribiori:2019xzp}. We stress that this is not the case for the supercurrent-squared flow satisfied by the $D=4$ supersymmetric Born-Infeld action.}

\section{Higher Dimensions and Connections to Amplitudes}
\label{concludingthoughts}

In this thesis, we have primarily focused on two-dimensional quantum field theories, where $\TT$ and related supercurrent-squared operators can be unambiguously defined at the quantum level. For $D>2$, there appears to be no complete argument showing  that any of the proposed operators 
$O_{T^2}^{[r]}$ of eq. \eqref{OT2r0}, including the holographic operator of \cite{Taylor:2018xcy, Hartman:2018tkw}, possess any particularly nice quantum properties.

However, despite the absence of a well-defined quantum operator in higher dimensions, the connection between non-linearly realized symmetries and our $D=4$, $\mathcal{N} = 1$ example suggests that operators of this form still have some special properties (at least at the classical level). It would be very interesting to understand whether any supersymmetric completion of our descendant operator $ {\sf  O}_{T^2}$ of \eqref{superTsquare2} could provide a well-defined deformation at the quantum level. Such an extension would necessarily involve contributions from other current-squared type operators which do not vanish on-shell. This possibility seems most promising in theories with extended, and perhaps even maximal, supersymmetry.

One interesting direction for future investigation concerns the relationship between $T \Tb$ deformations and amplitudes. In two dimensions, $T \Tb$ simply modifies the $S$-matrix of the undeformed theory by a CDD factor \cite{Dubovsky:2013ira}, but one might wonder about the $S$-matrices of higher-dimensional theories deformed by generalizations of $T \Tb$. One hint is that theories with non-linearly realized symmetries exhibit enhanced soft behavior -- indeed, in the case of non-linearly realized \textit{super}symmetry, there is a proof that such symmetries generically lead to constraints on the soft behavior of the $S$-matrix \cite{Kallosh:2016qvo}. This fact which has been applied to the Volkov-Akulov action \cite{Kallosh:2016lwj}, which also satisfies a $T \Tb$-like flow.

There are also examples involving purely bosonic theories. For instance, in four dimensions, the Dirac action is the unique Lorentz-invariant Lagrangian for a single scalar which is consistent with factorization, has one derivative per field, and exhibits soft degree $\sigma = 2$ for its scattering amplitudes \cite{Cheung:2014dqa}. Similarly, it has been shown that the Born-Infeld action for a vector can be fixed by demanding enhanced soft behavior in a particular multi-soft limit \cite{Cheung:2018oki}, which can be understood in the context of T-duality and dimensional reduction \cite{Elvang:2019twd}. Given the hints of a deeper relationship between supercurrent-squared deformations, non-linearly realized symmetries, and actions of Dirac or Born-Infeld type, it is natural to ask whether such deformations enhance the soft behavior of scattering amplitudes in a more general context.

\chapter{\texorpdfstring{$\TT$}{TT} and Non-Abelian Gauge Theory} \label{CHP:nonabelian}
In this chapter, we will define a theory for a non-abelian gauge field in two dimensions using the $\TT$ operator, following the discussion in ``A Non-Abelian Analogue of DBI from $\TT$'' \cite{Brennan:2019azg}.

Although we focus on the bosonic formulation of $\TT$ rather than the supercurrent-squared deformations developed in the preceding chapters, the gauge theory which we define here is compatible with maximal supersymmetry. This deformed theory shares some properties with the Dirac-Born-Infeld theory which describes gauge fields living on a brane, although it does not reduce to DBI even in the abelian case. We begin by reviewing some facts and motivation about brane physics.

\section{Background on Brane Physics} \label{brane_physics}

Imagine a $p$-brane embedded in an ambient $(D+1)$-dimensional Minkowski space-time. Any such brane spontaneously breaks the Poincar\'e symmetry of the ambient space-time:
\be\label{poincare}
ISO(D,1) \,\rightarrow\, ISO(p, 1). 
\ee
In particular, the breaking of translational symmetry guarantees the existence of $D-p$ universal scalar fields on the brane world-volume, collectively denoted $\phi$, which are Nambu-Goldstone (NG) bosons for the broken translations. The physics of these modes is governed by the Dirac action, 
\be\label{Dirac}
    S_{\rm Dirac}  = - T_p \int d^{p+1}\s \, \sqrt{- \det \left(\eta_{\m\n} + \p_\m \phi \p_\n\phi\right) },
\ee
with brane world-volume coordinates $\s$ and a single dimensionful parameter $T_p$. The form of this action is fixed by the broken Lorentz symmetries, which are non-linearly realized. There might also be a function of any additional non-universal scalar fields multiplying this form, which we will not consider here. This action can also equivalently be viewed as the Nambu-Goto action for the brane in static gauge.

The Born-Infeld action, on the other hand, defines a non-linear interacting extension of Maxwell theory with action: 
\be\label{BI_nonabelian}
    S_{\rm BI}  = - T_p \int d^{p+1}\s \,  \sqrt{- \det \left(\eta_{\m\n} + \alpha F_{\m\n}\right) }.
\ee
One of the striking physical differences between Born-Infeld theory and Maxwell theory is the existence of a critical electric field determined by the dimensionful parameter $\alpha$. In string theory, Born-Infeld theory describes the leading interactions for the gauge-field supported on a D-brane~\cite{Abouelsaood:1986gd}. In that context both $T_p$ and $\alpha$ are fixed in terms of the fundamental string tension $\alpha'$ with $\alpha = 2\pi\alpha'.$

The combined Dirac-Born-Infeld (DBI) action is a complete description of the physics of a single D-brane at leading order in string perturbation theory, and under the assumption that acceleration terms like $\p F$ or $\p^2 \phi$ are negligible: 
\be\label{DBI_nonabelian}
    S_{\rm DBI}  = - T_p \int d^{p+1}\s \, \sqrt{- \det \left( \eta_{\m\n} + \p_\m \phi \p_\n\phi+ \alpha F_{\m\n} \right) }.
\ee
For multiple coincident branes, the abelian gauge symmetry is replaced by a non-abelian symmetry, and the fields $(\phi, F)$ typically take values in the adjoint representation of the gauge group. We will not assume any particular representation for the scalar fields in this discussion. It is natural to pose the following long considered question: what might replace~\C{DBI_nonabelian} in the non-abelian theory? For the scalar fields appearing in the induced metric of the Dirac action~\C{Dirac}, one could easily imagine making the replacement
\be\label{replace}
\p_\m \phi \p_\n\phi\, \rightarrow \,\Tr\left(D_\m \phi D_\n\phi \right),
\ee
where $\phi$ is now matrix-valued and $D$ is an appropriate covariant derivative. For the Born-Infeld action of~\C{BI_nonabelian}, however, an interesting gauge-invariant replacement of this sort is not possible. In~\C{replace} $\Tr$ denotes the trace over gauge indices. When needed, we will use $\tr$ to denote the trace over Lorentz indices so that,
\be
\tr(F^2) = F_{\mu \nu} F^{\nu \mu}.
\ee
Indeed what one means by the Born-Infeld approximation, namely neglecting acceleration terms like $DF$, is ambiguous. Unlike the abelian case, 
\be
\comm{D_\m}{D_\n} = -iF_{\m\n},
\ee
so there is no clear cut way of truncating the full brane effective action by throwing out acceleration terms. 

With considerable hard work there is, however, some data known about brane couplings beyond the two derivative non-abelian kinetic terms $\Tr(F_{\m\n}F^{\m\n})$ at leading order in string perturbation theory. This information is very nicely summarized in the thesis~\cite{Koerber:2004ze}\ to which we refer for a more complete discussion. For comparative purposes with our analysis, we note that the known $F^4$  terms are correctly captured by a symmetrized trace prescription~\cite{Tseytlin:1997csa, Bergshoeff:2001dc}. Up to overall scaling, they are given by
\be\label{f4terms}
{\rm STr}\left( \tr F^4 - \frac{1}{4} (\tr F^2)^2 \right),
\ee
with
\be {\rm STr}\left(T_1 T_2 \ldots T_n \right) = \frac{1}{n!} \sum_{\rm \s \in S_n} \Tr(T_{\s(1)} T_{\s(2)} \ldots T_{\s(n)}). 
\ee
This prescription is known to fail for higher derivative terms. What is important for us is that~\C{f4terms} defines a single trace operator. 

To define a non-abelian analogue of the abelian DBI theory~\C{DBI_nonabelian}, our approach will be to $\TT$ deform a non-abelian gauge theory with scalar matter in two dimensions. In contrast to the preceding parts of this thesis, in this chapter we will restrict to bosonic theories for simplicity. A priori this approach has no connection to either brane physics or string theory. We will do this in steps by first recalling known results about deforming free scalars and Maxwell fields~\cite{Cavaglia:2016oda,Bonelli:2018kik, Conti:2018jho}, and then extending to charged matter and non-abelian gauge theories. Other than also involving an infinite collection of irrelevant operators, the reason this approach should be viewed as giving a non-abelian analogue of DBI is that the $\TT$ deformation of a free scalar with parameter $\lambda$ already gives the Dirac action, as we reviewed in Chapter \ref{CHP:TT}:
\be
S_\lambda = \int d^2\sigma\frac{1}{2 \lambda} \left( \sqrt{1 + 4 \lambda \dpsmall } - 1 \right) ~ .
\ee
This direct connection with brane physics is reason enough to suspect that $\TT$ applied to gauge-theory will give further insight into brane physics. 

The $\TT$ deformation of Maxwell theory, however, is already different from the Born-Infeld theory of~\C{BI_nonabelian}. This is the reason we call the $T\bar{T}$-deformed theory an analogue rather than a generalization of DBI; it does not reduce to DBI even in the case of abelian gauge theory. The couplings are not given by the square-root structure of a relativistic particle but rather by a hypergeometric function~\cite{Conti:2018jho}. This might not seem very exciting in two dimensions where pure gauge fields have no propagating degrees of freedom, but that is no longer the case when we add scalar fields, even in the abelian setting. For interesting recent discussions of $\TT$-deformed gauge theories, see~\cite{Santilli:2018xux, Ireland:2019vvj}. 

For the non-abelian theory defined using $\TT$, the $O(F^4)$ terms are already very different from what is known about non-abelian BI theory. Rather than involving a single trace operator like~\C{f4terms}, they involve double trace operators. The $\TT$-deformed theory is quite remarkable because it has the following properties:
\begin{itemize}
    \item The theory is compatible with supersymmetry~\cite{Baggio:2018rpv, Chang:2018dge, Jiang:2019hux, Chang:2019kiu, Coleman:2019dvf, Ferko:2019oyv, Jiang:2019trm}. Indeed, we have discussed at length in Chapters \ref{CHP:SC-squared-1} and \ref{CHP:SC-squared-2} that $\TT$ \textit{always} preserves degeneracies in energy levels like those implied by supersymmetry, and that one can often make the supersymmetry manifest via a superspace flow construction. In particular, if one $\TT$ deforms a maximally supersymmetric starting theory then this supersymmetry is preserved! 
    
    \item The theory is believed to exist at the quantum level, unlike DBI which is an effective theory with our present level of understanding. 
    
    \item The theory has a critical electric field like BI. 
\end{itemize}
This is already quite surprising in the abelian case with uncharged matter. Folklore suggests that some of these properties, like compatibility with maximal supersymmetry, should only have been true for DBI. 
Indeed there are no obvious reasons that the structures seen here should not emerge from string theory, either in a closed or open string setting. In fact, the $D=10$ space-time effective action for the type I/heterotic strings does contain a double trace $F^4$ term, which is required for anomaly cancelation in $D=10$, or more generally required by supersymmetry~\cite{Bergshoeff:1988nn}. In the heterotic string, the term arises at tree-level and takes the schematic form:
\be
S_{\rm het} \sim \int d^{10}x \sqrt{g} e^{-2\phi} \left(\Tr F^2 \right)^2 + \ldots.
\ee
In the dual type I frame, relevant for a brane picture, the same coupling arises from diagrams with Euler characteristic $-1$~\cite{Tseytlin:1995fy, Tseytlin:1995bi}. This leads us to suspect that the $\TT$ flow equation is connected with corrections to two-dimensional beta functions from higher orders in string perturbation theory.

As a final comment, we note in passing that a different way of deforming a $2d$ gauge theory is via holography. For instance, $2d$ JT gravity can be written in gauge theory variables as a BF theory, and one can then $\TT$-deform the $(0+1)$-dimensional dual to this BF theory. The bulk interpretation of such a deformation is explored in \cite{us:gravity}.

\section{Deforming Pure Gauge Theory}\label{sec:deforming_maxwell}

Before we go on to solve the $\TT$ equation to find a non-abelian analogue of the DBI action, we will first illustrate the above techniques by solving for the $\TT$-deformed Yang-Mills theory. That is, we begin with an undeformed Lagrangian of the form
\begin{align}
    k \mathcal{L}_0 &= \, F_{\mu \nu}^a F^{\mu \nu}_a = \, \Tr \left( F_{\mu \nu} F^{\mu \nu} \right) 
\end{align}
where we will often suppress the trace for convenience and simply write $F^2$ for $\Tr ( F^2 )$, so that
\begin{align}
    \mathcal{L}_0 = \frac{1}{k} F^2 .
    \label{undeformed_pure_gauge}
\end{align}

We retain an overall dimensionless constant $k$ in the Lagrangian; in most of the calculations that follow, which we will suppress factors of $k$ by setting $k=1$. The value of $k$ does not affect the equations of motion associated with the action (\ref{undeformed_pure_gauge}), but the sign of $k$ will be important for determining critical value of the field strength $F^2$. To see the maximum allowed electric field for the deformed theory in Minkowski signature, we will find that we must take $k < 0$ so that the undeformed action is positive (however, all of our other results are valid in either Minkowski or Euclidean signature). We will restore factors of $k$, replacing $F^2 \to \frac{1}{k} F^2$, when the sign is relevant.

Our goal is to find the deformed Lagrangian $\mathcal{L} ( \lambda ) = f ( \lambda , F^2 )$ which solves the flow equation
\be\label{flow}
\partial_\lambda \CL_\lambda={\rm det}\, T_{\mu\nu} ~ ,
\ee
with initial condition $\mathcal{L} ( 0 ) = \mathcal{L}_0$.

It is first convenient to rewrite this flow equation using the fact that $2 \times 2$ matrices $M$ satisfy
\begin{align}
    \det ( M ) = \frac{1}{2} \left( \left (\tr M \right)^2 - \tr \left( M^2 \right) \right) \, .
\end{align}
Thus we can write the $T \Tb$ flow equation in a manifestly Lorentz-invariant way as
\begin{align}\label{flow_invariant}
    \partial_\lambda \mathcal{L}_{\lambda} = \frac{1}{2} \left( \left( \tensor{T}{^\mu_\mu} \right)^2 - T^{\mu \nu} T_{\mu \nu} \right) .
\end{align}
Notice that the stress-energy tensor is a single-trace operator in the undeformed theory. In the $\lambda$ expansion, the leading order deformation of the Lagrangian is therefore automatically a double trace operator. Including further corrections in $\lambda$ will only generate higher order multi-trace operators. In particular -- as we show in Appendix \ref{flow_derivation} and discuss in more detail in Chapter \ref{CHP:new_chapter} -- a single-trace deformation like the leading $F^4$ terms of non-abelian DBI in~\C{f4terms} will never be generated from a $\TT$ flow beginning from an undeformed Lagrangian which is only a function of $F^2$.

The details of the calculation of the stress tensor components $T_{\mu \nu}^{(\lambda)}$ for the Lagrangian $\mathcal{L} ( \lambda , F^2 )$ are presented in Appendix \ref{flow_derivation}, where we find the $T \Tb$ operator for an arbitrary Lagrangian depending on a field strength $F_{\mu \nu}$ and a complex scalar $\phi$. We can set the scalar $\phi$ to zero in the result of that Appendix to find the flow equation for a pure gauge field Lagrangian. The result, using the shorthand notation $x = F^2$, is
\begin{align}
    \frac{d f}{d \lambda} = f(x)^2 - 4 f ( x ) x \frac{\partial f}{\partial x} + 4 x^2 \left( \frac{\partial f}{\partial x} \right)^2 .
    \label{flow_for_pure_gauge_theory}
\end{align}
Next we will present several methods for solving (\ref{flow_for_pure_gauge_theory}):
\begin{enumerate}
    \item by directly solving the flow equation \eqref{flow_invariant} in a series expansion;
    \item by writing the solution implicitly in terms of a complete integral;
    \item and by dualizing the field strength $F^2$ to a scalar, deforming, and dualizing back.
\end{enumerate}
The same three methods will also be applied to the case of a gauge field coupled to scalars in Section \ref{sec:DBI}.

\subsection{Series solution of flow equation}\label{sec:direct_flow_maxwell}

The differential equation (\ref{flow_for_pure_gauge_theory}) derived in the preceding subsection can be brought into a simpler form by refining our ansatz to
\begin{align}
    f ( \lambda, F^2 ) = F^2 g ( \lambda F^2 ) 
\end{align}
for some new function $g$. For convenience, we define the dimensionless variable $\chi = \lambda F^2 = \lambda x$. Then the function $g$ satisfies the differential equation
\begin{align}
    \frac{\partial g}{\partial \chi} = \left( g ( \chi ) + 2 \chi g'(\chi) \right)^2~ .
    \label{maxwell_ODE}
\end{align}
One can solve this differential equation by making a series ansatz of the form $g(\chi) = \sum_{n=0}^{\infty} c_n \chi^n$, determining the first several coefficients $c_n$. To order $\chi^6$, the function $g(\chi)$ is given by
\begin{align}
    g ( \chi ) = 1 + \chi + 3 \chi^2 + 13 \chi^3 + 68 \chi^4 + 399 \chi^5 + 2530 \chi^6 + \mathcal{O} ( \chi^7 ) .
\end{align}
To determine the generating function, one can refer to an encyclopedia of integer sequences \cite{hypergeometricSequence} to find that $g$ can be written as a generalized hypergeometric function,
\begin{align}
    g ( \chi ) = {}_{4} F_{3} \left( \frac{1}{2} , \frac{3}{4} , 1 , \frac{5}{4} ; \frac{4}{3} , \frac{5}{3} , 2 ; \frac{256}{27} \cdot \chi \right) ~. 
    \label{hypergeometric}
\end{align}
Thus the full solution for the deformed Lagrangian can be written as
\begin{align}
    \mathcal{L} ( \lambda ) &= F^2 \cdot {}_{4} F_{3} \left( \frac{1}{2} , \frac{3}{4} , 1 , \frac{5}{4} ; \frac{4}{3} , \frac{5}{3} , 2 ; \frac{256}{27} \cdot \lambda F^2 \right) \nonumber \\
    &= \frac{3}{4 \lambda} \left( {}_{3} F_2 \left( - \frac{1}{2} , - \frac{1}{4} , \frac{1}{4} ; \frac{1}{3} , \frac{2}{3} ; \frac{256}{27} \cdot \lambda F^2 \right) - 1 \right) .
    \label{two_forms_hypergeometric}
\end{align}
The functions on the first and second lines of (\ref{two_forms_hypergeometric}) are equivalent because of a hypergeometric functional identity. We will use the expression in the first line, written in terms of ${}_{4} F_3$ rather than ${}_3 F_2$, but we include the second expression to make contact with the work of \cite{Conti:2018jho}, where this expression was first derived. We also note that the function (\ref{two_forms_hypergeometric}) has appeared in an analogue of $T \Tb$ defined for $(0+1)$ dimensional theories \cite{Gross:2019ach}. 

\subsection{Implicit solution}\label{sec:implicit_flow_maxwell}

Later it will be useful to solve the $\TT$ flow equation using a different method. We begin with the differential equation (\ref{flow_for_pure_gauge_theory}), but this time we make the ansatz
\begin{align}
    f ( \lambda, F^2 ) = \frac{1}{\lambda} h ( \lambda F^2 ) .
\end{align}
As before, we define the dimensionless variable $\chi = \lambda F^2$. In terms of $h$, the differential equation becomes
\begin{align}
    4 \chi^2 \left( h' ( \chi ) \right)^2 - 4 \chi h ( \chi ) h' ( \chi ) - \chi h'(\chi) + h ( \chi )^2 + h ( \chi ) = 0 .
    \label{h_chi_intermediate}
\end{align}
Equation (\ref{h_chi_intermediate}) is quadratic in $h'(\chi)$, so we can solve to find
\begin{align}
    \frac{dh}{d \chi} = \frac{1 + 4 h ( \chi ) - \sqrt{1 - 8 h ( \chi ) } }{8 \chi} , 
    \label{h_deriv_separable}
\end{align}
where we have chosen the root which makes $h'(\chi)$ finite as $\chi \to 0$, assuming $\lim_{\chi \to 0} h ( \chi ) = 0$.

We may separate variables in (\ref{h_deriv_separable}) to write
\begin{align}
    \int \frac{8 \, dh}{1 + 4 h ( \chi ) - \sqrt{1 - 8 h ( \chi ) } } = \int \frac{d \chi}{\chi} .
    \label{sep_variables_pure_gauge}
\end{align}
The integrals can be evaluated in terms of logarithms; exponentiating both sides then yields
\begin{align}
    \chi = C \left( 1 - \sqrt{1 - 8 h} \right) \left( 3 + \sqrt{1 - 8 h} \right)^{3} . 
    \label{implicit_maxwell}
\end{align}
Equation (\ref{implicit_maxwell}) implicitly defines the solution $h ( \chi )$ to the $T \Tb$ flow equation via the roots of an algebraic equation. 

We note that, choosing $C = \frac{1}{256}$, equation (\ref{implicit_maxwell}) is consistent with the solution derived in the previous section. Recall that the two ansatzes we made here and in subsection (\ref{sec:direct_flow_maxwell}) are related by
\begin{align}
    f ( \lambda, F^2 ) = F^2 f ( \chi ) = \frac{1}{\lambda} h ( \chi ) , 
\end{align}
so $h ( \chi ) = \chi f ( \chi )$. Indeed, one can check that the function
\begin{align}
    h ( \chi ) = \chi \cdot {}_{4} F_{3} \left( \frac{1}{2} , \frac{3}{4} , 1 , \frac{5}{4} ; \frac{4}{3} , \frac{5}{3} , 2 ; \frac{256}{27} \cdot \chi \right) , 
\end{align}
satisfies the functional identity
\begin{align}
    \chi = \frac{1}{256} \left( 1 - \sqrt{1 - 8 h ( \chi ) } \right) \left( 3 + \sqrt{1 - 8 h ( \chi ) } \right)^{3} .
    \label{hypergeometric_as_root}
\end{align}
We therefore see that the hypergeometric (\ref{two_forms_hypergeometric}) obtained earlier is, in fact, an algebraic function that can be defined as a root of (\ref{hypergeometric_as_root}).\footnote{Other examples of hypergeometric functions which can be expressed algebraically include those on Schwarz's list \cite{Schwarz1873}, which is summarized \href{https://en.wikipedia.org/w/index.php?title=Schwarz\%27s_list\&oldid=933589284}{on Wikipedia}.}

\subsection{Solution via dualization}\label{sec:dualization_maxwell}
The above result can also be derived in a different way. The details of this procedure do not depend on the sign of our constant $k$ nor the signature, so we will set $k=1$ and take Minkowski signature for concreteness. The undeformed Lagrangian (\ref{undeformed_pure_gauge}) is then
\begin{align}
    \mathcal{L}_0 = \frac{1}{k} F_{\mu \nu} F^{\mu \nu} = - 2 F_{01}^2  ,
    \label{maxwell_undeformed_components}
\end{align}
which can be equivalently expressed by dualizing the field strength to a scalar, as in
\begin{align}
    \mathcal{L}_0 &= \frac{1}{2} \phi^2 + \phi \, \epsilon^{\mu \nu} F_{\mu \nu} \nonumber \\
    &= \frac{1}{2} \phi^2 + 2 \phi F_{01} .
    \label{maxwell_dualized}
\end{align}
The equation of motion for $\phi$ arising from (\ref{maxwell_dualized}) is
\begin{align}
    \frac{\delta \mathcal{L}_0}{\delta \phi} = \phi + 2 F_{01} = 0 ,
\end{align}
so $\phi = - 2 F_{01}$, and then replacing $\phi$ with its equation of motion yields
\begin{align}
    \mathcal{L}_0 = \frac{1}{2} \left( - 2 F_{01} \right)^2 + 2 \left( - 2 F_{01} \right) F_{01} = - 2 F_{01}^2 ,
\end{align}
which matches (\ref{maxwell_undeformed_components}). On the other hand, (\ref{maxwell_dualized}) is easy to $T \Tb$ deform. After coupling to gravity, one has
\begin{align}
    S_0 [ g ] = \left( \frac{1}{2} \int \sqrt{-g} \, \phi^2 \, d^2 x \right) + \left( \int \phi \, \epsilon^{\mu \nu} F_{\mu \nu} \, d^2 x \right) , 
\end{align}
where the second term is purely topological and thus independent of the metric. The undeformed Lagrangian, then, is a pure potential term $V ( \phi ) = \phi^2$ for the boson $\phi$. The solution to the $T \Tb$ flow equation at finite $\lambda$ for a general potential is well-known \cite{Cavaglia:2016oda, Bonelli:2018kik}; in this case, one finds
\begin{align}
    \mathcal{L} ( \lambda ) &= \frac{ \frac{1}{2} \phi^2 }{1 - \frac{\lambda}{2} \phi^2} + \phi \, \epsilon^{\mu \nu} F_{\mu \nu} \nonumber \\
    &= \frac{\phi^2}{2 - \lambda \phi^2} + 2 \phi F_{01} .
    \label{maxwell_deformed_phi}
\end{align}
We now integrate out $\phi$. The equation of motion for $\phi$ arising from (\ref{maxwell_deformed_phi}) is
\begin{align}
    0 &= \frac{\delta \mathcal{L} ( \lambda )}{\delta \phi} \nonumber \\
    &= 2 \phi + F_{01} \left( 2 - \lambda \phi^2 \right)^2 ,
\end{align}
or
\begin{align}
    F_{01} = - \frac{2 \phi}{ ( 2 - \lambda \phi^2 )^2 } .
    \label{deformed_sol_for_F12}
\end{align}
To proceed, we series expand (\ref{deformed_sol_for_F12}) in $\phi$ to find
\begin{align}
    F_{01} = - \frac{\phi}{2} - \frac{\lambda \phi^3}{2} - \frac{3 \lambda^2 \phi^5}{8} - \frac{\lambda^3 \phi^7}{4} - \frac{5 \lambda^4 \phi^9}{32} + \mathcal{O} \left( \phi^{11} \right) , 
\end{align}
and then apply the Lagrange inversion theorem to find a series expansion for $\phi$ in terms of $F_{01}$, yielding
\begin{align}
    \phi = - 2 F_{01} + 8 \lambda F_{01}^2 - 72 \lambda^2 F_{01}^5 + 832 \lambda^3 F_{01}^7 - 10880 \lambda^4 F_{01}^9 + \mathcal{O} \left( F_{01}^{11} \right) .
    \label{phi_from_inversion}
\end{align}
Substituting the expansion (\ref{phi_from_inversion}) into the action (\ref{maxwell_deformed_phi}), and expressing the result in terms of $F^2 = F_{\mu \nu} F^{\mu \nu} = - 2 F_{01}^2$, gives
\begin{align}
    \mathcal{L} ( \lambda ) = F^2 \left( 1 + \lambda F^2 + 3 \lambda F^4 + 13 \lambda^3 F^6 + 68 \lambda^4 F^8 + \cdots \right) . 
    \label{hypergeometric_series_approx}
\end{align}
The Taylor coefficients appearing in (\ref{hypergeometric_series_approx}) are precisely those of the hypergeometric (\ref{hypergeometric}). The procedure of iteratively solving (\ref{deformed_sol_for_F12}) for $\phi$ and substituting into (\ref{maxwell_deformed_phi}), therefore, reproduces
\begin{align}
    \mathcal{L} ( \lambda ) = F^2 \cdot {}_{4} F_{3} \left( \frac{1}{2} , \frac{3}{4} , 1 , \frac{5}{4} ; \frac{4}{3} , \frac{5}{3} , 2 ; \frac{256}{27} \cdot \lambda F^2 \right)
\end{align}
which matches the solution which we derived by different methods above.

This procedure -- dualizing the field strength $F^2$ to a scalar $\phi$, $\TT$ deforming the scalar action, and then dualizing back -- is closely related to an observation made in \cite{Conti:2018jho}. There the authors noted that, although the deformed Lagrangian (\ref{two_forms_hypergeometric}) is quite complicated, the corresponding Hamiltonian satisfies the simple relation
\begin{align}
    \mathcal{H}_\lambda = \frac{\mathcal{H}_0}{1 - \lambda \mathcal{H}_0} ,
\end{align}
where the Hamiltonian is a function of the conjugate momentum
\begin{align}
    \Pi^1 = \frac{\partial \mathcal{L}_\lambda}{\partial \dot{A}_1} .
\end{align}
The Legendre transform which converts the Lagrangian $\mathcal{L}_\lambda$ to the Hamiltonian $\mathcal{H}_\lambda$ is mathematically equivalent to the process of dualizing the field strength $F_{01}$ to a scalar $\phi$.

\subsection{Comparison to Born-Infeld}\label{sec:comp_BI}

For the moment, we specialize to the abelian case where the Born-Infeld action can be unambiguously defined. The Lagrangian (\ref{two_forms_hypergeometric}) differs from the Born-Infeld action in two dimensions. To order $\lambda^4$, our solution has the series expansion
\begin{align}
    \mathcal{L} ( \lambda ) = F^2 + \lambda F^4 + 3 \lambda^2 F^6 + 13 \lambda^3 F^8 + 68 \lambda^4 F^{10} + \mathcal{O} ( \lambda^5 ) .
    \label{series_expansion_compare_bi}
\end{align}
On the other hand, the Born-Infeld action (after normalizing the coefficient of $F^2$ to match (\ref{series_expansion_compare_bi}) at order $F^2$) has the expansion
\begin{align}
    \frac{1}{2 \lambda} \sqrt{1 + 4 \lambda F^2} = \frac{1}{2 \lambda} + F^2 - \lambda F^4 + 2 \lambda^2 F^6 - 5 \lambda^3 F^8 + 14 \lambda^4 F^{10} + \mathcal{O} ( \lambda^5 ) .
    \label{2d_BI_expansion}
\end{align}
Although the Taylor coefficients for the Born-Infeld action and the ``hypergeometric action'' differ, both exhibit a critical value for the electric field. In the case of Born-Infeld, this is obvious; replacing $F^2 = - 2 F_{01}^2$, we see that the action
\begin{align}
    \frac{1}{2 \lambda} \sqrt{1 - 8 \lambda F_{01}^2 }
\end{align}
is only real for
\begin{align}
    F_{01} < \frac{1}{\sqrt{8 \lambda}} .
    \label{critical_BI}
\end{align}
To see the critical electric field for the action $\mathcal{L} ( \lambda )$ defined in (\ref{hypergeometric}), it is most convenient to use the implicit form (\ref{hypergeometric_as_root}):
\begin{align}
    \lambda F^2 =  \frac{1}{256} \left( 1 - \sqrt{1 - 8 \lambda \mathcal{L} ( \lambda ) } \right) \left( 3 + \sqrt{1 - 8 \lambda \mathcal{L} ( \lambda ) } \right)^{3} .
\end{align}
The right side is maximized when $\mathcal{L} ( \lambda ) = \frac{1}{8 \lambda}$, where it takes the value $\frac{27}{256}$, which means that
\begin{align}
    F^2 < \frac{27}{256 \lambda} . 
\end{align}
Recall that our Lagrangian (\ref{undeformed_pure_gauge}) contained an overall constant to track signs; to restore factors of $k$, we replace $F^2 \to \frac{1}{k} F^2$. In Minkowski signature, we should take $k < 0$ so that $\mathcal{L}_0 = - \frac{1}{k} F_{\mu \nu} F^{\mu \nu} = - \frac{2}{k} F_{01}^2$ is positive. Letting $k=-1$, we find
\begin{align}
    F_{01} < \sqrt{ \frac{27}{512 \lambda} } , 
    \label{critical_hypergeometric}
\end{align}
which is a different critical value for the electric field than (\ref{critical_BI}).

However, pure Yang-Mills theory in two dimensions has no propagating degrees of freedom, so the difference between the expansions (\ref{series_expansion_compare_bi}) and (\ref{2d_BI_expansion}) does not have much physical effect (at least in infinite volume). To detect the difference between these theories, we should couple the gauge field to matter, as we do in section \ref{sec:DBI}.

\section{Non-Abelian Analogue of DBI}\label{sec:DBI}

In this section, we will consider an action for a Yang-Mills gauge field $F_{\mu \nu}^{a}$ coupled to a scalar $\phi$ in some representation of the gauge group. The undeformed Lagrangian is taken to be
\begin{align}
    \mathcal{L}_0 &= F_{\mu \nu}^{a} F^{\mu \nu}_a + | D \phi |^2 \nonumber \\
    &\equiv F^2 + | D \phi |^2 , 
    \label{coupled_undeformed}
\end{align}
where we again use the shorthand $F^2 = \Tr \left( F_{\mu \nu} F^{\mu \nu} \right)$. We have set the overall constant $k$, which appears in (\ref{undeformed_pure_gauge}), equal to $1$ for simplicity. If one were to retain this constant, however, one would need an extra minus sign in (\ref{coupled_undeformed}) in Minkowski signature.

In what follows, we will also define $x = F^2$ and $y = | D \phi |^2$ for convenience; here
\begin{align}
    | D \phi |^2 &= \left( D_\mu \phi \right) \left( D^\mu \phi \right)^\ast , \nonumber \\
    D_\mu &= \partial_\mu - i A_\mu , 
\end{align}
and gauge group indices will be suppressed.

At finite $\lambda$, we take a general ansatz of the form
\begin{align}
    \mathcal{L}_{\lambda} = f ( \lambda, x = F^2, y = | D \phi |^2 ) .
    \label{coupled_lagrangian_ansatz}
\end{align}
The stress tensor components $T_{\mu \nu}^{(\lambda)}$ for the Lagrangian (\ref{coupled_lagrangian_ansatz}), and the differential equation arising from (\ref{flow_invariant}), are worked out in Appendix \ref{flow_derivation}. The resulting partial differential equation, equation (\ref{main_pde}), is copied here for convenience:
\begin{align}
    \frac{d f}{d \lambda} =  f^2 - 4 f x \frac{\partial f}{\partial x} - 2 f y \frac{\partial f}{\partial y} + 4 x^2 \left( \frac{\partial f}{\partial x} \right)^2 + 4 x y \frac{\partial f}{\partial x} \frac{\partial f}{\partial y} .
\end{align}
Our goal in the following subsections will be to solve (\ref{main_pde}) by several methods, just as we did in the case of pure gauge theory.

\subsection{Series solution of flow equation}\label{sec:DBI_series}

We know that (\ref{main_pde}) reduces to the Dirac action, (\ref{turn_off_F}), when the gauge field is set to zero, and that it reduces to the hypergeometric action of Section \ref{sec:deforming_maxwell}, (\ref{turn_off_phi}), when the scalar is set to zero. Therefore, in the coupled case it is natural to make an ansatz of the form
\begin{align}
    f ( \lambda, x, y ) &= \frac{1}{2 \lambda} \left( \sqrt{1 + 4 \lambda y } - 1 \right) + {}_{3} F_4 \left( \frac{1}{2} , \frac{3}{4} , 1 , \frac{5}{4} ; \frac{4}{3} , \frac{5}{3} , 2 ; \frac{256}{27} \cdot \lambda x \right) \cdot x \nonumber \\
    &\qquad + \sum_{n=3}^{\infty} \sum_{k=1}^{n} c_{n, k} \lambda^{n-1} x^k y^{n-k} .
    \label{series_solution_coupled}
\end{align}
The sum on the final line of (\ref{series_solution_coupled}) allows for all possible couplings between $F^2$ and $| D \phi |^2$, with the appropriate power of $\lambda$ required by dimensional analysis. One can then determine the coefficients $c_{n, k}$ by plugging the ansatz (\ref{series_solution_coupled}) into (\ref{main_pde}) and solving order-by-order in $\lambda$. The result, up to coupled terms of order $\lambda^8$, is
\begin{align}
    f ( \lambda , x, y ) &= \frac{1}{4 \lambda} \left( \sqrt{1 + 16 \lambda y } - 1 \right) + {}_{3} F_4 \left( \frac{1}{2} , \frac{3}{4} , 1 , \frac{5}{4} ; \frac{4}{3} , \frac{5}{3} , 2 ; \frac{256}{27} \cdot \lambda x \right) \cdot x \nonumber \\
    &\quad - \lambda^2 x y^2 + \lambda^3 \left( 4 \ x y^3 - 4 \, x^2 y^2 \right) + \lambda^4 \left( 18 \, x^2 y^3 - 22 \, x^3 y^2 - 14 \, x y^4 \right) \nonumber \\
    &\quad + \lambda^5 \left( -140 \, x^4 y^2  + 104 \, x^3 y^3 - 65 \, x^2 y^4 + 48 \, x y^5 \right) \nonumber \\
    &\quad + \lambda^6 \left( -165 \, x y^6 + 220 \, x^2 y^5 - 364 \, x^3 y^4 + 680 \, x^4 y^3 - 969 \, x^5 y^2 \right) \nonumber \\
    &\quad + \lambda^7 \left( 572 \, x y^7 - 726 \, x^2 y^6 + 1120 \, x^3 y^5 - 2244 \, x^4 y^4 + 4788 \, x^5 y^3 - 7084 \, x^6 y^2 \right) \nonumber \\
    &\quad + \lambda^8 \Big( -2002 \, x y^8 + 2392 \, x^2 y^7 - 3160 \, x^3 y^6 + 5814 \, x^4 y^5 - 14630 \, x^5 y^4 + 35420 \,  x^6 y^3 \nonumber \\
    &\qquad \qquad - 53820 \, x^7 y^2 \Big) .
    \label{numerical_series_solution_coupled}
\end{align}
We were unable to find an closed-form expression for the function which generates the couplings (\ref{numerical_series_solution_coupled}). However, it is interesting to study the corrections in various approximations.

For instance, consider the coupled terms between $F^2$ and $|D \phi|^2$ to leading order in the variable $y = |D \phi|^2$. Retaining only the couplings in (\ref{numerical_series_solution_coupled}) proportional to $y^2$, one finds
\begin{align}\hspace{-10pt}
    f ( \lambda , x, y ) &= \frac{1}{4 \lambda} \left( \sqrt{1 + 16 \lambda y } - 1 \right) + {}_{3} F_4 \left( \frac{1}{2} , \frac{3}{4} , 1 , \frac{5}{4} ; \frac{4}{3} , \frac{5}{3} , 2 ; \frac{256}{27} \cdot \lambda x \right) \cdot x - \lambda^2 x y^2  - 4 \lambda^3  \, x^2 y^2 \nonumber \\
    &\quad - 2  \lambda^4 \, x^3 y^2 - 140 \lambda^5 \, x^4 y^2 - 969 \lambda^6  \, x^5 y^2 - 7084 \lambda^7 \, x^6 y^2   - 53820 \lambda^8 \, x^7 y^2 \nonumber \\ & \quad + \mathcal{O} \left( \lambda^9 , \lambda^3 x y^3 \right) .
\end{align}
These series coefficients resum into another hypergeometric function \cite{hypergeometricCoupledSequence},
\begin{align}\hspace{-30pt}
    f ( \lambda , x, y ) &= \frac{1}{4 \lambda} \left( \sqrt{1 + 16 \lambda y } - 1 \right) + {}_{3} F_4 \left( \frac{1}{2} , \frac{3}{4} , 1 , \frac{5}{4} ; \frac{4}{3} , \frac{5}{3} , 2 ; \frac{256}{27} \cdot \lambda x \right) \cdot x \nonumber \\
    &\quad - \lambda^2 x y^2 \, {}_{3} F_2 \left( \frac{1}{4} , \frac{1}{2} , \frac{3}{4} ; \frac{2}{3} , \frac{4}{3} ; \frac{256}{27} \cdot \lambda x \right) + \mathcal{O} \left( \lambda^3 x y^3 \right)  .
    \label{coupled_resummed}
\end{align}
Defining the hypergeometric appearing in the correction term as
\begin{align}
     g ( \chi ) = {}_{3} F_2 \left( \frac{1}{4} , \frac{1}{2} , \frac{3}{4} ; \frac{2}{3} , \frac{4}{3} ; \frac{256}{27} \cdot \chi \right) , 
\end{align}
one can show that $g$ satisfies the functional relation
\begin{align}
    \chi = \frac{g ( \chi ) - 1}{g(\chi)^4} .
    \label{first_correction_critical_field_strength}
\end{align}
The maximum of the function $\frac{g-1}{g^4}$ occurs when $g = \frac{4}{3}$, at which this function takes the maximal value of $\frac{27}{256}$. Therefore, the maximum value of $\chi$ for which the function (\ref{first_correction_critical_field_strength}) is defined is $\chi = \frac{27}{256}$, giving a critical field strength
\begin{align}
    F^2 < \frac{27}{256 \lambda} .
\end{align}
We note that this is the same value of the critical electric field as that in the uncoupled term involving ${}_3 F_4$ in (\ref{coupled_resummed}), which we saw in (\ref{critical_hypergeometric}) for the case of pure gauge theory. It is reasonable to expect that the value of the critical electric field is modified if one includes corrections to all orders in $|D \phi|^2$, as is the case for the Dirac-Born-Infeld action.

\subsection{Implicit solution}\label{sec:DBI_implicit}

We can instead solve \eqref{main_pde} in terms of a complete integral. First, we refine our ansatz to
\begin{align}
    f ( \lambda ) &= \frac{1}{\lambda} g ( \chi, \eta )\quad, \quad \chi = \lambda x~ ,~\eta = \lambda y~ .
\end{align}
After doing this, the differential equation becomes
\begin{align}
    0 = 4 \chi^2 \left( \frac{\partial g}{\partial \chi} \right)^2 + 4 \eta \chi \frac{\partial g}{\partial \chi} \frac{\partial g}{\partial \eta} - 4 \chi g \frac{\partial g}{\partial \chi} - \chi \frac{\partial g}{\partial \chi} - 2 \eta g \frac{\partial g}{\partial \eta} - \eta \frac{\partial g}{\partial \eta} + g^2 + g .
\end{align}
Making a change of variables to 
\begin{align}
    p &= \log ( \chi )\quad, \quad    q = \log ( \eta ) ~, 
\end{align}
and writing $g ( \chi, \eta ) = w( p , q )$,  the differential equation for $h$ becomes 
\begin{align}
    0 = 4 \left( \frac{\partial w}{\partial p} \right)^2 + 4 \frac{\partial w}{\partial p} \frac{\partial w}{\partial q} - \left( 4 w + 1 \right) \frac{\partial w}{\partial p} - \left( 2 w + 1 \right) \frac{\partial w}{\partial q} + w^2 + w~ .
    \label{final_coupled_flow}
\end{align}
This can be solved by consulting a handbook of partial differential equations (see, for instance, equation 15 in section 2.2.6 of \cite{polyanin}). For any partial differential equation of the form
\begin{align}
    0 = f_1 ( w ) \left( \frac{\partial w}{\partial x} \right)^2 + f_2 ( w ) \frac{\partial w}{\partial x} \frac{\partial w}{\partial y} + f_3 ( w ) \left( \frac{\partial w}{\partial y} \right)^2 + g_1 ( w ) \frac{\partial w}{\partial x} + g_2 ( w ) \frac{\partial w}{\partial y} + h ( w ) ,
    \label{general_pde_sol}
\end{align}
the solution $w ( x , y ) $ is given implicitly by the following complete integral:
\begin{align}
    C_3 &= C_1 x + C_2 y + \int \frac{2 F ( w ) \, dw }{ G ( w ) \pm \sqrt{ G ( w )^2 - 4 F(w) h(w) } } , \nonumber \\
    F ( w ) &= C_1^2 f_1 ( w ) + C_1 C_2 f_2 ( w ) + C_2^2 f_3 ( w ) , \nonumber \\
    G ( w ) &= C_1 g_1 ( w ) + C_2 g_2 ( w ) . 
\end{align}
Our equation (\ref{final_coupled_flow}) is precisely of the form (\ref{general_pde_sol}), after identifying the independent variables $p \sim x$, $q \sim y$, and with the following functions:
\begin{align}\begin{split}
    &f_1 = f_2= 4 \quad  , \quad f_3 = 0 \quad , \quad g_1 = - 4 w - 1~,\\
    &g_2 = - 2 w - 1 \quad , \quad h(w) = w^2 + w .
\end{split}\end{align}
Therefore, the functions $F$ and $G$ are
\begin{align}
    F ( w ) &= 4 C_1^2 + 4 C_1 C_2 , \nonumber \\
    G ( w ) &= C_1 \left( - 4 w - 1 \right) + C_2 \left( - 2 w - 1 \right) \nonumber \\
    &= \left( - 4 C_1 - 2 C_2 \right) w - \left( C_1 + C_2 \right) . 
\end{align}
Our solution, then, is
\begin{align}
    &C_3 = C_1 p + C_2 q  \\
    &\hspace{-15pt} + \int \frac{8 \left( C_1^2 + C_1 C_2 \right) \, d w }{ - \left( 4 C_1 + 2 C_2 \right) w - \left( C_1 + C_2 \right) - \sqrt{ \left( \left( 4 C_1 + 2 C_2 \right) w + \left( C_1 + C_2 \right) \right)^2 - 16 \left( C_1^2 + C_1 C_2 \right) \left( w^2 + w \right) } }~ .
    \label{coupled_implicit_solution} \nonumber
\end{align}
where we have taken the negative root in the denominator, appropriate if $C_1 + C_2 < 0$.

Choosing values of the constants $C_1, C_2, C_3$ in (\ref{coupled_implicit_solution}) gives an implicit relation for the function $w(p,q)$ which solves the $\TT$ flow equation. For instance, if we set $C_2 = 0$ and $C_1 = -1$, equation (\ref{coupled_implicit_solution}) becomes
\begin{align}
    C_3 + \log ( \chi ) = \int \frac{8 \, dw}{4 w + 1 - \sqrt{1 - 8 w} } , 
\end{align}
which reproduces the result (\ref{sep_variables_pure_gauge}) which we found in the case of pure gauge theory. In this sense, our implicit solution is a generalization of the technique of section \ref{sec:implicit_flow_maxwell} to the case where $| D \phi |^2 \neq 0$.

The integral appearing in (\ref{coupled_implicit_solution}) can be computed explicitly in terms of logarithms (or, equivalently, inverse hyperbolic tangents). The integral is of the form
\begin{align}
    I ( w ) = \int \frac{ \alpha \, dw }{- \beta w - \gamma - \sqrt{ \left( \beta w + \gamma \right)^2 - 2 \alpha \left( w^2 + w \right) } } , 
\end{align}
where
\begin{align}
    \alpha &= 8 ( C_1^2 + C_1 C_2 ) , \nonumber \\
    \beta &= 4 C_1 + 2 C_2 , \nonumber \\
    \gamma &= C_1 + C_2 .
\end{align}
The result can be written as
\begin{align}
    I ( w ) &= \frac{1}{2} \Bigg( (\gamma -\beta ) \tanh ^{-1}\left(\frac{\alpha  (w+1)+(\beta -\gamma ) (\gamma +\beta  w)}{(\beta -\gamma ) \sqrt{\gamma ^2+w^2 \left(2 \alpha +\beta ^2\right)+2 w (\alpha +\beta  \gamma )}}\right) \nonumber \\ 
    &\qquad -\gamma  \tanh ^{-1}\left(\frac{\gamma ^2+w (\alpha +\beta  \gamma )}{\gamma  \sqrt{\gamma ^2+w^2 \left(2 \alpha +\beta ^2\right)+2 w (\alpha +\beta  \gamma )}}\right) \nonumber \\
    &\qquad +\sqrt{2 \alpha +\beta ^2} \tanh ^{-1}\left(\frac{\alpha +2 \alpha  w +\beta  (\gamma +\beta  w)}{\sqrt{2 \alpha +\beta ^2} \sqrt{\gamma ^2+w^2 \left(2 \alpha +\beta ^2\right)+2 w (\alpha +\beta  \gamma )}}\right) \nonumber \\
    &\qquad + \left( \beta - \gamma \right)  \log (w+1)+\gamma  \log (w) \Bigg) .
\end{align}
Exponentiating both sides then gives
\begin{align}
    \exp \left( C_3 - C_1 p - C_2 q \right) = \exp \left( I ( w ) \right) .
    \label{implicit_w_solution}
\end{align}
After simplifying the exponentials of the inverse hyperbolic tangents in (\ref{implicit_w_solution}), the right side only involves rational functions and radicals. This relation, therefore, gives an algebraic equation in $w$ whose roots are the solution to the $T \Tb$ flow.

By construction, a function $w(p, q)$ which satisfies (\ref{implicit_w_solution}) solves the differential equation (\ref{final_coupled_flow}). However, this technique is more unwieldy than the direct series solution for generating Taylor coefficients. The main utility of this strategy is the conceptual result that the solution $w ( p, q )$ is, in principle, defined by the root of an equation which involves only radicals and quotients, as we saw for the pure gauge theory case in (\ref{hypergeometric_as_root}).

\subsection{Solution via dualization}\label{sec:DBI_dualization}

One can also apply the dualization technique of section (\ref{sec:deforming_maxwell}) to the coupled action. Begin with the undeformed action
\begin{align}
    \mathcal{L}_0 = | D \phi |^2 + F^2 ~ ,
\end{align}
where we put $k = 1$ since the sign will not affect this calculation. Exactly as before, this action is equivalent to
\begin{align}
    \mathcal{L}_0 = | D \phi |^2 + \frac{1}{2} \chi^2 + \chi \epsilon^{\mu \nu} F_{\mu \nu}~,
\end{align}
after integrating out the field $\chi$, although this form of the Lagrangian hides some complexity because the covariant derivative $D$ is now non-local in $\chi$. Ignoring this for the moment, we again note that the action coupled to a background metric is of the form
\begin{align}
    S_0 = \left( \int \sqrt{-g} \, d^2 x \, \left( | D \phi |^2 + \frac{1}{2} \chi^2 \right) \right) + \left( \int d^2 x \, \chi \epsilon^{\mu \nu} F_{\mu \nu} \right) ~.
    \label{coupled_action_dualized}
\end{align}
As far as the $T \Tb$ deformation is concerned, (\ref{coupled_action_dualized}) is simply the action of a complex scalar $\phi$ with a constant potential $V = \frac{1}{2} \chi^2$. The solution to the flow equation at finite $\lambda$ is \cite{Cavaglia:2016oda, Bonelli:2018kik}
\begin{align}
    \mathcal{L} ( \lambda ) = \frac{1}{2 \lambda} \sqrt{ \frac{ \left( 1 - \lambda \chi^2 \right)^2 }{ \left( 1 - \frac{1}{2} \lambda \chi^2 \right)^2 } + 2 \lambda \, \frac{ 2 |D \phi|^2 + \chi^2 }{1 - \frac{1}{2} \lambda \chi^2} } - \frac{1}{2 \lambda} \, \frac{1 - \lambda \chi^2}{1 - \frac{1}{2} \lambda \chi^2} + \chi \epsilon^{\mu \nu} F_{\mu \nu} ~. 
    \label{coupled_deformed_chi}
\end{align}
As in section (\ref{sec:deforming_maxwell}), one might hope to iteratively integrate out the auxiliary field $\chi$ in (\ref{coupled_deformed_chi}) in order to express the result in terms of $F^2$. The equation of motion for $\chi$ resulting from (\ref{coupled_deformed_chi}), after solving for $F_{01}$ (and assuming that $\lambda \chi^2 < 2$), is
\begin{align}
    F_{01} = \frac{\chi  \left(-|D \phi|^2 \lambda  \left(2-\lambda  \chi ^2\right)-\sqrt{2 |D \phi|^2 \lambda  \left(2-\lambda  \chi ^2\right)+1}-1\right)}{\left(2-\lambda  \chi ^2\right)^2 \sqrt{2 |D \phi|^2 \lambda  \left(2-\lambda  \chi ^2\right)+1}} ~.
    \label{coupled_F12_sol}
\end{align} 
Solving (\ref{coupled_F12_sol}) by series inversion to give $\chi$ as a function of $F_{01}$, then substituting back into (\ref{coupled_deformed_chi}), then determines the full $T \Tb$ deformed action. The result, up to order $F^8$ and using the shorthand $x = F^2, y = | D \phi |^2$, is
\begingroup
\allowdisplaybreaks
\begin{align}\hspace*{-50pt}
   & \mathcal{L} ( \lambda ) = \frac{\sqrt{1 + 4 \lambda y}-1}{2 \lambda } + \frac{2 x \left(\sqrt{1 + 4 \lambda y}+2 \lambda  y \left(\sqrt{1 + 4 \lambda y}+2\right)+1\right)}{\left(2 \lambda  y+\sqrt{1 + 4 \lambda y}+1\right)^2} \nonumber \\
    &\quad + \frac{16 \lambda x^2}{\left(2 \lambda  y+\sqrt{1 + 4 \lambda y}+1\right)^5} \cdot \Bigg[ 2 \lambda^3 y^3 \left(3 \sqrt{1 + 4 \lambda y}+14\right)+\lambda^2 y^2 \left(17 \sqrt{1 + 4 \lambda y}+31\right) \nonumber \\
    &\hspace{150pt} +2 \lambda  y \left(4 \sqrt{1 + 4 \lambda y}+5\right)+\sqrt{1 + 4 \lambda y}+1  \Bigg] \nonumber \\
    &\quad + \frac{128 \lambda^2 x^3}{\left(2 \lambda  y+\sqrt{1 + 4 \lambda y}+1\right)^8} \cdot \Bigg[ 4 \lambda^5 y^5 \left(13 \sqrt{1 + 4 \lambda y}+96\right)+2 \lambda^4 y^4 \left(183 \sqrt{1 + 4 \lambda y}+496\right) \nonumber \\
    &\hspace{150pt} +8 \lambda^3 y^3 \left(57 \sqrt{1 + 4 \lambda y}+101\right)+2 \lambda^2 y^2 \left(106 \sqrt{1 + 4 \lambda y}+145\right) \nonumber \\
    &\hspace{150pt} + 6 \lambda  y \left(7 \sqrt{1 + 4 \lambda y}+8\right)+3 \left(\sqrt{1 + 4 \lambda y}+1\right) \Bigg] \nonumber \\
    &\quad + \frac{ 1024 \lambda^3  x^4}{\left(2 \lambda  y+\sqrt{1 + 4 \lambda y}+1\right)^{11}} \cdot \Bigg[ 2 \lambda^7 y^7 \left(323 \sqrt{1 + 4 \lambda y}+3266\right) +\lambda^6 y^6 \left(8471 \sqrt{1 + 4 \lambda y}+30585\right) \nonumber \\
    &\hspace{150pt} +2 \lambda^5 y^5 \left(9879 \sqrt{1 + 4 \lambda y}+22795\right)+4 \lambda^4 y^4 \left(4589 \sqrt{1 + 4 \lambda y}+8019\right)  \nonumber \\
    &\hspace{150pt} +2 \lambda^3 y^3 \left(4244 \sqrt{1 + 4 \lambda y}+6093\right)+\lambda^2 y^2 \left(2083 \sqrt{1 + 4 \lambda y}+2577\right) \nonumber \\
    &\hspace{150pt} +26 \lambda  y \left(10 \sqrt{1 + 4 \lambda y}+11\right)+13 \left(\sqrt{1 + 4 \lambda y}+1\right)  \Bigg] ~ .
    \label{coupled_legendre_series}
\end{align}%
\endgroup
We have checked by explicit computation that the series expansion (\ref{coupled_legendre_series}) solves the flow equation (\ref{main_pde}) to order $x^4$.

\chapter{$\TT$ Deformations of Gauge Theories in Arbitrary Dimension} \label{CHP:new_chapter}
This chapter aims to extend the study of $\TT$ deformations of pure gauge theories -- which was discussed, in part, in Chapters \ref{CHP:nonlinear} and \ref{CHP:nonabelian} -- to general spacetime dimension. Because an unambiguous definition of the $\TT$ operator at the quantum level is available only in $d=2$, the treatment of this chapter will be totally classical. We will view stress tensor deformations as merely a mechanism for generating interesting one-parameter families of classical Lagrangians, rather than an irrelevant deformation by an operator which admits a definition by point-splitting.

Unlike the majority of this thesis, the current chapter contains new results which do not appear in any previous publications. I thank Savdeep Sethi and Gabriele Tartaglino-Mazzucchelli for useful discussions which prompted this work.\footnote{I am also grateful to Mukund Rangamani, Julio Virrueta, Shruti Paranjape, Brendan Halstead, and Luna Dole for fruitful conversations about extensions of these results to the $p$-form and non-abelian settings.}

\section{Motivation and $3d$ Example}\label{sec:motivation_and_3d}

In the preceding chapters, we have seen that interesting gauge theories can be built from $\TT$ flows in $d=2$ dimensions and in $d=4$ dimensions. In $d=2$, deforming the theory of a gauge field coupled to charged matter gives a quantum-mechanically well-defined theory which is compatible with maximal supersymmetry. This theory does not reduce to the Dirac-Born-Infeld (DBI) action, even in the case of an abelian gauge group, and the pure gauge part of the action is written in terms of a hypergeometric function. In $d=4$, deforming the theory of an abelian gauge field using a $\TT$-like operator gives the Born-Infeld action, although in this case we do not expect that the operator is well-defined by point-splitting without imposing extended supersymmetry. 



It is natural to ask whether interesting actions arise from deformations by stress tensor bilinears in other numbers of spacetime dimensions besides $d=2$ and $d=4$. Although the analysis for $d=3$ is a special case of the one that we will develop later in this chapter, we will include it explicitly here by way of motivating some of the ideas which appear in the general case. In particular, we will ask whether a Born-Infeld-type Lagrangian involving a square root can be obtained from deforming the $d=3$ Maxwell Lagrangian by any combination of stress tensor bilinears.

First consider a general Lagrangian depending on a field strength $F_{\mu \nu}$ and parameter $\lambda$. As we will prove later, in $d=3$ the only scalar which can be constructed from $F_{\mu \nu}$ is $F^2 = F^{\mu \nu} F_{\mu \nu}$. Thus an arbitrary Lagrangian can be written as
\begin{align}
    \mathcal{L} = f ( \lambda, F^2  ) .
\end{align}
For convenience, let $x = F^2$. We can now compute the stress-energy tensor by coupling to a background metric and varying with respect to the metric, where our conventions will be given in (\ref{hilbert_stress_defn}). One finds
\begin{align}
    T_{\mu \nu} = \eta_{\mu \nu} \, f - 4 \frac{\partial f}{\partial x} \tensor{F}{_\mu^\sigma} F_{\nu \sigma } .
\end{align}
We then compute the two combinations $T^{\mu \nu} T_{\mu \nu}$ and $\left( \tensor{T}{^\mu_\mu} \right)^2$. The first is
\begin{align}
    T^{\mu \nu} T_{\mu \nu} &= \left( \eta^{\mu \nu} \, f - 4 \frac{\partial f}{\partial x} \tensor{F}{^\mu^\sigma} \tensor{F}{^\nu_\sigma} \right) \left( \eta_{\mu \nu} \, f - 4 \frac{\partial f}{\partial x} \tensor{F}{_\mu^\rho} F_{\nu \rho} \right) \nonumber \\
    &= 3 f^2 - 8 f \frac{\partial f}{\partial x} F^2 + 16 \left( \frac{\partial f}{\partial x} \right)^2 F^{\mu \sigma} \tensor{F}{_\sigma^\nu} \tensor{F}{_\mu^\rho} F_{\rho \nu} .
\end{align}
The last term, $F^{\mu \sigma} \tensor{F}{_\sigma^\nu} \tensor{F}{_\mu^\rho} F_{\rho \nu}$, naively appears to be a new scalar quantity constructed from $F_{\mu \nu}$ which defines an independent quantity in the action. However, in three spacetime dimensions, one has the relation
\begin{align}
    F^{\mu \sigma} \tensor{F}{_\sigma^\nu} \tensor{F}{_\mu^\rho} F_{\rho \nu} = \frac{1}{2} \left( F_{\mu \nu} F^{\mu \nu} \right)^2 .
\end{align}
This is analogous to the fact that, in $d=2$ spacetime dimensions, one has $\tensor{F}{_\mu^\sigma} F_{\sigma \nu} = - \frac{1}{2} \eta_{\mu \nu} \left( F_{\alpha \beta} F^{\alpha \beta} \right)$. This was used in Appendix \ref{flow_derivation} to obtain the general flow equation for two-dimensional theories which was analyzed in Chapter \ref{CHP:nonabelian}; in both cases, we are exploiting the fact that in low spacetime dimension ($d=2, 3$), the number of independent scalars that can be constructed from $F_{\mu \nu}$ is small.

Replacing $F_{\mu \nu} F^{\mu \nu} = x$, we then find
\begin{align}
    T^{\mu \nu} T_{\mu \nu} = 3 f^2 - 8 f x \frac{\partial f}{\partial x} + 8 \left( \frac{\partial f}{\partial x} \right)^2 x^2 .
\end{align}
The second combination is
\begin{align}
    \left( \tensor{T}{^\mu_\mu} \right)^2 &= \left( 3 f - 4 F^2 \frac{\partial f}{\partial x}  \right)^2 = 9 f^2 - 24 f x \frac{\partial f}{\partial x} + 16 \left( \frac{\partial f}{\partial x} \right)^2 x^2 \, .
\end{align}
We would like to consider an arbitrary flow equation for the Lagrangian which is built out of these stress tensor bilinears,
\begin{align}
    \frac{\partial \mathcal{L}}{\partial \lambda} = a T^{\mu \nu} T_{\mu \nu} - b \left( \tensor{T}{^\mu_\mu} \right)^2 \, .
\end{align}
Of course, we can absorb an overall factor of $a$ into the normalization of $\lambda$ and then define the relative coefficient $- \frac{b}{a}$ as a new parameter $r$. So it suffices to consider flow equations of the form
\begin{align}\label{general_r_deformation}
    \frac{\partial \mathcal{L}}{\partial \lambda} = T^{\mu \nu} T_{\mu \nu} - r \left( \tensor{T}{^\mu_\mu} \right)^2 \, .
\end{align}
Using our expressions for the two stress tensor products derived above, the general flow equation is
\begin{align}\label{3d_flow}
    \frac{df}{d \lambda} &= T^{\mu \nu} T_{\mu \nu} - r \left( \tensor{T}{^\mu_\mu} \right)^2 \nonumber \\
    &= 3 \left( 1 - 3 r \right) f^2 - 8 \left( 1 - 3 r \right) f x \frac{\partial f}{\partial x} + 8 \left( 1 - 2 r \right) \left( \frac{\partial f}{\partial x} \right)^2 x^2 ,
\end{align}
This flow equation may have interesting solutions for various values of the relative coefficient $r$. For the moment, we will content ourselves by asking whether it has any solutions of Born-Infeld type. In three spacetime dimensions, one has
\begin{align}
    \sqrt{ - \det ( \eta_{\mu \nu} + \alpha F_{\mu \nu} ) } = \sqrt{1 + \frac{1}{2} \alpha^2 F^2} \, .
\end{align}
Viewing $\lambda$ as playing the role of $\alpha^2$, we ask whether we can ever get solutions of the form
\begin{align}\label{BI_ansatz_3d}
    f ( \lambda , x ) = \frac{a}{\lambda} \left( \sqrt{ 1 + b \lambda x } - 1 \right) \, , 
\end{align}
which by construction is proportional to the Maxwell action as $\lambda \to 0$.\footnote{\label{dimensional_analysis_footnote}We briefly comment on dimensional analysis. Ordinarily, the free action in $d$ spacetime dimensions is $S = - \frac{1}{4 g^2} \int d^d x \, F^{\mu \nu} F_{\mu \nu}$ and $F_{\mu \nu}$ has mass dimension $2$. However, it is convenient to re-scale $F \to \frac{1}{g} F$ and eliminate the prefactor of $\frac{1}{g^2}$ in the Lagrangian. In these conventions, the combination $x = F^2$ has mass dimension $d$ so the field strength $F_{\mu \nu}$ has mass dimension $\frac{d}{2}$. Similarly, the stress tensor $T_{\mu \nu}$ has mass dimension $d$ in any number of dimensions, so stress tensor bilinears like $\TT$ have mass dimension $2d$. Therefore, by dimensional analysis of the flow equation $\frac{\partial \mathcal{L}}{\partial \lambda} = \TT$, we see that $\lambda$ must have length dimension $d$. Thus in the case $d=3$ of interest here, $\lambda$ has length dimension $3$. This means that the combination $\lambda x$ under the square root in (\ref{BI_ansatz_3d}) is dimensionless, as required, and the overall factor of $\frac{1}{\lambda}$ has the same dimension as $F^2$, which is the appropriate dimension for a Lagrangian density.}

By using the ansatz (\ref{BI_ansatz_3d}) in the general flow equation
\begin{align}\label{general_pde_flow}
    \frac{d f}{d \lambda} = c_1 f^2 + c_2 f x \frac{\partial f}{\partial x} + c_3 \left( \frac{\partial f}{\partial x} \right)^2 x^2 \, ,
\end{align}
one can see that (\ref{general_pde_flow}) will have solutions of this square-root-type only if $c_3 = 0$ and $c_2 = - 2 c_1$, in which case the solution is
\begin{align}
    f ( \lambda, x ) = \frac{1}{2 \lambda c_1} \left( \sqrt{1 + 4 c_2 \lambda x} - 1 \right) \, , 
\end{align}
where $c_1, c_2$ are fixed by the initial condition, which in this case is $f(0, x) = - \frac{x}{4}$. However, even if we choose $r = \frac{1}{2}$ so that the unwanted final term in (\ref{3d_flow}) drops out, the differential equation for our three-dimensional flow is
\begin{align}
    \frac{d f}{d \lambda} = - \frac{3}{2} f^2 + 4 f x \frac{\partial f}{\partial x} . 
\end{align}
This does not have a square-root-type solution, since it is not of the form (\ref{general_pde_flow}) with $c_2 = - 2 c_1$; instead we have $c_2 = 4$ and $-2 c_1 = 3$.
Thus the Born-Infeld action cannot be obtained via any $\TT$-like deformation of the Maxwell action in three spacetime dimensions.

The above analysis raises several questions. Can the Born-Infeld action be realized as a $\TT$ flow in any other dimension $d > 4$? Can we obtain other interesting actions by applying $\TT$-like deformations to seed theories other than the free Maxwell action? For instance, perhaps we would like to begin with the Born-Infeld action as the seed theory, rather than as a desired output of deformation, and see what action is produced by the flow. What additional complications arise when deforming non-abelian gauge theories? For instance, can one define non-abelian analogues of BI in higher dimensions using $\TT$-like flows as well?

The aim of this chapter is to present a framework for answering questions like this by considering a $\TT$-like deformation of a quite general gauge theory in arbitrary spacetime dimension. Throughout this chapter, we will restrict to pure gauge theory without any charged matter. In the abelian case, we will be able to explicitly derive a flow equation for a family of $\TT$-type deformations of such general theories. Interestingly, the only case for which deformation of a free Maxwell theory yields a Lagrangian of Born-Infeld type occurs for $d=4$. We will also discuss the more subtle case of deformations for non-abelian gauge theories, where it is immediate that one can never obtain a deformed Lagrangian of Born-Infeld type because $\TT$ flows cannot generate terms with the appropriate trace structure.

\section{Deformations of General Abelian Gauge Theories}\label{sec:general_abelian}

In this section, we will restrict to the simpler case of a $U(1)$ gauge field $A_\mu$, which avoids complications from traces over group indices.

\subsection{Signature and Sign Conventions}

Unlike in the $3d$ example of the introductory section \ref{sec:motivation_and_3d}, for the remainder of this chapter we will work in Euclidean signature in $d$ spacetime dimensions. In this signature the field strength $F$ can be thought of as a $d \times d$ matrix regardless of the positioning of the indices, which are raised or lowered with $\delta_{\mu \nu}$.

Our conclusions will be valid for either Euclidean or Minkowski signature, so the choice of restricting to Euclidean signature is merely for simplicity; we did not feel that the additional generality of treating both cases merited the introduction of a constant $k = \pm 1$ and writing $\mathcal{L} = \frac{k}{4} F_{\mu \nu} F^{\mu \nu}$ as in Chapter \ref{CHP:nonabelian}. We can afford to be somewhat cavalier about the overall sign of the Lagrangian because our goal is to understand deformations by stress tensor bilinears, and since reversing the sign of the Lagrangian also flips the sign of $T_{\mu \nu}$, all bilinears in $T_{\mu \nu}$ are insensitive to the sign of $\mathcal{L}$. However, to avoid sign confusion and reiterate our conventions, we will state a few formulas here.

We will take the Euclidean action for free Maxwell theory to be
\begin{align}\label{S_euc}
    S_{\text{Euc}} = \frac{1}{4} \int d^d x \, F_{\mu \nu} F^{\mu \nu} \equiv \frac{1}{4} \int d^d x \, F^2 \, .
\end{align}
The factors of the coupling $g$ have been absorbed into the field strength, following the choice in footnote \ref{dimensional_analysis_footnote}, so that $F_{\mu \nu}$ now has mass dimension $\frac{d}{2}$. We emphasize that this is in contrast to the Lorentz signature action,
\begin{align}\label{S_lor}
    S_{\text{Lor}} = - \frac{1}{4} \int d^d x \, F_{\mu \nu} F^{\mu \nu} \, ,
\end{align}
which was only used in the introductory discussion of Section \ref{sec:motivation_and_3d} but will be mostly discarded in favor of (\ref{S_euc}) from here on.

In this chapter, expressions like $F^{2n}_{\mu \nu}$ (which we will often write with indices suppressed as $F^{2n}$) will always refer to the $d \times d$ matrix obtained by successively performing $2n$ matrix multiplications of $F_{\mu \nu}$ by itself, leaving the first and last indices free:
\begin{align}\label{F2n_def}
    F^{2n}_{\mu \nu} = \underbrace{F_{\mu \alpha_1} \tensor{F}{^{\alpha_1}_{\alpha_2}} \cdots \tensor{F}{^{\alpha_{2n-2}}_{\alpha_{2n-1}}} \tensor{F}{^{\alpha_{2n-1}}_\nu} }_{2n \text{ copies} } \, .
\end{align}
Likewise, $\tr ( M_{\mu \nu} ) = \delta^{\mu \nu} M_{\mu \nu}$ refers to the trace of any matrix $M_{\mu \nu}$ with the Kronecker delta.\footnote{As in Chapter \ref{CHP:nonabelian}, we use the notation $\tr$ for the trace over tensor indices and reserve $\Tr$ for the trace over gauge indices in the non-abelian setting.}

At the risk of being pedantic, we point out that this convention introduces a potentially confusing sign in expressions like $\tr ( F^2 )$. In our conventions, the matrix $F^2_{\mu \nu}$ is defined according to (\ref{F2n_def}) as
\begin{align}
    F^2_{\mu \nu} = F_{\mu \rho} \tensor{F}{^\rho_\nu} \, , 
\end{align}
and therefore its trace is
\begin{align}
    \tr ( F^2_{\mu \nu} ) = \delta^{\mu \nu} F_{\mu \rho} \tensor{F}{^\rho_\nu}  = F^{\nu \rho} F_{\rho \nu} \, .
\end{align}
In particular, this means that
\begin{align}
    \tr ( F^2 ) = F^{\nu \rho} F_{\rho \nu} = - F^{\rho \nu} F_{\rho \nu} = - F^2 \, , 
\end{align}
using the notation $F^2 = F^{\mu \nu} F_{\mu \nu}$ as above and in previous chapters. For this reason, in our notation the Euclidean action $S_{\text{Euc}}$ can be written as
\begin{align}
    S_{\text{Euc}} = \frac{1}{4} \int d^d x \, F^2 \,  = - \frac{1}{4} \int d^d x \, \tr ( F^2 ) \, .
\end{align}
Our convention for the Hilbert stress energy tensor is
\begin{align}\label{hilbert_stress_defn}
    T_{\mu \nu} = \frac{-2}{\sqrt{g}} \frac{\delta S}{\delta g^{\mu \nu}} = g_{\mu \nu} \mathcal{L} - 2 \frac{\partial \mathcal{L}}{\partial g^{\mu \nu}} \, .
\end{align}
For example, the stress tensor associated with the Euclidean Lagrangian (\ref{S_euc}) is
\begin{align}\label{euc_stress_tensor}
    T_{\mu \nu}^{\text{Euc}} = \frac{1}{4} g_{\mu \nu} F_{\alpha \beta} F^{\alpha \beta} - F_{\mu \alpha} \tensor{F}{_\nu^\alpha} \, ,
\end{align}
while that for the Lorentzian Lagrangian (\ref{S_lor}) is
\begin{align}
    T_{\mu \nu}^{\text{Lor}} = -\frac{1}{4} g_{\mu \nu} F_{\alpha \beta} F^{\alpha \beta} + F_{\mu \alpha} \tensor{F}{_\nu^\alpha} = - T_{\mu \nu}^{\text{Euc}} \, .
\end{align}
So for instance (\ref{euc_stress_tensor}) agrees with the Maxwell stress tensor (\ref{maxwell_lag_nonlinear_chapter}) in Chapter \ref{CHP:nonlinear}. But as we have emphasized, the stress tensor bilinears which enter into our $\TT$-like deformations are identical in the two cases:
\begin{align}
    \left( T^{\text{Euc}} \right)_{\mu \nu} \left( T^{\text{Euc}} \right)^{\mu \nu} &= \left( T^{\text{Lor}} \right)_{\mu \nu} \left( T^{\text{Lor}} \right)^{\mu \nu} \, , \nonumber \\
    \tensor{\left( T^{\text{Euc}}\right)}{^\mu_\mu} \tensor{\left( T^{\text{Euc}}\right)}{^\nu_\nu} &= \tensor{\left( T^{\text{Lor}}\right)}{^\mu_\mu} \tensor{\left( T^{\text{Lor}}\right)}{^\nu_\nu} \, .
\end{align}
One might object that, although the \emph{deformations} constructed from bilinears in $T_{\mu \nu}$ are invariant under reversing the sign of $\mathcal{L}$, the \emph{solution} to flow equations with initial condition $\mathcal{L}_0$ might differ since $\mathcal{L}_0$ disagrees by a sign in the two signatures. However, this ambiguity can be absorbed into a re-scaling of the flow parameter $\lambda$. Since all of the analysis of this chapter will hold for either sign of $\lambda$, we are free to modify a flow equation of the form
\begin{align}
    \frac{\partial \mathcal{L}}{\partial \lambda} = T^{\mu \nu} T_{\mu \nu} - r \left( \tensor{T}{^\mu_\mu} \right)^2
\end{align}
by sending $\lambda \to \lambda' = - \lambda$, so that
\begin{align}
    \frac{\partial \left( - \mathcal{L} \right)}{\partial \lambda'} = T^{\mu \nu} T_{\mu \nu} - r \left( \tensor{T}{^\mu_\mu} \right)^2 \, .
\end{align}
This operation allows us to map a solution to a flow equation with given initial condition $\mathcal{L}_0$ to a solution with initial condition $-\mathcal{L}_0$.

Of course, we have seen in the preceding chapters that the two signs of the $\TT$ flow parameter $\lambda$ lead to very different behavior. Consider the example of deforming a $2D$ CFT. For a ``good sign'' choice, all of the finite-volume energies in the deformed theory remain real, at least for sufficiently small $\lambda$. The other ``bad sign'' choice generically leads to a deformed theory where all but finitely many of the energies are complex. One might therefore say that we should be more careful about making arguments which require reversing the sign of $\lambda$, since this can lead to pathologies in the quantum theory.

However, as we mentioned above, in this chapter we are concerned exclusively with flow equations for classical Lagrangians. Indeed, since we work in arbitrary dimension, our deforming operators are not even known to exist at the quantum level. Thus it is perfectly acceptable to treat $\lambda$ as a formal variable labeling a one-parameter family of Lagrangians and treat either sign of this variable as equally valid.

This concludes the (somewhat tedious) discussion of our sign manifesto and conventions for the present chapter. We now return to the more interesting task of obtaining a flow equation for an arbitrary abelian gauge theory in $d$ dimensions.

\subsection{Enumeration of Trace Structures}

We must first parameterize the independent scalar quantities which can be built from $F_{\mu \nu}$. This is obviously a dimension-dependent question. For instance, in $d=4$, every scalar constructed from $F_{\mu \nu}$ can be expressed in terms of two combinations
\begin{align}\label{4d_scalar_combinations}
    x_1 = F_{\mu \nu} F^{\nu \mu} = \tr ( F^2 ) \, , \qquad x_2 = F^{\mu \sigma} \tensor{F}{_\sigma^\nu} \tensor{F}{_\nu^\rho} F_{\rho \mu} = \tr ( F^4 ) \, .
\end{align}
%
Equivalently, if we express the $4d$ Euclidean field strength tensor as
\begin{align}
    F_{\mu \nu} = \begin{bmatrix} 0 & E_x & E_y & E_z \\ - E_x & 0 & -B_z & B_y \\ - E_y & B_z & 0 & -B_x \\ -E_z & - B_y & B_x & 0 \end{bmatrix}_{\mu \nu} \, , 
\end{align}
then the two invariants satisfy
\begin{align}
    x_1 &= - 2 \left( \vec{E}^2 + \vec{B}^2 \right) \, , \nonumber \\
    x_2 &= \frac{1}{2} x_1^2 - 4 \left( \vec{E} \cdot \vec{B} \right)^2 \, .
\end{align}
The fact that all scalars can be written in terms of $x_1$ and $x_2$ might be more familiar if phrased in Lorenztian signature, where the two Lorentz invariants that can be constructed from the field strength tensor in $d = 4$ are $\vec{E}^2 - \vec{B}^2$ and $\vec{E} \cdot \vec{B}$, which are related to $F \wedge \ast F$ and $F \wedge F$ respectively.

To begin the enumeration of independent scalars, we first recall a very basic fact which follows from the Cayley-Hamilton theorem: every matrix satisfies its own characteristic equation. That is, any $d \times d$ matrix $M$ obeys a condition of the form
\begin{align}\label{cayley_hamilton}
    p ( M ) = M^d + c_{d-1} M^{d-1} + \cdots + c_1 M + ( - 1 )^d \det ( M ) I_d = 0 \, ,
\end{align}
where $I_d$ is the $d \times d$ identity matrix. The coefficients $c_i$ are functions only of the traces $\mathrm{tr} \big( M^l \big)$ for $l = 1, \cdots, d$. Explicitly, they are given by
\begin{align}\label{ci_defn}
    c_i = \sum_{\{ k_l \}} \, \prod_{l=1}^{d} \,  \frac{(-1)^{k_l + 1}}{l^{k_l} k_l!} \left( \mathrm{tr} \, \big( M^l \big) \, \right)^{k_l} \, ,
\end{align}
where the sum runs over all sets of non-negative integers $k_l$ which satisfy
\begin{align}
    \sum_{l=1}^{d} l k_l = d - i \, .
\end{align}
The condition (\ref{cayley_hamilton}) implies a limit on the number of independent trace structures that a matrix can have, as we now review.\footnote{Another way to think about this fact is that each $\tr ( M^k )$ is the sum $\lambda_1^k + \cdots + \lambda_d^k$ where the $\lambda_i$ are eigenvalues of $M$. Knowing the first $d$ such sums gives $d$ equations, which is sufficient to determine all $d$ of the eigenvalues. From this data, any trace $\tr ( M^n )$ can be computed. For antisymmetric matrices the eigenvalues come in pairs $\pm \lambda$, except in odd dimensions where there is one unpaired zero eigenvalue.}

Suppose that we are given all of the traces
\begin{align}
    \tr ( M ) \, , \, \tr \big( M^2 \big) \, , \, \cdots \, , \, \tr \big( M^d \big) \, .
\end{align}
We claim that all higher traces $\tr \big( M^n \big)$ for $n > d$ can then be determined. To see this, first note that $\det ( M )$ is also known, since we can take the trace of equation (\ref{cayley_hamilton}) and then solve for $\det ( M )$ in terms of the known traces:
\begin{align}\label{det_solved}
    \det ( M ) = \frac{(-1)^{d+1}}{d} \left( \tr ( M^d ) + c_{d-1} \tr ( M^{d-1} ) + \cdots + c_1 \tr ( M ) \right) \, .
\end{align}
On the other hand, we can first multiply (\ref{cayley_hamilton}) by $M^{n - d}$ and then trace to find that
\begin{align}\label{tr_Mn_solved}
    \tr ( M^{n} ) = - \left( c_{d-1} \tr ( M^{n-1} ) + \cdots + c_1 \tr ( M^{d - n + 1 } ) + ( -1 )^d \det ( M ) \tr ( M^{d-n} ) \right) \, .
\end{align}
Since all the coefficients $c_i$ and $\det ( M )$ are known, we have therefore solved for $\tr ( M^n )$ in terms of the traces $\tr \big( M^k \big)$ for $k < n$. By induction, this shows that all traces $\tr ( M^n )$ for $n > d$ can be determined from the known traces $\tr ( M^i )$ for $i = 1, \cdots, d$.

The upshot of this theorem is that, in $d$ spacetime dimensions, there are only $\lfloor \frac{d}{2} \rfloor$ independent trace structures $\mathrm{tr} \big( F^k \big)$ for an antisymmetric $2$-tensor $F_{\mu \nu}$. The reason for the floor function is because of a slight difference between the cases of even and odd dimensions. For odd $d$, the determinant of any antisymmetric matrix vanishes, as does the trace of any odd power of the matrix. So in this case, if $d = 2 k +1$, the only independent trace structures are
\begin{align}
    \tr ( F^2 ) \, , \tr ( F^4 ) \, , \cdots \, , \tr F^{2k} \, .
\end{align}
Therefore there are $k = \frac{d-1}{2}$ independent quantities. In even dimension $d = 2k$, where the determinant of $F$ may be non-zero, the independent trace structures are
\begin{align}
    \tr ( F^2 ) \, , \tr ( F^4 ) \, , \cdots \, , \tr F^{2k} \, .
\end{align}
so there are $k = \frac{d}{2}$ independent quantities. Therefore, for the general case, we need $\lfloor \frac{d}{2} \rfloor$ trace structures in order to express an arbitrary scalar constructed from $F_{\mu \nu}$.

\subsection{Deforming a General Lagrangian}

In the previous subsection, we have seen that a general scalar quantity built from an antisymmetric $2$-form $F_{\mu \nu}$ in $d$ dimensions can be expressed as a function of the independent traces $\tr ( F^2 )$, $\tr ( F^4 )$, $\cdots$, $F^{2k}$ where $k = \lfloor \frac{d}{2} \rfloor$. In particular, the Lagrangian is a scalar and therefore takes this form, so the most general classical theory of an abelian gauge field can be written
\begin{align}\label{general_lagrangian_gauge}
    \mathcal{L} ( F ) = \mathcal{L} \left( \tr ( F^2 ) , \tr ( F^4 ) , \cdots, \tr ( F^{2k} ) \right) \, ,
\end{align}
For convenience, we will define a set of variables $x_i$ defined by
\begin{align}
    x_i = \tr ( F^{2i} ) \, , 
\end{align}
for $i = 1, \cdots, k$. We now wish to compute the stress tensor associated with the Lagrangian (\ref{general_lagrangian_gauge}) so that we can construct the $\TT$ deformation. Using our convention (\ref{hilbert_stress_defn}) for the Hilbert stress tensor and $g_{\mu \nu} = \delta_{\mu \nu}$ in Euclidean signature, one has
\begin{align}
    T_{\mu \nu} = \delta_{\mu \nu} \mathcal{L} - 2 \sum_{i=1}^{k} \frac{\partial \mathcal{L}}{\partial x_i} \cdot \frac{\delta x_{i}}{\delta g^{\mu \nu}} \Bigg\vert_{g = \delta} \, .
\end{align}
Restoring metric factors, each $x_i$ has the expansion
\begin{align}
    x_{i} = g^{\alpha_0 \beta_0} g^{\alpha_1 \beta_1} \cdots g^{\alpha_{2i-1} \beta_{2i-1}}  F_{\beta_0 \alpha_1} F_{\beta_1 \alpha_2} \cdots F_{\beta_{2i-2} \alpha_{2i-1}} F_{\beta_{2i-1} \alpha_0} \, .
\end{align}
Computing the metric derivative thus yields $2i$ separate terms, one for each factor of $g^{\alpha \beta}$:
\begin{align}\label{xi_metric_derivative}
    \frac{\delta x_{i}}{\delta g^{\mu \nu}} &= g^{\alpha_1 \beta_1} \cdots g^{\alpha_{2i-1} \beta_{2i-1}}  F_{\nu \alpha_1} F_{\beta_1 \alpha_2} \cdots F_{\beta_{2i-2} \alpha_{2i-1}} F_{\beta_{2i-1} \mu} \nonumber \\
    &\quad + g^{\alpha_0 \beta_0} g^{\alpha_2 \beta_2} \cdots g^{\alpha_{2i-1} \beta_{2i-1}}  F_{\beta_0 \mu} F_{\nu \alpha_2} \cdots F_{\beta_{2i-2} \alpha_{2i-1}} F_{\beta_{2i-1} \alpha_0}  \nonumber \\
    &\quad + \cdots + g^{\alpha_0 \beta_0} g^{\alpha_1 \beta_1} \cdots g^{\alpha_{2i-2} \beta_{2i-2}}  F_{\beta_0 \alpha_1} F_{\beta_1 \alpha_2} \cdots F_{\beta_{2i-2} \mu} F_{\nu \alpha_0} \, .
\end{align}
Each of the terms appearing in (\ref{xi_metric_derivative}) can be brought into an identical form via a re-labeling of indices. This must have been true on general grounds, since each term in (\ref{xi_metric_derivative}) is obtained by removing one metric factor $g^{\alpha_k \beta_k}$ and replacing the corresponding indices as $\alpha_k \to \mu$, $\beta_k \to \nu$. But since the entire expression $x_i$ is invariant under cyclic permutations of $k$, each such term is identical and proportional to $F_{\mu \nu}^{2i}$.

However, one can also prove this claim explicitly. In each term, we simply push the factor of $F$ with the $\nu$ index all the way to the left, push the $F$ with the $\mu$ index to the right, and then cyclically permute the indices. For example, in the second term we have
\begin{align}
    &g^{\alpha_0 \beta_0} g^{\alpha_2 \beta_2} \cdots g^{\alpha_{2i-1} \beta_{2i-1}}  F_{\beta_0 \mu} F_{\nu \alpha_2} F_{\beta_2 \alpha_3} \cdots F_{\beta_{2i-2} \alpha_{2i-1}} F_{\beta_{2i-1} \alpha_0} \nonumber \\
    &\quad = g^{\alpha_0 \beta_0} g^{\alpha_2 \beta_2} \cdots g^{\alpha_{2i-1} \beta_{2i-1}}  F_{\nu \alpha_2} F_{\beta_2 \alpha_3} \cdots F_{\beta_{2i-2} \alpha_{2i-1}} F_{\beta_{2i-1} \alpha_0} F_{\beta_0 \mu} \nonumber \\ 
    &\quad = g^{\alpha_0 \beta_0} g^{\alpha_1 \beta_1} \cdots g^{\alpha_{2i-2} \beta_{2i-2}}  F_{\nu \alpha_1} F_{\beta_1 \alpha_2} \cdots F_{\beta_{2i-3} \alpha_{2i-2}} F_{\beta_{2i-2} \alpha_{0}} F_{\beta_{0} \mu} \nonumber \\
    &\quad = g^{\alpha_1 \beta_1} \cdots g^{\alpha_{2i-2} \beta_{2i-2}} g^{\alpha_{2i-1} \beta_{2i-1}}  F_{\nu \alpha_1} F_{\beta_1 \alpha_2} \cdots F_{\beta_{2i-3} \alpha_{2i-2}} F_{\beta_{2i-2} \alpha_{2i-1}} F_{\beta_{2i-1} \mu} \, . 
\end{align}
In going from the first step to the second, we have simply commuted $F_{\beta_0 \mu}$ and $F_{\nu \alpha_2}$. To go to the third step, we relabeled each $\alpha^k$ and $\beta^k$ to $\alpha^{k-1}$ and $\beta^{k-1}$ for $k = 2 , \cdots, 2i-1$. In the final step, we relabeled $\alpha_0, \beta_0$ to $\alpha_{i-1}, \beta_{i-1}$. The final expression is manifestly equivalent to the term on the first line of (\ref{xi_metric_derivative}).

We note that this re-labeling procedure requires us to commute factors of $F_{\beta \alpha}$ past one another. In the non-abelian context, to be investigated in Section \ref{sec:general_def_nonabelian}, the field strengths have non-trivial commutators so that this calculation is more subtle.

Using a similar cyclic manipulation in each of the $2i$ terms to bring them all into the same form, we obtain
\begin{align}
    \frac{\delta x_{i}}{\delta g^{\mu \nu}} &= 2i g^{\alpha_1 \beta_1} \cdots g^{\alpha_{2i-1} \beta_{2i-1}}  F_{\nu \alpha_1} F_{\beta_1 \alpha_2} \cdots F_{\beta_{2i-2} \alpha_{2i-1}} F_{\beta_{2i-1} \mu} \equiv 2i F^{2i}_{\mu \nu} \, ,
\end{align}
where again we use the notation
\begin{align}
    F^{2i}_{\mu \nu} = g^{\alpha_1 \beta_1} \cdots g^{\alpha_{2i-1} \beta_{2i-1}}  F_{\mu \alpha_1} F_{\beta_1 \alpha_2} \cdots F_{\beta_{2i-2} \alpha_{2i-1}} F_{\beta_{2i-1} \nu} \, .
\end{align}
Therefore, the stress tensor for a general Lagrangian is
\begin{align}
    T_{\mu \nu} = \delta_{\mu \nu} \mathcal{L} - 4 \sum_{i=1}^{k} i \frac{\partial \mathcal{L}}{\partial x_i} F^{2i}_{\mu \nu} \, .
\end{align}
Next we compute the appropriate bilinears. We have
\begin{align}
    T^{\mu \nu} T_{\mu \nu} &= \left( \delta^{\mu \nu} \mathcal{L} - 4 \sum_{i=1}^{k} i \frac{\partial \mathcal{L}}{\partial x_i} F^{2i, \mu \nu} \right) \left( \delta_{\mu \nu} \mathcal{L} - 4 \sum_{j=1}^{k} j \frac{\partial \mathcal{L}}{\partial x_j} F^{2j}_{\mu \nu}  \right) \nonumber \\
    &= \mathcal{L}^2 d - 8 \mathcal{L} \sum_{i=1}^{k} i \frac{\partial \mathcal{L}}{\partial x_i} \tr ( F^{2i} ) + 16 \sum_{i, j=1}^{k} i j \frac{\partial \mathcal{L}}{\partial x_i} \frac{\partial \mathcal{L}}{\partial x_j} F^{2i, \mu \nu} F_{\mu \nu}^{2j} \, .
\end{align}
Meanwhile, the trace is
\begin{align}
    \tensor{T}{^\mu_\mu} = \mathcal{L} d - 4 \sum_{i=1}^{k} i \frac{\partial \mathcal{L}}{\partial x_i} \tr ( F^{2i} ) \, ,
\end{align}
and its square is
\begin{align}
    \left( \tensor{T}{^\mu_\mu} \right)^2 = \mathcal{L}^2 d^2 - 8 \mathcal{L} d \sum_{i=1}^{k} i \frac{\partial \mathcal{L}}{\partial x_i} \tr ( F^{2i} ) + 16 \sum_{i, j=1}^{k} i j \frac{\partial \mathcal{L}}{\partial x_i} \frac{\partial \mathcal{L}}{\partial x_j} \tr ( F^{2i} ) \tr ( F^{2j} ) \, .
\end{align}
As we argued in the introduction, in order to treat flow equations by a general linear combination of stress tensor bilinears, it is enough to consider flow equations of the form
\begin{align}\label{general_r_flow_new_conventions}
    \frac{\partial \mathcal{L}}{\partial \lambda} = T^{\mu \nu} T_{\mu \nu} - r \left( \tensor{T}{^\mu_\mu} \right)^2 \, .
\end{align}
Deforming the Lagrangian by this linear combination gives
\begin{align}\label{flow_general_exact}
    \frac{\partial \mathcal{L}}{\partial \lambda} &= ( d - r d^2 ) \mathcal{L}^2 - 8 \mathcal{L} ( 1 - r d ) \sum_{i=1}^{k} i \frac{\partial \mathcal{L}}{\partial x_i} x_i + 16 \sum_{i, j=1}^{k} i j \frac{\partial \mathcal{L}}{\partial x_i} \frac{\partial \mathcal{L}}{\partial x_j} \left( x_{i+j} - r x_i x_j \right) \, .
\end{align}
Although (\ref{flow_general_exact}) is exact, it is not expressed as a closed flow equation using only the set of independent trace structures $x_1 , \cdots , x_k$. This is because term involving $x_{i+j} = \tr ( F^{2(i+j)} )$ introduces dependence on traces $\tr ( F^{ 2 n} )$ for integers $n > k$. These traces are not independent of the lower traces, as we have argued, and must therefore be re-expressed in terms of the other independent trace structures using the Cayley-Hamilton theorem.

It is unwieldy, though possible, to perform this procedure and write a closed flow equation using only the independent trace structures $x_i$ by using equations (\ref{det_solved}), (\ref{tr_Mn_solved}), and the definition (\ref{ci_defn}) of the $c_i$. However, we will prefer to leave this implicit and only perform the necessary substitutions as needed in special cases later in this chapter.

Despite requiring the implicit definition of higher traces in terms of lower ones, (\ref{flow_general_exact}) reveals some general structure about these $\TT$-like flows in arbitrary dimension, such as the dependence on $r$ and $d$ and the introduction of dependence on higher traces via the final term. For instance, one can see immediately that the flow equation simplifies for certain choices of $r$ such as $r = \frac{1}{d}$, for which some of the terms on the right vanish. One can also see an interesting structure in the large $d$ limit. If we retaining only the leading term on the right side as $d \to \infty$, the flow equation reduces to
\begin{align}
    \frac{\partial \mathcal{L}}{\partial \lambda} = - r d^2 \mathcal{L}^2 \, ,
\end{align}
which has the simple solution
\begin{align}\label{large_d_solution}
    \mathcal{L} ( \lambda ) = \frac{\mathcal{L}_0}{1 + d^2 \mathcal{L}_0 r \lambda } \, .
\end{align}
For negative $\mathcal{L}_0$, such as $\mathcal{L}_0 = - \frac{1}{4} F^2$ in Lorentzian signature, or for positive $\mathcal{L}_0$ but negative $\lambda$ or $r$, we see that (\ref{large_d_solution}) has a singularity at finite $\lambda$. This is reminiscent of the structure we saw in equation (\ref{pole_potential_solution}) of Chapter \ref{CHP:SC-squared-2}, namely
\begin{align}
    V ( \lambda , \phi ) = \frac{V ( 0, \phi ) }{1 - \lambda V ( 0 , \phi ) } \, .
\end{align}
In both cases, it appears that the $\TT$ deformation has led to a pole in the Lagrangian for certain field configurations. As we stressed in Chapter \ref{CHP:SC-squared-2}, this is a purely classical result, and one cannot necessarily draw inferences about the structure of the quantum mechanical theory from this observation. However, this structure suggests the possibility of a modification of the vacuum structure in the theory -- one might naively expect such an infinite potential barrier to cut off different sectors of the Hilbert space from one another. It would be interesting to study this possibility in greater detail, perhaps in a simplified quantum mechanical (i.e. $(0+1)$-dimensional) example.
%
%
%

Finally, we comment on one additional manipulation of (\ref{flow_general_exact}) that may be useful in obtaining series solutions for certain flows. The Lagrangian may always be expanded as a power series in the variables $x_k = \tr ( F^{2k} )$. We recall from footnote \ref{dimensional_analysis_footnote} that, in our conventions, the field strength $F_{\mu \nu}$ has mass dimension $\frac{d}{2}$ and thus $x_k$ has mass dimension $kd$. Likewise, $\lambda$ has length dimension $d$ so combinations like $\lambda^k x_k$ are dimensionless. We will now consider the combinations of terms that can appear in the power series expansion of the Lagrangian, subject to the usual requirement that each term have mass dimension $d$.

The coefficient of $x_1 = F^2$, the free kinetic term, already has mass dimension $d$ and thus appears with no powers of $\lambda$. Additional terms must be suppressed by powers of $\lambda$; for instance, $\lambda x_2$ and $\lambda x_1^2$ have the appropriate dimension to appear in the Lagrangian. More generally, a term with $2m$ factors of $F_{\mu \nu}$ must appear multiplying $\lambda^{m-1}$. The total mass dimension of a term $x_1^{n_1} x_2^{n_2} \cdots x_k^{n_k}$ is $\left( n_1 + 2 n_2 + \cdots + k n_k \right) d$, and must therefore appear with a factor of $\lambda^{n_1 + 2 n_2 + \cdots + k n_k - 1}$. Thus the most general expansion of the Lagrangian is
\begin{align}
    \mathcal{L} = \sum_{N} c_N \lambda^{  n_1 + 2 n_2 + \cdots + k n_k - 1 } x_1^{n_1} x_2^{n_2} \cdots x_{k}^{n_k} \, .
\end{align}
Here $N$ represents a multi-index of $k$ non-negative integers $n_1, \cdots, n_k$ (not all zero). If we are beginning from an undeformed theory which is the ordinary Maxwell Lagrangian, the initial condition $\mathcal{L} ( \lambda = 0 ) = - \frac{1}{4} x_1$ fixes the coefficient $N_0 = \{ 1, 0, 0, \cdots, 0 \}$ to be $c_{N_0} = - \frac{1}{4}$.

Such an expansion simplifies the problem of finding a series solution to our master equation (\ref{flow_general_exact}), since now the task is reduced to determining each of the coefficients $c_N$. For instance, the left side of the flow equation is now
\begin{align}\label{master_lambda_deriv}
    \frac{\partial \mathcal{L}}{\partial \lambda} = \sum_{N > N_0} c_N \left( n_1 + 2 n_2 + \cdots + k n_k - 1  \right) \lambda^{  n_1 + 2 n_2 + \cdots + k n_k - 2 } x_1^{n_1} x_2^{n_2} \cdots x_{k}^{n_k} \, ,
\end{align}
where now the notation $\sum_{N > N_0}$ means to sum over all multi-indices $N = \{ n_1, \cdots , n_k \}$ except for $\{ 1 , 0 , \cdots , 0 \}$. Likewise, each of the $x_i$ derivatives appearing in the flow equation can be simply evaluated as
\begin{align}\label{master_x_deriv}
    \frac{\partial \mathcal{L}}{\partial x_j} = \sum_{N \, , \, n_j > 0} c_N \, n_j \,\lambda^{  n_1 + 2 n_2 + \cdots + k n_k - 1 } x_1^{n_1} x_2^{n_2} \cdots x_j^{n_j -1} \cdots x_{k}^{n_k} \, ,
\end{align}
where similarly $\sum_{N, n_j>0}$ means to sum over all multi-indices $N = \{ n_1, \cdots , n_k \}$ where $n_j$ is at least $1$.

After eliminating all of the higher trace structures $x_n$ for $n>k$ in the master flow equation (\ref{flow_general_exact}) using Cayley-Hamilton, one can then substitute (\ref{master_lambda_deriv}) on the left and (\ref{master_x_deriv}) for each $x_i$ derivative on the right. The result, though perhaps complicated, is in principle a perturbative expansion that allows us to solve for the coefficients $c_N$ order-by-order in $\lambda$.

\subsection{Low-Dimensional Examples}

To illustrate the general formalism developed in the preceding subsection, here we will develop the flow equations in $d=2$ and $d=3$. These cases are especially simple since there is only one independent trace structure $x_1$, so we need only express $x_2$ in terms of $x_1$ using the Cayley-Hamilton theorem. Since we have studied these examples before, this will also provide a usesful check of whether the formalism recovers previous results.

\subsubsection{\uline{Two Dimensions}}

In two dimensions, the Cayley-Hamilton theorem reads
\begin{align}
    M^2 - ( \tr ( M ) ) M + \det ( M ) I_2 = 0 \, .
\end{align}
For an antisymmetric matrix $M = F$, then, we have $\tr ( F ) = 0$ and $F^2 = - \det ( F ) I_2$. Taking traces, one has
\begin{align}
    \tr ( F^2 ) = - 2 \det ( F ) \, .
\end{align}
If we first multiply by $F^2$ and then take traces, we instead find
\begin{align}
    \tr ( F^4 ) = - \det ( F ) \tr ( F^2 ) = \frac{1}{2} \left( \tr ( F ) \right)^2 \, .
\end{align}
In terms of our variables $x_i = \tr ( F^{2i} )$, this means that
\begin{align}
    x_2 = \frac{1}{2} x_1^2 \, .
\end{align}
Using this, our general flow equation (\ref{flow_general_exact}) with $d=2$ becomes
\begin{align}
    \frac{\partial \mathcal{L}}{\partial \lambda} = 2 \left( 1 - 2 r \right) \mathcal{L}^2 - 8 \mathcal{L} ( 1 - 2 r ) x_1 \frac{\partial \mathcal{L}}{\partial x_1} + 8 \left( 1 - 2 r \right) x_1^2 \left( \frac{\partial \mathcal{L}}{\partial x_1} \right)^2  \, .
\end{align}
This should match the result that we derived in Chapter \ref{CHP:nonabelian}. Indeed, if $r=1$ the flow equation reduces to
\begin{align}\label{master_flow_2d_example}
    \frac{\partial \mathcal{L}}{\partial \lambda} = - 2 \mathcal{L}^2 + 8 \mathcal{L} x_1 \frac{\partial \mathcal{L}}{\partial x_1} - 8 x_1^2 \left( \frac{\partial \mathcal{L}}{\partial x_1} \right)^2  \, .
\end{align}
We may now multiply both sides by $-\frac{1}{2}$ and absorb this factor into a re-scaling of $\lambda$ on the left side, which gives
\begin{align}\label{rescaled_2d_example}
    \frac{\partial \mathcal{L}}{\partial \lambda} = \mathcal{L}^2 - 4 \mathcal{L} x_1 \frac{\partial \mathcal{L}}{\partial x_1} + 4 x_1^2 \left( \frac{\partial \mathcal{L}}{\partial x_1} \right)^2  \, .
\end{align}
This matches (\ref{maxwell_pde_intermediate}), as expected. We note that we have re-scaled $\lambda$ by a negative constant, which reverses the sign; this corresponds to the fact that our convention for the ``good sign'' of $\lambda$ is opposite that in Chapter \ref{CHP:nonabelian}. We will see a similar re-scaling by a constant $a=-8$ when comparing to Born-Infeld in the next section.

\subsubsection{\uline{Three Dimensions}}

In the case of $d=3$, the Cayley-Hamilton result (\ref{cayley_hamilton}) gives the identity
\begin{align}
    M^3 - \left( \tr ( M ) \right) M^2 + \frac{1}{2} \left( \left( \tr ( M ) \right)^2 - \tr ( M^2 ) \right) M - \det ( M ) I_3 = 0 \, .
\end{align}
Let $M=F$ be antisymmetric. In odd dimensions, the determinant of any antisymmetric matrix vanishes, and we manifestly also have $\tr ( M ) = 0$, so the condition becomes
\begin{align}\label{M_inter}
    F^3 - \frac{1}{2} \left( \tr ( F^2 ) \right) F = 0 \, .
\end{align}
We may multiply through by $F$ and take the trace to find
\begin{align}
    \tr ( F^4 ) = \frac{1}{2} \left( \tr ( F^2 ) \right)^2 \, ,
\end{align}
which gives the same condition $x_2 = \frac{1}{2} x_1^2$ which we found in $d=2$. The flow equation, however, will differ because it explicitly depends on $d$. The master equation (\ref{flow_general_exact}) becomes
\begin{align}\label{3d_flow_exact}
    \frac{\partial \mathcal{L}}{\partial \lambda} = 3 (1 - 3 r ) \mathcal{L}^2 - 8 \mathcal{L} ( 1 - 3 r ) x_1 \frac{\partial \mathcal{L}}{\partial x_1} + 8 ( 1 - 2 r ) x_1^2 \left( \frac{\partial \mathcal{L}}{\partial x_1} \right)^2  \, .
\end{align}
This matches the result (\ref{3d_flow}) which we derived earlier.

\section{Comparison to Born-Infeld}

In this section, we will examine the flow equation for gauge theories in general dimension to see whether any such flow can yield the Born-Infeld action from the Maxwell Lagrangian as the seed theory, besides the known case in $d=4$. 

\subsection{Perturbative Comparison in $\lambda$}\label{sec:deformation_bi_dgeq6}

It will be sufficient to check whether Born-Infeld-type actions can satisfy a $\TT$-like flow equation to second order in $\lambda$. To make this comparison, we first expand the Born-Infeld Lagrangian to the first few orders. First note that, since $\det ( M ) = \det ( M^T )$ for any matrix $M$, we have
\begin{align}
    \det \left( \delta_{\mu \nu} + k F_{\mu \nu} \right) = \det \left( \delta_{\nu \mu} + k F_{\nu \mu} \right) = \det \left( \delta_{\mu \nu} - k F_{\mu \nu} \right) \, .
\end{align}
Therefore, we can rewrite
\begin{align}
    \sqrt{ \det \left( \delta_{\mu \nu} + k F_{\mu \nu} \right) } &= \Big[ \det \left( \delta_{\mu \nu} + k F_{\mu \nu} \right) \det \left( \delta_{\alpha \beta} - k F_{\alpha \beta} \right) \Big]^{\frac{1}{4}} \nonumber \\
    &= \left[ \det \left( \delta_{\mu \nu} - k^2 F^2_{\mu \nu} \right) \right]^{\frac{1}{4}} \, .
\end{align}
On the other hand, one has $\det ( A ) = \exp \left( \tr ( \log ( A ) ) \right)$ for any matrix $A$, so
\begin{align}\label{bi_expansion_intermediate}
    \left( \det \left( \delta_{\mu \nu} - k^2 F^2_{\mu \nu} \right) \right)^{\frac{1}{4}} = \exp \left( \frac{1}{4} \tr \left( \log \left( \delta_{\mu \nu} - k^2 F^2_{\mu \nu} \right) \right) \right) \, .
\end{align}
In order to compare the Born-Infeld action to $\TT$-flows up to order $\lambda^2$, we must Taylor-expand (\ref{bi_expansion_intermediate}) to order $k^6$. This gives
\begin{align}\label{bi_expansion_final_k}
    \sqrt{ \det \left( \delta_{\mu \nu} + k F_{\mu \nu} \right) } &= 1 - \frac{k^2}{4} \tr ( F^2 ) + \frac{k^4}{4} \left( \frac{1}{8} \left( \tr ( F^2 ) \right)^2 - \frac{1}{2} \tr ( F^4 ) \right) \nonumber \\
    &\qquad- \frac{k^6}{6} \left( \frac{1}{64} \left( \tr ( F^2 ) \right)^3 - \frac{3}{16} \tr ( F^2 ) \tr ( F^4 ) + \frac{1}{2} \tr ( F^6 ) \right) + \mathcal{O} \left( k^8 \right) \, .
\end{align}
%
%
Following the conventions in footnote \ref{dimensional_analysis_footnote}, where the gauge coupling $g$ has been absorbed into the field strength $F$, the combination $\sqrt{\lambda} F_{\mu \nu}$ is dimensionless. Thus the appropriate form of a Born-Infeld type Lagrangian which could potentially arise from a $\TT$ flow is
\begin{align}\label{BI_flow_ansatz}
    \mathcal{L}_\lambda = \frac{1}{a \lambda} \left( - 1 + \sqrt{ \det \left( \delta_{\mu \nu} + \sqrt{a \lambda}  F_{\mu \nu} \right) } \right) , 
\end{align}
where $a$ is arbitrary and the coefficient $\sqrt{a \lambda}$ in front of $F_{\mu \nu}$ is chosen so that the Lagrangian reduces to $\mathcal{L}_0 = - \frac{1}{4} \tr ( F^2 ) = \frac{1}{4} F^2$ as $\lambda \to 0$. Using (\ref{bi_expansion_final_k}) with $k = \sqrt{a \lambda}$, then, the expansion of $\mathcal{L}_\lambda$ to order $\lambda^2$ is
\begin{align}\label{BI_expansion_order_lambdasquare}
    \mathcal{L}_\lambda &= - \frac{1}{4} \tr ( F^2 ) + \frac{a \lambda}{4} \left( \frac{1}{8} \left( \tr ( F^2 ) \right)^2 - \frac{1}{2} \tr ( F^4 ) \right) \nonumber \\
    &\qquad- \frac{a^2 \lambda^2}{6} \left( \frac{1}{64} \left( \tr ( F^2 ) \right)^3 - \frac{3}{16} \tr ( F^2 ) \tr ( F^4 ) + \frac{1}{2} \tr ( F^6 ) \right) + \mathcal{O} \left( \lambda^3 \right) \, .
\end{align}
Let's now compare this to the Lagrangian one would get using the master flow equation (\ref{flow_general_exact}). Beginning with the initial condition $\mathcal{L}_0 = - \frac{1}{4} x_1$, we find
\begin{align}\label{leading_correction_maxwell_general_rd}
    \frac{\partial \mathcal{L}}{\partial \lambda} \Big\vert_{\lambda = 0} &= \frac{1}{16} ( d - r d^2 ) x_1^2 - \frac{1}{2} x_1^2 ( 1 - r d ) + \left( x_2 - r x_1^2 \right) \, \nonumber \\
    &= x_2 + x_1^2 \left( - \frac{1}{2} + \frac{d}{16} - \left( 1 - \frac{d}{2} + \frac{d^2}{16} \right) r \right) \, .
\end{align}
Note that, when $d = 4$, the quantity multiplying $r$ in the parentheses of (\ref{leading_correction_maxwell_general_rd}) vanishes, and the coefficient of $x_1^2$ simply becomes $- \frac{1}{4}$. Assuming that $d \neq 4$, in order for the coefficients to match the order $\lambda$ expansion of Born-Infeld we need
\begin{align}
    a = - 8 \, , \qquad r = \frac{1}{d - 4} \, .
\end{align}
This scaling by $a = - 8$ agrees with the factor of $\frac{1}{8}$ appearing in (\ref{one_eighth_maxwell}). The overall sign difference arises from the fact that our convention for the ``good sign'' of $\lambda$ in this chapter is the opposite of that which we have previously used in two dimensions, for instance in Chapter \ref{CHP:nonabelian}. One can see this by comparing equation (\ref{general_r_flow_new_conventions}) to equation (\ref{flow_invariant}). For the present purposes, it is sufficient to assume $\lambda < 0$ so that $a \lambda$ is positive.

Thus, to order $\lambda$, the deformed Lagrangian is
\begin{align}
    \mathcal{L} = - \frac{1}{4} x_1 + \lambda x_2 - \frac{\lambda}{4} x_1^2 + \mathcal{O} ( \lambda^2 ) \, .
\end{align}
Repeating the process by plugging this deformed Lagrangian into the master equation and extracting the $\mathcal{O} ( \lambda^2 )$ piece, we find
\begin{align}\label{order_lambdasquare_def_lag}
    \mathcal{L} \Big\vert_{\lambda^2} = \frac{\lambda^2}{2} \left( \frac{- 12 + d - 32 r + 12 r d - r d^2}{8} \, x_1^3 + \frac{20 - d + 32 r - 12 r d + r d^2}{2} \, x_1 x_2 - 16 x_3 \right) \, .
\end{align}
Substituting the required value of $r = \frac{1}{d-4}$, this becomes
\begin{align}
    \mathcal{L} \Big\vert_{\lambda^2} = \lambda^2 \left( - \frac{1}{4} x_1^3 + 3 x_1 x_2 - 8 x_3 \right) \, .
\end{align}
The coefficients of $x_1^3, x_1 x_2, x_3$ in this term are $- \frac{1}{4}, 3,$ and $-8$, respectively, which do not match the coefficients in (\ref{BI_expansion_order_lambdasquare}) when $a = - 8$, which are $- \frac{1}{6}$, $2$, and $- \frac{16}{3}$. Thus the Born-Infeld action can never be obtained by a flow involving stress tensor bilinears in any number of dimensions where $\tr ( F^6 )$ is independent of lower traces, i.e. $d \geq 6$.

In $d < 6$ dimensions, $x_3$ can be expressed in terms of $x_1$ and $x_2$ so checking the coefficients at order $\lambda^2$ requires eliminating $x_3$ using the Cayley-Hamilton theorem. We have already checked most of the cases $d < 6$; in particular, $d = 2$ gives the hypergeometric action discussed in Chapter \ref{CHP:nonabelian}, we saw that $d=3$ does not give Born-Infeld in Section \ref{sec:motivation_and_3d}, and we know from Chapter \ref{CHP:nonlinear} that $\TT$ in four dimensions \emph{does} generate a Born-Infeld type action. The only remaining case to check (besides $d=1$, which is trivial) is $d=5$ spacetime dimensions, where $\tr ( F^2 )$ and $\tr ( F^4 )$ are the independent trace structures. We perform this explicit check in the next subsection.

\subsection{Deformation in $d=5$}

We will now handle one edge case which is not included in the preceding analysis, namely $d=5$. In this case, the expression (\ref{order_lambdasquare_def_lag}) for the $\mathcal{O} ( \lambda^2 )$ term in the $\TT$ flow is formally correct. However, since the trace $x_3 = \tr ( F^6 )$ is not independent of the lower $x_i$ in five dimensions, we must eliminate it in favor of lower traces using the Cayley-Hamilton theorem. In five dimensions, (\ref{cayley_hamilton}) reads
\begin{align}
    &M^5 - ( \tr ( M ) ) M^4 + \frac{1}{2} \left( \left( \tr ( M ) \right)^2 - \tr ( M^2 ) \right) M^3 \nonumber \\
    &\quad - \frac{1}{6} \left( \left( \tr ( M ) \right)^3 - 3 \tr ( M^2 ) \tr ( M ) + 2 \tr ( M^3 ) \right) M^2 \nonumber \\
    &\quad + \frac{1}{24} \left( \tr ( M )^4 + 8 \tr (M^3) \tr ( M ) + 3 \left( \tr ( M^2 ) \right)^2 - 6 \tr ( M^2 ) \left( \tr ( M ) \right)^2 - 6 \tr (M^4) \right) M \nonumber \\
    &\quad - \det ( M ) I_5 = 0 \, .
\end{align}
Letting $M = F$ be antisymmetric, which implies that $\tr ( F ) = \tr ( F^3 ) = \det ( F ) = 0$, this becomes
\begin{align}\label{5d_f_cayley}
    F^5 - \frac{1}{2} \tr ( F^2 ) F^3 + \frac{1}{24} \left( 3 \left( \tr \big( F^2 \big) \right)^2 - 6 \tr \big( F^4 \big) \right) F = 0 \, .
\end{align}
Multiplying both sides of (\ref{5d_f_cayley}) by $F$ and tracing, one finds
\begin{align}
    \tr ( F^6 ) = \frac{1}{2} \tr ( F^2 ) \tr ( F^4 ) - \frac{1}{24} \left( 3 \left( \tr \big( F^2 \big) \right)^2 - 6 \tr \big( F^4 \big) \right) \tr ( F^2 ) \, ,
\end{align}
or in terms of the variables $x_i$,
\begin{align}\label{5d_x3_condition}
    x_3 = \frac{3}{4} x_1 x_2 - \frac{1}{8} x_1^3 \, .
\end{align}
%
%
This means that the order $\lambda^2$ term arising from the $\TT$ flow is
\begin{align}\label{5d_tt_lambdasquare}
    \mathcal{L} \Big\vert_{\lambda^2} = \frac{3 \lambda^2}{4} \left( x_1^3 - 4 x_1 x_2 \right) \, .
\end{align}
We can compare this to the $\mathcal{O} ( \lambda^2 )$ term in the expansion (\ref{BI_expansion_order_lambdasquare}) of Born-Infeld, after substituting the constraint (\ref{5d_x3_condition}), which yields
\begin{align}\label{5d_bi_lambdasquare}
    \mathcal{L} \Big\vert_{\lambda^2} = \frac{\lambda^2}{2} \left( x_1^3 - 4 x_1 x_2 \right) \, .
\end{align}
Since (\ref{5d_tt_lambdasquare}) and (\ref{5d_bi_lambdasquare}) differ by an overall multiplicative factor, we conclude that the Born-Infeld action is not generated by a $\TT$ flow in five dimensions. This completes the proof that the only dimension in which the Born-Infeld action arises from a $\TT$-like deformation of the free Maxwell Lagrangian, for any choice of $r$, is in $d = 4$.

\subsection{Deformation in $d=4$}

Although we have already seen in Chapter \ref{CHP:nonlinear} that $\TT$ deformation of the $4$-dimensional Maxwell action yields the Born-Infeld action, it is a useful check of our general formalism to see how this emerges from the master flow equation (\ref{flow_general_exact}). In particular, since we have shown that this flow equation does \emph{not} yield Born-Infeld in any $d > 4$, it is instructive to see how the numerical factors appearing in the Cayley-Hamilton theorem conspire to give a deformed action of square-root type in $d = 4$.

As we mentioned around equation (\ref{4d_scalar_combinations}), in four spacetime dimensions the two independent scalars that can be constructed from $F_{\mu \nu}$ are
\begin{align}
    x_1 = F_{\mu \nu} F^{\nu \mu} = \tr ( F^2 ) \, , \qquad x_2 = F^{\mu \sigma} \tensor{F}{_\sigma^\nu} \tensor{F}{_\nu^\rho} F_{\rho \mu} = \tr ( F^4 ) \, .
\end{align}
However, our master flow equation (\ref{flow_general_exact}) will introduce dependence on two additional trace structures, $x_3 = \tr ( F^{6} )$ and $x_4 = \tr ( F^{8} )$. First we must express these in terms of lower trace structures. For a general $4 \times 4$ matrix $M$, the Cayley-Hamilton theorem (\ref{cayley_hamilton}) reads
\begin{align}\label{4d_cayley_hamilton}
    0 &= M^4 - \left( \tr ( M ) \right) M^3 + \frac{1}{2} \left( \left( \tr ( M ) \right)^2 - \tr ( M^2 ) \right) M^2  \nonumber \\
    &\quad - \frac{1}{6} \left( \left( \tr ( M ) \right)^3 - 3 \tr ( M^2 ) \tr ( M ) + 2 \tr ( M^3 ) \right) M + \det ( M ) I_4 \, .
\end{align}
If we specialize to the case where $M = F$, which is antisymmetric and thus has vanishing trace (as does $F^3$), we have
\begin{align}\label{F_4d_cayley_hamilton}
    0 = F^4 - \frac{1}{2} \tr ( F^2 ) F^2  + \det ( F ) I_4 \, .
\end{align}
By taking traces, we can solve for the determinant as
\begin{align}
    \det ( F ) = \frac{1}{8} \left( \tr \big( F^2 \big) \right)^2  - \frac{1}{4} \tr ( F^4 ) \, .
\end{align}
Similarly, if we first multiply (\ref{F_4d_cayley_hamilton}) by $F^2$ and then take traces, we find
\begin{align}
    \tr ( F^6 ) = \frac{1}{2} \tr ( F^2 ) \tr ( F^4 ) - \det ( F ) \tr ( F^2 ) \, ,
\end{align}
and if we first multiply (\ref{F_4d_cayley_hamilton}) by $F^4$ and then take traces, one obtains
\begin{align}
    \tr ( F^8 ) = \frac{1}{2} \tr ( F^2 ) \tr ( F^6 ) - \det ( F ) \tr ( F^4 ) \, .
\end{align}
Expressed in terms of $x_i = \tr ( F^{2i} )$, these equations can be solved to find
\begin{align}\label{4d_x3_and_x4}
    x_3 = - \frac{1}{8} x_1 \left( x_1^2 - 6 x_2 \right) \, , \qquad x_4 = - \frac{1}{16} \left( x_1^4 - 4 x_1^2 x_2 - 4 x_2^2 \right) \, . 
\end{align}
The numerical factors appearing in (\ref{4d_x3_and_x4}) work out such that the general argument that a $\TT$ flow disagrees with Born-Infeld at order $\lambda^2$, presented in Section \ref{sec:deformation_bi_dgeq6} for $d \geq 6$, fails in four dimensions. This was already suggested by the term at leading order in $\lambda$, where the coefficients automatically agreed in $d = 4$ and required $r = \frac{1}{d-4}$ otherwise. However, we can also check that the $\mathcal{O} ( \lambda^2 )$ piece works. Substituting in $d=4$ in equation (\ref{order_lambdasquare_def_lag}), one finds
\begin{align}
    \mathcal{L}_{\TT} \Big\vert_{\lambda^2} = - \frac{\lambda^2}{2} \left( x_1^3 - 8 x_1 x_2 + 16 x_3 \right) \, , 
\end{align}
and using our expression (\ref{4d_x3_and_x4}) for $x_3$, this is
\begin{align}
    \mathcal{L}_{\TT} \Big\vert_{\lambda^2} = \frac{\lambda^2}{2} \left( x_1^3 - 4 x_1 x_2 \right) \, , 
\end{align}
On the other hand, when $a = - 8$, the $\mathcal{O} ( \lambda^2 )$ term in (\ref{BI_expansion_order_lambdasquare}) is
\begin{align}
    \mathcal{L}_{BI} \Big\vert_{\lambda^2} &= - \frac{64 \lambda^2}{6} \left( \frac{1}{64} \left( \tr ( F^2 ) \right)^3 - \frac{3}{16} \tr ( F^2 ) \tr ( F^4 ) + \frac{1}{2} \Tr ( F^6 ) \right) \nonumber \\
    &= \frac{\lambda^2}{2} \left( x_1^3 - 4 x_1 x_2 \right) \, .
\end{align}
So in this case, there is an exact match at order $\lambda^2$. In fact, using the expressions (\ref{4d_x3_and_x4}), we can see how the all-orders flow equation emerges, which has a Born-Infeld solution as we saw in Chapter \ref{CHP:nonlinear}. Beginning from the master flow equation (\ref{flow_general_exact}) and letting $d=4$, $\mathcal{L} = \mathcal{L} ( \lambda , x_1, x_2 )$, then eliminating $x_3$ and $x_4$, we have
%
\begin{align}\label{4d_ode_master_flow}
    \frac{\partial \mathcal{L}}{\partial \lambda} &= \left( 4 - 16 r \right) \mathcal{L}^2 - 8 \left( 1 - 4 r \right) \mathcal{L} \left( x_1 \frac{\partial \mathcal{L}}{\partial x_1} + 2 x_2 \frac{\partial \mathcal{L}}{\partial x_2} \right) + 16\left( \frac{\partial \mathcal{L}}{\partial x_1} \right)^2 \left( x_2 - r x_1^2 \right) \nonumber \\
    &\quad + 16 \left( - \frac{1}{2} \frac{\partial \mathcal{L}}{\partial x_1} \frac{\partial \mathcal{L}}{\partial x_2} x_1 \left( x_1^2 - 6 x_2 + 8 r x_2 \right) + \left( \frac{\partial \mathcal{L}}{\partial x_2} \right)^2 \left( - \frac{1}{4} x_1^4 + x_1^2 x_2 + ( 1 - 4 r ) x_2^2 \right) \right) \, .
\end{align}
%
%
Although this differential equation is quite complicated, it simplifies if we make the ansatz
\begin{align}
    \mathcal{L} = \frac{1}{2 \alpha \lambda} \left( - 1 + \sqrt{ 1 - \alpha \lambda x_1 + \lambda^2 \left( \beta x_2 + \gamma x_1^2 \right) }  \right) \, .
\end{align}
With this ansatz, the differential equation (\ref{4d_ode_master_flow}) reduces to a set of algebraic consistency conditions for the parameters $\alpha, \beta, \gamma, r$, whose solution is
\begin{align}
    \alpha = - 4 \, , \quad \beta = - 16 \, , \quad \gamma = 8 \, , \quad r = \frac{1}{2} \, .
\end{align}
Therefore, the Lagrangian
\begin{align}
    \mathcal{L} = - \frac{1}{8 \lambda} \left( - 1 + \sqrt{ 1 + 4 \lambda x_1 + \lambda^2 \left( 8 x_1^2 - 16 x_2 \right) }  \right) \,
\end{align}
is an exact solution to the flow equation (\ref{flow_general_exact}) in $d = 4$, as expected from the analysis of Chapter \ref{CHP:nonlinear}. This confirms that our formalism reproduces the known result in this case.

\section{Born-Infeld-Type Seed Theories}

One advantage of the general formalism of Section \ref{sec:general_abelian} is that it allows us to more uniformly treat $\TT$-flows with different initial conditions across various dimensions. As an application, we will be motivated by the deformation of a seed theory which is not the free Maxwell action, but rather the Born-Infeld action.

We will focus on the low-dimensional examples $d=2, 3$, since here there is only a single trace structure $x_1$ to worry about. In these cases, the seed theory which we wish to deform depends on $x_1$ via the functional form
\begin{align}\label{BI_type_seed}
    \mathcal{L}_0 = \mathcal{L}_0 \left( \sqrt{ 1 + c \alpha' x_1} \right) \, .
\end{align}
where $\alpha'$ is dimensionful and $c$ is a numerical constant. Deforming such a theory is more difficult than deforming the free Maxwell theory $\mathcal{L}_0 = - \frac{1}{4} x_1$, since in the latter case, the only dimensionful quantities in the problem are $\lambda$ and $x_1$. We exploited this simplicity in previous analyses by making an ansatz of the form
\begin{align}\label{one_var_ansatz}
    \mathcal{L} ( x_1, \lambda ) = \frac{1}{\lambda} f ( \lambda x_1 ) \, ,
\end{align}
which allowed us to convert the $\TT$ flow equation into an ordinary differential equation in the dimensionless variable $\lambda x_1$.

This strategy is not available when deforming a seed theory of the type (\ref{BI_type_seed}), since now there are three dimensionful quantities $x_1, \alpha', \lambda$. This means that an ansatz of the form (\ref{one_var_ansatz}) is not sufficient. In two dimensions (where $\alpha'$ and $\lambda$ have the same dimension, according to our conventions), we would need a more complicated ansatz such as
\begin{align}\label{two_var_ansatz}
    \mathcal{L} ( x_1, \alpha' , \lambda ) = \frac{1}{\lambda} f \left( \lambda x_1 , \frac{\alpha'}{\lambda} \right) \, ,
\end{align}
which leads to a partial differential equation rather than an ordinary differential equation. 

Although the promotion from an ODE to a PDE complicates the analysis, this complication is not special to a seed theory of the form (\ref{BI_type_seed}). The additional difficulty merely comes from the presence of a second dimensionful constant $\alpha'$ in the undeformed theory. For this reason, although we are motivated by the deformation of Born-Infeld theories like (\ref{BI_flow_ansatz}) (which take the form (\ref{BI_type_seed}) in $d=2$ and $d=3$), we can in fact analyze flow equations beginning from a general seed theory of the form
\begin{align}
    \mathcal{L}_0 = \mathcal{L}_0 ( \alpha', x_1 ) \, .
\end{align}
By making an ansatz of the form (\ref{two_var_ansatz}), we can treat the deformation of all such seed theories simultaneously. The solution to this general problem reduces to the solution of the flow equation for a Born-Infeld-type seed theory in the special case (\ref{BI_type_seed}).

\subsection{Two Dimensions}

As a first example, we will study this problem in two spacetime dimensions. Here there is a preferred value $r = 1$ for the relative coefficient in our deformation, which defines the usual two-dimensional $\TT$ operator; for simplicity, we will take this value $r=1$ in this subsection. As we saw in (\ref{master_flow_2d_example}), in this case our master formula (\ref{flow_general_exact}) with $\mathcal{L} = f ( \lambda, x_1 )$ reduces to
\begin{align}
    \frac{df}{d \lambda} = - 2 f^2 + 8 f x_1 \frac{\partial f}{\partial x_1} - 8 x_1^2 \left( \frac{\partial f}{\partial x_1} \right)^2 \, .
\end{align}
In order to match the conventions of Chapter \ref{CHP:nonabelian}, we will multiply both sides by $- \frac{1}{2}$ and absorb this factor into a re-scaling of $\lambda$ on the left side as in (\ref{rescaled_2d_example}), so that the flow equation becomes
\begin{align}
    \frac{df}{d \lambda} = f^2 - 4 f x_1 \frac{\partial f}{\partial x_1} + 4 x_1^2 \left( \frac{\partial f}{\partial x_1} \right)^2 \, .
\end{align}
As we described above, rather than the initial condition $f(\lambda = 0) = \frac{1}{4} x_1$, we are motivated by a class of initial conditions includes the seed theory
\begin{align}\label{BI_initial_condition_one}
    f ( \lambda = 0 ) = \frac{1}{2 \alpha'} \left( \sqrt{ 1 + \alpha' F^2 } -1 \right) \, .
\end{align}
On dimensional grounds, the solution to the flow equation beginning from an initial condition of this form can be expressed as
\begin{align}
    f ( x_1, \alpha', \lambda ) = \frac{1}{\lambda} h \left( \lambda x_1 , \frac{\alpha'}{\lambda} \right) \, .
\end{align}
In the limit as the ratio $\frac{\alpha'}{\lambda}$ goes to $0$, the undeformed theory reduces to the free Maxwell action, and the deformed theory is then the usual hypergeometric action. Changing variables to $\xi = \lambda x_1, \eta = \frac{\alpha}{\lambda}$, the differential equation becomes
\begin{align}
    h^2 + \eta \frac{\partial h}{\partial \eta} + \left( 1 - 4 \xi \frac{\partial h}{\partial \xi} \right) h - \xi \left( 1 - 4 \xi \frac{\partial h}{\partial \xi} \right) \frac{\partial h}{\partial \xi} = 0 \, .
\end{align}
It is again convenient to switch to $\log$ variables of the form
\begin{align}
    p = \log ( \xi ) \, , \qquad q = \log ( \eta ) \, , 
\end{align}
and write $h ( \xi , \eta ) = w ( p, q )$, so that
\begin{align}\label{2d_bi_deform_flow}
    w^2 + w + \frac{\partial w}{\partial q} + 4 \left( \frac{\partial w}{\partial q} \right)^2 - \left( 1 + 4 w \right) \frac{\partial w}{\partial p} = 0 \, .
\end{align}
Perhaps surprisingly, this equation is of the same form as (\ref{final_coupled_flow}) in Chapter \ref{CHP:nonabelian}, which arose for a gauge field coupled to matter. In fact, (\ref{2d_bi_deform_flow}) can be solved using the complete integral from \cite{polyanin} as was used to solve (\ref{final_coupled_flow}).

We will review this solution technique here for convenience. Any partial differential equation which can be written as
\begin{align}
    0 = f_1 ( w ) \left( \frac{\partial w}{\partial x} \right)^2 + f_2 ( w ) \frac{\partial w}{\partial x} \frac{\partial w}{\partial y} + f_3 ( w ) \left( \frac{\partial w}{\partial y} \right)^2 + g_1 ( w ) \frac{\partial w}{\partial x} + g_2 ( w ) \frac{\partial w}{\partial y} + h ( w ) ,
    \label{general_pde_sol_new_chapter}
\end{align}
has a solution $w ( x , y ) $ is given implicitly via
\begin{align}
    C_3 &= C_1 x + C_2 y + \int \frac{2 F ( w ) \, dw }{ G ( w ) \pm \sqrt{ G ( w )^2 - 4 F(w) h(w) } } , \nonumber \\
    F ( w ) &= C_1^2 f_1 ( w ) + C_1 C_2 f_2 ( w ) + C_2^2 f_3 ( w ) , \nonumber \\
    G ( w ) &= C_1 g_1 ( w ) + C_2 g_2 ( w ) . 
\end{align}
To match (\ref{2d_bi_deform_flow}) onto the form (\ref{general_pde_sol_new_chapter}), we identify the independent variables as $p \to x$, $q \to y$, and we define
\begin{align}\begin{split}
    &f_1 = f_2 = 0 \, , \quad f_3 = 4 \, , \quad g_1 = - ( 1 + 4 w ) \, , \\
    &g_2 = 1 \quad , \quad h(w) = w^2 + w .
\end{split}\end{align}
Therefore, the functions $F$ and $G$ are
\begin{align}
    F ( w ) &= 4 C_2^2 , \nonumber \\
    G ( w ) &= C_2 - C_1 \left( 1 + 4 w \right) \, .
\end{align}
The solution via complete integral can hence be written as
\begin{align}\label{new_chapter_complete_integral_soln}
    C_3 = C_1 p + C_2 q + \int \frac{8 C_2^2 \, dw}{C_2 - C_1 ( 1 + 4 w ) \pm \sqrt{ \left( C_2 - C_1 ( 1 + 4 w ) \right)^2 - 16 C_2^2 \left( w^2 + w \right) } } \, .
\end{align}
As with the corresponding complete integral from Chapter \ref{CHP:nonabelian}, the integral on the right side of (\ref{new_chapter_complete_integral_soln}) can be evaluated in terms of square roots and inverse hyperbolic tangents (or equivalently logarithms), although the explicit expression may not be especially illuminating. The integration constants $C_1, C_2, C_3$ are then determined by the initial conditions; for instance, in the case of a Born-Infeld seed theory one would impose the initial condition (\ref{BI_initial_condition_one}) and that the implicit solution reduce to (\ref{hypergeometric_as_root}) when $\eta = 0$. 

We will also record a particular special solution to the flow equation, parameterized by constants $c_1$ and $c_2$ which are not the same as the capital constants $C_1, C_2$ above. This is a special case of (\ref{new_chapter_complete_integral_soln}) where the integral simplifies. Restoring $\xi, \eta$, the implicit expression
\begin{align}\label{special_soln}
    ( 1 + c_1 ) \log ( \xi ) + \log ( \eta ) + c_2 &= c_1 \log \left( \sqrt{ c_1^2 - 8 ( c_1 + 1 ) ( c_1 + 2 ) h ( \xi, \eta ) } - c_1 \right) \nonumber \\
    &\qquad + ( 4 + 3 c_1 ) \log \left( 4 + 3 c_1 + \sqrt{ c_1^2 - 8 ( 1 + c_1 ) ( 2 + c_1 ) h ( \xi, \eta ) } \right) \, , 
\end{align}
defines a function $h (\xi, \eta )$ which is a particular solution to this flow equation. The form of the square roots inside the logarithms of (\ref{special_soln}) is reminiscent of those in Chapter \ref{CHP:nonabelian}.

\subsection{Three Dimensions}

For completeness, we will also study the corresponding problem in three dimensions, although it is similar to the two-dimensional case but with different numerical factors. We saw in (\ref{3d_flow_exact}) that the master flow equation in three spacetime dimensions becomes
\begin{align}
    \frac{\partial \mathcal{L}}{\partial \lambda} = 3 (1 - 3 r ) \mathcal{L}^2 - 8 \mathcal{L} ( 1 - 3 r ) x_1 \frac{\partial \mathcal{L}}{\partial x_1} + 8 ( 1 - 2 r ) x_1^2 \left( \frac{\partial \mathcal{L}}{\partial x_1} \right)^2  \, .
\end{align}
As before, we will make an ansatz of the form
\begin{align}
    \mathcal{L} = f ( x_1, \beta, \lambda ) = \frac{1}{\lambda} h \left( \lambda x_1 , \frac{\beta}{\lambda} \right) \, .
\end{align}
We now use the symbol $\beta$ rather than $\alpha'$ to avoid confusion since $\alpha'$ is typically taken to have length dimension $2$ in any spacetime dimension $d$ (as it is the square of the string length), whereas in our conventions, $\lambda$ has length dimension $d$ so $\frac{\alpha'}{\lambda}$ is not dimensionless in $d=3$. Thus in this case, one can think of $\beta = \left( \alpha' \right)^{3/2}$ and use this to form a dimensionless ratio $\frac{\beta}{\lambda}$.

Making the same change of variables
\begin{align}
    \xi = \lambda x_1 \, , \qquad \eta = \frac{\beta}{\lambda} \, ,
\end{align}
as in two dimensions, we arrive at the differential equation
\begin{align}
    \left( 9 r - 3 \right) h^2 + h \left( -1 + 8 ( 1 - 3 r ) \xi \frac{\partial h}{\partial \xi} \right) + \xi \frac{\partial h}{\partial \xi} \left( 1 + 8 ( 2 r - 1 ) \xi \frac{\partial h}{\partial \xi} \right) - \eta \frac{\partial h}{\partial \eta} = 0 \, .
\end{align}
We can once again switch to log variables $p = \log ( \xi ) , q = \log ( \eta )$ to find
\begin{align}
    \left( 9 r - 3 \right) w^2 - w + \frac{\partial w}{\partial p} + 8 ( 2 r - 1 ) \left( \frac{\partial w}{\partial q} \right)^2 + ( 8 - 24 r ) w \frac{\partial w}{\partial p} - \frac{\partial w}{\partial q} = 0 \, .
\end{align}
This PDE is still of the general form (\ref{general_pde_sol_new_chapter}), so it may be solved with the same method of writing the solution implicitly using a complete integral. However, we will not belabor the point further by writing the corresponding formulas.

This concludes our discussion of deformations of more general seed theories, including those of Born-Infeld type, in $d=2$ and $d=3$. The question of the corresponding deformations in $d \geq 4$ is more interesting. In that case, since there are more trace structures involved, one would need to begin with an ansatz of the form
\begin{align}
    \mathcal{L} = \frac{1}{\lambda} h \left( \lambda x_1 , \lambda^2 x_2 \, , \cdots , \lambda^k x_k , \frac{\alpha'}{\lambda} \right) \, ,
\end{align}
which produces a partial differential equation in $k+1$ variables. Barring some miraculous simplification, the complete integral techniques described here will not work in such cases, since this method applies only to partial differential equations in two variables.

\section{Deformations of Non-Abelian Gauge Theories}\label{sec:general_def_nonabelian}

In this section, we turn to the deformation of non-abelian gauge theories. We will have much less to say in this context, instead limiting ourselves to a few general observations and pointing out the complications which make this case more difficult.

We now assume that the field strength $F_{\mu \nu} = F_{\mu \nu}^{a} T_a$ carries an additional gauge index, where we will always use Greek letters for spacetime indices and Latin letters for gauge indices. We will assume that the Killing-Cartan form has been diagonalized and the generators $T^a$ of our Lie algebra are scaled so that $\Tr ( T^a T^b ) = \delta^{ab}$, and write
\begin{align}\label{Fsquare_nonabelian}
    F^2 = \Tr \left( F_{\mu \nu} F^{\mu \nu} \right) = \sum_a F_{\mu \nu}^a F^{\mu \nu a} \, ,
\end{align}
where the capital trace $\Tr$ refers to the trace over gauge indices.

The presence of additional indices means that we can build more inequivalent combinations of field strengths than in the abelian setting. For instance, in abelian gauge theories the two scalars involving four field strengths are
\begin{align}
    x_1^2 = \left( F_{\mu \nu} F^{\mu \nu} \right)^2 \, , \qquad x_2 = \tensor{F}{^\mu_\rho} F^{\rho \sigma} F_{\sigma \nu} \tensor{F}{^\nu_\mu} \, .
\end{align}
However, in a non-abelian theory, we can construct at least $6$ inequivalent combinations:
\begin{align}\label{extra_possibilities_nonabelian}
    x_{1, 1}^2 &= \left( \Tr \left( F_{\mu \nu} F^{\mu \nu} \right) \right)^2 = F_{\mu \nu}^a F^{\mu \nu}_a F_{\alpha \beta}^{b} F^{\alpha \beta}_b \, , \nonumber \\ 
    x_{2, 1} &= \Tr \left( F_{\mu \nu} F^{\alpha \beta} \right) \Tr \left( F^{\mu \nu} F_{\alpha \beta} \right) = F^{\mu \nu}_a F_{\mu \nu}^b F_{\alpha \beta}^a  F^{\alpha \beta}_b  \, , \nonumber \\
    x_{2,2} &=\Tr \left( F_{\mu \nu} F^{\mu \nu} F_{\alpha \beta} F^{\alpha \beta} \right) = F_{\mu \nu a} F^{\mu \nu}_b F_{\alpha \beta c} F^{\alpha \beta}_d \Tr \left( T^a T^b T^c T^d \right) \nonumber \\
    x_{2, 3} &= \Tr \left( \tensor{F}{^\mu_\rho} F^{\rho \sigma} \right) \Tr \left( F_{\sigma \nu} \tensor{F}{^\nu_\mu} \right) = \tensor{F}{^\mu_\rho^a} F^{\rho \sigma}_a F_{\sigma \nu}^b \tensor{F}{^\nu_\mu_b} \, , \nonumber \\ 
    x_{2, 4} &= \Tr \left( \tensor{F}{^\mu_\rho} F^{\sigma \nu} \right) \Tr \left( \tensor{F}{_\nu^\rho} \tensor{F}{_\mu_\sigma} \right) = \tensor{F}{^\mu_\rho_a} F^{\sigma \nu a}  \tensor{F}{_\nu^\rho^b} \tensor{F}{_\mu_\sigma_b} \, , \nonumber \\ 
    x_{2, 5} &= \Tr \left( \tensor{F}{^\mu_\rho} F^{\rho \sigma} F_{\sigma \nu} \tensor{F}{^\nu_\mu} \right) = \tensor{F}{^\mu_\rho_a} F^{\rho \sigma}_b F_{\sigma \nu c} \tensor{F}{^\nu_\mu_d} \Tr \left( T^a T^b T^c T^d \right) \, .
\end{align}
%
%
Four of the combinations in (\ref{extra_possibilities_nonabelian}) are ``double-trace'' in the sense that they are the products of two factors, each of which involves a trace over gauge indices. These double-trace terms, in this case, only depend on traces of products of two generators $T^a T^b$, which have been normalized to $\delta^{ab}$. However, two terms are ``single-trace'' expressions which involve only one trace over gauge indices, and therefore depend on the trace of a product of four generators.

It is clear that the flow equation (\ref{flow_general_exact}) which we developed in the abelian case is not sufficient here. Whereas a general scalar constructed from an abelian field strength $F_{\mu \nu}$ could be expressed in terms of a fairly limited set of variables $x_1, \cdots, x_k$, in the non-abelian setting one seems to need a larger set of variables $x_{i, j}$, where the $j$ index distinguishes between terms with inequivalent trace structures over gauge indices. For instance, we have introduced a (fairly \emph{ad hoc}) example of such a naming convention in (\ref{extra_possibilities_nonabelian}): one might use the notation $\big( x_{1,1} \big)^2$ for the expression in the first line, then call the remaining five expressions $x_{2, 1}, \cdots, x_{2, 5}$, respectively. Note that, in the abelian setting, the objects in first three lines of (\ref{extra_possibilities_nonabelian}) are all equivalent and equal to $x_1^2$ whereas the expressions in the final three lines are all proportional to $x_2$. Although the terms $x_{2, 1}$ and $x_{2,2}$ reduce to $x_1^2$ for abelian gauge groups, in the non-abelian case they cannot be written as the square of any scalar quantity, so we have chosen to label them as versions of $x_2$.

Although one might be able to write a version of the master flow equation (\ref{flow_general_exact}) in the case of a non-abelian theory, keeping track of all inequivalent $x_{i, j}$, it seems that this analysis would become very involved unless one developed some more clever way to organize the various trace structures. We will not attempt such an analysis in the present work. However, we can already make an obvious, but general, observation: if a seed Lagrangian is only a function of the combination $F^2 = \Tr ( F_{\mu \nu} F^{\mu \nu} )$, then a $\TT$-like flow will never introduce dependence on combinations involving a trace of four or more field strengths. For instance, if we begin with a Lagrangian $\mathcal{L} = f ( x )$, where $x = F^2$, then
\begin{align}
    T_{\mu \nu} = \delta_{\mu \nu} \mathcal{L} - 4 \frac{\partial f}{\partial x} \Tr \left( F_{\mu \rho} \tensor{F}{^\rho_\nu} \right) \, ,
\end{align}
and therefore any bilinears in the stress tensor like $T^{\mu \nu} T_{\mu \nu}$ or $\left( \tensor{T}{^\mu_\mu} \right)^2$ will involve only double-trace combinations. To leading order, then, the $\TT$-deformed Lagrangian will have terms which involve products of two traces over gauge indices, with two factors of the field strength inside of each trace, but it will contain no terms which are written as a single trace with four field strengths inside the trace. This statement generalizes to higher orders: that is, if a given Lagrangian $\mathcal{L}$ contains terms which involve traces of at most $2n$ generates $T^a$, then the stress tensor $T_{\mu \nu} = \frac{-2}{\sqrt{g}} \frac{\delta \mathcal{L}}{\delta g^{\mu \nu}}$ also involves traces of at most $2n$ generators, and therefore bilinears constructed from $T_{\mu \nu}$ will contain traces of no more than $2n$ generators (although they will involve \emph{products} of such traces).

It follows that, in the non-abelian context, a $\TT$-like deformation of free Yang-Mills theory can never yield the non-abelian Born-Infeld action. As we saw in equation (\ref{f4terms}) of Chapter \ref{CHP:nonabelian}, the $F^4$ terms in non-abelian DBI are purely single-trace, whereas $\TT$ deformation of a seed theory $\mathcal{L}_0 \sim F^2$ will yield only double-trace terms at order $\lambda^2$. This deformation, therefore, can never agree with Born-Infeld.




\appendix

\chapter{Details of Supercurrent and Flow Equation Computations} \label{details}
\section{Derivation of \texorpdfstring{$(1,1)$}{(1, 1)} Flow Equation}\label{11_appendix}

Here we will show some steps of the calculation which leads to the partial differential equation (\ref{potential_pde}) defining the supercurrent-squared deformation of a $(1,1)$ free theory with a potential. By setting the superpotential $h$ to zero, this calculation also reproduces the PDE (\ref{free_pde}) which describes deformations of the free theory.

We would like to consider what happens when we deform the superspace Lagrangian $\mathcal{A}^{(0)} = D_+ \Phi D_- \Phi + h( \Phi )$, according to the flow equation (\ref{ttbar_general_flow}),
\begin{align*}
	\frac{\partial}{\partial t} \mathcal{A}^{(t)} = \mathcal{T}_{+++}^{(t)} \mathcal{T}_{---}^{(t)} - \mathcal{T}_{--+}^{(t)} \mathcal{T}_{++-}^{(t)}  .
\end{align*}
It will help to introduce some shorthand: we define $A = D_+ \Phi D_- \Phi$ so that $\mathcal{A}^{(0)} = A$, and let $x = t \partial_{++} \Phi \partial_{--} \Phi$ and $y  = t \left( D_+ D_- \Phi \right)^2$ as before. Also define the dimensionful combinations 
\begin{align} 
X = \partial_{++} \Phi \partial_{--} \Phi = \frac{x}{t}, \qquad Y = \left( D_+ D_- \Phi \right)^2 = \frac{y}{t}.
\end{align}
Our ansatz for the superspace Lagrangian at finite $t$ will be $\mathcal{A}^{(t)} = F ( x , y ) A + h(\Phi)$.

With this ansatz, some of the terms in (\ref{final_ttbar_general}) will not contribute to the right side of (\ref{ttbar_general_flow}). For instance, the terms $\frac{\delta \mathcal{A}}{\delta D_+ D_- \Phi} D_{\pm} \partial_{\pm} \Phi$ will be proportional to $D_+ \Phi D_- \Phi = A$. However, every term in the superspace supercurrent is proportional to $D_+ \Phi$, $D_- \Phi$, or $D_+ \Phi D_- \Phi$. Therefore, when we construct a bilinear in $\mathcal{T}$, any term containing $D_+ \Phi D_- \Phi$ will not contribute because it can only appear multiplying another term which contains at least one of $D_{\pm} \Phi$, which vanishes because $\left( D_{\pm} \Phi \right)^2 = 0$.

For our special ansatz, we will re-write the components of $\mathcal{T}$ keeping only terms which contribute to bilinears,
\begin{align}
\begin{split}
	\mathcal{T}_{++-} &\sim \partial_{++} \Phi \frac{\delta \mathcal{A}}{\delta D_+ \Phi} + \partial_{++} \Phi D_+ \left(  \frac{\delta \mathcal{A}}{\delta \partial_{++} \Phi} \right) - \frac{1}{2} \partial_{++} \Phi D_- \left( \frac{\delta \mathcal{A}}{\delta D_+ D_- \Phi} \right) - D_+ \mathcal{A}  , \label{final_ttbar_equiv} \\	
    \mathcal{T}_{+++} &\sim \partial_{++} \Phi \frac{\delta \mathcal{A}}{\delta D_- \Phi} + \partial_{++} \Phi D_- \left(  \frac{\delta \mathcal{A}}{\delta \partial_{--} \Phi} \right) + \frac{1}{2} \partial_{++} \Phi D_+ \left( \frac{\delta \mathcal{A}}{\delta D_+ D_- \Phi} \right) ,  \\
    \mathcal{T}_{---} &\sim \partial_{--} \Phi \frac{\delta \mathcal{A}}{\delta D_+ \Phi} +  \partial_{--} \Phi D_+ \left( \frac{\delta \mathcal{A}}{\delta \partial_{++} \Phi} \right) - \frac{1}{2} \partial_{--} \Phi D_- \left( \frac{\delta \mathcal{A}}{\delta D_+ D_- \Phi} \right)  ,  \\
    \mathcal{T}_{--+} &\sim \partial_{--} \Phi \frac{\delta \mathcal{A}}{\delta D_- \Phi} + \partial_{--} \Phi  D_- \left( \frac{\delta \mathcal{A}}{\delta \partial_{--} \Phi} \right) + \frac{1}{2} \partial_{--} \Phi D_+ \left( \frac{\delta \mathcal{A}}{\delta D_+ D_- \Phi} \right) - D_- \mathcal{A} . 
\end{split}
\end{align}
The terms are
\begin{align*}
	D_+ \mathcal{A} &\sim F D_+ A + h'(\Phi) D_+ \Phi, \\
	D_- \mathcal{A} &\sim F D_- A + h'(\Phi) D_- \Phi, \\
    \partial_{++} \Phi \frac{\delta \mathcal{A}}{\delta D_+ \Phi} &\sim F \partial_{++} \Phi D_- \Phi, \\
    \partial_{++} \Phi \frac{\delta \mathcal{A}}{\delta D_- \Phi} &\sim - F \partial_{++} \Phi D_+ \Phi , \\
    \partial_{--} \Phi \frac{\delta \mathcal{A}}{\delta D_+ \Phi} &\sim F \partial_{--} \Phi D_- \Phi ,\\ 
    \partial_{--} \Phi \frac{\delta \mathcal{A}}{\delta D_- \Phi} &\sim - F \partial_{--} \Phi D_+ \Phi , \\
    \partial_{++} \Phi D_+ \left( \frac{\delta \mathcal{A}}{\delta \partial_{++} \Phi} \right) &\sim X \frac{\partial F}{\partial X} D_+ A ,\\
    \partial_{++} \Phi D_- \left( \frac{\delta \mathcal{A}}{\delta \partial_{--} \Phi} \right) &\sim \left( \partial_{++} \Phi \right)^2 \frac{\partial F}{\partial X} D_- A ,\\
    \partial_{--} \Phi D_+ \left( \frac{\delta \mathcal{A}}{\delta \partial_{++} \Phi} \right) &\sim \left( \partial_{--} \Phi \right)^2 \frac{\partial F}{\partial X} D_+ A ,\\
    \partial_{--} \Phi D_- \left( \frac{\delta \mathcal{A}}{\delta \partial_{--} \Phi} \right) &\sim X \frac{\partial F}{\partial X} D_- A , \\
	\frac{1}{2} \partial_{++} \Phi D_+ \left( \frac{\delta \mathcal{A}}{\delta D_+ D_- \Phi} \right) &\sim \sqrt{Y} \partial_{++} \Phi \frac{\partial F}{\partial Y} \cdot D_+ A, \\
	\frac{1}{2} \partial_{--} \Phi D_+ \left( \frac{\delta \mathcal{A}}{\delta D_+ D_- \Phi} \right) &\sim \sqrt{Y} \partial_{--} \Phi \frac{\partial F}{\partial Y} \cdot D_+ A , \\
    - \frac{1}{2} \partial_{++} \Phi D_- \left( \frac{\delta \mathcal{A}}{\delta D_+ D_- \Phi} \right) &\sim - \sqrt{Y} \partial_{++} \Phi \frac{\partial F}{\partial Y} \cdot D_- A ,\\
    - \frac{1}{2} \partial_{--} \Phi D_- \left( \frac{\delta \mathcal{A}}{\delta D_+ D_- \Phi} \right) &\sim - \sqrt{Y} \partial_{--} \Phi \frac{\partial F}{\partial Y} \cdot D_- A , 
\end{align*}
where $\sim$ means ``equal modulo terms which are proportional to $D_+ \Phi D_- \Phi$,'' since any products involving these terms will contain two nilpotent factors and thus vanish.

The first piece of supercurrent-squared is
\begin{align}
\begin{split}
	\mathcal{T}_{++|+} \mathcal{T}_{--|-} &= \left( - F \partial_{++} \Phi D_+ \Phi + \left( \partial_{++} \Phi \right)^2 \frac{\partial F}{\partial X} D_- A + \sqrt{Y} \partial_{++} \Phi \frac{\partial F}{\partial Y} \cdot D_+ A \right) \\
    &\times \left( F \partial_{--} \Phi D_- \Phi +  \left( \partial_{--} \Phi \right)^2 \frac{\partial F}{\partial X} D_+ A - \sqrt{Y} \partial_{--} \Phi \frac{\partial F}{\partial Y} D_- A \right), \\
    &= - F^2 X A - F X \frac{\partial F}{\partial X} \partial_{--} \Phi D_+ \Phi D_+ A + F X \frac{\partial F}{\partial X} \partial_{++} \Phi D_- A D_- \Phi \\
    &+ X^2 \left( \frac{\partial F}{\partial X} \right)^2 D_- A D_+ A + F \frac{\partial F}{\partial Y} \sqrt{Y} X D_+ A D_- \Phi \\
    &+ F X \sqrt{Y} \frac{\partial F}{\partial Y} D_+ \Phi D_- A - Y X \left( \frac{\partial F}{\partial Y} \right)^2 D_+ A D_- A .
\end{split}
\end{align}
The second piece is
\begin{align}
\begin{split}
	\mathcal{T}_{++|-} \mathcal{T}_{--|+} &= \left( F \partial_{++} \Phi D_- \Phi + \left( X \frac{\partial F}{\partial X} - F \right) D_+ A - G' D_+ \Phi - \sqrt{Y} \partial_{++} \Phi \frac{\partial F}{\partial Y} D_- A \right) \\
    &\times \left(  - F \partial_{--} \Phi D_+ \Phi + \left( X \frac{\partial F}{\partial X} - F \right) D_- A - G' D_- \Phi + \sqrt{Y} \partial_{--} \Phi \frac{\partial F}{\partial Y} \cdot D_+ A \right), \\
    &= F^2 X A + F \left( X \frac{\partial F}{\partial X} - F \right) \partial_{++} \Phi D_- \Phi D_- A + F X \sqrt{Y} \frac{\partial F}{\partial Y} D_- \Phi D_+ A \\
    &+ F X \sqrt{Y} \frac{\partial F}{\partial Y} D_- A D_+ \Phi - F \left( X \frac{\partial F}{\partial X} - F \right) \partial_{--} \Phi D_+ A D_+ \Phi \\
    &+ \left( X \frac{\partial F}{\partial X} - F \right)^2 D_+ A D_- A - Y X \left( \frac{\partial F}{\partial Y} \right)^2 D_- A D_+ A \\
    &- G' \left( X \frac{\partial F}{\partial X} - F \right) D_+ \Phi D_- A + \left( G' \right)^2 D_+ \Phi D_- \Phi - G' \sqrt{Y} \partial_{--} \Phi \frac{\partial F}{\partial Y} D_+ \Phi D_+ A \\
    &- G' \left( X \frac{\partial F}{\partial X} - F \right) D_+ A D_- \Phi + G' \sqrt{Y} \partial_{++} \Phi \frac{\partial F}{\partial Y} D_- A D_- \Phi . 
    \label{potential_ttbar}
\end{split}
\end{align}
Using the definitions $A = D_+ \Phi D_- \Phi$, $X = \partial_{++} \Phi \partial_{--} \Phi$, and $\sqrt{Y} = D_+ D_- \Phi$, we see that the products appearing in the above bilinears can be simplified as follows:
\begin{align}
\begin{split}
	D_+ \Phi D_+ A &= D_+ \Phi D_+ \left( D_+ \Phi D_- \Phi \right) = D_+ \Phi D_+ D_+ \Phi D_- \Phi = A \partial_{++} \Phi , \\
    D_+ \Phi D_- A &= D_+ \Phi D_- \left( D_+ \Phi D_- \Phi \right) = D_+ \Phi D_- D_+ \Phi D_- \Phi = - A \sqrt{Y} , \\
    D_- \Phi D_+ A &= D_- \Phi D_+ \left( D_+ \Phi D_- \Phi \right) = - D_- \Phi D_+ \Phi D_+ D_- \Phi = A \sqrt{Y} , \\
    D_- \Phi D_- A &= D_- \Phi D_- \left( D_+ \Phi D_- \Phi \right) = - D_- \Phi D_+ \Phi D_- D_- \Phi = A \partial_{--} \Phi , \\
    D_+ A D_- A &= \left( \partial_{++} \Phi D_- \Phi - D_+ \Phi \sqrt{Y} \right) \left( - \sqrt{Y} D_- \Phi - \partial_{--} \Phi D_+ \Phi \right) = (X + Y) A .
\end{split}
\end{align}
So after simplifying,
\begin{align}
\begin{split}
	\mathcal{T}_{++|+} \mathcal{T}_{--|-} &= - F^2 X A - 2 F X^2 \frac{\partial F}{\partial X} A - X^2 \left( \frac{\partial F}{\partial X} \right)^2 A (X + Y) - 2 F \frac{\partial F}{\partial Y} Y X A , \label{sc-square-terms-potential} \\
	\mathcal{T}_{++|-} \mathcal{T}_{--|+} &= F^2 X A + 2 F X \left( X \frac{\partial F}{\partial X} - F \right) A + 2 F X Y \frac{\partial F}{\partial Y} A + \left( X \frac{\partial F}{\partial X} - F \right)^2 (X + Y) A \\
	&\quad + \left( \left( h' \right)^2 + 2 h' \sqrt{Y} \left( X \frac{\partial F}{\partial X} - F \right) - 2 \sqrt{Y} X h' \frac{\partial F}{\partial Y} \right) A .
\end{split}
\end{align}
In particular, we see that every term appearing in (\ref{sc-square-terms-potential}) is proportional to $A = D_+ \Phi D_- \Phi$. This means that the deformation only generates a change in the first term of our ansatz $\mathcal{A}^{(t)} = F D_+ \Phi D_- \Phi + h ( \Phi )$, but it does not source any change in the potential. This justifies our choice of ansatz which leaves the potential as $h(\Phi)$ rather than allowing a more general function $G(t, \Phi)$ with $G(0, \Phi) = h(\Phi)$.

Adding the contributions gives,
\begin{align}
\begin{split}
	\mathcal{T}_{++|+} \mathcal{T}_{--|-} +  \mathcal{T}_{++|-} \mathcal{T}_{--|+} &= \Big[ \left( Y - X \right) F^2 - 2 F X ( X + Y ) \frac{\partial F}{\partial X} + 2 h' \sqrt{Y} \left( X \frac{\partial F}{\partial X} - F \right) \\
	&\quad - 2 \sqrt{Y} X h' \frac{\partial F}{\partial Y} + \left( h' \right)^2 \Big] A .
\end{split}
\end{align}
Setting this deformation equal to $\frac{\partial}{\partial t} \mathcal{A}^{(t)}$, and multiplying both sides by $t$ to convert dimensionlful variables $X$ and $Y$ into their dimensionless counterparts $x$ and $y$, gives our final result (\ref{potential_pde}), 
\begin{align}
	x \frac{\partial}{\partial x} F + y \frac{\partial}{\partial y} F &= \left( y - x \right) F^2 - 2 F x ( x + y ) \frac{\partial F}{\partial x} + \left( h' \right)^2 + 2 h' \sqrt{y} \left( x \frac{\partial F}{\partial x} - F \right) \cr & - 2 \sqrt{y} x h' \frac{\partial F}{\partial y} . 
	\label{pde_in_appendix}
\end{align}
We were unable to find a closed-form solution to (\ref{potential_pde}) in the general case. However, we can find the solution in a few special cases. If $y=0$, (\ref{pde_in_appendix}) reduces to
\begin{align}
    x F'(x) = - x \left( F(x)^2 + 2 F(x) F'(x) x \right) ,
\end{align}
which is solved by the Dirac-type ansatz $F(x) = \frac{\sqrt{1+4x}-1}{2x}$. If $x=0$, equation (\ref{pde_in_appendix}) is solved by $F(y) = \frac{1}{1-y}$. If $y = - x$, the second term on the right side of (\ref{pde_in_appendix}) drops out and the solution is $F(x) = \frac{1}{1+2x}$. See also Section \ref{sec:on-shell}, where we show that an exact solution can be found in the case where $h=0$ if we are willing to impose part of the superspace equations of motion.

\section{Derivation of \texorpdfstring{$(0,1)$}{(0, 1)} Flow Equation}\label{01-appendix}

Next we would like to write down a partial differential equation, similar to (\ref{free_pde}) in the $(1,1)$ case, which determines the Lagrangian deformed by the $(0,1)$ supercurrent-squared operator at finite $t$.

Define the three combinations of fields 
\begin{align} 
x = t \partial_{++} \Phi \partial_{--} \Phi, \qquad  y = t \left( D_+ \Psi_- \right)^2, \qquad z = t D_+ \Phi D_+ \partial_{--} \Phi,
\end{align}
and their dimensionful counterparts $X = \frac{x}{t}$, $Y = \frac{y}{t}$, $Z = \frac{z}{t}$. Our ansatz for the Lagrangian at finite $t$ is
\begin{align}
    \mathcal{A}^{(t)} = F ( x, y, z) \left( D_+ \Phi \partial_{--} \Phi + \Psi_- D_+ \Psi_- \right) + F_{2, -} ( x, y, z ) \left( \Psi_- D_+ \Phi \right) . 
\end{align}
Since the function $F_{2,-}$ is fermionic, it actually contains several different functions since we may combine the fields $\Phi, \Psi_-$ and derivatives in a few independent ways to obtain a fermionic function. We will expand $F_{2,-}$ as follows:
\begin{align}
    F_{2,-} = G ( x, y, z ) D_+ \Psi_- D_+ \partial_{--} \Phi + H ( x, y, z ) \partial_{++} \Phi \partial_{--} \Psi_- + J (x, y, z) \partial_{--} \Phi \partial_{++} \Psi_- .
\end{align}
Altogether our ansatz for the deformed Lagrangian is, 
\begin{align}
    \mathcal{A}^{(t)} &= F \left( D_+ \Phi \partial_{--} \Phi + \Psi_- D_+ \Psi_- \right) \nonumber \\
    &\quad + \left( G D_+ \Psi_- D_+ \partial_{--} \Phi + H \partial_{++} \Phi \partial_{--} \Psi_- + J \partial_{--} \Phi \partial_{++} \Psi_- \right) \left( \Psi_- D_+ \Phi \right) . 
    \label{01-reduced-ansatz}
\end{align}
This is a $(0,1)$ superspace Lagrangian with the functional dependence
\begin{align}
    \mathcal{A} = \mathcal{A} \left( \Phi, \Psi_-, D_+ \Phi, D_+ \Psi_-, \partial_{\pm \pm} \Phi, \partial_{\pm \pm} \Psi_-, D_+ \partial_{--} \Phi \right) . 
\end{align}
Following the procedure of section (\ref{section:supercurrent-squared}), we can consider a transformation $x^{\pm \pm} \to x^{\pm \pm} + a^{\pm \pm}$ and extract the components of conserved currents. In this case, they are
\begin{align}
\begin{split}
    \mathcal{T}_{++++} &= \partial_{++} \Phi \frac{\delta \mathcal{A}}{\delta \partial_{--} \Phi} + \partial_{++} \Psi_- \frac{\delta \mathcal{A}}{\delta \partial_{++} \Psi_-} - \partial_{++} \Phi D_+ \left( \frac{\delta \mathcal{A}}{\delta D_+ \partial_{--} \Phi} \right) , \\
    \mathcal{T}_{++--} &= \partial_{--} \Phi \frac{\delta \mathcal{A}}{\delta \partial_{--} \Phi} + \partial_{--} \Psi_- \frac{\delta \mathcal{A}}{\delta \partial_{--} \Psi_-} - \partial_{--} \Phi D_+ \left( \frac{ \delta \mathcal{A}}{\delta D_+ \partial_{--} \Phi} \right) - \mathcal{A} , \\
    \mathcal{S}_{++-} &= \partial_{++} \Phi \frac{\delta \mathcal{A}}{\delta D_+ \Phi} + D_+ \left( \partial_{++} \Phi \frac{\delta \mathcal{A}}{\delta \partial_{++} \Phi} - \mathcal{A} \right) + \partial_{++} \Psi_- \frac{\delta \mathcal{A}}{\delta D_+ \Psi_-}  \cr & + \left( \partial_{--} \partial_{++} \Phi \right) \frac{\delta \mathcal{A}}{\delta D_+ \partial_{--} \Phi} , \\
    \mathcal{S}_{---} &= \partial_{--} \Phi \frac{\delta \mathcal{A}}{\delta D_+ \Phi} + D_+ \left( \partial_{--} \Phi \frac{\delta \mathcal{A}}{\delta \partial_{++} \Phi} \right) + \partial_{--} \Psi_- \frac{\delta \mathcal{A}}{\delta D_+ \Psi_-} + \partial_{--}^2 \Phi \frac{\delta \mathcal{A}}{\delta D_+ \partial_{--} \Phi} .
\end{split}
\end{align}
Next we will compute each of these contributions. As before, we will drop terms which are proportional to $\Psi_- D_+ \Phi$, since every term in $\mathcal{S}$ and $\mathcal{T}$ is proportional to either $\Psi_-$ or to $D_+ \Phi$, so any terms involving both of these nilpotent factors will not contribute to bilinears. We will also introduce the shorthand $A = D_+ \Phi \partial_{--} \Phi$ and $B = \Psi_- D_+ \Psi_-$.

Doing this, we see that:
\begingroup
\allowdisplaybreaks
\begin{align}
    \partial_{++} \Phi \frac{\delta \mathcal{A}}{\delta \partial_{--} \Phi} &\sim F  D_+ \Phi \partial_{++} \Phi + \frac{\partial F}{\partial x} \left( \partial_{++} \Phi \right)^2 \left( A + B \right) , \label{01-reduced-components} \cr
    \partial_{--} \Phi \frac{\delta \mathcal{A}}{\delta \partial_{--} \Phi} &\sim F A + x \frac{\partial F}{\partial x} \left( A + B \right) , \cr
    \partial_{++} \Phi \frac{\delta \mathcal{A}}{\delta D_+ \Phi} &\sim \partial_{++} \Phi \frac{\partial F}{\partial z} \left( D_+ \partial_{--} \Phi \right) \left( A + B \right) + F x \cr
    &\quad - \partial_{++} \Phi \left( G D_+ \Psi_- D_+ \partial_{--} \Phi + H \partial_{++} \Phi \partial_{--} \Psi_- + J \partial_{--} \Phi \partial_{++} \Psi_- \right) \Psi_- , \cr
    \partial_{--} \Phi \frac{\delta \mathcal{A}}{\delta D_+ \Phi} &\sim \partial_{--} \Phi \frac{\partial F}{\partial z} \left( D_+ \partial_{--} \Phi \right) + F \left( \partial_{--} \Phi \right)^2 \cr
    &\quad - \partial_{--} \Phi \left( G D_+ \Psi_- D_+ \partial_{--} \Phi + H \partial_{++} \Phi \partial_{--} \Psi_- + J \partial_{--} \Phi \partial_{++} \Psi_- \right) \Psi_-, \cr
    \partial_{++} \Psi_- \frac{\delta \mathcal{A}}{\delta \partial_{++} \Psi_-} & \sim 0 , \cr
    \partial_{--} \Psi_- \frac{\delta \mathcal{A}}{\delta \partial_{--} \Psi_-} &\sim 0 , \cr
    D_+ \left( \partial_{++} \Phi \frac{\delta \mathcal{A}}{\delta \partial_{++} \Phi} \right) &\sim D_+ \left( x \frac{\partial F}{\partial x} ( A + B ) \right) + \Big( x \frac{\partial G}{\partial x} D_+ \Psi_- D_+ \partial_{--} \Phi + x \frac{\partial H}{\partial x} \partial_{++} \Phi \partial_{--} \Psi_- \cr
    &\hspace{50pt}  + x \frac{\partial J}{\partial x} \partial_{--} \Phi \partial_{++} \Psi_- \Big) \cdot \Big( D_+ \Psi_- D_+ \Phi - \Psi_- \partial_{++} \Phi \Big),  \cr
    D_+ \left( \partial_{--} \Phi \frac{\delta \mathcal{A}}{\delta \partial_{++} \Phi} \right) &\sim D_+ \left( \left( \partial_{--} \Phi \right)^2 \frac{\partial F}{\partial x} ( A + B ) \right) + \Big( \frac{\partial G}{\partial x} \left( \partial_{--} \Phi \right)^2 D_+ \partial_{--} \Phi \cr
    &+ \frac{\partial H}{\partial x} \partial_{--} \Phi \partial_{--} \Psi_- + \frac{\partial J}{\partial x} \left( \partial_{--} \Phi \right)^3 \partial_{++} \Psi_- \Big) \Big( D_+ \Psi_- D_+ \Phi - \Psi_- \partial_{++} \Phi \Big), \cr
    \partial_{++} \Phi D_+ \left( \frac{\delta \mathcal{A}}{\delta D_+ \partial_{--} \Phi} \right) &\sim \partial_{++} \Phi \frac{\partial F}{\partial z} \left( \partial_{++} \Phi \Psi_- D_+ \Psi_- - D_+ \Phi \left( D_+ \Psi_- \right)^2 \right), \cr
    \partial_{--} \Phi D_+ \left( \frac{\delta \mathcal{A}}{\delta D_+ \partial_{--} \Phi} \right) &\sim \partial_{--} \Phi \frac{\partial F}{\partial z} \left( \partial_{++} \Phi \Psi_- D_+ \Psi_- - D_+ \Phi \left( D_+ \Psi_- \right)^2 \right), \cr
    \partial_{++} \Psi_- \frac{\delta \mathcal{A}}{\delta D_+ \Psi_-} &\sim \partial_{++} \Psi_- \left( \frac{\partial F}{\partial y} ( A + B ) + F \Psi_- \right) , \cr
    \partial_{--} \Psi_- \frac{\delta \mathcal{A}}{\delta D_+ \Psi_-} &\sim \partial_{--} \Psi_- \left( \frac{\partial F}{\partial y} ( A + B ) + F \Psi_- \right) , \cr
    \partial_{--} \partial_{++} \Phi \frac{\delta \mathcal{A}}{\delta D_+ \partial_{--} \Phi} &\sim \partial_{--} \partial_{++} \Phi  \left( \frac{\partial F}{\partial z} D_+ \Phi (A + B) \right) \sim 0 , \cr
    \partial_{--}^2 \Phi \frac{\delta \mathcal{A}}{\delta D_+ \partial_{--} \Phi} &\sim \partial_{--}^2 \Phi \left( \frac{\partial F}{\partial z} D_+ \Phi ( A + B ) \right) \sim 0 , \cr
    D_+ \mathcal{A} &\sim \left( \frac{\partial F}{\partial x} D_+ x + \frac{\partial F}{\partial y} D_+ y + \frac{\partial F}{\partial z} D_+ z \right) ( A + B ) + F \left( D_+ A + D_+ B \right) \cr
    &+ \left( G D_+ \Psi_- D_+ \partial_{--} \Phi + H \partial_{++} \Phi \partial_{--} \Psi_- + J \partial_{--} \Phi \partial_{++} \Psi_- \right) \cr
    &\quad \cdot \left( D_+ \Psi_- D_+ \Phi - \Psi_- \partial_{++} \Phi \right).
\end{align}
\endgroup
We will argue that the coupled differential equations for $F, G, H$, and $J$ resulting from (\ref{01-reduced-components}) are consistent. This will be the case if they do not source any additional combinations of fields that do not appear in the ansatz (\ref{01-reduced-ansatz}).

The only thing that could spoil consistency is a $D_+ x$ term, since
\begin{align}
    D_+ x = \left( D_+ \partial_{++} \Phi \right) \partial_{--} \Phi + \partial_{++} \Phi D_+ \partial_{--} \Phi .
\end{align}
We have already allowed for dependence on $D_+ \partial_{--} \Phi$ in our Lagrangian, but terms proportional to $D_+ \partial_{++} \Phi$ are forbidden. We will show that, in the $\mathcal{S} \cdot \mathcal{T}$ deformation resulting from (\ref{01-reduced-components}), all $D_+ x$ terms drop out.

Tracking only the $D_+ x$ terms in bilinears, the supercurrent components are
\begin{align}
\begin{split}
    \mathcal{S}_{++-} &\sim x D_+ \left( \frac{\partial F}{\partial x} \right) \left( A + B \right) + \ldots , \\
    \mathcal{T}_{++--} &\sim x \frac{\partial F}{\partial x} ( A + B ) - F B - \partial_{--} \Phi \frac{\partial F}{\partial z} D_+ \left( D_+ \Phi \Psi_- D_+ \Psi_- \right) , \\
    \mathcal{S}_{---} &\sim \left( \partial_{--} \phi \right)^2 D_+ \left( \frac{\partial F}{\partial x} \right) ( A + B ) + \ldots , \\
    \mathcal{T}_{++++} &\sim F D_+ \Phi \partial_{++} \Phi + \frac{\partial F}{\partial x} \left( \partial_{++} \Phi \right)^2 ( A + B ) - \partial_{++} \Phi \frac{\partial F}{\partial z} D_+ \left( D_+ \Phi \Psi_- D_+ \Psi_- \right) ,
\end{split}
\end{align}
where $\ldots$ indicates terms that are not proportional to $D_+ x$ or $D_+ \left( \frac{\partial F}{\partial x} \right)$.

The relevant contributions in the deformation are
\begin{align}
\begin{split}
  &  \hspace{-20pt} \mathcal{T}_{++++} \mathcal{S}_{---} - \mathcal{T}_{++--} \mathcal{S}_{++-} \sim \left( \partial_{--} \Phi \right)^2 D_+ \left( \frac{\partial F}{\partial x} \right) ( A + B ) \cdot \left( F D_+ \Phi \partial_{++} \Phi + \frac{\partial F}{\partial x} \left( \partial_{++} \Phi \right)^2 (A + B ) \right) \\
    &\quad - x D_+ \left( \frac{\partial F}{\partial x} \right) (A + B ) \cdot \left( x \frac{\partial F}{\partial x} ( A + B ) - F B \right) + \ldots,
    \\
    &= \left( \partial_{--} \Phi \right)^2 D_+ \left( \frac{\partial F}{\partial x} \right) ( F B D_+ \Phi \partial_{++} \Phi ) - x D_+ \left( \frac{\partial F}{\partial x} \right) ( - F A B ) + \ldots , 
\end{split}
\end{align}
where we have used the fermionic nature of $A$ and $B$ so $A^2 = B^2 = (A+B)^2 = 0$. However in the last line, we recognize that $\left( \partial_{--} \Phi \right)^2 D_+ \Phi \partial_{++} \Phi = x A$, since $x = \partial_{++} \Phi \partial_{--} \Phi$ and $A = D_+ \Phi \partial_{--} \Phi$, so
\begin{align}
\begin{split}
    \mathcal{T}_{++++} \mathcal{S}_{---} - \mathcal{T}_{++--} \mathcal{S}_{++-} &\sim x D_+ \left( \frac{\partial F}{\partial x} \right) F B A + x D_+ \left( \frac{\partial F}{\partial x} \right) F A B = 0 , 
\end{split}
\end{align}
and thus the problematic $D_+ \left( \frac{\partial F}{\partial x} \right)$ terms do not contribute.

\section{The \texorpdfstring{$\cS$}{S}-multiplet in Components}
\label{components-S}

In this appendix, we provide the component expansion of the superfields of the $\cS$-multiplet introduced in section \ref{section-S-multiplet}. The results presented below are equivalent to the results first obtained in \cite{Dumitrescu:2011iu} up to  differences in notation.

The constraints~\eqref{conservation-S} 
are solved in terms of component fields by, 
\bea
\label{comp-S}
{\cal S}_{\pm\pm}  
&= &
 j_{\pm\pm} - i \q^\pm S_{\pm\pm\pm} 
 - i \q^\mp \left( S_{\mp\pm\pm} \mp 2 \sqrt 2 i \bar \rho_\pm\right) 
 - i \qb^\pm \bar S_{\pm\pm\pm} \cr
&& 
- i \qb^\mp \left(\bar S_{\mp\pm\pm} \pm 2 \sqrt2 i \rho_\pm\right) 
- \q^\pm \qb^\pm T_{\pm\pm\pm\pm} 
+ \q^\mp \qb^\mp \left(A \mp { \frac{k + k'}{2}}\right) \cr
&& 
+ i \q^+ \q^- \bar Y_{\pm\pm} + i \qb^+ \qb^- Y_{\pm\pm} \pm i \q^+ \qb^- \bar G_{\pm\pm} \mp i \q^- \qb^+ G_{\pm\pm} 
\cr
&& 
\mp \hf \q^+ \q^- \qb^\pm \pa_{\pm\pm} S_{\mp\pm\pm} \mp \hf \q^+ \q^- \qb^\mp \pa_{\pm\pm}
\left(S_{\pm\mp\mp} \pm 2 \sqrt 2 i \bar \rho_\mp\right)\cr
&&
 \mp \hf \qb^+ \qb^- \q^\pm \pa_{\pm\pm} \bar S_{\mp\pm\pm} \mp \hf \qb^+ \qb^-\q^\mp \pa_{\pm\pm} 
 \left( \bar S_{\pm\mp\mp} \mp 2 \sqrt 2  i \rho_\mp\right)\cr
&& + {\frac{1}{4}} \q^+ \q^- \qb^+ \qb^- \pa_{\pm\pm}^2 j_{\mp\mp}~.
\eea
Let us introduce the usual useful combinations: 
$y^{\pm\pm} = x^{\pm\pm} - \frac{i}{2}  \q^\pm\qb^\pm$ and $\tilde y^{\pm\pm} = x^{\pm\pm} \mp \frac{i}{2} \q^\pm \qb^\pm$. 
The chiral superfields~$\chi_\pm$ are
\begin{align}
\label{comp-chi}
\chi_+ &=
 - i \lambda_+(y) - i \q^+ \bar G_{++}(y) +  \q^- \left(E(y) + {\frac{k}{2}}\right) + \qb^- C^{(-)}  + \q^+ \q^- \pa_{++} \bar \lambda_-(y) \, , \\
\chi_- &=
 - i \lambda_-(y) - \q^+ \left(\bar E(y) - {\frac{k}{2}}\right) + i \q^- G_{--}(y) - \qb^+ C^{(+)}- \q^+ \q^- \pa_{--} \bar \lambda_+(y)~,\\
\lambda_\pm &= \pm\bar S_{\mp\pm\pm} + \sqrt 2 i \rho_\pm~,
\\
E &= \hf \left( \Theta - A \right)+ {\frac{i}{4}} \left( \pa_{++} j_{--} - \pa_{--} j_{++}\right)~,\\
0&= \pa_{++} G_{--} - \pa_{--} G_{++} ~,
\end{align}
and the twisted-(anti-)chiral superfields~$\cY_\pm$ are given by
\begin{align}
\label{comp-Y}
\cY_+ &= \sqrt 2 \rho_+ (\bar {\tilde y}) +  \q^- \left( F(\bar {\tilde y}) 
+ {\frac{k'}{2}}\right) - i \qb^+ Y_{++}(\bar {\tilde y}) - \qb^- C^{(-)}+ \sqrt2 i \q^- \bar \q^+ \pa_{++} \rho_-(\bar {\tilde y}) \, , 
\\
\cY_- &= \sqrt 2 \rho_-(\tilde y) -  \q^+ \left( F(\tilde y) - {\frac{k'}{2}}\right) + \qb^+ C^{(+)}
- i \qb^- Y_{--}(\tilde y) + \sqrt 2 i \q^+ \qb^- \pa_{--} \rho_+ (\tilde y)~,\\
F &= - \hf \left( \Theta + A\right) - {\frac{i}{4}} \left(\pa_{++} j_{--} + \pa_{--} j_{++}\right)~,\\
0&= \pa_{++} Y_{--} - \pa_{--} Y_{++} ~.
\end{align}
For the FZ-multiplet defined by the constraints \eqref{FZ-2_(2,2)}, the $\cS$-multiplet reduces to a set of $4+4$ real independent
component fields described by the $j_{\pm\pm}$ $U(1)_A$ axial conserved $R$-symmetry current  
($\pa_{++} j_{--} - \pa_{--} j_{++}=0$). In addition, there is a complex scalar field $v(x)$,
see eq.~\eqref{FZ-3},
together with the independent supersymmetry current and energy momentum tensor:
\bsubeq
\bea
 S_{\pm\pm\pm}(x)
&:=&
 ~~i  D_{\pm}\cJ_{\pm\pm}(\z)|_{\q=0}
~,\\
\bar{S}_{\pm\pm\pm}(x)
&:=&
- i  \Db_{\pm}\cJ_{\pm\pm}(\z)|_{\q=0}
~,\\
 S_{\mp\pm\pm}(x)
&:=&
- i  D_{\mp}\cJ_{\pm\pm}(\z)|_{\q=0}
=
\pm i  \Db_{\pm}\overline{\cV}(\z)|_{\q=0}
~,
\\
\bar{S}_{\mp\pm\pm}(x)
&:=&
~~ i  \Db_{\mp}\cJ_{\pm\pm}(\z)|_{\q=0}
=\mp i  D_{\pm}\cV(\z)|_{\q=0}
~,
\\
T_{\pm\pm\pm\pm}(x)
&:=&
~~\hf{[}D_\pm,\Db_\pm{]}\cJ_{\pm\pm}(\z)|_{\q=0}
~,
\\
\Theta(x)
&:=&
-\hf{[}D_+,\Db_+{]}\cJ_{--}(\z)|_{\q=0}
=-\hf{[}D_-,\Db_-{]}\cJ_{++}(\z)|_{\q=0}
\non\\
&=&-\hf D_+D_-\cV(\z)|_{\q=0}
+\hf \Db_+\Db_-\overline{\cV}(\z)|_{\q=0}
~.
\eea
\esubeq
For the FZ-multiplet, the following relation holds:
\bea
{\cal J}_{\pm\pm}  
&= & j_{\pm\pm} 
- i \q^\pm S_{\pm\pm\pm} 
 - i \qb^\pm \bar S_{\pm\pm\pm} 
+ i \q^\mp S_{\mp\pm\pm}
+ i \qb^\mp \bar S_{\mp\pm\pm} 
\cr
&&
- \q^\pm \qb^\pm T_{\pm\pm\pm\pm} + \q^\mp \qb^\mp \Theta
+ i   \q^+ \q^- \pa_{\pm\pm}\bar v + i  \qb^+ \qb^- \pa_{\pm\pm} v
\cr
&& \mp \hf \q^+ \q^- \qb^\pm \pa_{\pm\pm} S_{\mp\pm\pm} 
\pm \hf \q^+ \q^- \qb^\mp \pa_{\pm\pm}S_{\pm\mp\mp}
\cr
&&
 \mp \hf \qb^+ \qb^- \q^\pm \pa_{\pm\pm} \bar S_{\mp\pm\pm} 
 \pm \hf \qb^+ \qb^-\q^\mp \pa_{\pm\pm}  \bar S_{\pm\mp\mp} 
\cr
&&
 + {\frac{1}{4}} \q^+ \q^- \qb^+ \qb^- \pa_{\pm\pm}^2 j_{\mp\mp}
 ~.
\eea
Moreover, the chiral superfields~$\chi_\pm$ are set to zero 
and the twisted-(anti-)chiral superfields~$\cY_\pm=D_\pm \cV$ are given by
\bsubeq
\bea
\cY_+ &=&   i\bar S_{-++} (\bar {\tilde y}) +  \q^- \, 
G(\bar {\tilde y}) 
- i \qb^+ \pa_{++}v(\bar {\tilde y}) 
+ \q^- \bar \q^+ \pa_{++}  \bar S_{+--}(\bar {\tilde y})~,
\\
\cY_- &=& - i \bar S_{+--}(\tilde y) -  \q^+ G(\tilde y) 
- i \qb^- \pa_{--}v(\tilde y) 
+ \q^+ \qb^- \pa_{++}  \bar S_{---}(\tilde y)~,
\\
G &=& - \Theta - {\frac{i}{2}} \pa_{++} j_{--}
~.
\eea
\esubeq

\section{Details of the $(2,2)$ FZ Multiplet Calculation}\label{appendix:supercurrent_calculation}

In this appendix, we compute the fields $\mathcal{J}_{\pm \pm}$ and $\sigma$ appearing in
 the FZ-multiplet for Lagrangians of a chiral superfield $\Phi$ with the general form
\begin{align}
    \mathcal{L}_0 = \left( \int d^4 \theta \, \mathcal{A} ( \Phi, D_{\pm} \Phi , D_+ D_- \Phi, \partial_{\pm \pm} \Phi , \text{c.c.} ) \right) 
    + \left( \int d^2 \theta \, W ( \Phi ) \right) + \left( \int d^2 \thetab \, \Wb ( \Phib ) \right) ~, 
    \label{general_susy_lag}
\end{align}
where ``\text{c.c.}'' indicates dependence on the conjugates $\Phib, \Db_{\pm} \Phib, \Db_+ \Db_- \Phib$, and 
$\partial_{\pm \pm} \Phib$. To do this, we will minimally couple the theory to supergravity using the old-minimal supergravity 
formulation and extract the currents which couple to the metric superfield $H^{\pm \pm}$ and the chiral compensator $\sigma$. 
The minimal coupling prescription involves promoting $\mathcal{L}_0$ to%
\footnote{Conforming to 
notation of \cite{Grisaru:1994dm,Grisaru:1995dr,Grisaru:1995kn,Grisaru:1995py,Gates:1995du}, 
in this section we will sometimes 
use the index notations $\a=+,-$ and $m=++,--$.}
\begin{align}
    \mathcal{L}_0 \longrightarrow \mathcal{L}_{\text{SUGRA}} &= \left( \int d^4 \theta \, E^{-1} \,
     \mathcal{A} ( \bbPhi, \nabla_{\pm} \bbPhi , \nabla_+ \nabla_- \bbPhi, \nabla_{\pm \pm} \bbPhi , \text{c.c.} ) \right)
      \nonumber \\
    &\quad 
    + \left(
    \int d^2 \theta \, \mathcal{E}^{-1} \, W ( \bbPhi ) 
    \right) 
    + \left( 
    \int d^2 \thetab \, \overline{\mathcal{E}}^{-1} \, 
    \Wb ( \bbPhib ) 
    \right)~ .
    \label{general_lag_coupled}
\end{align}
Here $\nabla_{\pm}$ is the derivative which is covariant with respect to the full local supergravity gauge group, 
$E^{-1}$ is the full superspace measure, $\mathcal{E}^{-1}$ is the chiral measure, and $\bbPhi$ is the covariantly 
chiral version of the chiral superfield $\Phi$---that is, $\nablab_{\pm} \bbPhi = 0$ whereas $\Db_{\pm} \Phi = 0$.

Expressions for these supercovariant derivatives and measures have been worked out in a series of papers
 \cite{Grisaru:1994dm,Grisaru:1995dr,Grisaru:1995kn,Grisaru:1995py,Gates:1995du} from which we will import the results
that we need for our analysis. 
To leading order in $H^m$, 
the linearized inverse superdeterminant of the supervielbein is
\begin{align}
    E^{-1} = 1 - \left[ \Db_+ , D_+ \right] H^{++} - \left[ \Db_- , D_- \right] H^{--} \, ,
\end{align}
while the chiral measure is given by
\begin{align}
    \mathcal{E}^{-1} &= e^{- 2 \sigma } \left( 1 \cdot e^{iH^m\overset{\leftarrow}{\pa_m}} \right)
    =1-2\s+i(\pa_m H^m)
    +\cdots
    ~ ,
\end{align}
where the ellipsis are terms of higher-order in $H^m$ and $\s$.
The covariantly chiral superfield $\bbPhi$ is related to the ordinary chiral superfield $\Phi$ by
\begin{align}
    \bbPhi = e^{i H^m\partial_m} \Phi = \Phi + i \left( H^{++} \partial_{++} + H^{--} \partial_{--} \right) \Phi + \mathcal{O} ( H^2 ) ~.
    \label{cov_chiral}
\end{align}
The spinor supercovariant derivatives $\nabla_{\pm}$ are
\begin{align}
    \nabla_\alpha = E_\alpha + \Omega_{\alpha} M + \Gamma_{\alpha} \Mb + \Sigma_{\alpha} N~ ,
    \label{supercovariant_deriv_grisaru}
\end{align}
where $M$ and $N$ are linear combinations of the Lorentz, $U(1)_V$, and $U(1)_A$ generators which act on spinors as
\bsubeq\begin{align}
    [ M , \psi_{\pm} ] &= \pm \frac{1}{2} \psi_{\pm}~,
    & [ M , \psib_{\pm} ] &= 0
    ~ ,\\
    [ \Mb , \psib_{\pm} ] &= \pm \frac{1}{2} \psib_{\pm}~,
    &  [ \Mb, \psi] &= 0~ , \\
    [ N , \psi_{\pm} ] &= - \frac{i}{2} \psi_{\pm}
    ~,& [N, \psib_{\pm}] &= + \frac{i}{2} \psib_{\pm}~ . 
\end{align}
\esubeq
The spinor inverse of the 
supervielbein $E_\alpha=E_\alpha{}^M\pa_M$,
and the structure group connections
$\Omega_{\alpha}$, $\Gamma_{\alpha}$, and $\Sigma_{\alpha}$ can be expressed to linear order in terms of the metric 
superfield $H^{\pm \pm}$ and an unconstrained complex
scalar compensator $S$. In the case of old-minimal supergravity,
the unconstrained superfield $S$ is related to the chiral compensator $\sigma$ by
\begin{align}
    S &= \sigma - \frac{i}{2} \partial_m H^m - \frac{1}{2} \left[ \Db_+ , D_+ \right] H^{++} - \frac{1}{2} \left[ \Db_- , D_- \right] H^{--}~ ,
    \label{S_sigma_contraint}
\end{align}
to linear order.
In the following analysis we will first obtain expressions for the supercovariant derivatives in terms of 
$S=S(H^m,\s)$, 
and use \eqref{S_sigma_contraint} 
to give them in terms of $H^m$ and $\sigma$. 

The spinorial inverse of the supervielbein is given at first order in the prepotentials by
\be  
E_\pm 
=
(1+\Sb)D_\pm 
+ i ( D_\pm H^m ) \partial_m 
- 2 \left( \Db_\mp D_\pm H^{\mp\mp} \right) D_\mp 
~,
\label{spin_viel}
\ee 
together with their complex conjugates.
Meanwhile, the connections $\Omega_\alpha$, $\Gamma_\alpha$, and $\Sigma_\alpha$ can be written to leading order as
\bsubeq
  \label{gammas_sigmas_omegas}
\bea
    \Gamma_\pm 
    &=&
    \pm2 D_\pm \left( S + \Sb \right) 
    \mp 2 D_\mp \Db_\mp D_\pm H^{\mp\mp} 
    ~,\\
    \Sigma_\pm &=& 
    - 2 i D_\pm \Sb + 2 i D_\mp \Db_\mp D_\pm H^{\mp\mp} 
    ~,
    \\
    \Omega_\pm &=& \mp 2 D_\mp \Db_\mp D_\pm H^{\mp\mp} 
    ~ .
\eea\esubeq
Using (\ref{supercovariant_deriv_grisaru}), the vielbeins (\ref{spin_viel}), and the expression (\ref{cov_chiral}) for $\bbPhi$, 
we find the supercovariant derivatives
\begin{align}    \label{linear_first_supercov_deriv}
    \nabla_\pm \bbPhi &=
    \left( 1 + \Sb \right) D_\pm \Phi 
    + 2 i ( D_\pm H^m ) \partial_m \Phi 
    + i H^m ( D_\pm \partial_m \Phi ) 
    - 2 \left( \Db_\mp D_\pm H^{\mp\mp} \right) D_\mp \Phi
    ~ , \\
     \nablab_\pm \bbPhib 
    &=
    \left( 1 + S \right) \Db_\pm \Phib 
    - 2 i \left( \Db_\pm  H^m \right) \partial_m \Phib 
    - i H^m ( \Db_\pm \partial_m \Phib ) 
    - 2 \left( \Db_\pm D_\mp H^{\mp\mp} \right) \Db_\mp  \Phib  \, .
\end{align}
To compute the second supercovariant derivatives
acting on $\bbPhi$ and $\bbPhib$, we must include the contributions from 
$\Omega_{\alpha}$, $\Gamma_{\alpha}$, $\Sigma_{\alpha}$, and their conjugates. One finds
\begin{align}
    \nablab_+ \nabla_+ \bbPhi &= i ( 1 + S + \Sb ) \partial_{++} \Phi     
    - 2 ( \Db_+ \Db_- D_+ H^{--} ) D_- \Phi
+ 2 i ( \Db_+ D_+ H^m ) \partial_m \Phi  
    \non\\
    &\quad 
    - H^m \partial_{++} \partial_m \Phi 
    + 2 \left( \Db_+ ( S + \Sb ) + \Db_- D_- \Db_+ H^{--} \right) D_+ \Phi
    ~,  \\
    \nabla_+ \nabla_- \bbPhi &= 
    ( 1 + 2 \Sb )  D_+ D_- \Phi 
    + 2 i ( D_+ D_- H^m ) \partial_m \Phi 
    - 2 i ( D_- H^m )  D_+ \partial_m \Phi 
    \non\\
    &\quad
    + 2 i ( D_+ H^m )  D_- \partial_m \Phi
    + i H^m  D_+ D_- \partial_m \Phi  
    - 2 ( D_+ \Db_+ D_- H^{++} ) D_+ \Phi 
    \non\\
    &\quad
    + 2 \left( D_- \Db_- D_+ H^{--} \right) D_- \Phi 
    ~,  \\
    \nablab_- \nabla_- \bbPhi 
    &=
    i ( 1 + S + \Sb ) \partial_{--} \Phi 
    - 2 ( \Db_- \Db_+ D_- H^{++} ) D_+ \Phi
    + 2 i ( \Db_- D_- H^m ) \partial_m \Phi 
      \non\\
    &\quad 
    - H^m \partial_{--} \partial_m \Phi
    + 2 \left( \Db_- ( S + \Sb ) + \Db_+ D_+ \Db_- H^{++} \right) D_- \Phi
    ~,
\end{align}
together with their complex conjugates.
Armed with these expressions, we can linearize the supergravity couplings in (\ref{general_lag_coupled}). 
First let us consider the contribution from the D-term. We would like to extract the terms proportional to $H^{\pm \pm}$ and $\sigma$ in
\begin{align}
    \mathcal{L} =&
    ~
     \int d^4 \theta \, E^{-1} \, \mathcal{A} ( \bbPhi, \nabla_{\pm} \bbPhi , \nabla_+ \nabla_- \bbPhi, \nabla_{\pm \pm} \bbPhi , \text{c.c.} ) 
    ~, \nonumber \\
    &\sim 
    \int d^4 \theta \Big( H^{\alpha \alphad} [ D_{\alpha}, \Db_{\alphad} ] \mathcal{A} 
    + i \frac{\partial \mathcal{A}}{\partial \Phi} H^m \partial_m \Phi + \left( \nabla_\alpha \bbPhi - D_\alpha \Phi \right) 
     \frac{\partial \mathcal{A}}{\partial \nabla_{\alpha} \Phi}
      \nonumber \\
    &
    + \frac{\partial \mathcal{A}}{\partial \nabla_{+} \nabla_{-} \Phi} \left( \nabla_+ \nabla_- \bbPhi - D_+ D_- \Phi \right) 
    + \frac{\partial \mathcal{A}}{\partial \nabla_m \Phi} \left( \nabla_m \bbPhi - \partial_m \Phi \right) + \text{c.c.} \Big) 
    ~,
\end{align}
where $\nabla_{\pm \pm} = - i \left\{ \nabla_{\pm} , \nablab_{\pm} \right\}$. Doing so, we see that the currents which couple to 
$H^{\pm \pm}$ are
\begin{align}\hspace{-50pt}
    \mathcal{J}_{++} &=
    [D_+, \Db_+] \Bigg[\, 
    \frac{1}{2} \mathcal{A}
    - \frac{1}{2}   \frac{\partial \mathcal{A}}{\partial \nabla_{-} \Phi}  D_- \Phi
    - \frac{1}{2} \frac{\partial \mathcal{A}}{\partial \nabla_{+} \Phi} D_+ \Phi
    +  \frac{\partial \mathcal{A}}{\partial \nabla_{+} \nabla_{-} \Phi}D_+ D_- \Phi 
    +  \frac{\partial \mathcal{A}}{\partial \nabla_{++} \Phi}\partial_{++} \Phi 
    \nonumber \\
    &\hspace{60pt}   
    + 2 i \Db_- \left(\frac{\partial \mathcal{A}}{\partial \nabla_{--} \Phi}  D_- \Phi\right) 
    + 2 i  \Db_+ \left(\frac{\partial \mathcal{A}}{\partial \nabla_{++} \Phi}D_+ \Phi \right) 
    + \frac{\partial \mathcal{A}}{\partial \nabla_{--} \Phi}
    \partial_{--} \Phi \,\Bigg]
 \nonumber \\
    &\quad + i \Bigg[\,
    \frac{\partial \mathcal{A}}{\partial \Phi} \partial_{++} \Phi 
    + \frac{1}{2} \partial_{++} \left(\frac{\partial \mathcal{A}}{\partial \nabla_{-} \Phi} D_- \Phi  \right)
    - \partial_{++} \left(\frac{\partial \mathcal{A}}{\partial \nabla_+ \nabla_- \Phi} D_+ D_- \Phi \right) 
    -\frac{\partial \mathcal{A}}{\partial \nabla_{-} \Phi}  D_- \partial_{++} \Phi  \nonumber \\
    &\hspace{30pt} 
    - 2 D_- \left(\frac{\partial \mathcal{A}}{\partial \nabla_{-} \Phi} \partial_{++} \Phi\right)
    + \frac{1}{2} \partial_{++} \left(\frac{\partial \mathcal{A}}{\partial \nabla_{+} \Phi}D_+ \Phi  \right)  
    - \frac{\partial \mathcal{A}}{\partial \nabla_+ \Phi}D_+ \partial_{++} \Phi
    - 2 D_+ \left( \frac{\partial \mathcal{A}}{\partial \nabla_+ \Phi} \partial_{++} \Phi\right) 
    \nonumber \\
    &\hspace{30pt} 
    - 2 D_- D_+ \left(  \frac{\partial \mathcal{A}}{\partial \nabla_{+} \nabla_{-} \Phi}\partial_{++} \Phi \right) 
    + 2 D_- \left(  \frac{\partial \mathcal{A}}{\partial \nabla_{+} \nabla_{-} \Phi} D_+ \partial_{++} \Phi  \right)
    \nonumber \\
    &\hspace{30pt} 
    - 2 D_+ \left(  \frac{\partial \mathcal{A}}{\partial \nabla_{+} \nabla_{-} \Phi}D_- \partial_{++} \Phi \right)
    + \frac{\partial \mathcal{A}}{\partial \nabla_{+} \nabla_{-} \Phi} D_+ D_- \partial_{++} \Phi 
    + 2 i D_+ \Db_+ \left(\frac{\partial \mathcal{A}}{\partial \nabla_{++} \Phi} \partial_{++} \Phi  \right) 
    \nonumber \\
    &\hspace{30pt} 
    + 2 i D_- \Db_- \left(  \frac{\partial \mathcal{A}}{\partial \nabla_{--} \Phi}\partial_{++} \Phi \right) 
    + \frac{\partial \mathcal{A}}{\partial \nabla_{++} \Phi}\partial_{++}^2 \Phi 
    +\frac{\partial \mathcal{A}}{\partial \nabla_{--} \Phi}\partial_{--} \partial_{++} \Phi \,\Bigg]
    \nonumber \\
    &\quad 
    + 2 \Bigg[ \,- D_- \Db_+ \left( \frac{\partial \mathcal{A}}{\partial \nabla_{-} \Phi}D_+ \Phi \right) 
    - D_- \Db_+ D_+ \left(  \frac{\partial \mathcal{A}}{\partial \nabla_{+} \nabla_{-} \Phi}D_+ \Phi \right)
    \nonumber \\
    &\hspace{40pt} 
    + i D_- \Db_+ \Db_- \left(   \frac{\partial \mathcal{A}}{\partial \nabla_{--} \Phi}D_+ \Phi \right)  
    - i \Db_- D_+ \Db_+ \left(  \frac{\partial \mathcal{A}}{\partial \nabla_{--} \Phi}D_- \Phi \right) \Bigg] 
    \nonumber \\
    &\quad + \text{ c.c. } ~,
    \label{general_jpp}
\end{align}
and
\begin{align}\hspace{-50pt}
    \mathcal{J}_{--} &= [D_-, \Db_-] \Bigg[\,
    \frac{1}{2} \mathcal{A} 
    - \frac{1}{2}  \frac{\partial \mathcal{A}}{\partial \nabla_{-} \Phi}D_- \Phi
    - \frac{1}{2} \frac{\partial \mathcal{A}}{\partial \nabla_{+} \Phi} D_+ \Phi 
    +  \frac{\partial \mathcal{A}}{\partial \nabla_{+} \nabla_{-} \Phi} D_+ D_- \Phi
    + \frac{\partial \mathcal{A}}{\partial \nabla_{++} \Phi} \partial_{++} \Phi 
    \nonumber \\
    &\hspace{60pt}  
    + 2 i \Db_- \left(  \frac{\partial \mathcal{A}}{\partial \nabla_{--} \Phi}D_- \Phi  \right) 
    + 2 i  \Db_+ \left(   \frac{\partial \mathcal{A}}{\partial \nabla_{++} \Phi}D_+ \Phi \right) 
    + \frac{\partial \mathcal{A}}{\partial \nabla_{--} \Phi}\partial_{--} \Phi 
    \,\Bigg] \nonumber \\
    &\quad + i \Bigg[\,
     \frac{\partial \mathcal{A}}{\partial \Phi} \partial_{--} \Phi 
     + \frac{1}{2} \partial_{--} \left(   \frac{\partial \mathcal{A}}{\partial \nabla_{-} \Phi} D_- \Phi \right) 
     - \partial_{--} \left( \frac{\partial \mathcal{A}}{\partial \nabla_+ \nabla_- \Phi}D_+ D_- \Phi  \right)  
     -  \frac{\partial \mathcal{A}}{\partial \nabla_{-} \Phi} D_- \partial_{--} \Phi
     \nonumber \\
    &\hspace{30pt} 
- 2 D_- \left(  \frac{\partial \mathcal{A}}{\partial \nabla_{-} \Phi} \partial_{--} \Phi\right)
+ \frac{1}{2} \partial_{--} \left(   \frac{\partial \mathcal{A}}{\partial \nabla_{+} \Phi} D_+ \Phi \right)  
- \frac{\partial \mathcal{A}}{\partial \nabla_+ \Phi}D_+ \partial_{--} \Phi
- 2 D_+ \left(  \frac{\partial \mathcal{A}}{\partial \nabla_+ \Phi} \partial_{--} \Phi\right) 
\nonumber \\
    &\hspace{30pt} 
    - 2 D_- D_+ \left(  \frac{\partial \mathcal{A}}{\partial \nabla_{+} \nabla_{-} \Phi} \partial_{--} \Phi\right) 
    + 2 D_- \left(  \frac{\partial \mathcal{A}}{\partial \nabla_{+} \nabla_{-} \Phi}D_+ \partial_{--} \Phi  \right) 
    \nonumber \\
    &\hspace{30pt} 
    - 2 D_+ \left(\frac{\partial \mathcal{A}}{\partial \nabla_{+} \nabla_{-} \Phi}D_- \partial_{--} \Phi\right)
    +\frac{\partial \mathcal{A}}{\partial \nabla_{+} \nabla_{-} \Phi} D_+ D_- \partial_{--} \Phi 
    + 2 i D_+ \Db_+ \left(  \frac{\partial \mathcal{A}}{\partial \nabla_{++} \Phi}\partial_{--} \Phi \right)  
    \nonumber \\
    &\hspace{30pt}
    + 2 i D_- \Db_- \left(  \frac{\partial \mathcal{A}}{\partial \nabla_{--} \Phi} \partial_{--} \Phi\right) 
    + \frac{\partial \mathcal{A}}{\partial \nabla_{++} \Phi} \partial_{++} \partial_{--} \Phi 
    +  \frac{\partial \mathcal{A}}{\partial \nabla_{--} \Phi} \partial_{--}^2 \Phi\,  \Bigg] 
    \nonumber \\
    &\quad + 2 \Bigg[\,
     -D_+ \Db_- \left(   \frac{\partial \mathcal{A}}{\partial \nabla_{+} \Phi} D_- \Phi\right) 
     + D_+ \Db_- D_- \left(  \frac{\partial \mathcal{A}}{\partial \nabla_{+} \nabla_{-} \Phi} D_- \Phi  \right) 
     \nonumber \\
    &\hspace{40pt} 
    + i D_+ \Db_- \Db_+ \left(  \frac{\partial \mathcal{A}}{\partial \nabla_{++} \Phi}  D_- \Phi\right) 
    - i \Db_+ D_- \Db_- \left(  \frac{\partial \mathcal{A}}{\partial \nabla_{++} \Phi}D_+ \Phi  \right) \Bigg] \nonumber \\
    &\quad + \text{ c.c. }~ ,
    \label{general_Jmm}
\end{align}
where $+\text{c.c.}$ means to add the complex conjugates of \emph{all} preceding terms (including the real quantity 
$\frac{1}{2} [ D_{\pm} , \Db_{\pm} ] \mathcal{A}$ for which the complex conjugate merely removes the factor of $\frac{1}{2}$).

The field $\mathcal{V}$ which appears in our deformation (\ref{FZTTbar}) receives two contributions, one from the D-term coupling 
which depends only on $\mathcal{A}$, and one from the F-term coupling which depends only on the superpotential $W$. 
Adding them, we find
\bea
    \mathcal{V}
    &=&
    \Db_+ \Db_- \Bigg[\,
    -\frac{\partial \mathcal{A}}{\partial \nabla_{\alpha} \Phi}  D_{\alpha} \Phi 
    + 2 \frac{\partial \mathcal{A}}{\partial \nabla_+ \nabla_- \Phi}D_+ D_- \Phi 
    + \frac{\partial \mathcal{A}}{\partial \nabla_{m} \Phi} \partial_{m} \Phi 
    +  \frac{\partial \mathcal{A}}{\partial \nabla_{m} \Phib}\partial_{m} \Phib
    \non\\
    &&~~~~~~~~~~
    + 2 i \Db_{+} \left(\frac{\partial \mathcal{A}}{\partial \nabla_{++} \Phi}  D_+ \Phi \right) 
    + 2 i D_{+} \left(  \frac{\partial \mathcal{A}}{\partial \nabla_{++} \Phib} \Db_{+} \Phib\right)
    \non\\
    &&~~~~~~~~~~
    + 2 i \Db_{-} \left( \frac{\partial \mathcal{A}}{\partial \nabla_{--} \Phi} D_- \Phi \right) 
    + 2 i D_{-} \left( \frac{\partial \mathcal{A}}{\partial \nabla_{--} \Phib}\Db_{-} \Phib  \right)
    \Bigg]
    \non\\
    &&
    + 2 W ( \Phi )~ .
\eea

\section{On-Shell Simplification of Chiral Scalar Theories}\label{appendix:on-shell}

In this appendix, we prove the following claim which applies to $(2, 2)$ theories of a chiral superfield of the form considered in Chapter \ref{CHP:SC-squared-2}: one can drop all terms which involve products of $(D_+ D_- \Phi)$ or $(\Db_+ \Db_- \Phib)$ and the four-fermion term $|D\Phi|^4 = D_+ \Phi \Db_+ \Phib D_- \Phi \Db_- \Phib$ 
when the equations of motion are satisfied. This is the $(2,2)$ analogue of the statement that terms involving products of $y = t ( D_+ D_- \Phi )^2$ and the two-fermion term $D_+ \Phi D_- \Phi$ can be dropped on-shell in $(1,1)$ theories, as shown in Section \ref{sec:on-shell}.

To see this for the $(2,2)$ models we consider,  
it suffices to consider a superspace Lagrangian of
the form
\begin{align}
    \mathcal{L} &= \int d^4 \theta \, \mathcal{A}
     \left( \Phi, D_{\pm} \Phi, D_+ D_- \Phi, \partial_{\pm \pm} \Phi , \text{c.c.} \right) \nonumber \\
    &= \int d^4 \theta \, \left( K ( \Phi , \Phib ) + f ( x, \xb, y ) |D\Phi|^4 \right) ~,
\end{align}
which has the superspace equation of motion
\begin{align}
    \Db_+ \Db_- K_{\Phi} = \Db_+ \Db_- \Bigg(
     D_{\alpha} \left[ \frac{\partial ( f |D\Phi|^4 ) }{\partial D_{\alpha} \Phi} \right]
     - D_+ D_- \left[ \frac{\partial ( f |D\Phi|^4 )}{\partial D_+ D_- \Phi} \right] 
     - \partial_{m} \left[\frac{\partial ( f |D\Phi|^4 )}{\partial ( \partial_{m} \Phi ) } \right] \Bigg)
    \label{superspace_eom}
\end{align}
for $\Phi$, and the conjugate equation of motion for $\Phib$. 
If we multiply (\ref{superspace_eom}) on both sides by the four-fermion term 
$|D\Phi|^4 = D_+ \Phi \Db_+ \Phib D_- \Phi \Db_- \Phib$ then any term containing $(D_\pm\Phi)$ and
$(\Db_\pm\overline{\Phi})$
fermions in (\ref{superspace_eom}) will vanish by nilpotency. On the left, the only surviving term is 
$K_{\Phi \Phib} \Db_+ \Db_- \Phib$, while on the right we get contributions from the first and second terms:
\begin{align}
     K_{\Phi \Phib} \left(\Db_+ \Db_- \Phib \right) |D\Phi|^4 = 
     \left( \Db_+ \Db_- \Phib \right) |D\Phi|^4
     \Bigg(  \lambda  \Db_+ \Db_- \left[ \frac{\partial f}{\partial y}  ( \partial_{--} \Phib ) ( \partial_{++} \Phib ) \right]
  - \left( \frac{x + \xb}{\lambda} \right) f\Bigg)  \, .
\end{align}
On collecting terms, the previous equation turns into
\begin{align}
  \left( \Db_+ \Db_- \Phib \right) |D\Phi|^4 \left\{
     K_{\Phi \Phib} + \left( \frac{x + \xb}{\lambda} \right) f 
     - \lambda  \Db_+ \Db_- \left[ \frac{\partial f}{\partial y}  ( \partial_{--} \Phib ) ( \partial_{++} \Phib ) \right] \right\}  
     = 0 ~. 
     \label{C4}
\end{align}
The parenthesis multiplying 
$( \Db_+ \Db_- \Phib )|D\Phi|^4$ 
in the previous expression
does not vanish in general,
at least for 
$\lambda$ small enough.
Then for \eqref{C4} to be satisfied,  
the equation
\begin{align}
    \left( \Db_+ \Db_- \Phib \right) |D\Phi|^4 = 0  
\end{align}
has to hold
when the equations of motion are satisfied. This justifies our claim in section \ref{subsec:kahler}
 that we may drop all terms involving the product $y |D\Phi|^4$ in the deformation, assuming we restrict to on-shell configurations.
 
 \section{On-Shell Simplification of Born-Infeld-Type Theories}
\label{appendix:EoMBI}

This appendix is devoted to deriving the on-shell relation \eqref{on-shell_susy-condition}.
We are going to prove this holds for an action of the form \eqref{GeneralBI}.
Let us start by considering the following Lagrangian
\bea
\cL&=&
\frac{1}{4}\int d^2\q\,W^2
+\frac{1}{4}\int d^2\bar{\q}\,\overbar{W}^2
+\int d^2\q d^2\qb\,W^2\overbar{W}^2\,\O\big[D^2W^2,\Db^2\overbar{W}^2\big]
~.
\label{N=1_nl}
\eea
We recall that $W_\a$ and $\overbar{W}_\ad$ satisfy the Bianchi identity
$D^\a W_\a=\Db_\ad \overbar{W}^\ad$, whose solution is given in terms of a real but otherwise unconstrained scalar prepotential superfield $V$:
$W_\a=-1/4\,\Db^2D_\a V$
and
$\overbar{W}_\ad=-1/4\,D^2\Db_\ad V$.
It is a straightforward calculation to derive the ${\rm EOM}$ 
by varying the action \eqref{N=1_nl} with respect to the prepotential $V$.
The ${\rm EOM}$ reads
\begin{align}
    0&=
-D^\a W_\a
+\frac{1}{2} D^\a\Db^2\Big(W_\a \overbar{W}^2\O\Big)
+\frac{1}{2} \Db_\ad D^2\Big(W^2\overbar{W}^\ad\O\Big)
\nonumber \, , 
\\&\quad 
+\frac{1}{2}D^\a\Big{[}  W_\a\Db^2D^2\Big(W^2\overbar{W}^2\frac{\pa\O}{\pa(D^2W^2)}\Big)\Big{]}
+\frac{1}{2} \Db_\ad\Big{[}  \overbar{W}^\ad \Big(D^2\Db^2W^2\overbar{W}^2\frac{\pa\O}{\pa(\Db^2\overbar{W}^2)}\Big)\Big{]} \, .
\label{EOM-N=1_nl}
\end{align}
Because of the constraint that $W_\a W_\b W_\g=0$ and its complex conjugate,
multiplying eq.~\eqref{EOM-N=1_nl} by $W^2\overbar{W}^2$ and using the  
${\rm EOM}$ gives the following condition
\bea
&&
W^2\overbar{W}^2(D^\a W_\a)\Big(1+f(\O)\Big)
=0
~,
\eea
where the functional $f(\O)$ is given by
\bea
f(\O)
&:=&
-\frac{1}{2}(\Db^2\overbar{W}^2+D^2W^2)\O
\non\\
&&
-\frac{1}{2}\Bigg{[}
(D^2W^2)(\Db^2\overbar{W}^2)\frac{\pa\O}{\pa(D^2W^2)}
+(D^2W^2)(\Db^2\overbar{W}^2)\frac{\pa\O}{\pa(\Db^2\overbar{W}^2)}
\Bigg{]}
~.
\eea
This implies
\bea
&&
W^2\overbar{W}^2(D^\a W_\a)=0
~,
\eea
which is precisely condition \eqref{on-shell_susy-condition}.

\section{Derivation of General Flow Equation for Gauge Field and Scalars}\label{flow_derivation}

In this appendix, we will obtain the flow equation of a sufficiently general Lagrangian for all cases involving gauge fields and scalars that are of interest in Chapter \ref{CHP:nonabelian}.

Consider a general $\lambda$-dependent Lagrangian for a complex scalar $\phi$ and field strength $F$:
\begin{align}
    \mathcal{L} = f ( \lambda, F^2, | D \phi |^2 ) , 
\end{align}
For convenience, we will also define $x = F^2$ and $y = | D \phi |^2$. As in the main body of Chapter \ref{CHP:nonabelian}, $D$ is the gauge-covariant derivative and the field strength $F$ need not be abelian; we use the shorthand
\begin{align}
    F^2 = F_{\mu \nu}^a F^{\mu \nu}_a = \Tr \left( F^2 \right) ,
\end{align}
and we will suppress gauge group indices in what follows.

We can now compute the stress-energy tensor by coupling to a background metric and varying with respect to the metric, which gives 
\begin{align}
    T_{\mu \nu}^{(\lambda)} &= \eta_{\mu \nu} \, f - 4 \frac{\partial f}{\partial x} \tensor{F}{_\mu^\sigma} F_{\sigma \nu} - 2 \frac{\partial f}{\partial y} D_{\mu} \phi D_\nu \phib \nonumber \\
    &= \eta_{\mu \nu} \, f + 2 \frac{\partial f}{\partial x} \eta_{\mu \nu} F^2 - 2 \frac{\partial f}{\partial y} D_{\mu} \phi D_\nu \phib ~, 
\end{align}
where we have used that $\tensor{F}{_\mu^\sigma} F_{\sigma \nu} = - \frac{1}{2} \eta_{\mu \nu} \left( F_{\alpha \beta} F^{\alpha \beta} \right)$ in two dimensions.

The determinant of $T$ is then expressed in terms of the combinations
\begin{align}
    T^{\mu \nu} T_{\mu \nu} &= \left( \eta^{\mu \nu} \, f + 2  \eta^{\mu \nu} F^2 \frac{\partial f}{\partial x} - 2  D^{\mu} \phi D^\nu \phib\frac{\partial f}{\partial y} \right) \left( \eta_{\mu \nu} \, f + 2 \eta_{\mu \nu} F^2 \frac{\partial f}{\partial x}  - 2  D_{\mu} \phi D_\nu \phib \frac{\partial f}{\partial y}\right) \nonumber \\
    &= 2 f^2 - 8  F^2 f \frac{\partial f}{\partial x}  - 4 \dps f \frac{\partial f}{\partial y}  + 8 F^4 \left( \frac{\partial f}{\partial x} \right)^2  + 8 F^2 \dps \frac{\partial f}{\partial x} \frac{\partial f}{\partial y}   + 4 \dpf \left( \frac{\partial f}{\partial y} \right)^2 \nonumber \\
    &= 2 f^2 - 8 x  f \frac{\partial f}{\partial x}  - 4 y f \frac{\partial f}{\partial y}  + 8 x^2  \left( \frac{\partial f}{\partial x} \right)^2  + 8 x y \frac{\partial f}{\partial x} \frac{\partial f}{\partial y}  + 4 y^2 \left( \frac{\partial f}{\partial y} \right)^2 ~,
\end{align}
and
\begin{align}
    \left( \tensor{T}{^\mu_\mu} \right)^2 &= \left( 2 f - 4 F^2 \frac{\partial f}{\partial x}  - 2 \dps \frac{\partial f}{\partial y}  \right)^2 \nonumber \\
    &= 4 f^2 - 16 F^2 f \frac{\partial f}{\partial x}  - 8 \dps f \frac{\partial f}{\partial y}  + 16 F^4 \left( \frac{\partial f}{\partial x} \right)^2  + 4 \dpf \left( \frac{\partial f}{\partial y} \right)^2  + 16 F^2 \frac{\partial f}{\partial x} \frac{\partial f}{\partial y}  \dps \nonumber \\
    &= 4 f^2 - 16 x f \frac{\partial f}{\partial x}  - 8 y f \frac{\partial f}{\partial y}  + 16 x^2 \left( \frac{\partial f}{\partial x} \right)^2  + 16 x y \frac{\partial f}{\partial x} \frac{\partial f}{\partial y}  + 4 y^2 \left( \frac{\partial f}{\partial y} \right)^2 ~ .
\end{align}
Using these, we can write the $\TT$ operator as
\begin{align}
    \det ( T ) &= \frac{1}{2} \left( \left( \tensor{T}{^\mu_\mu} \right)^2 - T^{\mu \nu} T_{\mu \nu} \right) \nonumber \\
    &= f^2 - 4 f x \frac{\partial f}{\partial x} - 2 f y \frac{\partial f}{\partial y} + 4 x^2 \left( \frac{\partial f}{\partial x} \right)^2 + 4 x y \frac{\partial f}{\partial x} \frac{\partial f}{\partial y} ~,
\end{align}
and hence the $\TT$-flow equation as
\begin{align}
    \frac{d f}{d \lambda} =  f^2 - 4 f x \frac{\partial f}{\partial x} - 2 f y \frac{\partial f}{\partial y} + 4 x^2 \left( \frac{\partial f}{\partial x} \right)^2 + 4 x y \frac{\partial f}{\partial x} \frac{\partial f}{\partial y}~.
    \label{main_pde}
\end{align}
This is the main differential equation of interest which we will study in Chapter \ref{CHP:nonabelian}.  Although we have used $\eta_{\mu \nu}$ for the metric, these results are valid either in Minkowski signature or in Euclidean signature -- replacing $\eta_{\mu \nu}$ with $\delta_{\mu \nu}$ in the intermediate steps of these calculations does not affect our final result (\ref{main_pde}).

In the case where we turn off the field strength, setting $x = 0$, this differential equation becomes
\begin{align}
    \frac{d f}{d \lambda} = f^2 - 2 f y \frac{\partial f}{\partial y} ~ .
    \label{scalars_pde_intermediate}
\end{align}
Imposing the boundary condition that $f ( \lambda = 0 ) = | D \phi |^2$, we find
\begin{align}
    f ( \lambda, \dps ) = \frac{1}{2 \lambda} \left( \sqrt{1 + 4 \lambda \dps } - 1 \right) ~ .
    \label{turn_off_F}
\end{align}
On the other hand, in the case where we turn off the scalars (setting $\dps = 0$), the differential equation (\ref{main_pde}) becomes
\begin{align}
    \frac{d f}{d \lambda} = f^2 - 4 f x \frac{\partial f}{\partial x} + 4 x^2 \left( \frac{\partial f}{\partial x} \right)^2 .
    \label{maxwell_pde_intermediate}
\end{align}
which has the solution
\begin{align}
    f ( \lambda, F^2 ) = F^2 \cdot {}_{3} F_4 \left( \frac{1}{2} , \frac{3}{4} , 1 , \frac{5}{4} ; \frac{4}{3} , \frac{5}{3} , 2 ; \frac{256}{27} \cdot \lambda F^2 \right) ~ . 
    \label{turn_off_phi}
\end{align}
%

\makebibliography

\end{document}